\documentclass [a4paper,11pt]{article}
\pdfoutput=1
\usepackage[english]{babel}
\usepackage{graphicx,amsmath,amsfonts,amssymb,slashed,upgreek,color,caption,amscd,cite,cancel}
\usepackage{cite}
\usepackage{framed}
\usepackage{multirow}
\usepackage{graphicx,subfigure}
\usepackage{multicol}
\usepackage{a4wide}
\usepackage{fullpage}
\usepackage[utf8]{inputenc} % for accents ...
\usepackage[english]{babel} % to know your using this language
\usepackage{color} % to color text or boxes
\usepackage{epstopdf}
\usepackage{float} %for the placement of images...
\usepackage[usenames,dvipsnames,svgnames,table]{xcolor}
\usepackage{multicol}
\usepackage{amsmath, amsfonts, amssymb, amsthm} % maths features
\usepackage{bm} % bold ex : greek letters in math mode
\usepackage{mathrsfs} % whrite caligraph letters in mathmode
\usepackage{relsize} % larger fonts in math mode \mathlarger{}
\usepackage{multirow}% for multirow in tables
\usepackage{pdfpages} %include pdf pages
\usepackage{url}
\definecolor{Myblue}{rgb}{0.,0.,0.8}
\usepackage{hyperref}
\hypersetup{colorlinks,citecolor=Myblue,linkcolor=black,filecolor=black,urlcolor=black}

\def\gsp{\eta}

\def\cb{\mathcal{C}}
\def\wb{w_{\rm eff}}
\def\R{{}^{(3)}\!R}
\def\mps{M_{\rm pl}^2}
\def\mp{M_{\rm pl}}

\def\gn{G_{\rm N}}
\def\fs{f\sigma_8} 
\def\omo{\Omega_{m,0}}

\def\m22{\mu_2^2}

\def\l33{\Lambda^3_3}

\newcommand{\vv}{\vskip 2mm}

\graphicspath{{figs/}}

\begin{document}
%**************************************************************************************************
%												PAPER
%**************************************************************************************************

\begin{center}
\Large{\textbf{Diagnostic of Horndeski Theories}} \\[1cm]

\large{Louis Perenon, Christian Marinoni and Federico Piazza}
\\[0.5cm]

\small{
\textit{Aix Marseille Univ, Universit\'e de Toulon, CNRS, CPT, Marseille, France. \\
}
}

\vspace{0.5cm}
%\today

\end{center}

\vspace{2cm}

\begin{abstract}
We study the effects of Horndeski models of dark energy on the  observables of the large-scale structure  in the late time universe. A novel classification into  {\it Late dark energy}, {\it Early dark energy} and {\it Early modified gravity} scenarios is proposed, according to whether such models predict deviations from the standard paradigm persistent at early time in the matter domination epoch. We discuss the physical imprints left by each specific class of models  on the effective Newton constant $\mu$, the gravitational slip parameter $\eta$, the light deflection parameter $\Sigma$ and the growth function $\fs$ and demonstrate that a convenient way to dress a complete portrait of  the viability of  the Horndeski accelerating mechanism is via two, redshift-dependent,  diagnostics: the  $\mu(z)-\Sigma(z)$ and the $\fs(z)-\Sigma(z)$ planes. If future, model-independent, measurements  point to either $\Sigma-1<0$ at redshift zero or $\mu-1<0$ with $\Sigma-1>0$ at high redshifts or $\mu-1>0$ with $\Sigma-1<0$ at high redshifts, Horndeski theories are effectively ruled out.  If $\fs$ is measured to be larger than expected in a $\Lambda$CDM model at $z>1.5$  then Early dark energy models are definitely ruled out. On the opposite case,  Late dark energy models are rejected by data if $\Sigma<1$, while,  if  $\Sigma>1$, only Early modifications of gravity provide a viable framework to interpret  data.
\end{abstract}

\newpage 
\tableofcontents

\vspace{.5cm}

\section{Introduction}\label{sec_1}

Current and future observations aiming at understanding the nature of cosmic acceleration offer the unique possibility of testing predictions of general relativity (GR) on scales well beyond those of the solar system, where GR has received its most impressive confirmations.  Upcoming galactic surveys such as DES \cite{DES}, Euclid \cite{Euclid,Amendola:2016saw,Taddei:2016iku}, DESI \cite{DESI}, LSST \cite{LSST}, WFIRST \cite{WFIRST} and SKA \cite{SKA,Bull:2015lja,Camera:2015fsa} are expected to provide unprecedented datasets with which to investigate, in an accurate way, how structures form and grow, and how light rays bend in the presence of local gravitational potentials. Anticipating interesting signals of non-standard gravity that could be potentially detected by such future surveys of the large-scale structure (LSS)  of the universe is a crucial task. 

Any deviation from the standard $\Lambda$CDM paradigm will imply some anomalous relation among the curvature perturbation $\Psi$, the Newtonian potential $\Phi$ and the comoving density contrast of non relativistic matter $\Delta$. These effects can be encoded in time and scale modifications to the effective Newton's constant parameter $\mu$ and to the gravitational slip parameter $\gsp$~\cite{Pogosian:2010tj}. The former quantity describes how fluctuations of the matter fields interact in the universe, while the latter encapsulates non-standard relation between the Newtonian potential $\Phi$ (time-time part of the metric fluctuations) and the curvature potential $\Psi$ (space-space part). From $\mu$ and $\eta$ one can derive a further parameter, $\Sigma$, of more direct relevance for lensing surveys~\cite{Song:2010fg,Simpson:2012ra}. $\Sigma$ relates the matter over-density with the lensing (or Weyl) potential $\Phi_+ = (\Phi+\Psi)/2$. Another convenient quantity to describe the gravitational clustering of matter is the product of the linear growth factor $f$ and the $rms$ density fluctuations on  a scale of $8h^{-1}$ Mpc  ($f \sigma_8$). This quantity, which can be optimally estimated from the analysis of the redshift space distortions induced by the large-scale,  coherent, in-falling(/out-flowing) of matter into(/out of)  high(/low)  density regions, is another key quantity turning galaxy redshift surveys into gravity probes.

While any observed deviation would represent a major discovery in itself, it is important to understand what type of signals are implied by concrete alternatives to the standard model and interpret them in terms of fundamental theoretical proposals. 
In particular, theories containing one extra scalar degree of freedom and leading to equations of motion of at most second order---\emph{Horndeski theories}~\cite{Horndeski:1974wa,Deffayet:2009mn}---despite the freedom in the choice of their free functions and the richness of their potential phenomenology, have proven to share common features and universal behaviours. The exclusion of pathologies and instabilities imposes tight constraints and well defined patterns for the time scaling of relevant observables of the LSS in the universe~\cite{Piazza:2013pua,Perenon:2015sla,Perenon:2016itr}. For instance, it was pointed out in~\cite{Perenon:2015sla} that the linear growth rate of Horndeski theories is systematically lower, at low redshift, than the value predicted by the  standard $\Lambda$CDM model. In~\cite{Song:2010rm} it was shown that Brans-Dicke theories, clustering and interacting dark energy models follow characteristic paths in the $\mu $ - $\Sigma$ plane. It has also been argued~\cite{Pogosian:2016pwr} that Horndeski theories are expected to display a systematic sign agreement in the $\mu $ - $\Sigma$ plane across all cosmic epochs.

Investigating the existence of further general patterns displayed by LSS  observables is the main goal of the present work. To this purpose, we present a complete study of the Horndeski phenomenology that generalizes in many respects that presented in~\cite{Perenon:2015sla}. First of all, we explore  accelerating cosmologies in which the presence of dark energy is not confined to the late times ~\cite{Wetterich:1987fm,Ratra:1987rm,Caldwell:1997ii,Hebecker:2000zb,Doran:2001rw,Wetterich:2001jw,Bean:2001wt,Wetterich:2004pv,Doran:2006kp,PhysRevD.83.123504,Reichardt:2011fv,Tsujikawa:2013fta,Sievers:2013ica,Archidiacono:2014msa,Pettorino:2013ia,Shi:2015tje,Pu:2014goa,Brax:2013fda,Lima:2016npg }.  At first sight, this is counter intuitive, as the acceleration is a recent phenomenon and there is no need to invoke dark energy effects at early times.  Truth  is that, although such effects are not needed, not to say wanted, they are allowed within the context of Horndeski theories,  therefore they must be thoroughly  investigated and systematised. In particular, we find convenient to highlight three novel possibilities of increasing generality.
\begin{itemize}
\item \emph{Late-time dark energy} {\bf (LDE)}: \ This is the reference class of models (explored at length in~\cite{Perenon:2015sla}), in which both the dark energy momentum tensor and the possible modifications of gravity (\emph{i.e.} the non-minimal gravitational couplings) become negligible at early times. 
\item \emph{Early dark energy} {\bf (EDE)}: \ In these scenarios dark energy can contribute  to the total energy momentum tensor even at early times, while non-minimal gravitational couplings are kept as a late-time phenomenon.
\item \emph{Early modified gravity} {\bf (EMG)}: \  Horndeski theories in their full generality. Not only does dark energy always contribute to the total energy momentum tensor, but modified gravity effects are also persistent at early times, during matter domination. 
\end{itemize}

On top of singling out the specific phenomenological features of the  Horndeski sub-classes listed above, in this work we extend the analysis of~\cite{Perenon:2015sla} by including different background expansion histories than the  $\Lambda$CDM model. Beside an effective equation of state parameter $w=-1$ (roughly, the value preferred by current observations, \emph{e.g.}~\cite{Betoule:2014frx,Ade:2015xua,Aubourg:2014yra}), we also consider models  with $w = -0.9$ and $w = -1.1$.  In~\cite{Perenon:2015sla} it was found that viability priors do impose tight constraints and well defined patterns for the time scaling of relevant observables of the LSS in the universe. As such, viability criteria  can be effectively used to complement data and observational information  in  statistical inferences~\cite{Salvatelli:2016mgy}.
Testing the consequences of relaxing some of these restrictions is also a goal of the present study.

The main results of the paper are recapped  in Figure~\ref{fig_diag} as exclusion regions in the parameter space of LSS observables. In the $\mu$-$\Sigma$ plane we highlight regions where the eventual presence of  data would rule out the entire class of Horndeski  theories. On the other hand, specific regions in the $f\sigma_8$-$\Sigma$ allow to rule out specific subclasses  of models (LDE and/or EDE) presented above. A ``complete diagnostic" of Horndeski theories is presented  in the other figures of the paper.

The paper's structure is  as follows: in Sec.~\ref{sec_2} we introduce the formalism adopted for the description of the background cosmic evolution, the non-minimal gravitational couplings and their relations with the LSS observables. In Sec.~\ref{sec_4} we present our results in the case of a cosmic expansion history identical to that of $\Lambda$CDM, for the three classes of models described above. In Sec.~\ref{sec_4bis} we verify the robustness of our conclusion by considering different equations of state for dark energy and by relaxing some of our viability conditions. The synthesis of our results as well as some digressions on future prospects are in Sec.~\ref{sec_5}.

\section{Formalism: the effective theory of dark energy}\label{sec_2}

The effective field theory of dark energy (EFT of DE)~\cite{Creminelli:2008wc,Gubitosi:2012hu,Gleyzes:2013ooa,Bloomfield:2012ff,Bloomfield:2013efa,Piazza:2013coa,Gergely:2014rna} proves a very powerful mean to explore the cosmological implications of Horndeski theories (see~\cite{Frusciante:2013zop,Hu:2013twa,Raveri:2014cka,Frusciante:2016xoj} 
for a numerical implementation of this formalism and~\cite{Gleyzes:2014qga,Gleyzes:2014rba,Gleyzes:2015pma,Gleyzes:2015rua,D'Amico:2016ltd} for generalizations to beyond-Horndeski and/or to models with non-minimally coupled dark matter). For such theories, the action up to second order in cosmological perturbations displays six functions of cosmic time,
\begin{equation}
\begin{split}
S \ = \ & \  S_m[g_{\mu \nu},\psi_i] \ \ + \int \! d^4x \sqrt{-g} \, \frac{M^2(t)}{2} \, \left[R \, -\,  2 \lambda(t) \, - \, 2 \cb(t) g^{00} \, \right. \\
 &  \left.-\,\mu_2^2(t) (\delta g^{00})^2 \, -\, \mu_3(t) \, \delta K \delta g^{00} + \,  \epsilon_4(t) \left(\delta K^\mu_{ \nu} \, \delta K^\nu_{ \mu} -  \delta K^2  +  \frac{\!\!\R\,   \delta g^{00}}{2}\right) \right] \;, 
\end{split}
\label{action}
\end{equation}
$M(t)$ is the ``bare planck mass", $\cb(t)$ and $\lambda(t)$ are the contributions of the scalar field to the background energy momentum tensor, and $\mu_2^2(t)$, $\mu_3(t)$ and $\epsilon_4(t)$ are non-minimal couplings\footnote{The coupling function $\mu_i$ have the dimension of mass and of order Hubble. $\mu_3$ appears in cubic galileon and Horndeski-3 Lagrangians whereas $\epsilon_4$ is a dimensionless order one function characterising galileon-Horndeski 4 and 5 Lagrangians.}. 

The Brans-Dicke subset of theories is characterized by a time-varying bare Plank mass $M(t)$, while all other non-minimal couplings are set to zero. A measure of the deviations from general relativity within the Brans-Dicke sector are more usefully defined by
\begin{equation}\label{mu1}
\mu_1 = \frac{d\,{\rm ln}\,M^2(t)}{dt} \;.
\end{equation}
One apparent feature of the above action is that the second line is quadratic in cosmological perturbations ($\delta K_{\mu \nu}$ being the perturbation of the extrinsic curvature of the hypersurfaces at constant scalar field value, and $\!\!\R$ their intrinsic curvature. We refer the reader \emph{e.g.} to the review~\cite{Piazza:2013coa} for more details), which means the first three operators uniquely govern the evolution of the background. Indeed, by varying the first line with respect to the metric one obtains the two equations
\begin{align}\label{candlambda}
\cb & =  \dfrac12 \left( H\mu_1-\dot\mu_1-\mu_1^2 \right) - \dot H -\dfrac{\rho_m}{2M^2} \ , \\
\lambda & =  \dfrac12 \left( 5H\mu_1+\dot\mu_1+\mu_1^2 \right) + \dot H +3H^2-\dfrac{\rho_m}{2 M^2 } \ ,
\end{align}
where a dot means a derivative with respect to cosmic time, $H(t)=\dot{a}/a$ is the Hubble parameter and only pressureless non-relativistic matter of energy density $\rho_m $ has been assumed. Note that $\mu_1$ is the only non-minimal coupling entering the background evolution. When $\mu_1$ vanishes and $M = \mp$, the above equations are particularly transparent, $\cal C$ and $\lambda$ play the role of the kinetic and potential term of a quintessence field respectively. 

\subsection{Setting the background}

The homogeneous background expansion history is characterised by the Hubble rate $H(t)$. We focus on models with a constant effective equation of state $\wb$, \emph{i.e.} those with a Hubble rate that scales as a function of the redshift as
\begin{equation} \label{h}
\frac{H^2(z)}{H_0^2} = \omo (1+z)^3 + \left(1-\omo\right)(1+z)^{3(1+\wb)}\, ,
\end{equation}
where $\omo$ is the present fractional matter density.  Without direct relation with the EFT action~\eqref{action}, the above expression descends from a ``standard" Friedmann equation 
\begin{equation} \label{frieddd}
H^2 = \frac{1}{3 M_{\rm Pl}^2}(\rho_m + \rho_D^{\rm eff}),
\end{equation}
with $\rho_D^{\rm eff}\propto a^{-3 (w_{\rm eff} +1)}$. As known, observations  suggest $\omo\sim 0.3$, $\wb\sim -1$ ~\cite{Ade:2013zuv,Ade:2015xua} and severely constrain any deviations from a $\Lambda$CDM expansion history~\cite{Algoner:2016nio}.
\vv
In summary,  our theories  are defined, in their background and perturbation sectors, by two parameters and four functions of the time,
\begin{equation} \label{couplings}
\left\{\Omega_{m,0}, \ \wb, \ \mu_1(t),\ \mu^2_2(t), \ \mu_3(t),\ \epsilon_4(t) \right\} \,.
\end{equation}

To characterise the evolution of the late time universe, we find convenient to use the fractional matter density of the background reference model~\eqref{h} calculated at any epoch, $x$, as our time variable. Expressed as a function of the redshift it reads 
\begin{equation}
\label{xdef}
x\ = \ \frac{\omo}{\omo + (1- \omo) (1+ z)^{3 \wb}}\, .
\end{equation}
and its present day value is  $x(t_0)\equiv x_0= \Omega_{m,0}$.

\subsection{Classifying dark energy models}\label{edeclass}

In general, it is natural to \emph{define} as in~\cite{Gubitosi:2012hu} the energy density of dark energy through the equation
\begin{equation} \label{fried}
H^2 = \frac{1}{3 M^2}(\rho_m + \rho_D)\, .
\end{equation}
We note that, as opposed to the definition given e.g. in~\cite{Gleyzes:2014rba}, here $\rho_D$ does not depend on the matter fields, as much as, since we work in the Jordan frame, $\rho_m$ is not a functional of the scalar field $\phi$. 
We have now the instruments to define the three different general types of behaviour for dark energy at early times:
 
 \begin{itemize}
\item \emph{Late-time dark energy} {\bf (LDE)}: \ This is the minimal model, in which all effects of dark energy are confined to late times. Not only do non minimal couplings ($\mu_1,\ \mu^2_2, \ \mu_3,\ \epsilon_4$) go to zero at early times---which in the case of the coupling $\mu_1$ implies $M$ going to a constant---but also the dark energy density $\rho_D$ becomes a sub-dominant component for $t\rightarrow 0$. In other words, the energy density of non relativistic particles $\rho_m$ must saturate the Friedmann equations at early time. By comparing~\eqref{frieddd} and~\eqref{fried} this means that $M^2/M_{\rm Pl}^2\rightarrow 1$. In summary, 
\begin{equation} \label{LDEdef}
{\rm \bf LDE} : \left\{\frac{M^2}{\mps} \rightarrow  1 \;,\; \frac{\mu_1}{H} \rightarrow 0 \;,\; \frac{\mu^2_2}{H^2} \rightarrow 0 \;,\; \frac{\mu_3}{H}\rightarrow 0 \;,\; \epsilon_4 \rightarrow 0 \right\}\underset{x \rightarrow 1} \;\;.
\end{equation}

The above defined class of models  generalises the set explored by \cite{Perenon:2015sla}. Indeed, we  allow  the coupling $\mu_2^2$ to be nonzero  over most of  cosmic history, {\it i.e.} for $x\ne 1$,  and  we also consider  expansion histories  different from that of a $\Lambda$CDM model, {\it i.e.} $\wb\ne -1$. 

\item \emph{Early dark energy} {\bf (EDE)}: \ The dark energy contributes  to the total energy momentum tensor even at early times, when, however, all non-minimal couplings vanish. The only way this is possible is for dark energy to acquire the same equation of state as dark matter early on, so that it becomes indistinguishable from the latter as long as the background evolution is concerned, 
\begin{equation} \label{EDEdef}
{\rm \bf EDE} : \left\{\frac{M^2}{\mps} \rightarrow  {\rm const.} \;,\; \frac{\mu_1}{H}  \rightarrow 0 \;,\; \frac{\m22}{H^2} \rightarrow 0 \;,\; \frac{\mu_3}{H}\rightarrow 0 \;,\; \epsilon_4 \rightarrow 0 \right\}\underset{x \rightarrow 1} \;\;\;.
\end{equation} 
A caveat must be issued regarding the use we make of the adjective ``early". Our study is oblivious of the radiation dominated epoch. Therefore ``early" for us means always well after equivalence, say, at $z \simeq 100$,  but well before the onset of acceleration, at $z \simeq 1$. Accordingly, the early non-standard scenarios we are considering evade Big-Bang Nucleosynthesis (BBN) constraints.\footnote{ In principle, since the  background expansion is  fixed to that of $\Lambda$CDM, the time at which  neutrinos decouple and  the neutron-proton fluid exits from equilibrium is not  modified in our  non-standard gravitational scenarios.  However, the  time elapsed from this epoch ($T \sim 0.8$ Mev) to that when  BBN begins ($ T \sim 0.1$ Mev), which regulates neutron decays and account for  the final  neutron-to-proton ratio  available for nucleosynthesis, critically depends on the value of the Newton constant. }

\item \emph{Early modified gravity} {\bf (EMG)}: \ This is the most general case. Here we allow also the asymptotic value of the non-minimal couplings at early times to be different from zero.
\begin{equation} \label{EMGdef}
{\rm \bf EMG} : \left\{\frac{M^2}{\mps} \rightarrow  {\rm const.} \;,\; \frac{\mu_1}{H}  \rightarrow 0 \;,\; \frac{\m22}{H^2} \rightarrow {\rm const.} \;,\; \frac{\mu_3}{H}\rightarrow  {\rm const.} \;,\; \epsilon_4 \rightarrow {\rm const.} \right\}\underset{x \rightarrow 1} \;\;.
\end{equation} 
\end{itemize}
Note that the Brans-Dicke non-minimal coupling $\mu_1$ needs a special attention due to its link with $M^2$ (see eq.~\eqref{mu1}): for any asymptotic value of $\mu_1$ different than zero, $M^2$ would tend to either zero or plus infinity, corresponding to infinite or zero gravitational coupling respectively. We thus restrict to the cases when $\mu_1 \rightarrow 0$. Note that recent observational bounds~\cite{Ade:2015rim} constraining the amount of dark energy at early times do not apply here, because the EFT of DE allows us to explore modified gravity models which only gives rise to modifications in the perturbed sector while keeping the background evolution to that of the standard model $\Lambda$CDM. Allowing modifications of gravity also deep in matter domination without altering the background evolution is the novelty of the scenarios EDE and EMG.

 The imprints of LDE, EDE and EMG  that can be revealed through the analysis of cosmological observables is discussed in Sec.~\ref{sec_4}. In appendix \ref{cov} we show how these scenarios can be implemented in a simplified covariant theory. 
 
\subsection{Extracting observables}

Extracting observables of the perturbation sector in modified gravity (MG) theories is mostly straightforward on cosmic comoving Fourier modes well below the non-linear limit and well above the DE sound horizon. In this regime, linear theory  and the \emph{quasi-static approximation}~\cite{Noller:2013wca,Sawicki:2015zya} can be trusted. The latter allows one to neglect time derivatives of scalar and metric fluctuations over spatial derivatives. Moreover, in our framework, the extra scalar degree of freedom has the purpose of sourcing cosmic acceleration thus its mass must be of order Hubble or lighter. Therefore, Fourier modes close to the Hubble scale would be the only ones affected by these mass scales. Given that surveys of the LSS observe modes generally deep inside the Hubble horizon, one can thus neglect any scale-dependence in our observables. These observables can be schematically split into two types, the ones linked to the growth of matter perturbations and the ones sensitive to the gravitational potentials. Let us briefly present them, for which we have adopted the following convention for the perturbed metric in Newtonian gauge:

\begin{equation}\label{newtonian}
ds^2 = - (1+ 2 \Phi) dt^2 + a^2 (1 - 2 \Psi) \delta_{i j} d x^i d x^j \;.
\end{equation}

\begin{itemize}

\item[$\diamond$] \emph{Effective gravitational coupling} ($\bm\mu$): \ In most MG theories it is possible to compile a part of the modifications of gravity in an observer-friendly quantity, an effective gravitational coupling $\mu$. It is defined through the Poisson equation, $-\frac{k^2}{a^2}\Phi=4\pi \mu \gn\rho_m \delta_m$. In~\cite{Perenon:2015sla} it was shown that the Newton constant $\gn$ of an EFT model is defined by: 
\begin{equation} \label{gn}
\gn \ = \ \frac{1}{8 \pi  M^2(x_0)[1+\epsilon_4(x_0)]^2} \, .
\end{equation} 
\vskip2mm
The effective gravitational constant in the EFT formulation then yields: 
\begin{equation} \label{mu}
\mu \ = \  \left ( \frac{M(x_0) [1+\epsilon_4(x_0)]}{ M[1+\epsilon_4]} \right )^2  \ \frac{a_0}{b_0} \;,
\end{equation}
where 
\begin{align} \label{a0}
a_0 &= 2 {\cb}  +  \mathring{\mu}_3  - 2 \dot H \epsilon_4 + 2 H \mathring{\epsilon}_ 4 + 2 (\mu_1 + \mathring{\epsilon}_4)^2 \;, \nonumber\\
b_0 &= 2 {\cb} + \mathring{\mu}_3  - 2 \dot H \epsilon_4 + 2 H \mathring{\epsilon}_ 4 + 2 \dfrac{(\mu_1 + \mathring{\epsilon}_ 4) (\mu_1 - \mu_3)}{1+\epsilon_4} - \dfrac{(\mu_1 - \mu_3)^2}{2 (1+\epsilon_4)^2}  \;,
\end{align}
and
\begin{align}
\mathring{\mu}_3  \ &\equiv \ \dot \mu_3 + \mu_1 \mu_3 + H \mu_3 \; ,\\
\mathring{\epsilon}_4  \ &\equiv \ \dot \epsilon_4 + \mu_1 \epsilon_4 + H \epsilon_4 \;.
\end{align}

\item[$\diamond$] \emph{Growth function} ($\bm\fs$): \ The effective gravitational constant is naturally part of the source term in the evolution of the linear density perturbations of matter $\delta$ : \begin{equation}\label{ed_delta}
\ddot{\delta}+2H\dot{\delta}-4\pi \mu \gn \rho_m\delta=0	\;.
\end{equation}
The $\delta$ variable is of difficult observability. However its second statistical moment,   the $rms$ of linear density fluctuations on the characteristic  scale $R=8 Mpc/h$, $\sigma_8$,  and  its logarithmic derivative with respect to the scale factor of the universe,  the  linear growth rate  $f$,  can be combined in an observable quantity
($\fs$) which is minimally affected my observational biases.\footnote{We predict the amplitude of the present-day value of the $rms$ density fluctuations in a given EFT model of gravity, by  rescaling the Planck best fitting value $\sigma_8(x_0)$ as follows: $\sigma^{EFT}_8(x)=\frac{D_{+}^{EFT}(x)}{D_{+}^{Planck}(x_0)} \sigma_{8}^{Planck}(x_0)$,  where $D_{+}$ is the growing mode of linear matter density perturbations.}

\item[$\diamond$] \emph{Gravitational slip parameter} ($\bm\gsp$): \ Gathering modifications of gravity in an effective gravitational constant does not suffice to model all deviations from GR. The Poisson equation must be supplemented with an equation for the gravitational slip parameter, namely the quantity sensitive to differences between the two gravitational potentials, $\gsp\equiv \Psi/\Phi$ . In the EFT of DE it yields as a function of the couplings: 
\begin{equation} \label{postn}
\gsp \ = 1 - \frac{c_0}{a_0} \; ,
\end{equation}
where
\begin{equation} \label{c0}
c_0 = (\mu_1 + \mathring{\epsilon}_ 4) (\mu_1 + \mu_3 + 2  \mathring{\epsilon}_ 4) - \epsilon_4 (2 \cb +  \mathring{\mu}_3 - 2 \dot H \epsilon_4 + 2 H \mathring{\epsilon}_ 4)\; .
\end{equation}
Note that $\mu$ and $\gsp$ share the same term $a_0$. The implication of this constraint will  be discussed in Sec.~\ref{sec_4bis}. 

\item[$\diamond$] \emph{Light deflection parameter} ($\bm\Sigma$): \ In general, observations probing  the gravitational potentials,  such as weak lensing measurements,  are not directly  sensitive to the gravitational slip parameter but to the light deflection parameter $\Sigma$. In GR, as for $\gsp$, it is equal to 1. In MG, it is not necessarily and is defined through the equation $-\frac{k^2}{a^2}(\Phi+\Psi)=8\pi \Sigma(t,k)\gn\rho_m \delta_m $. It can be expressed straightforwardly as a combination of $\mu$ and $\gsp$
\begin{equation} \label{postn}
\Sigma=\frac{\mu}{2}\,(1+\gsp) \; .
\end{equation}
This theoretical degeneracy between observables of the perturbed sector will be instrumental in understanding specific predictions of Horndeski theories.

\end{itemize}

\subsection{Viability criteria}\label{stabcond}

Although asymptotic behaviours of the EFT functions can be changed, not all their possible time scalings are permitted. A healthy theory must indeed fulfil a set of stability conditions: it must not be affected by ghosts, nor by gradient instabilities. Furthermore, along the arguments detailed in~\cite{Adams:2006sv}, we will not allow superluminal propagation speeds for either scalar or tensor modes. On top of these theoretical requirements, we should exclude models that are already ruled out by current observations. 
As for the choice of the background expansion rate, which we describe via  the effective Hubble rate 
 \eqref{h}, we exploit  current limits available in the perturbed sector of the universe. Notably, the local value of gravitational waves of EFT of DE models has been recently constrained leading to a bound on the value of $\epsilon_4(x_0) \sim 10^{-2}$~\cite{Jimenez:2015bwa}, thus we simply set its present value to 0 for simplicity. In summary,
\vv
\noindent{\it Stability of the theory},
\begin{align}\label{conditions}
& A = (\cb +2\mu_2^2)(1+\epsilon_4)+\frac34(\mu_1-\mu_3)^2\ge 0 \qquad\qquad\qquad\qquad\qquad\qquad\qquad\quad\;\, \text{ghost free}, \\ \label{conditions2}
& B = b_0 \ge 0 \qquad\qquad\qquad\qquad\qquad\qquad\qquad\qquad\qquad\qquad\qquad\qquad\qquad\qquad\quad \text{gradient,}
\end{align}
\noindent{\it Subluminal propagation speeds},
\begin{align}\label{cs}
& c_s^2 = \frac{B}{A} \leqslant 1  \qquad\qquad\qquad\qquad\qquad\qquad\qquad\qquad\qquad\qquad\qquad\qquad\qquad\;\;\;\, \text{scalar modes},\\ \label{ct}
& c_T^2 =\dfrac{1}{1+\epsilon_4} \leqslant 1  \qquad\qquad\qquad\qquad\qquad\qquad\qquad\qquad\qquad\qquad\qquad\qquad\;\;\;\, \text{tensor modes}, 
\end{align}
\noindent{\it Observational requirement (compatibility with current constraints)},
\begin{align}
\epsilon_4(x=x_0)=0 \;.
\end{align}

Since they  impose tight constraints on the functional behaviour of relevant observables of the LSS, viability criteria can be effectively used to complement data and observational information  in  statistical inferences \cite{Salvatelli:2016mgy}. The dependence of our conclusions on the requirement of sub-luminal propagation speeds will be assessed in Sec.~\ref{sec_viab}.
As already discussed in Section \ref{edeclass} the models we are considering are ineffective in describing cosmic evolution at such early epoch as those where  BBN could be used to constrain them. However, on the opposite end, \emph{ i.e.} today,  
Lunar ranging tests have put constraints on the variation of Newton's constant, at around $\dot{\gn}/ \gn<0.02 H_0$ (see \cite{Uzan:2010pm} for a detailed review). Since we are not considering EFT operators beyond the linear level, it is difficult to predict how non linearities would affect the definition of $G_N$ for our models. It is however misleading to draw  the conclusion that the coupling  $\mu_1$ would end up being severely constrained. Indeed variations in the Planck mass could still be relatively large, although   appropriately counter-balanced by the specific timescaling of $\epsilon_4$ (and so $\dot{c_T}$, see eq.\eqref{gn}). 
Accordingly,   we do  not  consider the Lunar ranging  constraint as an additional viability criteria in our study. We just warn the reader that  Horndeski models passing this constraint would constitute a subsample of the whole set of healthy theories  considered  in this study.

\section{General predictions on LSS observables} \label{sec_4}

In this section, we explore the space of theories following the protocol elaborated  by \cite{Perenon:2015sla}: the non-minimal couplings are expanded in power series of $(x-x_0)$ up to order 2 (see Appendix~\ref{param}), where each coefficient is randomly chosen within the window $[-1,1]$ with a flat uniform prior. This is enough to cover all the rich phenomenology arising in our EFT models. A pre-factor $(1-x)$ in the expansion is either switched on or off depending on the DE scenario, $i.e$ whether a non-minimal coupling needs to vanish at early times or not (see eqs.~\eqref{LDEdef},~\eqref{EDEdef} and~\eqref{EMGdef} for the conditions imposed in the various scenarios). The initial time we consider numerically is set to $z_i=100$, where radiation is already sub-dominant. It is for example the time where initial conditions are set for the integration of growth observables. We reject the theories that do not pass the viability conditions from early times until today. In addition, to lie within the range of applicability of the quasi-static approximation only models with $c_s^2>0.1$ are kept~\cite{Sawicki:2015zya}. With this procedure we randomly generate $10^4$ viable EFT models of each DE scenario.

For what concerns the class of LDE theories, an important generalization with respect to~\cite{Perenon:2015sla} is that we also consider here the coupling $\m22$ not to be equal to zero at all times. The latter, although not entering the expressions of the relevant observables, controls the effectiveness of the no-ghost stability condition and thus generally relaxes the selection processes. We then study the implications of early dark energy scenarios (EDE and EMG).

\subsection{LDE scenario}

\begin{figure}[H]
\begin{center}
\begin{flushleft} LDE : \end{flushleft} 
   \includegraphics[scale=0.27]{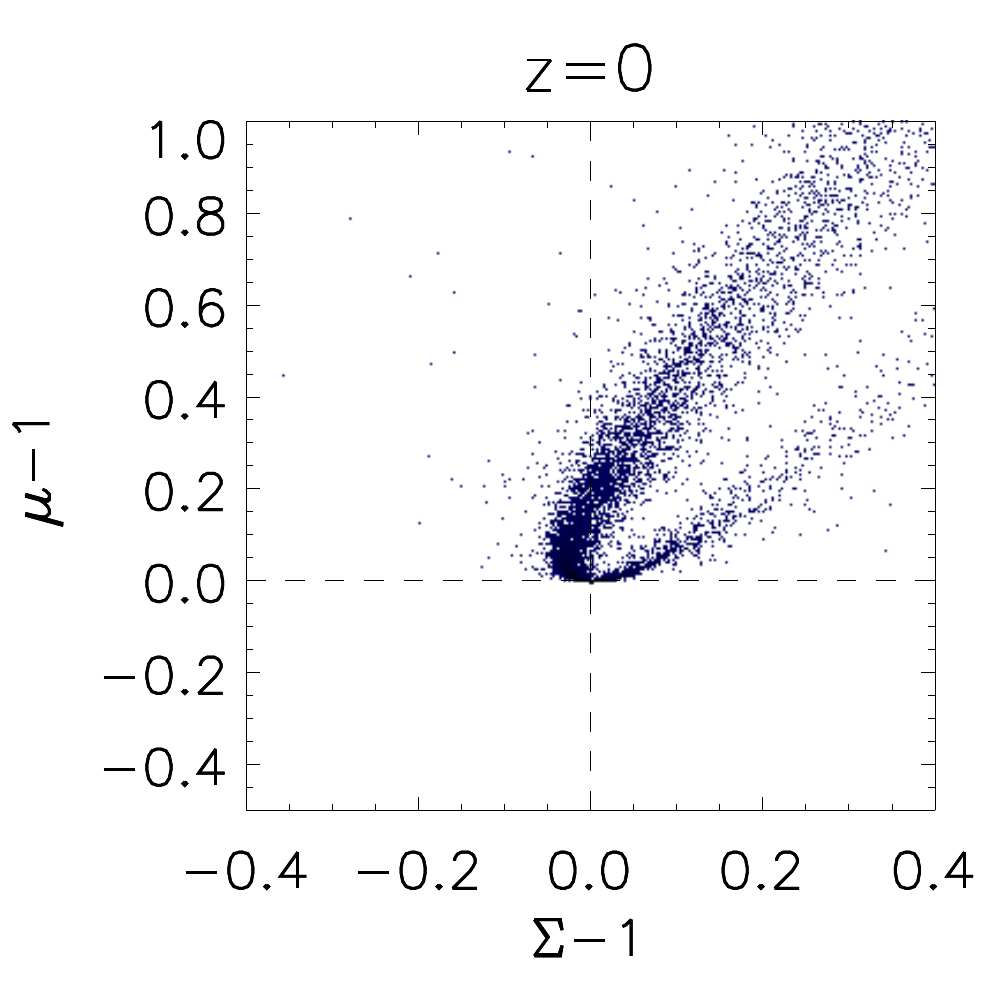}  \hskip-4mm
   \includegraphics[scale=0.27]{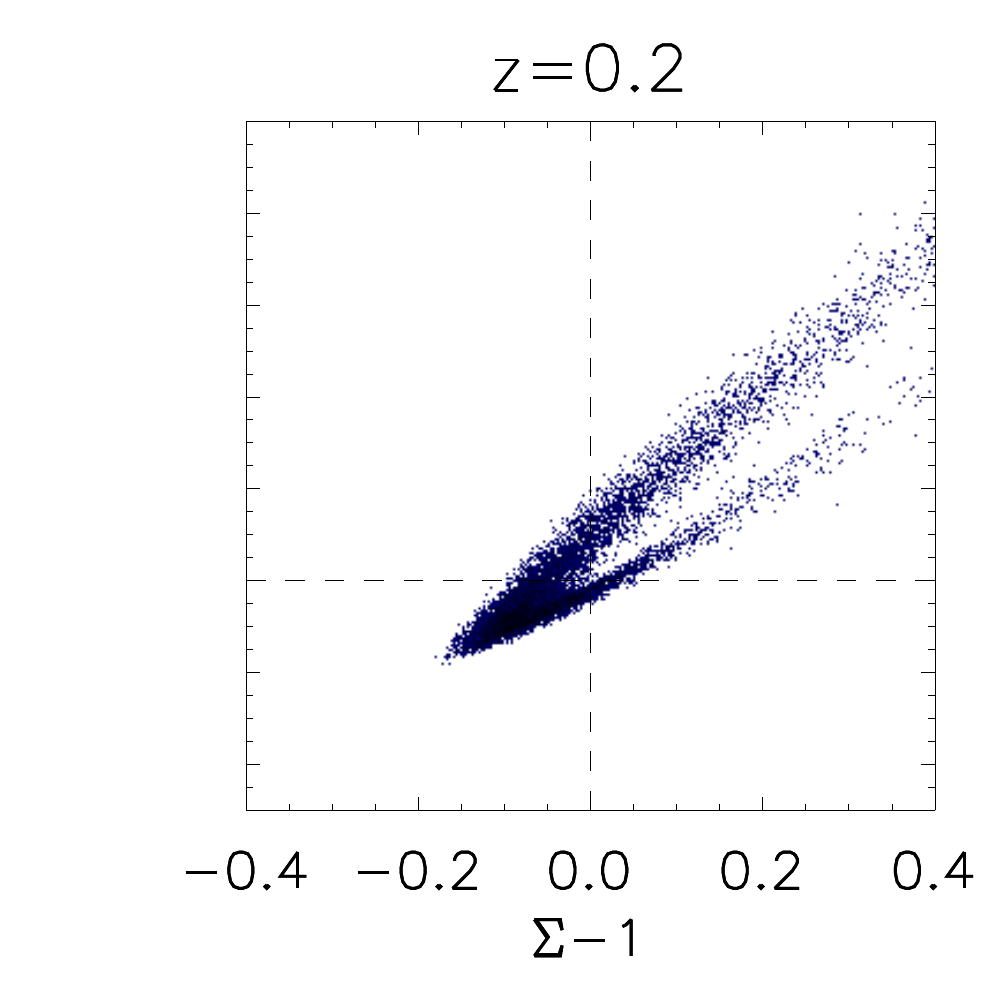}  \hskip-4mm
   \includegraphics[scale=0.27]{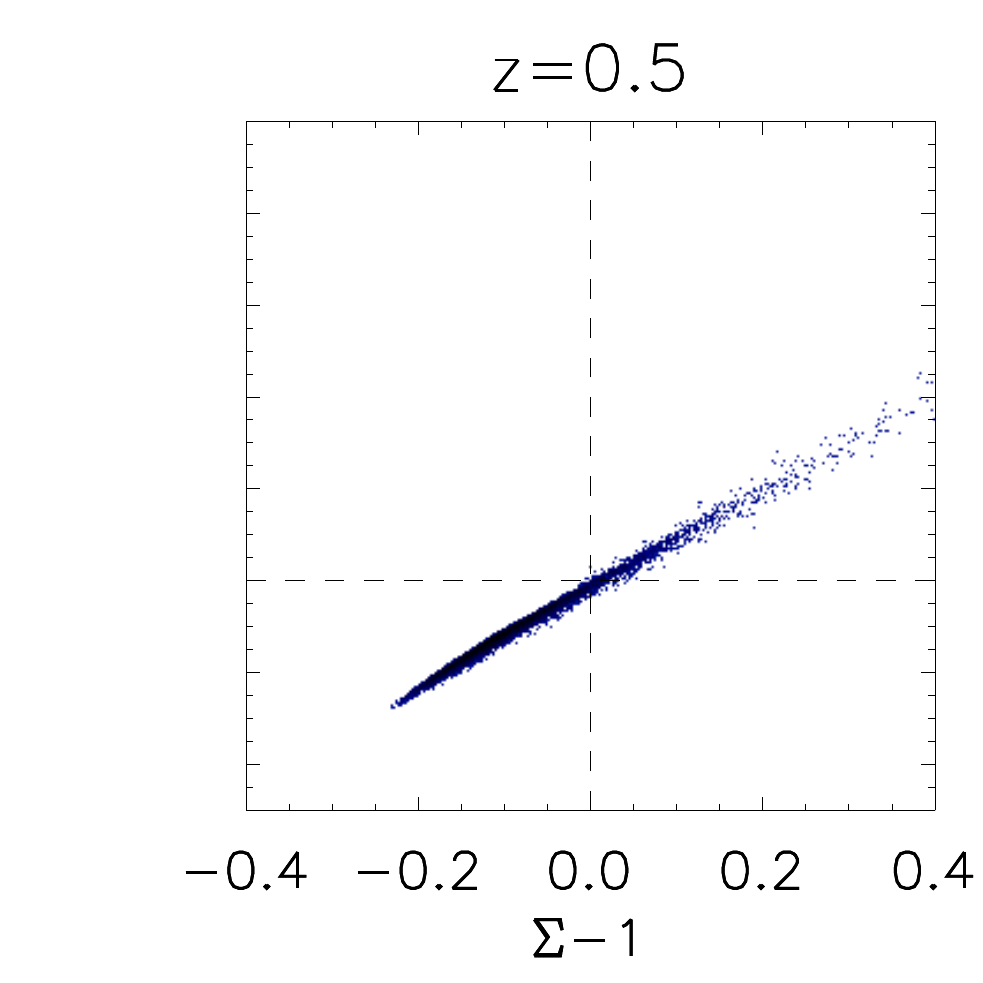}  \hskip-4mm 
   \includegraphics[scale=0.27]{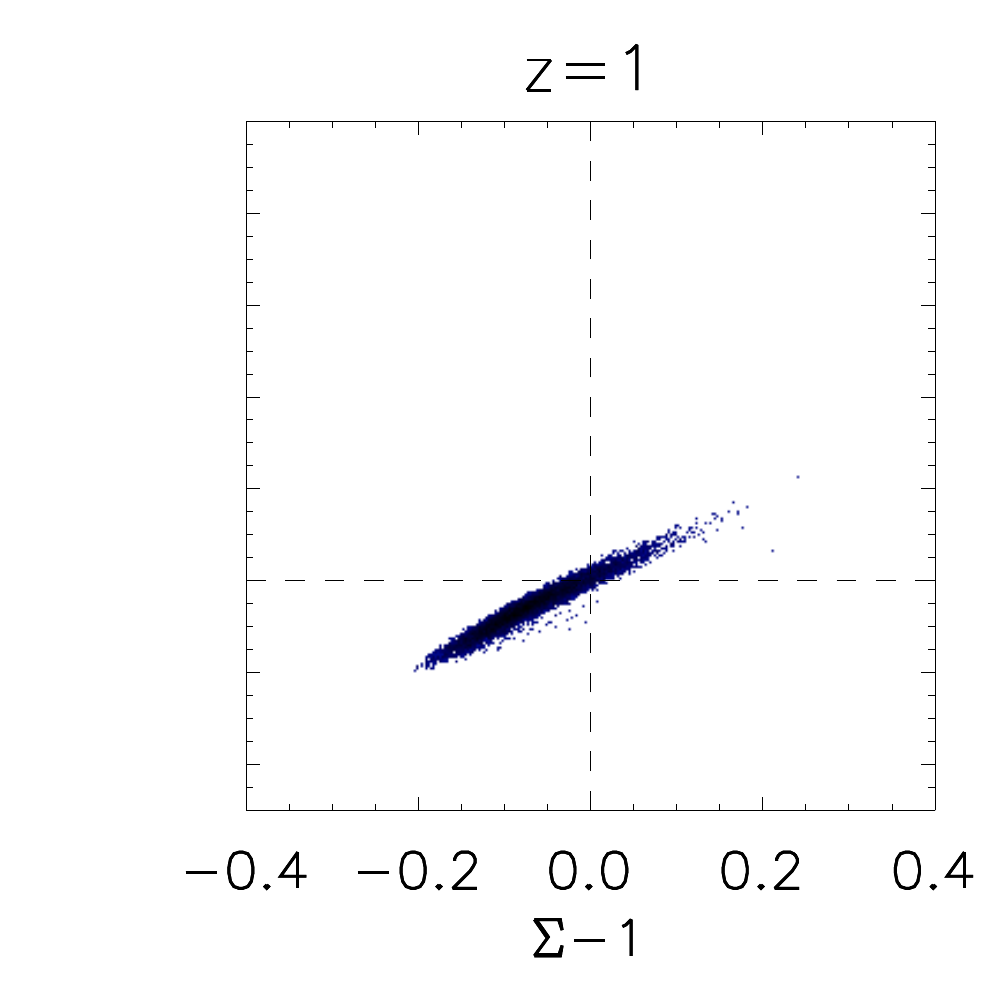}  \hskip-4mm
   \includegraphics[scale=0.27]{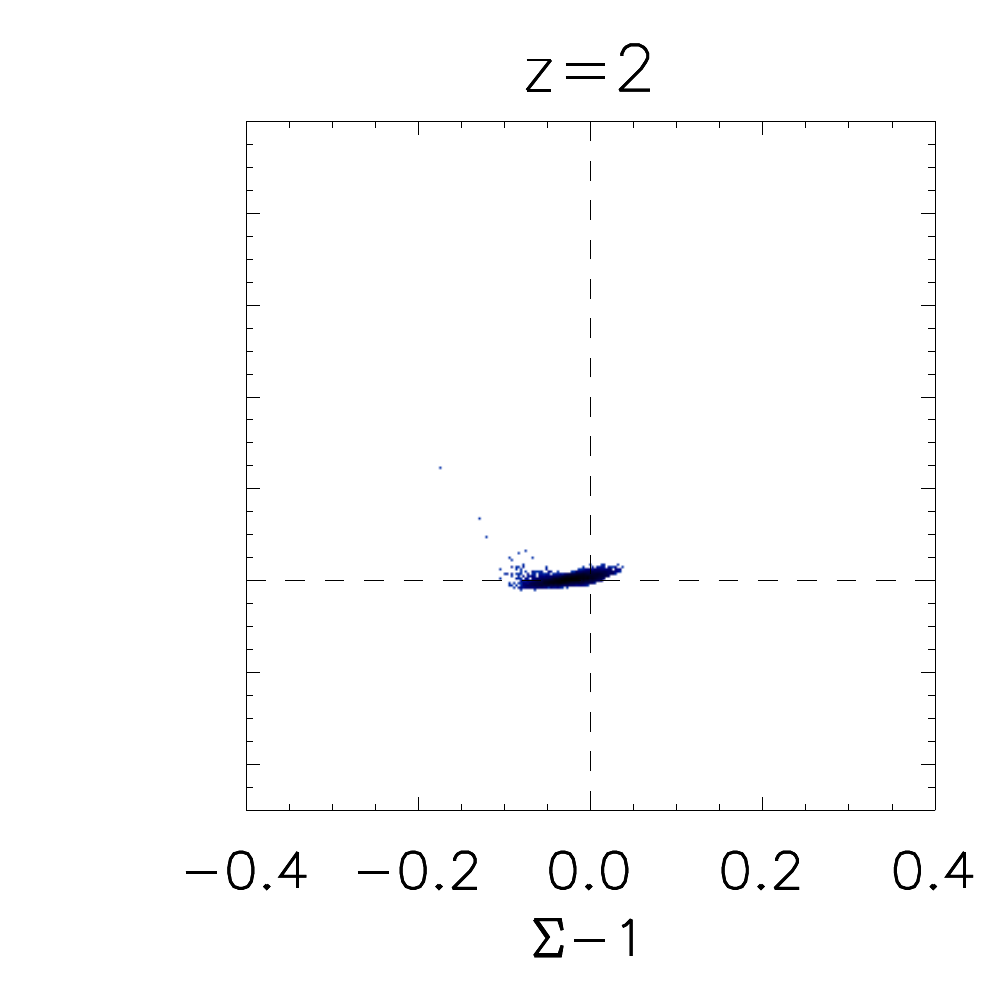}  \hskip-4mm
   \includegraphics[scale=0.27]{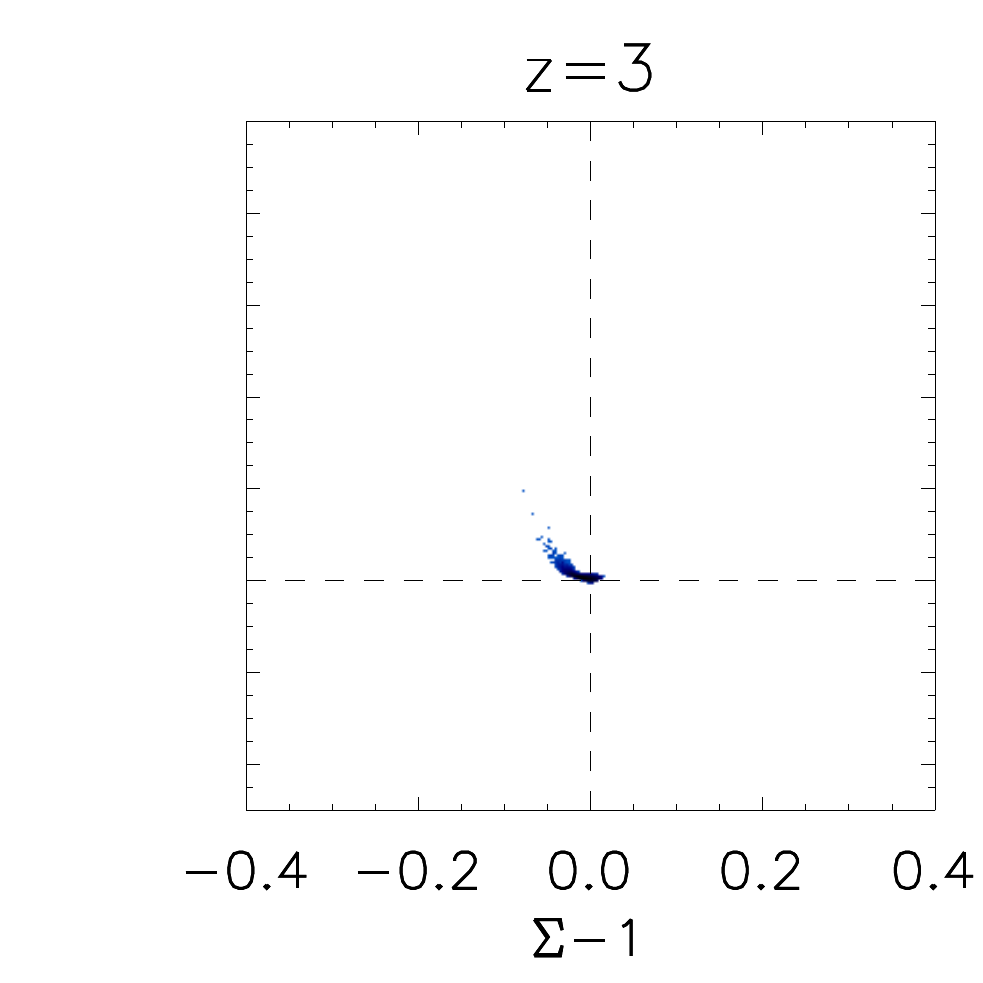}
   \vskip-2mm
   \includegraphics[scale=0.27]{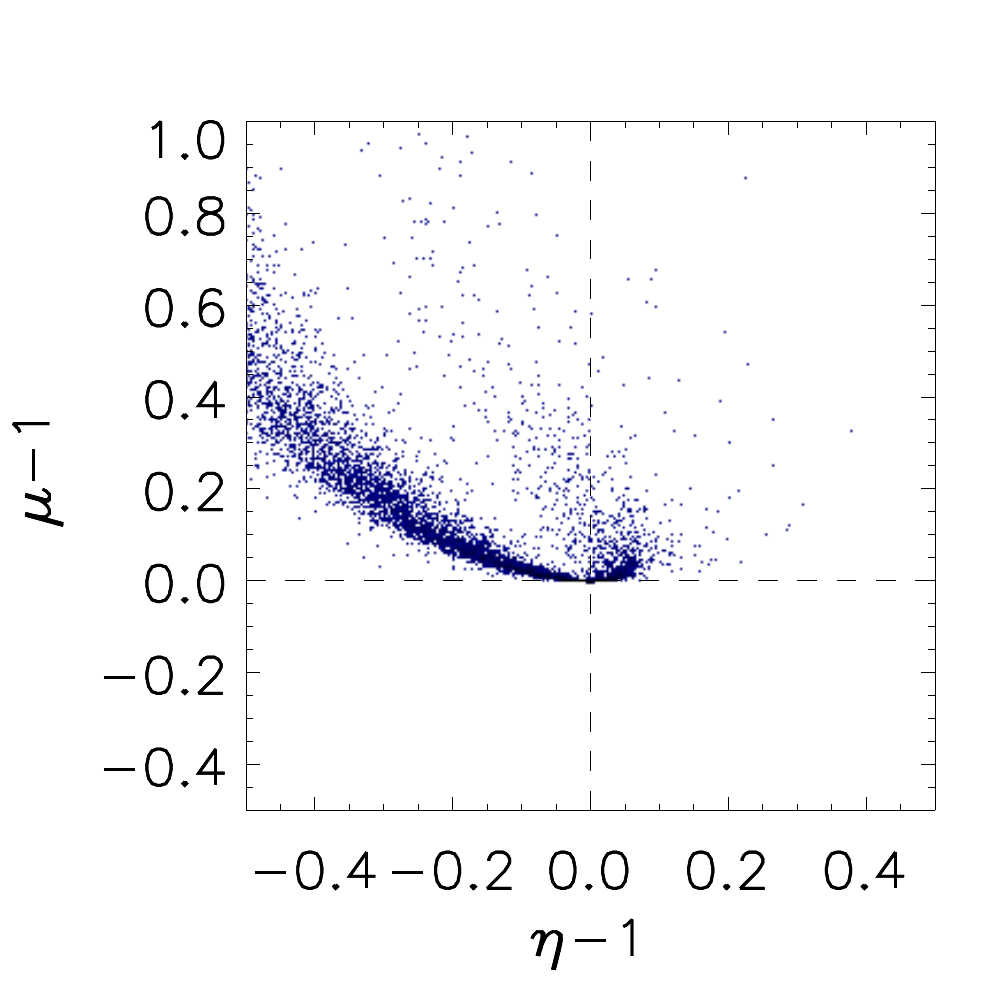}  \hskip-4mm
   \includegraphics[scale=0.27]{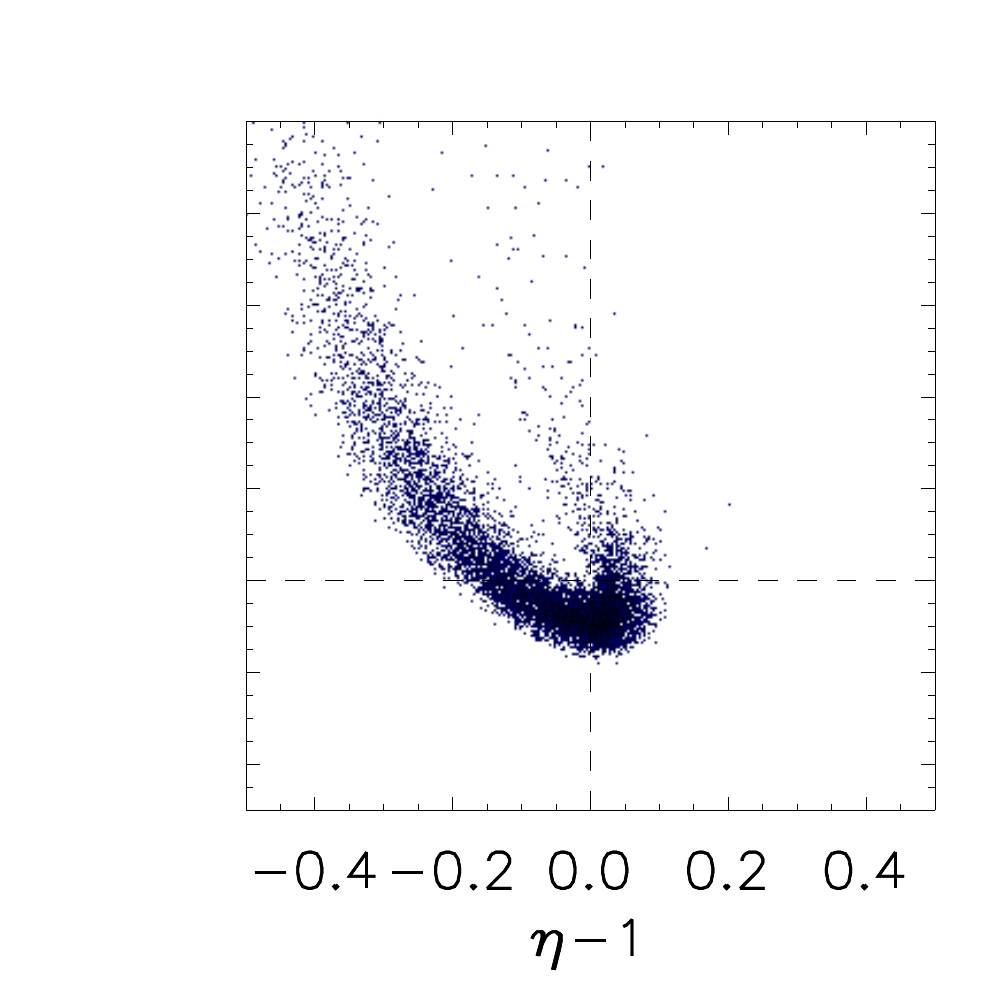}  \hskip-4mm
   \includegraphics[scale=0.27]{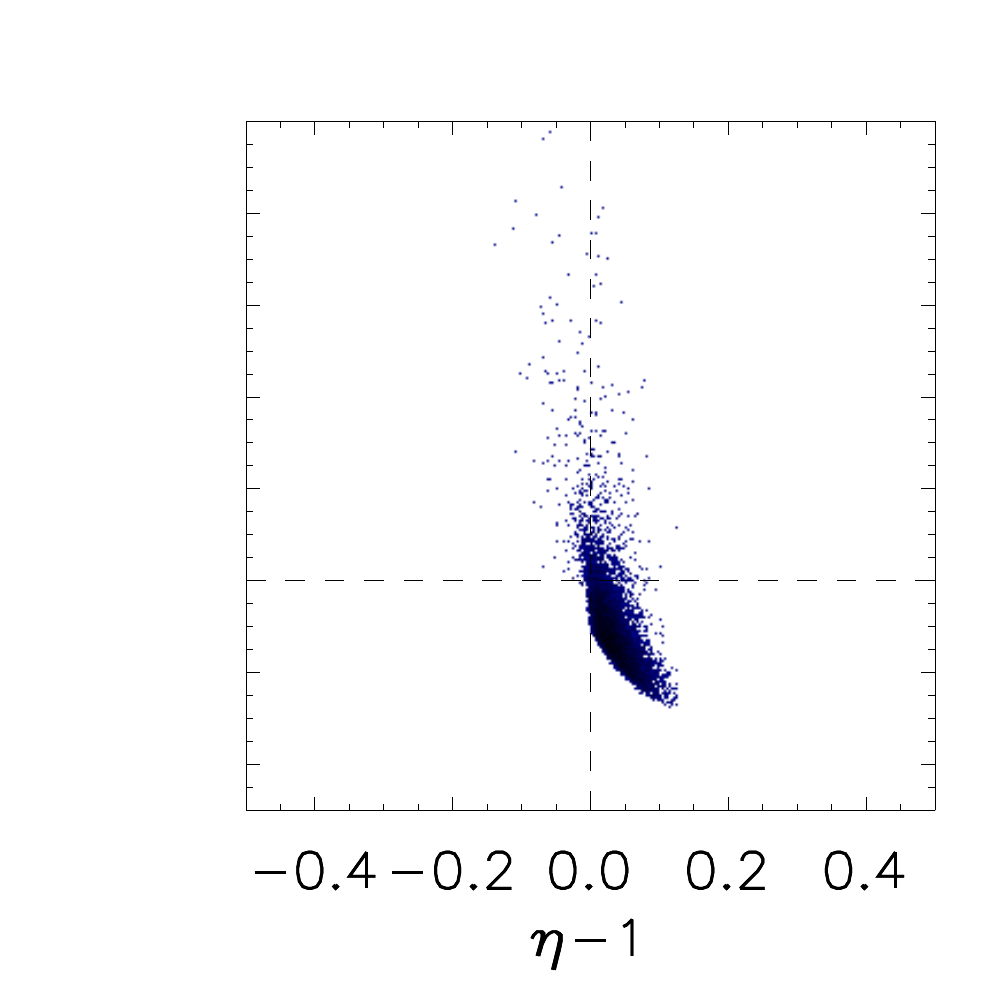}   \hskip-4mm
   \includegraphics[scale=0.27]{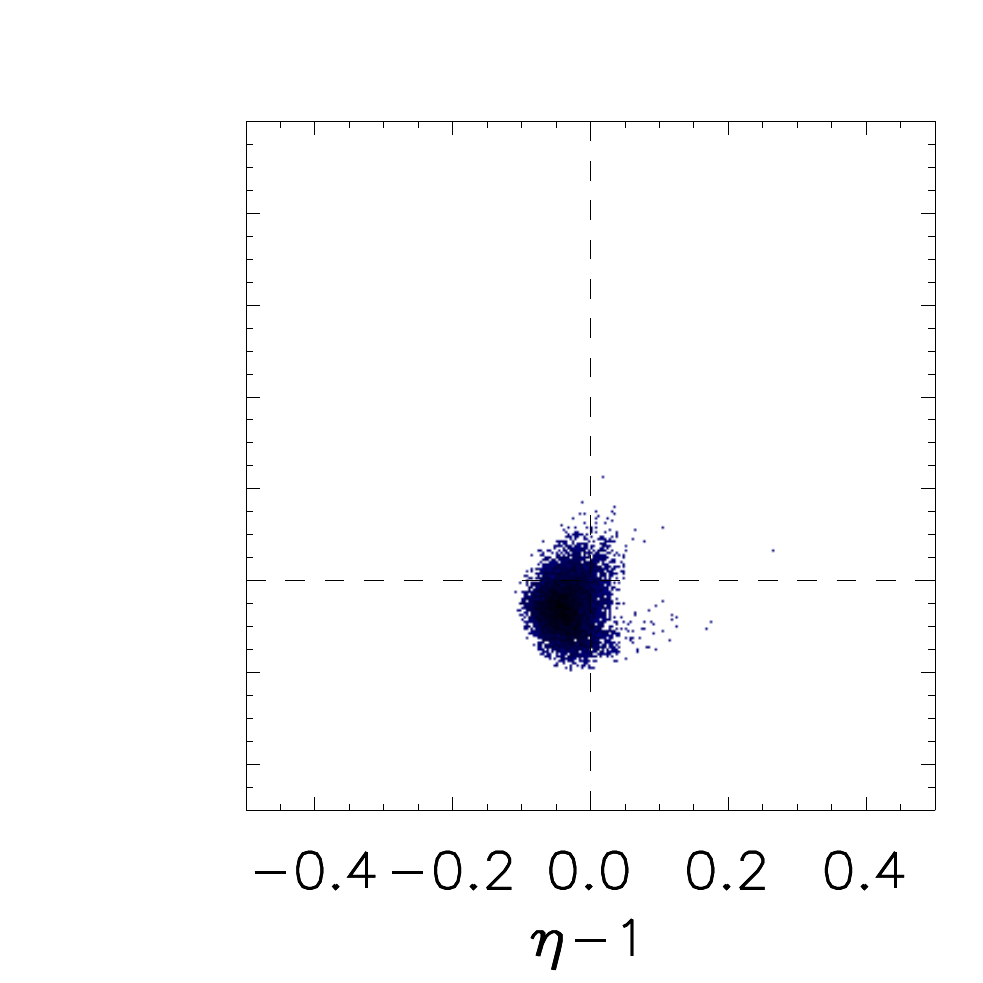}  \hskip-4mm
   \includegraphics[scale=0.27]{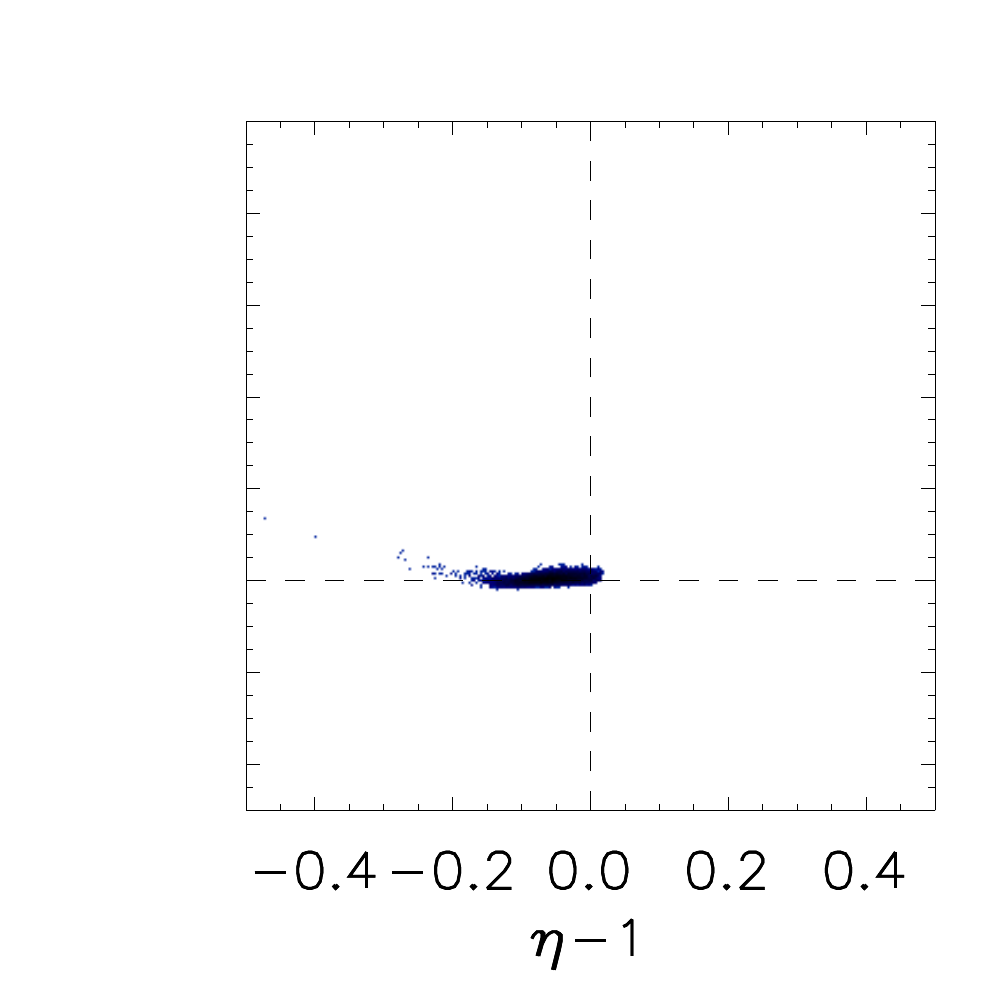}  \hskip-4mm
   \includegraphics[scale=0.27]{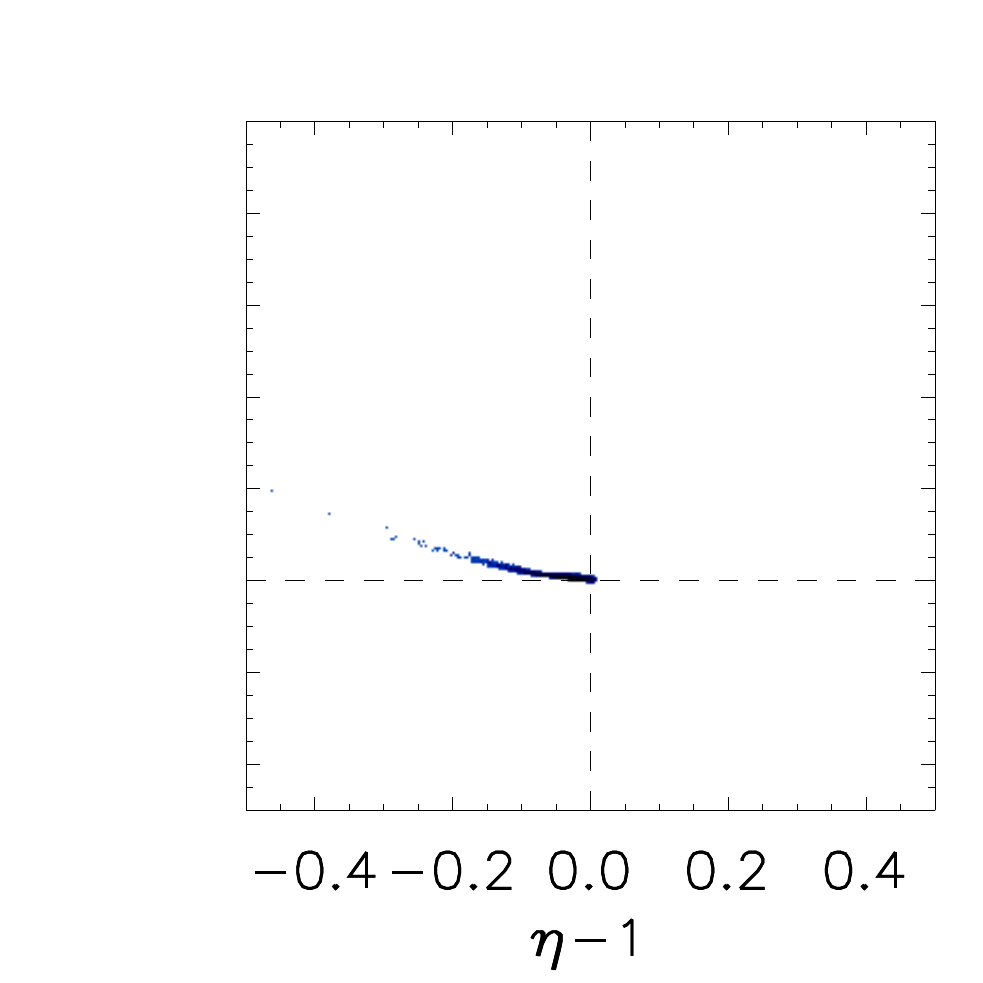}  
   \vskip-2mm
   \includegraphics[scale=0.27]{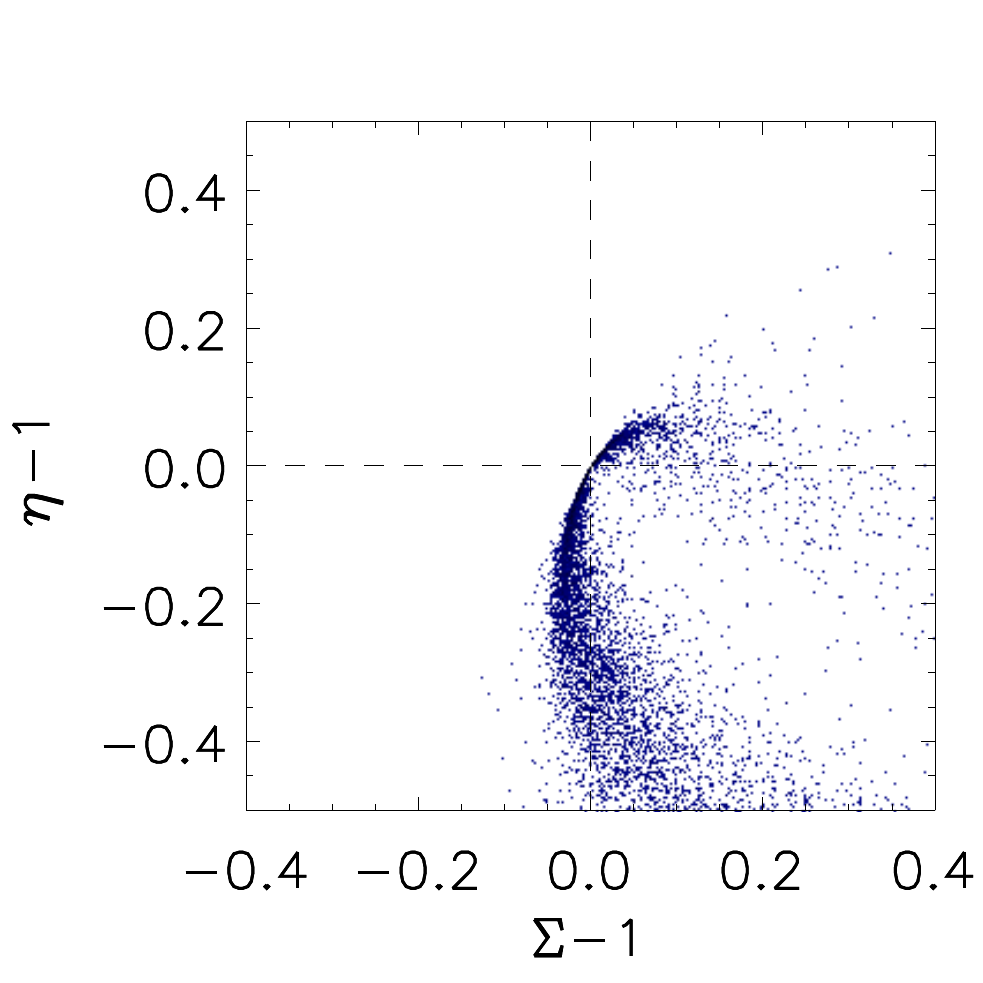}  \hskip-4mm
   \includegraphics[scale=0.27]{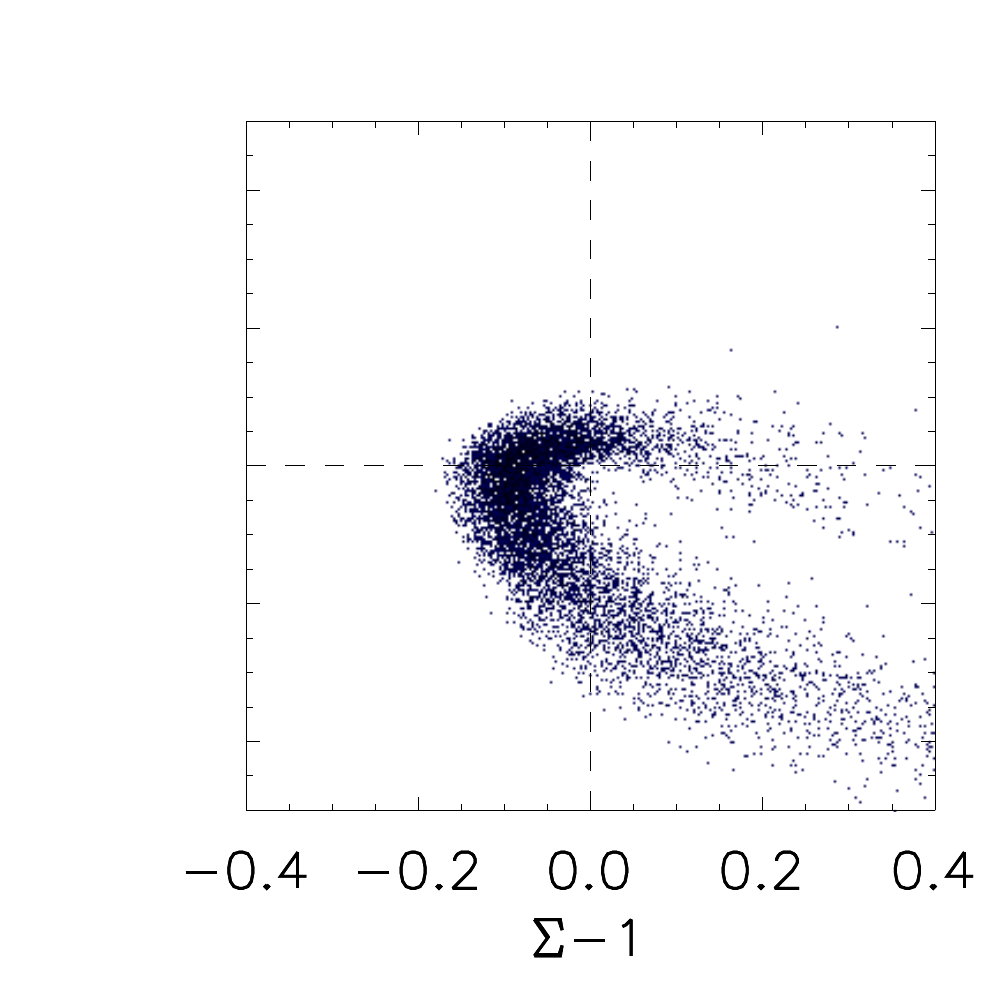}  \hskip-4mm
   \includegraphics[scale=0.27]{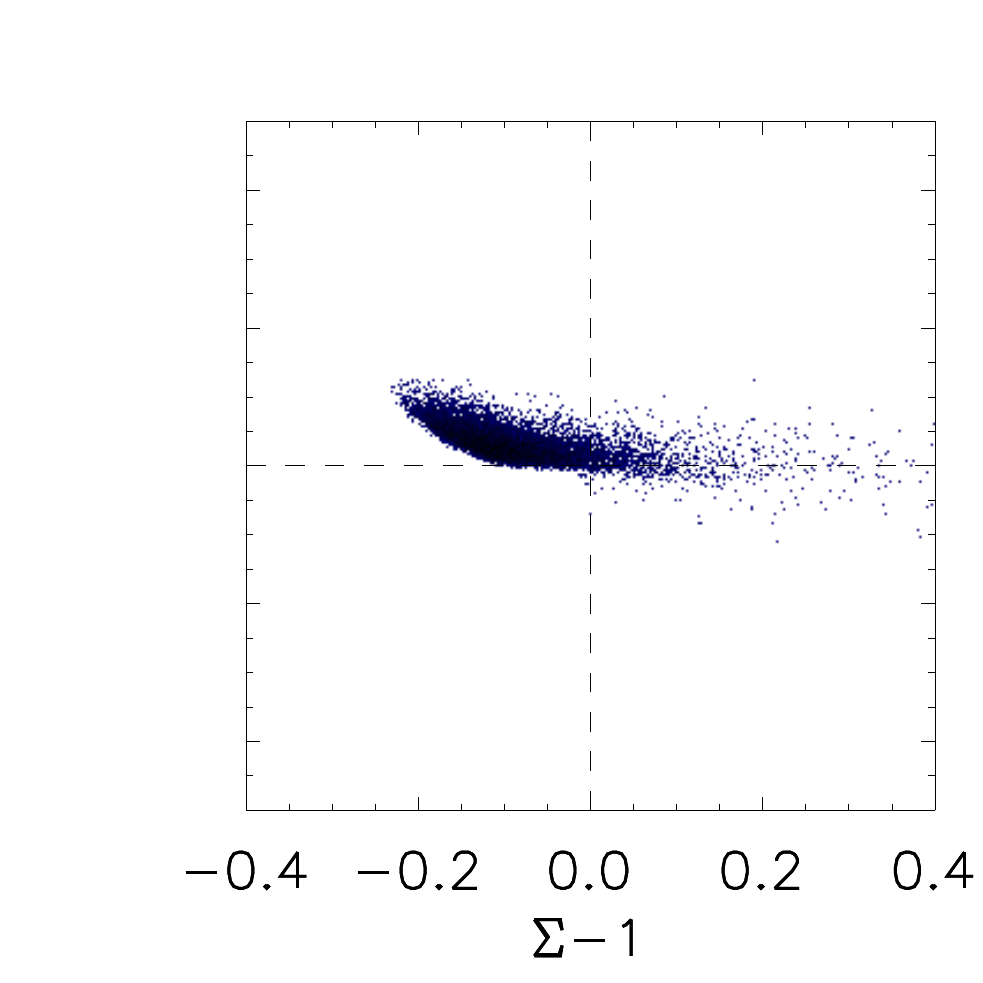}   \hskip-4mm
   \includegraphics[scale=0.27]{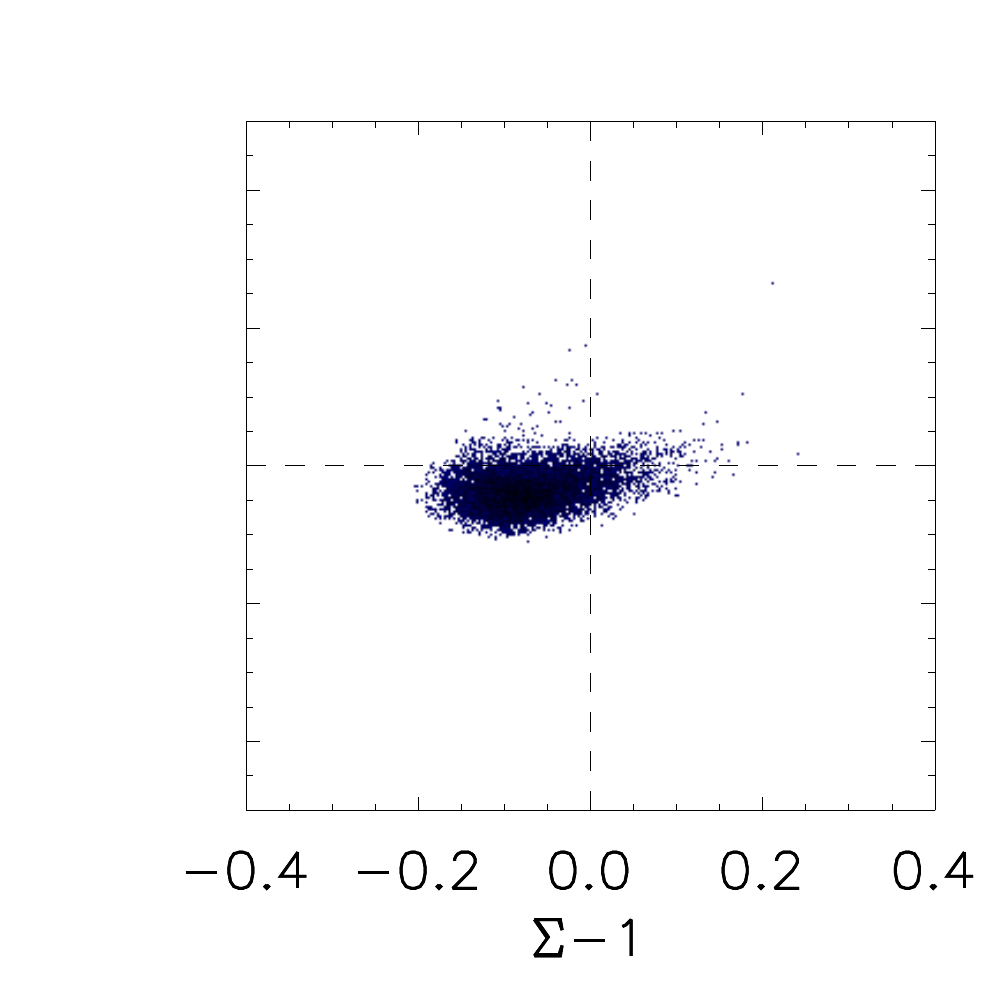}  \hskip-4mm
   \includegraphics[scale=0.27]{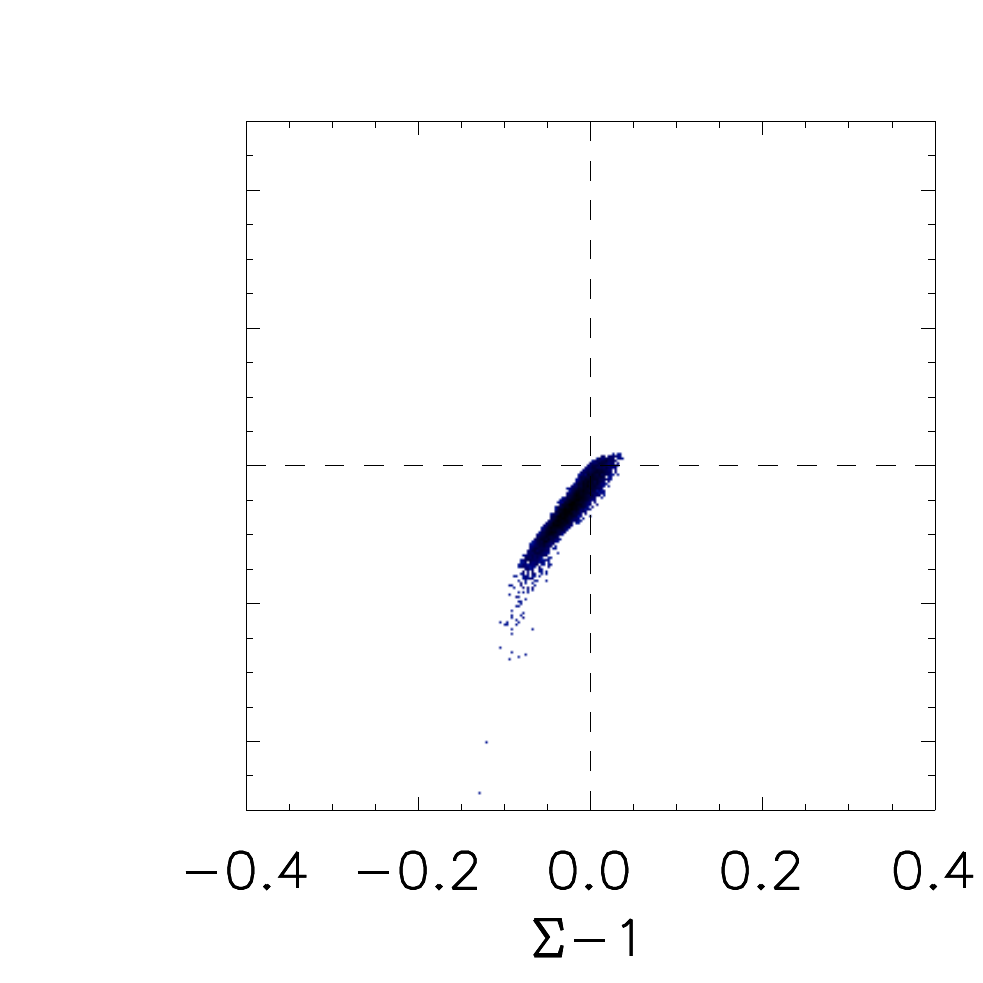}  \hskip-4mm
   \includegraphics[scale=0.27]{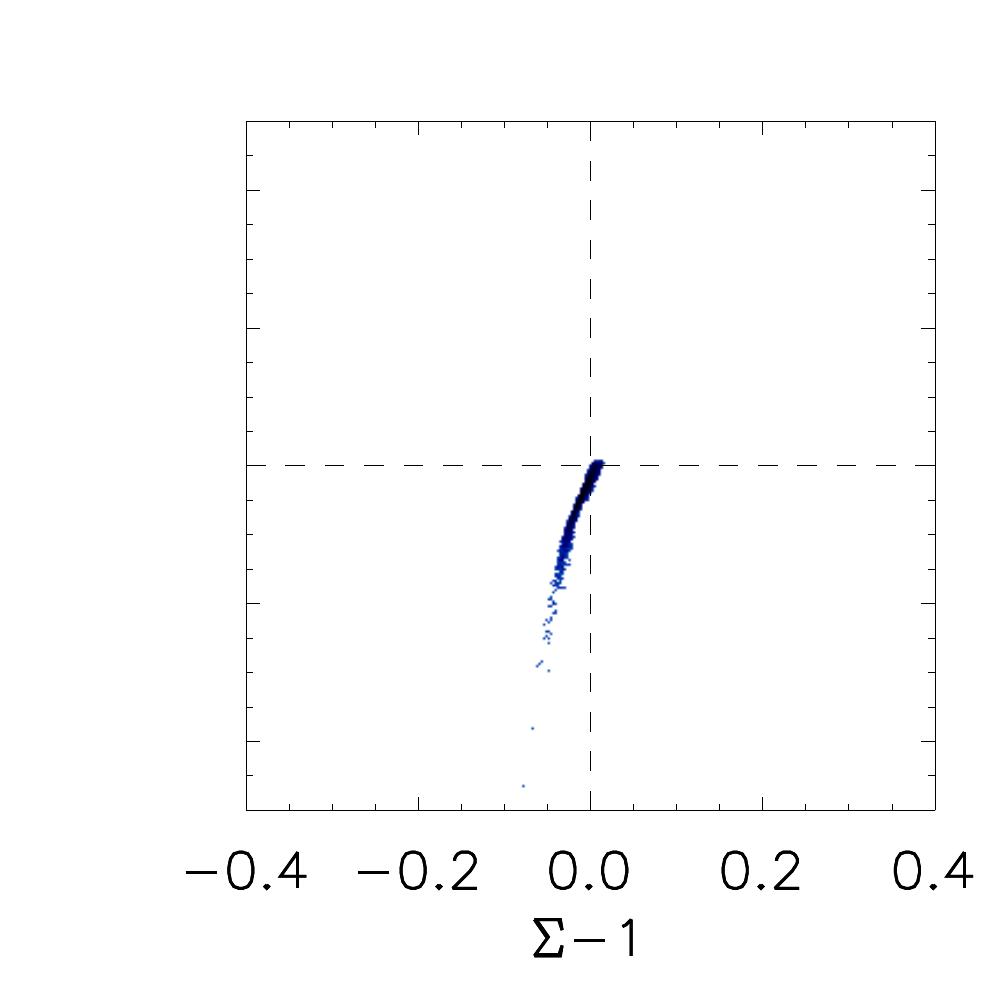}  
   \vskip-2mm
   \includegraphics[scale=0.27]{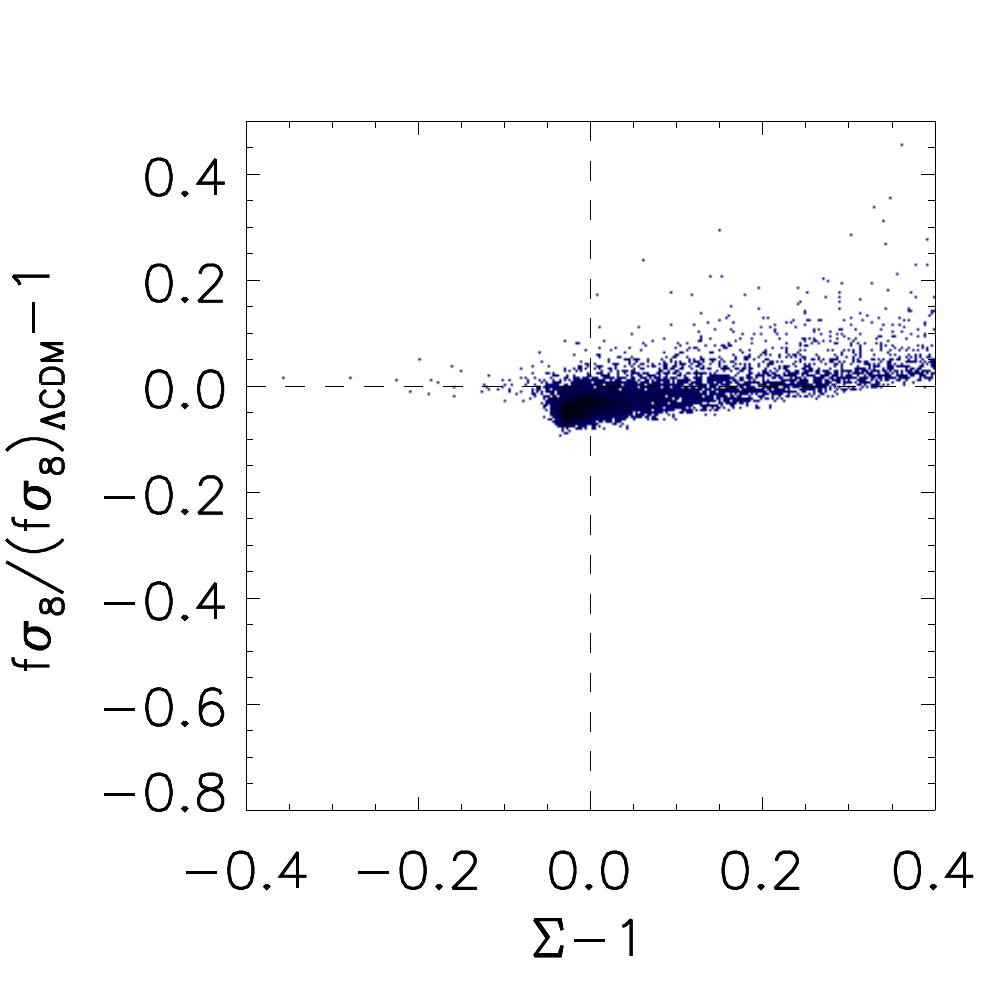}  \hskip-4mm
   \includegraphics[scale=0.27]{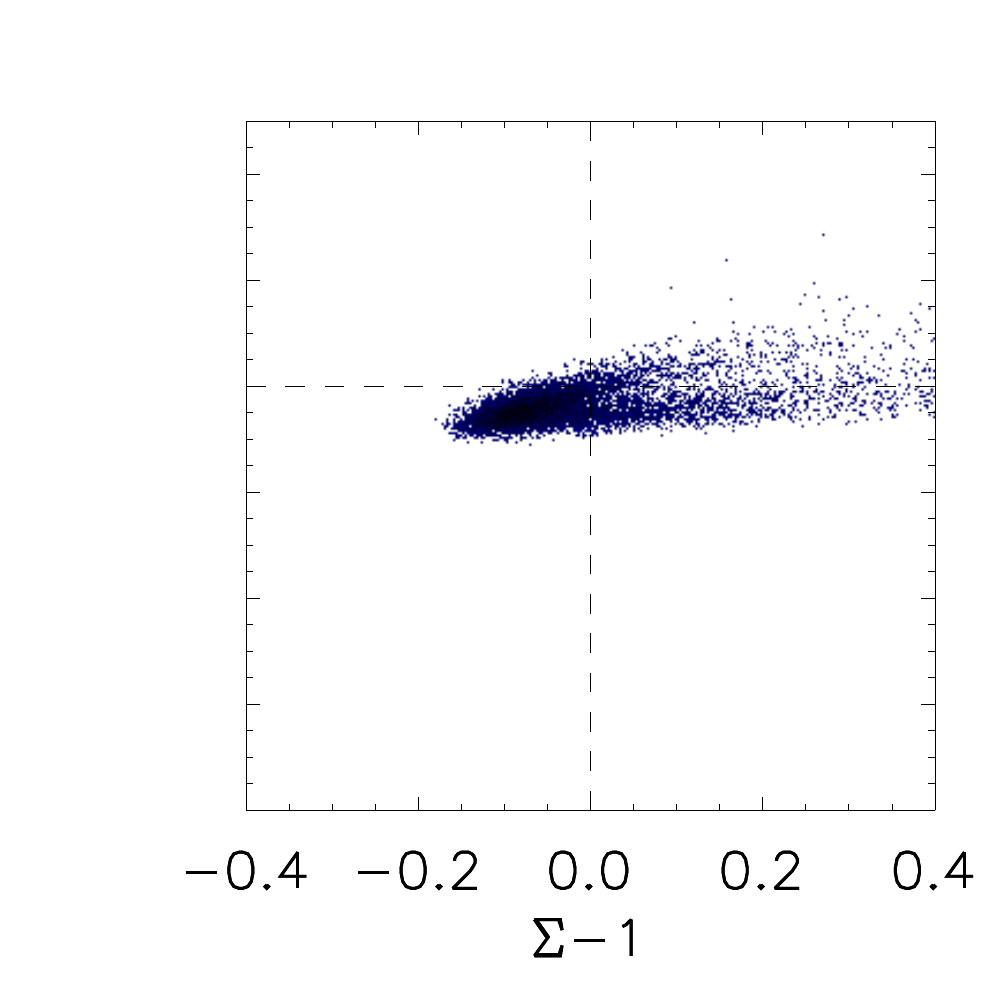}  \hskip-4mm
   \includegraphics[scale=0.27]{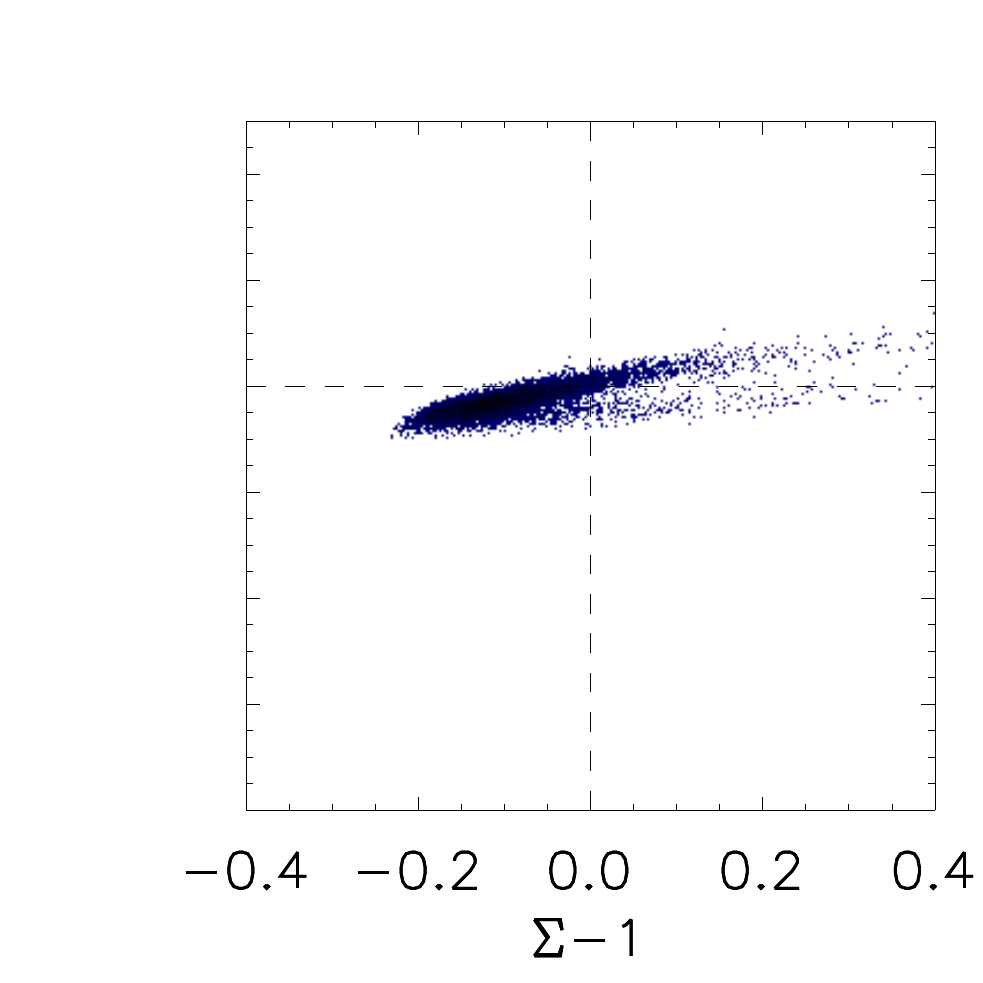}  \hskip-4mm 
   \includegraphics[scale=0.27]{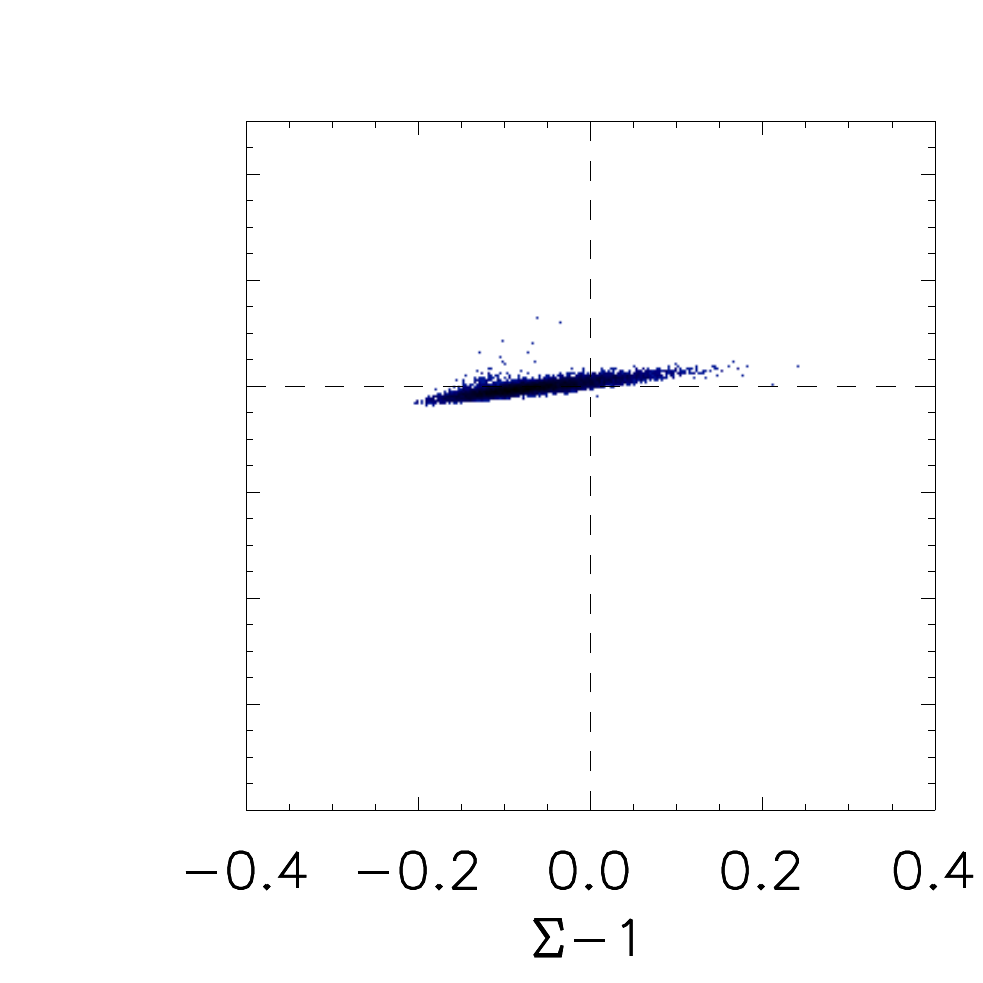}  \hskip-4mm
   \includegraphics[scale=0.27]{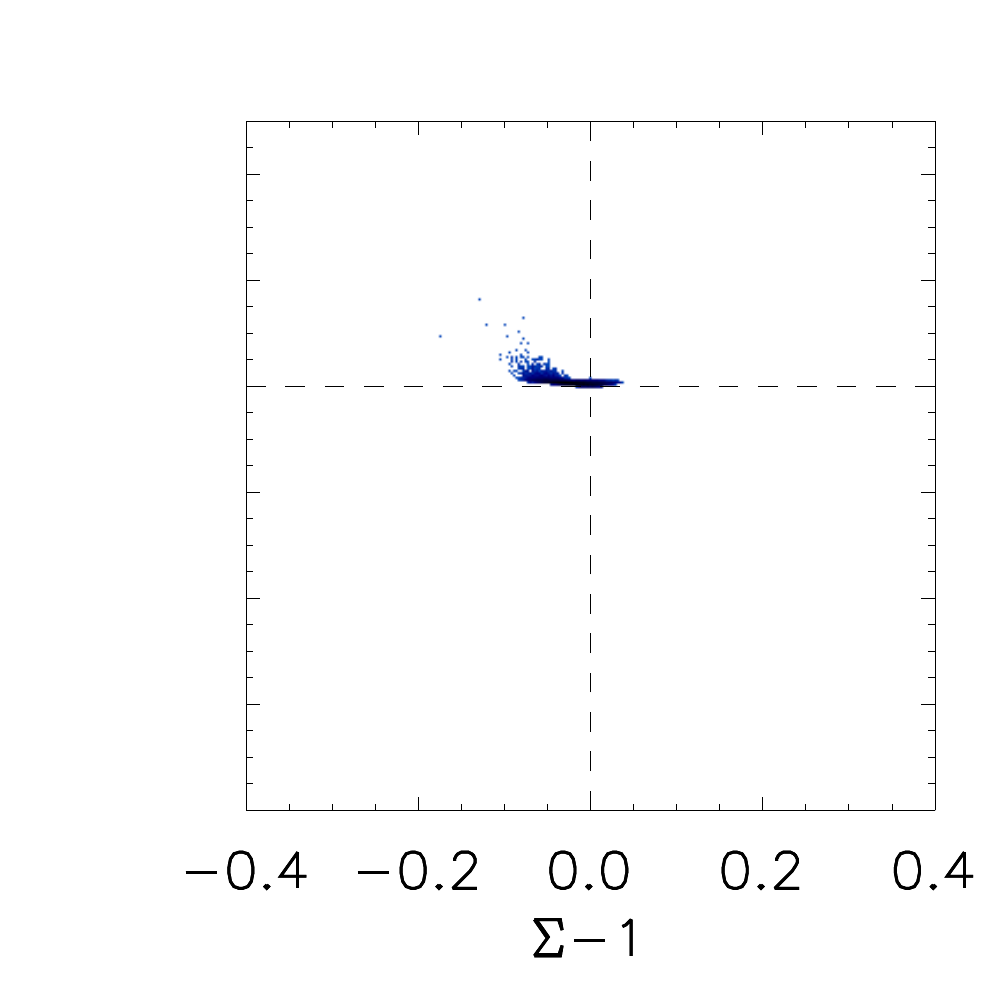}  \hskip-4mm
   \includegraphics[scale=0.27]{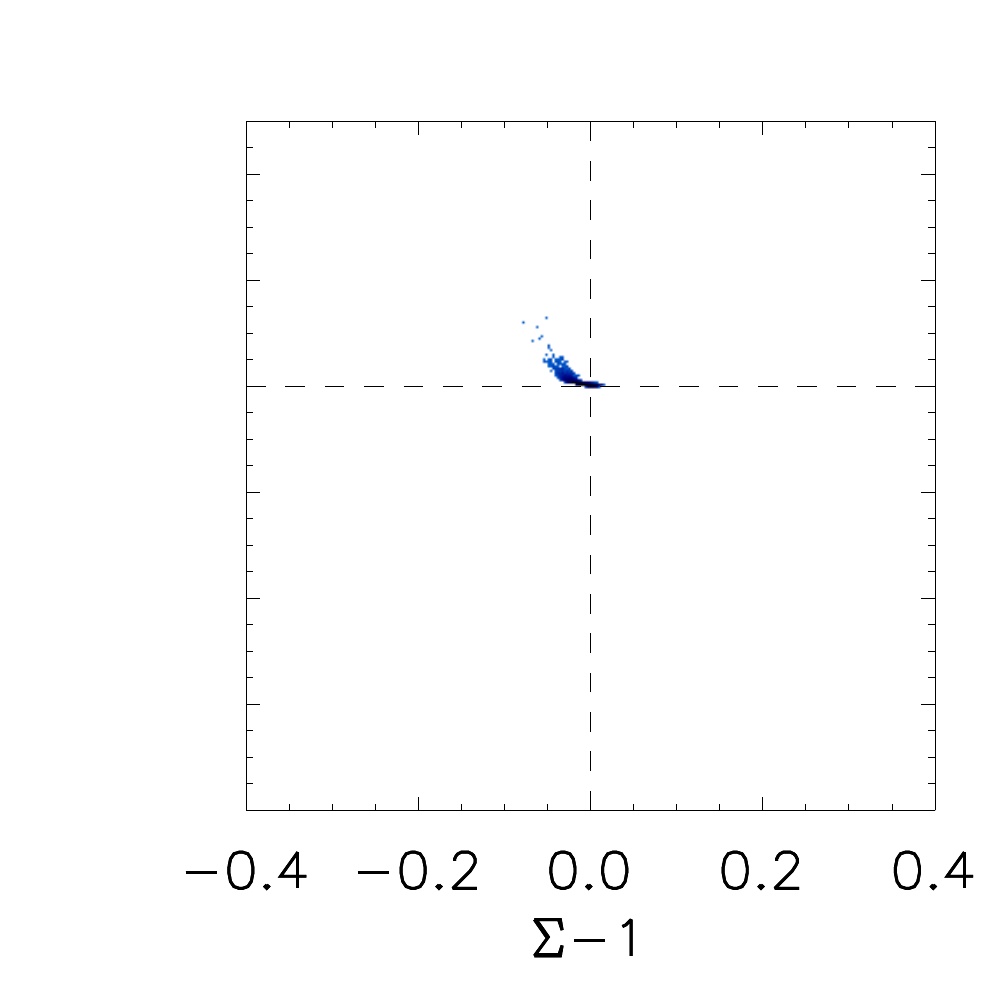}  
   \caption{The correlations between  $\mu$, $\gsp$, $\Sigma$ and $\fs$ is  displayed at several redshift epochs, from left to right $z=0$, $0.2$, $0.5$, $1$, $2$, $3$, for $10^4$ EFT models in the LDE scenario. The background evolution has been set to match that of a flat $\Lambda$CDM model.  The $\Lambda$CDM prediction corresponds to the intersection of the two dashed lines. The gray scale highlights the density of points.}
   \label{fig_lde} 
\end{center}
\end{figure}

A definite feature emerging from inspecting the first row of panels in figure~\ref{fig_lde}, is the peculiar \emph{S-shape} redshift evolution of the effective Newton constant $\mu(z)$ in LDE models. Notably, one has $\mu\geqslant 1$ at both late ($z \sim 0$) and early epochs ($z>2$), while power is suppressed in the interval $0.5<z<1$, in the sense that in most theories $\mu$ is found to be less than 1. This extremely constrained functional behaviour was already noticed by \cite{Perenon:2015sla} and confirmed by~\cite{Tsujikawa:2015mga}, although for a more restricted class of  Horndeski models.
 
The subset of models displaying $\mu>1$ in the interval $0.5<z<1$, despite having small size relative to the entire set of simulated models, has not strictly zero measure as was previously found in \cite{Perenon:2015sla}. This is the consequence of switching on the non-minimal coupling $\m22$ which, in the present analysis, it is allowed to vary freely in the interval $[-1,1]$. Indeed, affecting the sound speed, and more precisely the no-ghost stability condition~\eqref{conditions}, this parameter induces non-negligible back reactions on the LSS observables. From \cite{Perenon:2015sla} it was understood that the period of weaker gravity in $\mu(z)$ at intermediate redshifts was induced by the $1/M^2$ component (see eq.\eqref{mu}). Our current study reveals that switching on $\m22$ is the necessary condition for LDE models to exhibit $M^2/\mps < 1$ in a stable way and therefore a subset of theories with $\mu>1$ at intermediate redshifts, $i.e$ stronger gravity and also deeper gravity potentials than the standard model. Since under these conditions light should bend more on average, it does not come as a surprise that models exhibiting $\mu>1$ in $z\in [0.5,1]$ also display $\gsp>0$, $i.e$ $\Phi>\Psi$, or $\Sigma>0$, as the inspection of the second and third row of Figure~\ref{fig_lde} shows.
\vv
The bounded evolution history of $\mu$ has major implications for the growth of structures, as captured by the $\fs$ observable. Indeed the effective Newton constant is part of the source term in the equation used to compute the growth factor $f$: 
\begin{equation}\label{lgrowth}
3 w_{\rm eff}(1-x)x f'(x)+f(x)^2+ \left [  2- \frac{3}{2} \left( w_{\rm eff} (1-x) +1 \right )  \right ] f(x) =\frac{3}{2}x \, \mu \, \gn \, ,
\end{equation}
and also affects the amplitude of $\sigma_8$, the  $r.m.s$ of the matter density fluctuations. Characteristic features of $\mu$ at time $x$ will be seen time translated at later epochs, $i.e$ lower $x$, in the $\fs$ evolution since the effective source term in eq.~\eqref{lgrowth} is $x\,\mu$ and since $\sigma_8(z)$ is an integral quantity summed from the past (here $z_i=100$) until $z$. 

\begin{figure}[H]
\begin{center}
\begin{flushleft}   \vskip-2mm EDE : \end{flushleft}
   \includegraphics[scale=0.27]{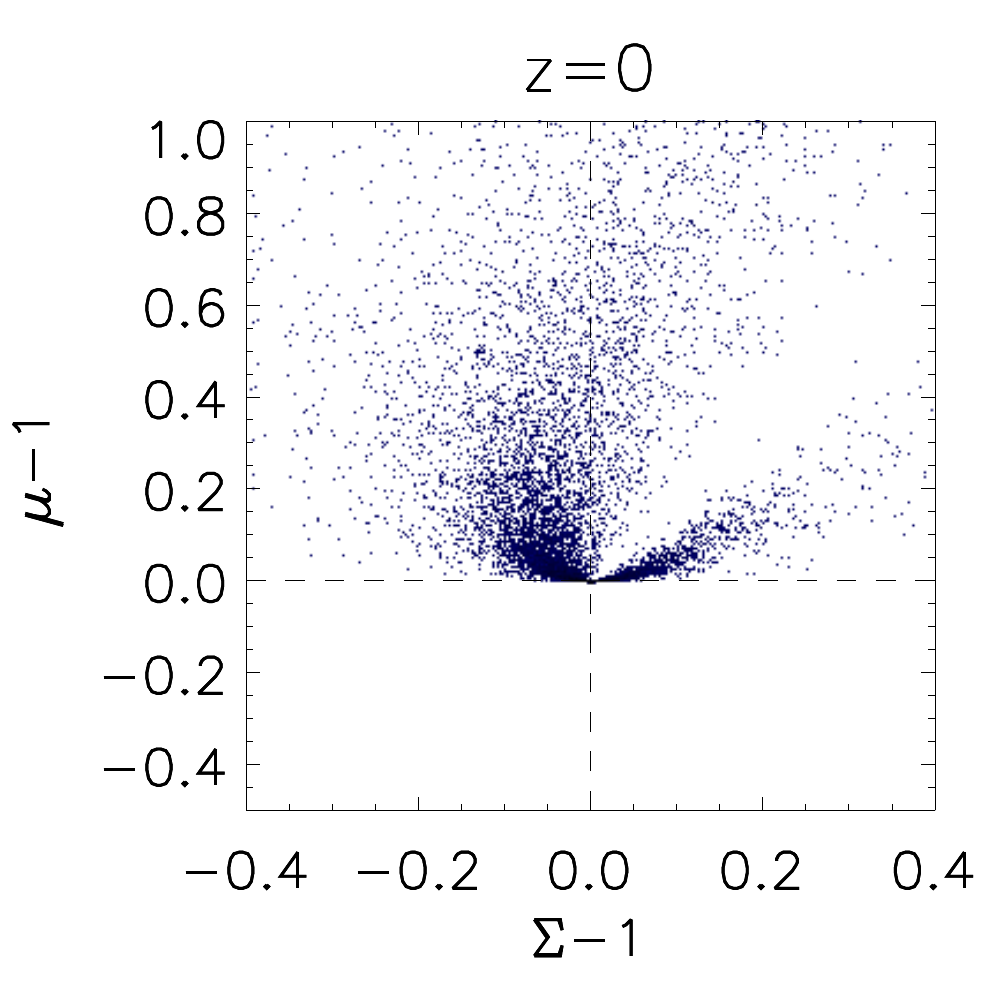}  \hskip-4mm
   \includegraphics[scale=0.27]{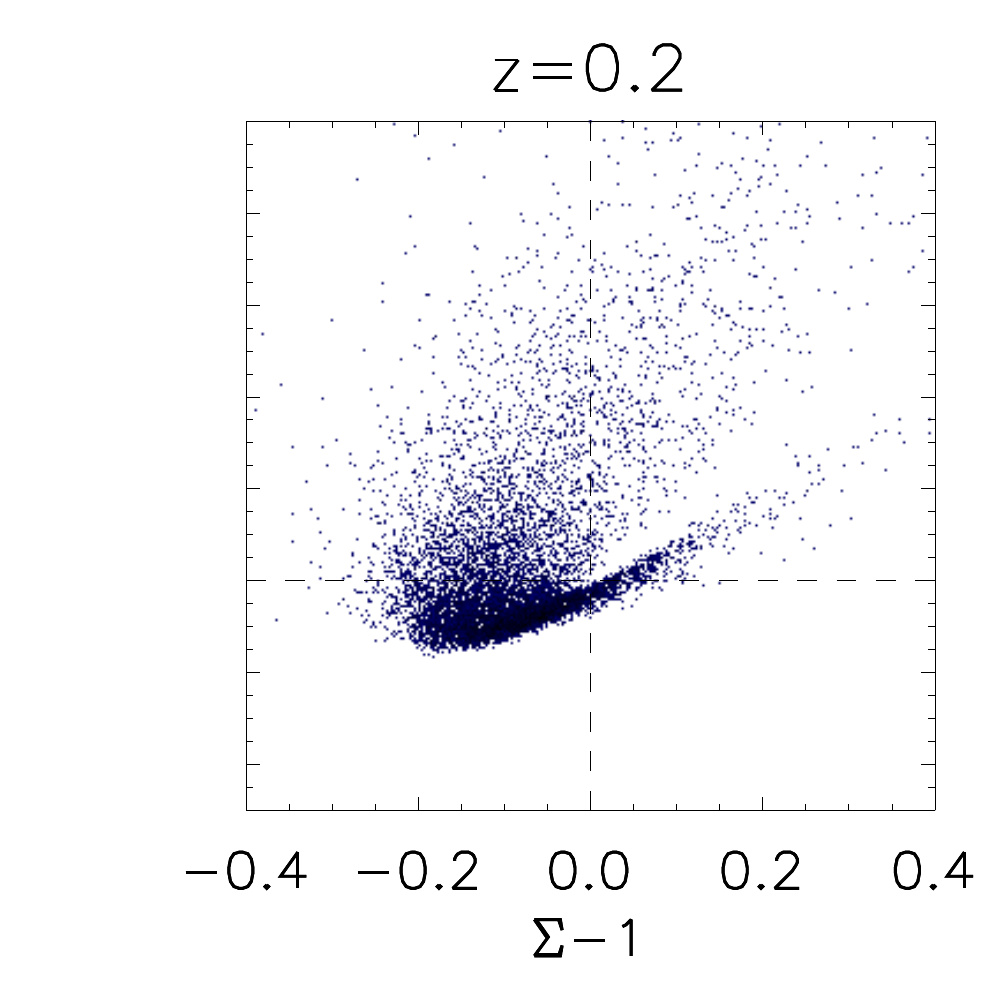}  \hskip-4mm
   \includegraphics[scale=0.27]{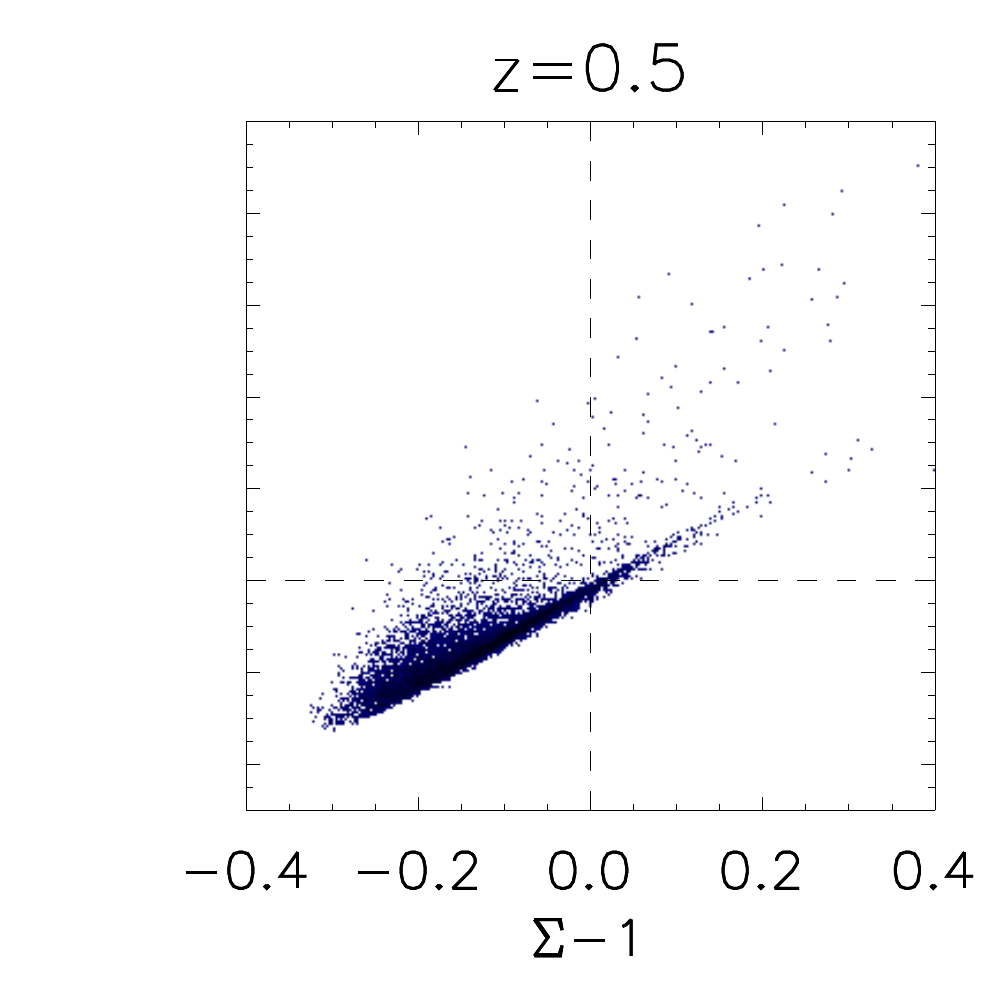}  \hskip-4mm 
   \includegraphics[scale=0.27]{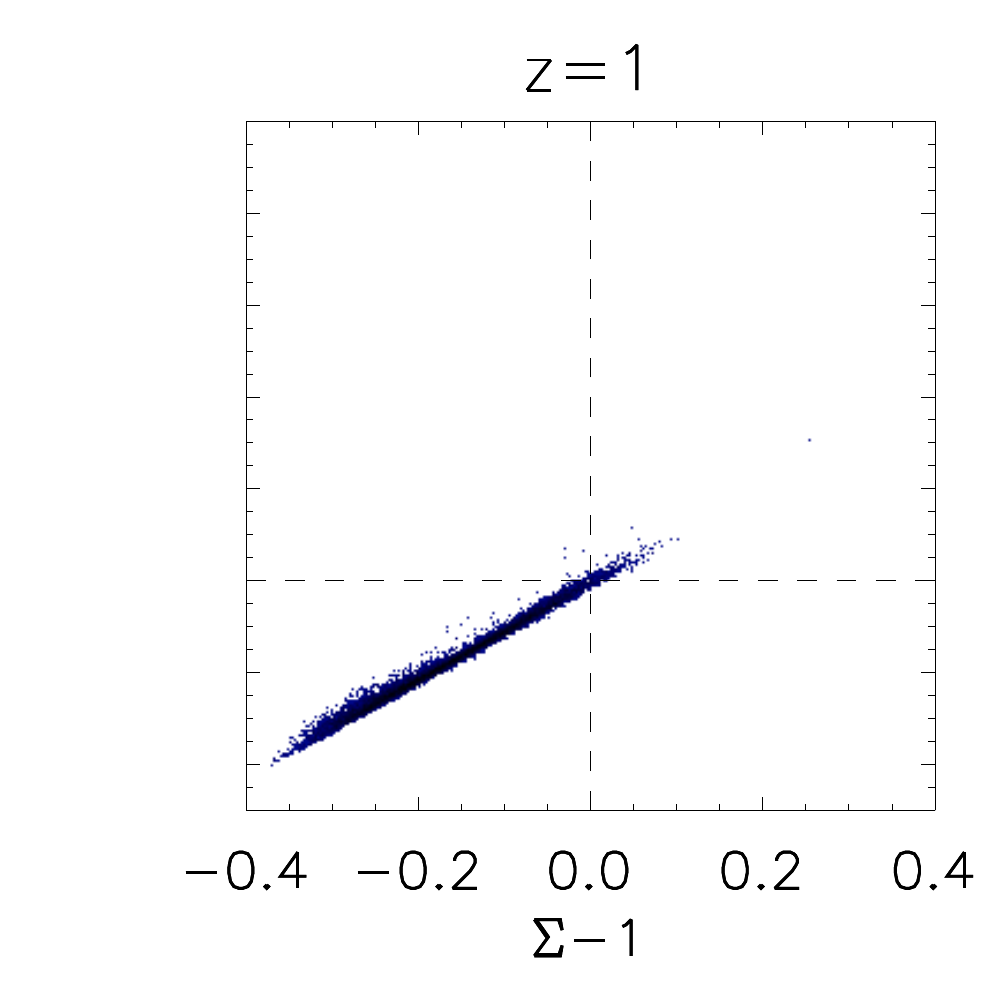}  \hskip-4mm
   \includegraphics[scale=0.27]{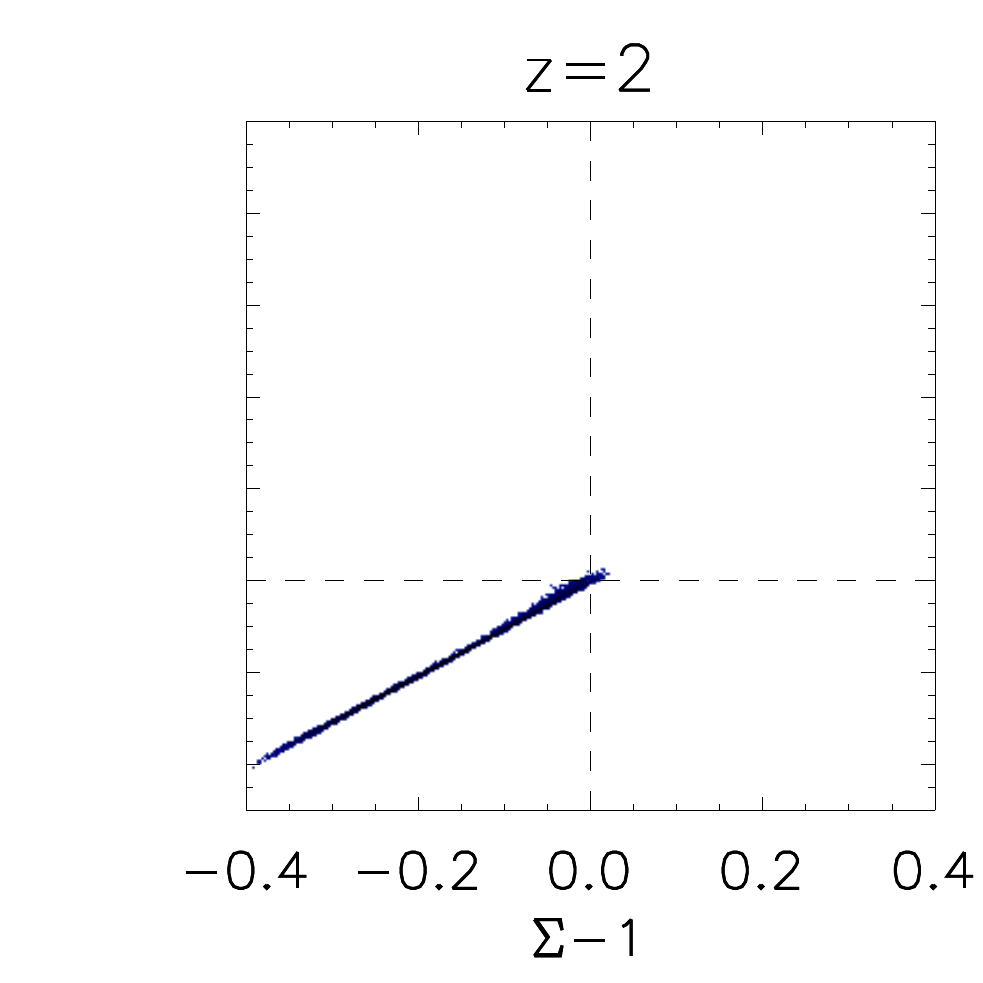}  \hskip-4mm
   \includegraphics[scale=0.27]{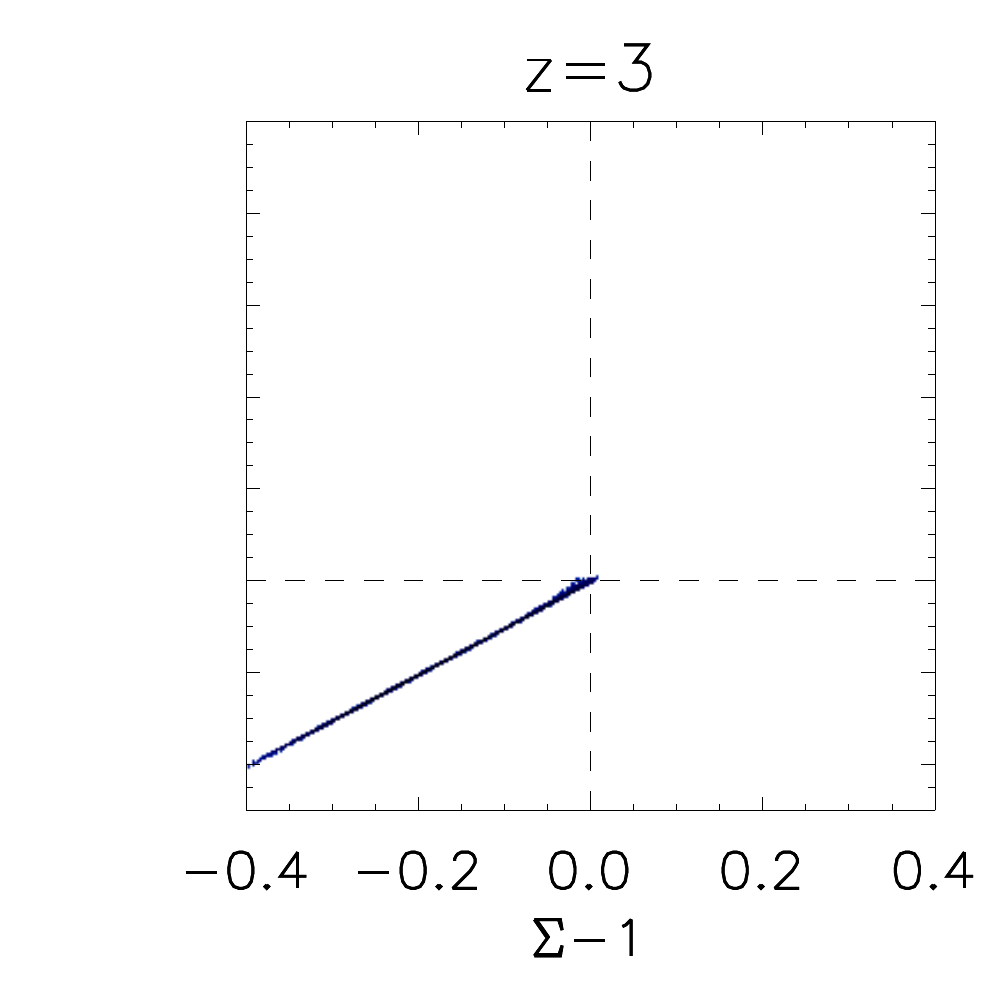}
   \vskip-2mm
   \includegraphics[scale=0.27]{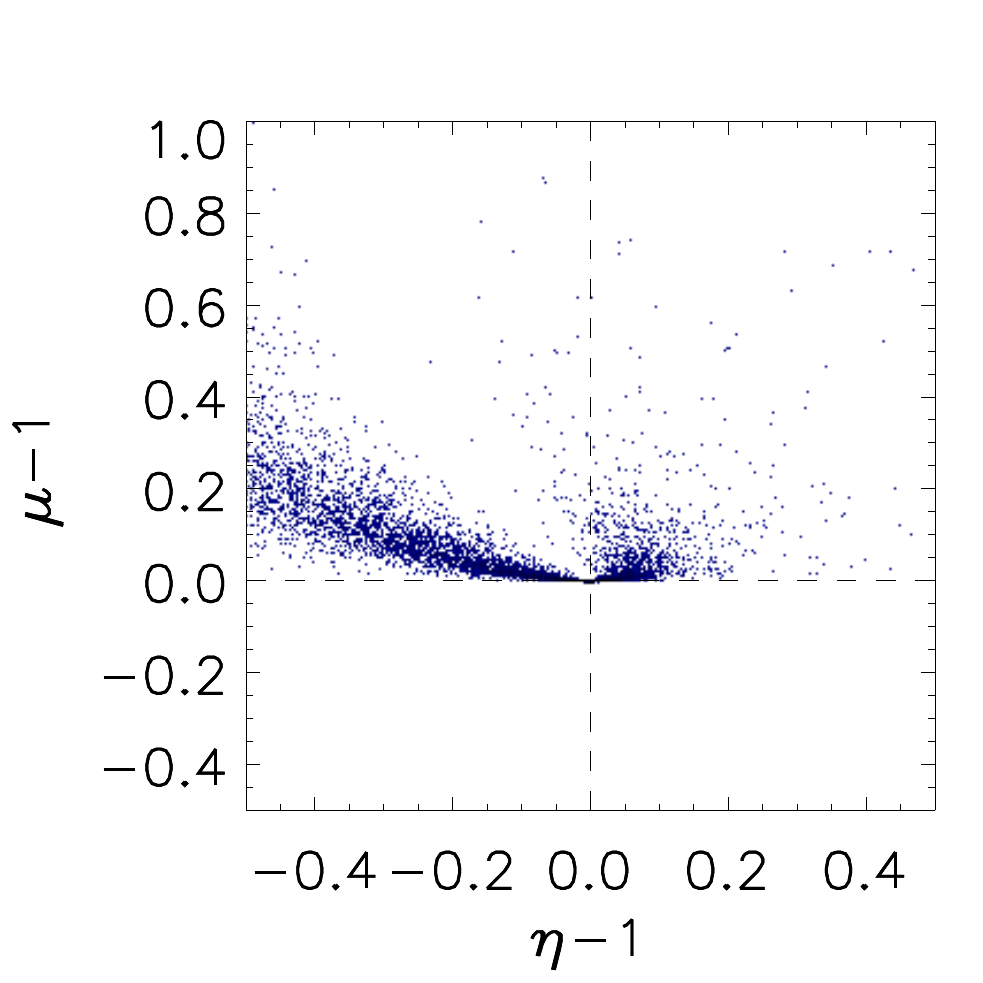}  \hskip-4mm
   \includegraphics[scale=0.27]{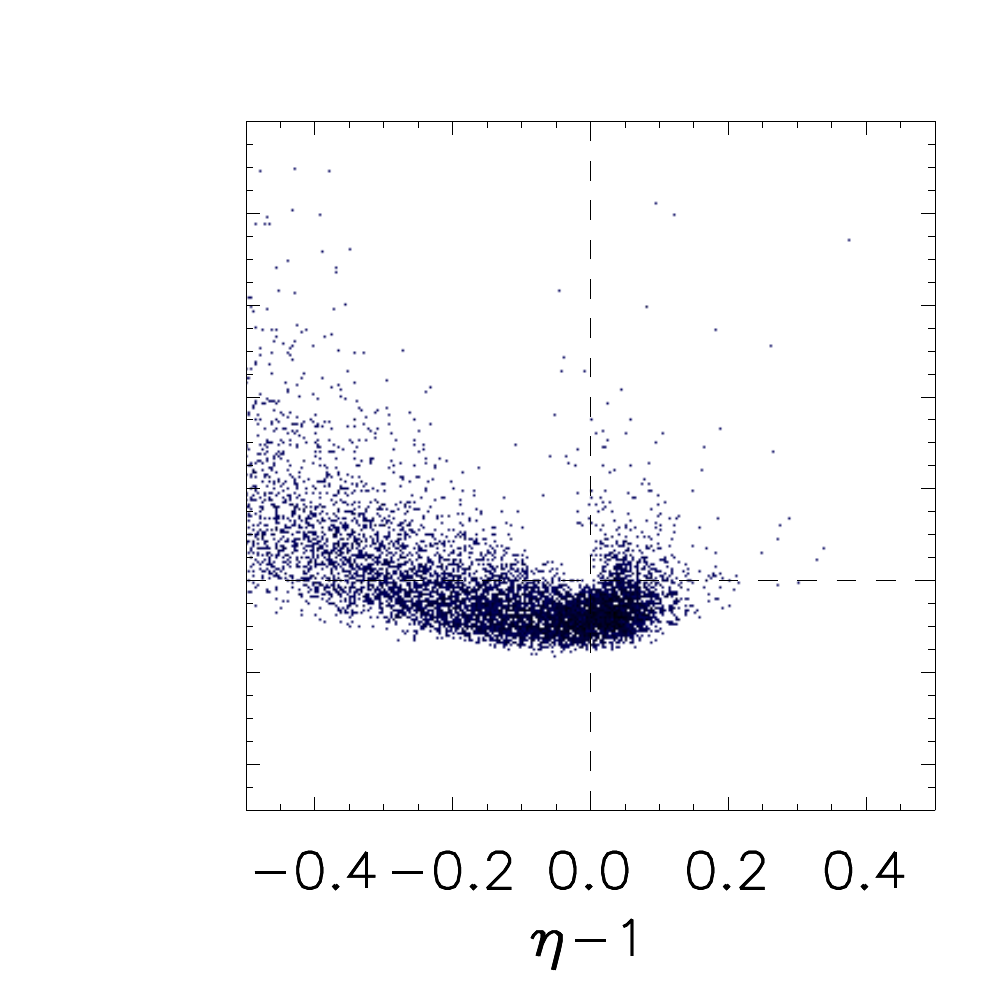}  \hskip-4mm
   \includegraphics[scale=0.27]{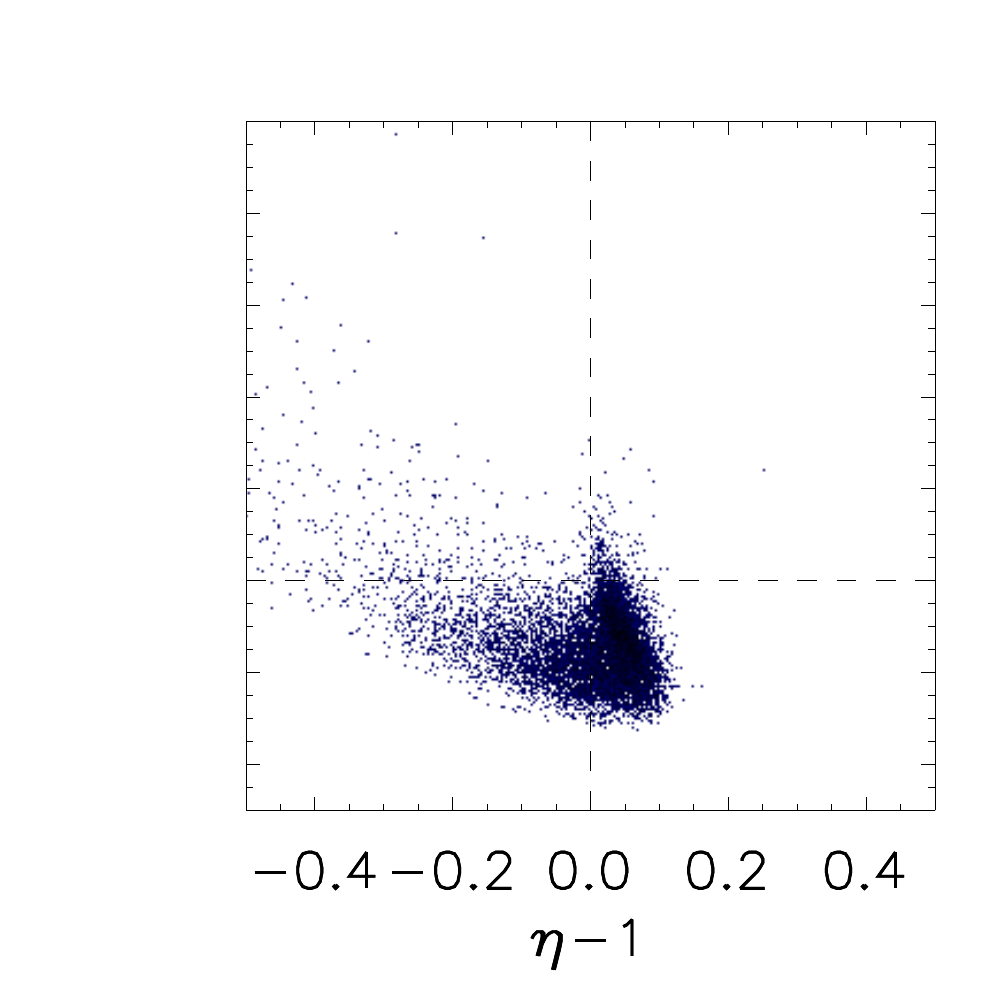}  \hskip-4mm 
   \includegraphics[scale=0.27]{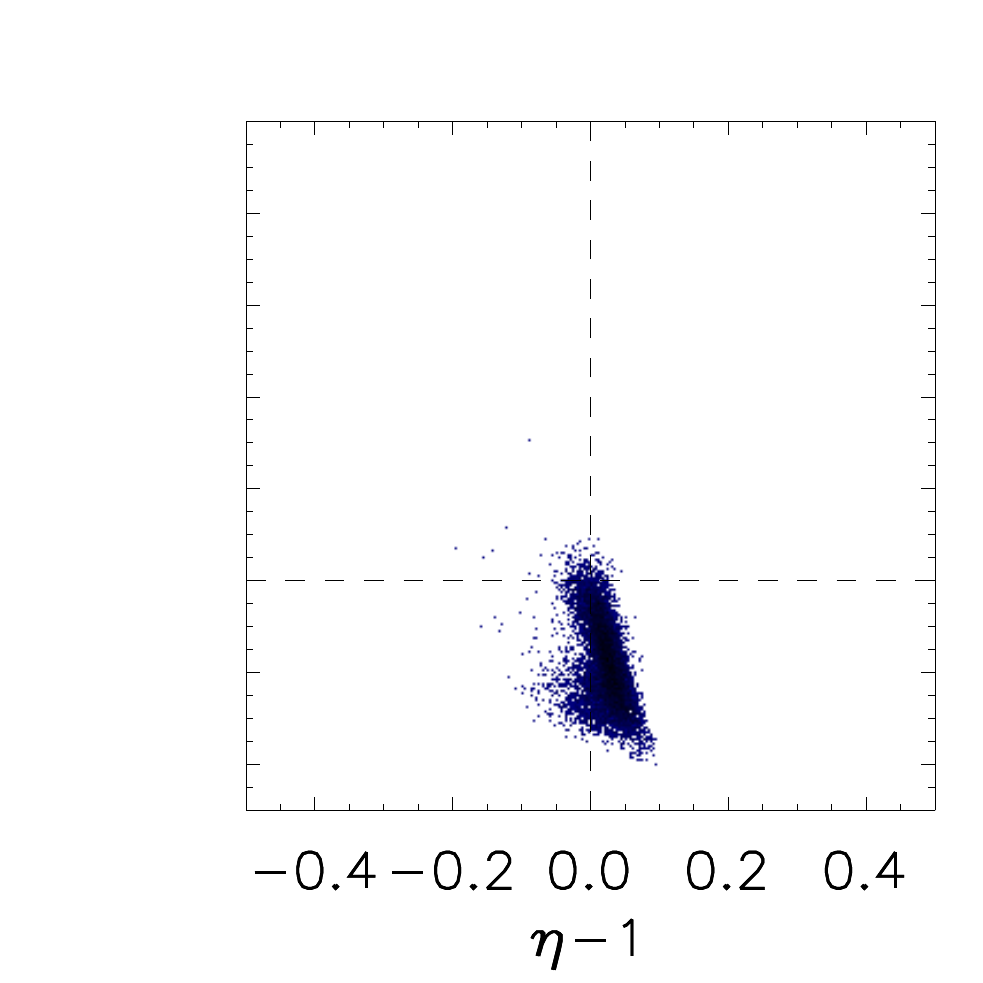}  \hskip-4mm
   \includegraphics[scale=0.27]{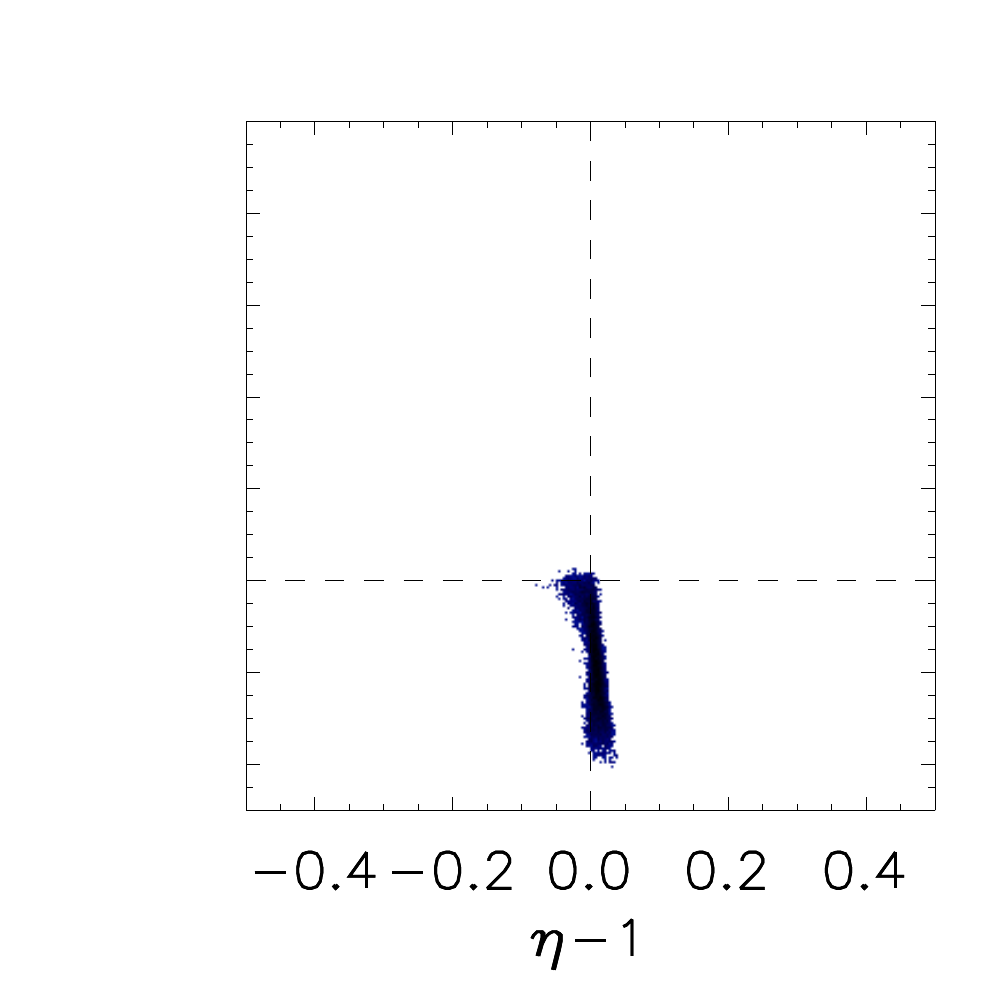}  \hskip-4mm
   \includegraphics[scale=0.27]{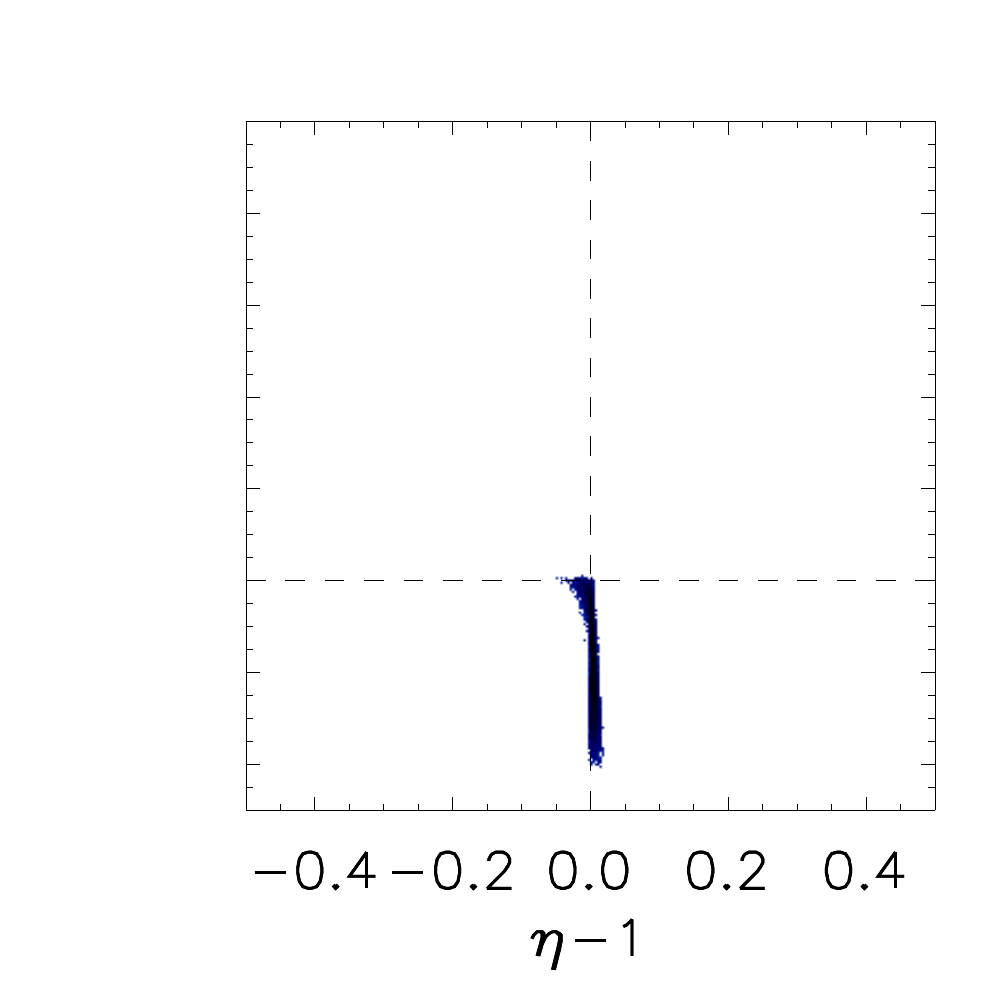}
   \vskip-2mm
   \includegraphics[scale=0.27]{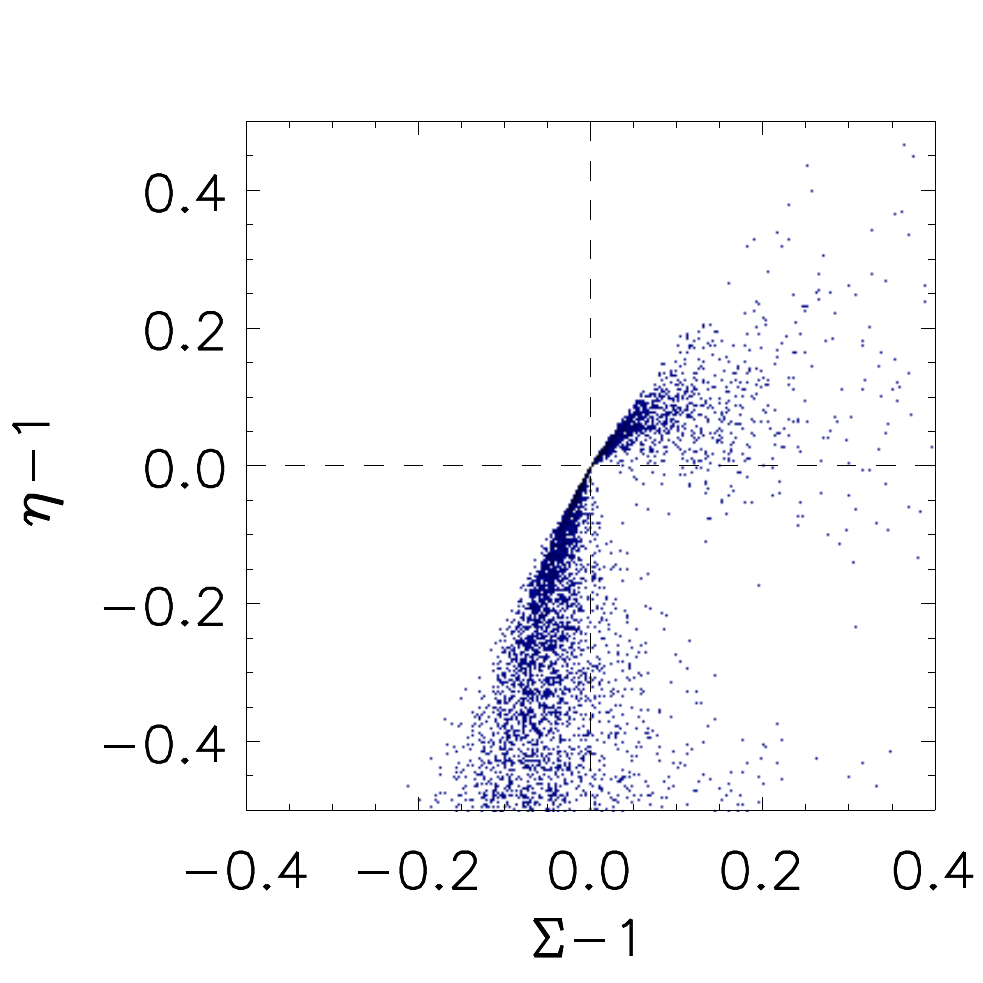}  \hskip-4mm
   \includegraphics[scale=0.27]{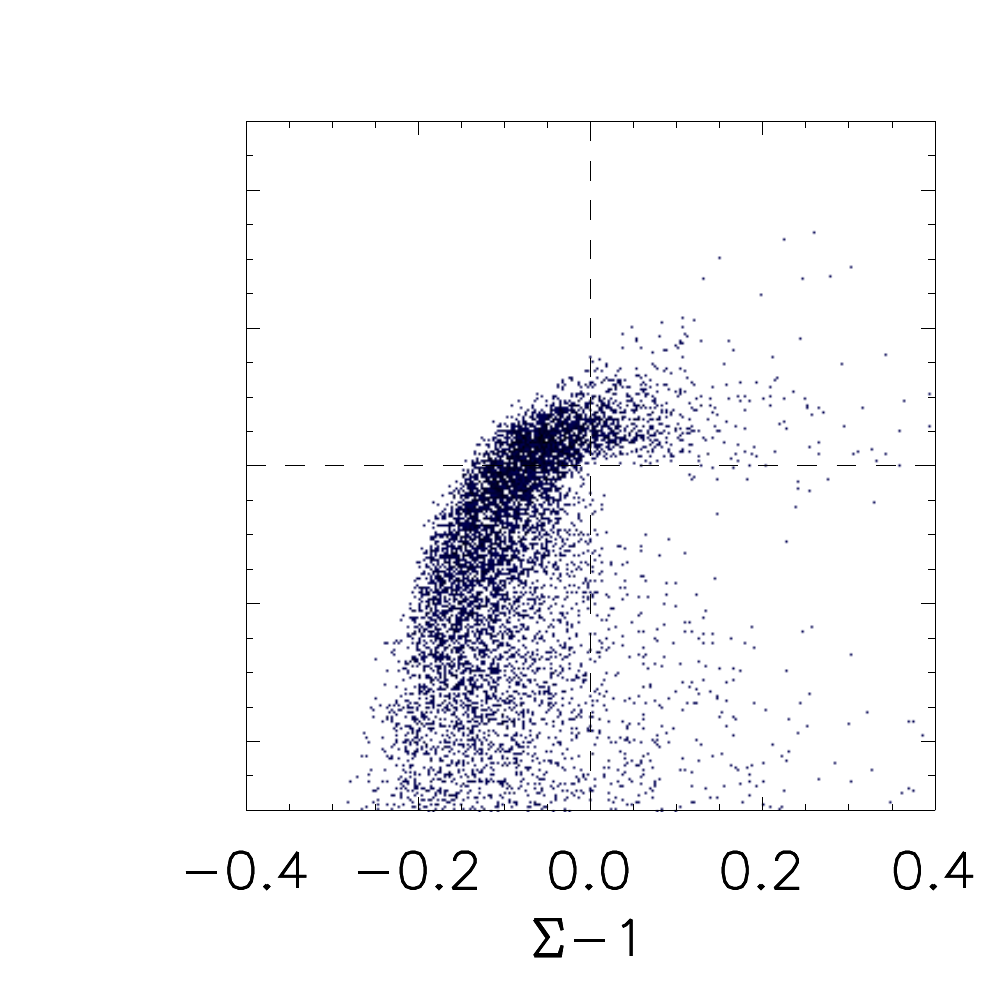}  \hskip-4mm
   \includegraphics[scale=0.27]{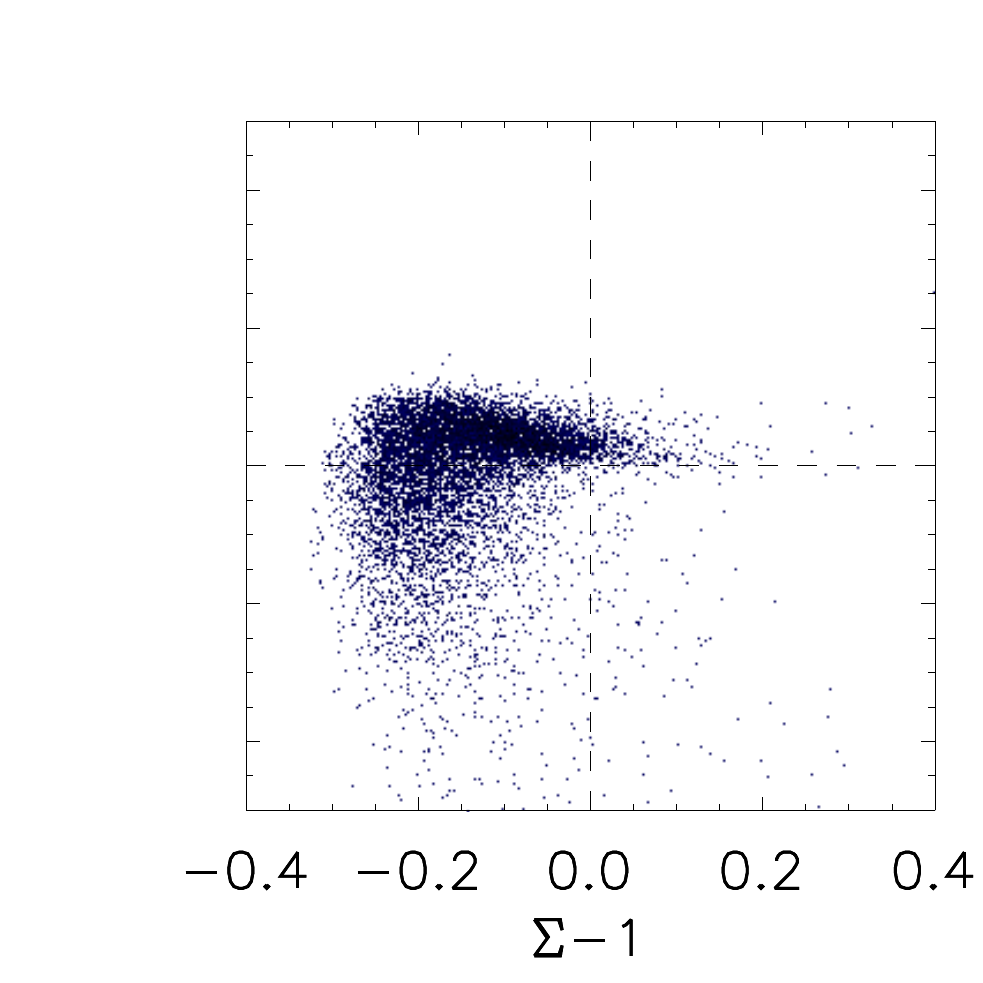}   \hskip-4mm
   \includegraphics[scale=0.27]{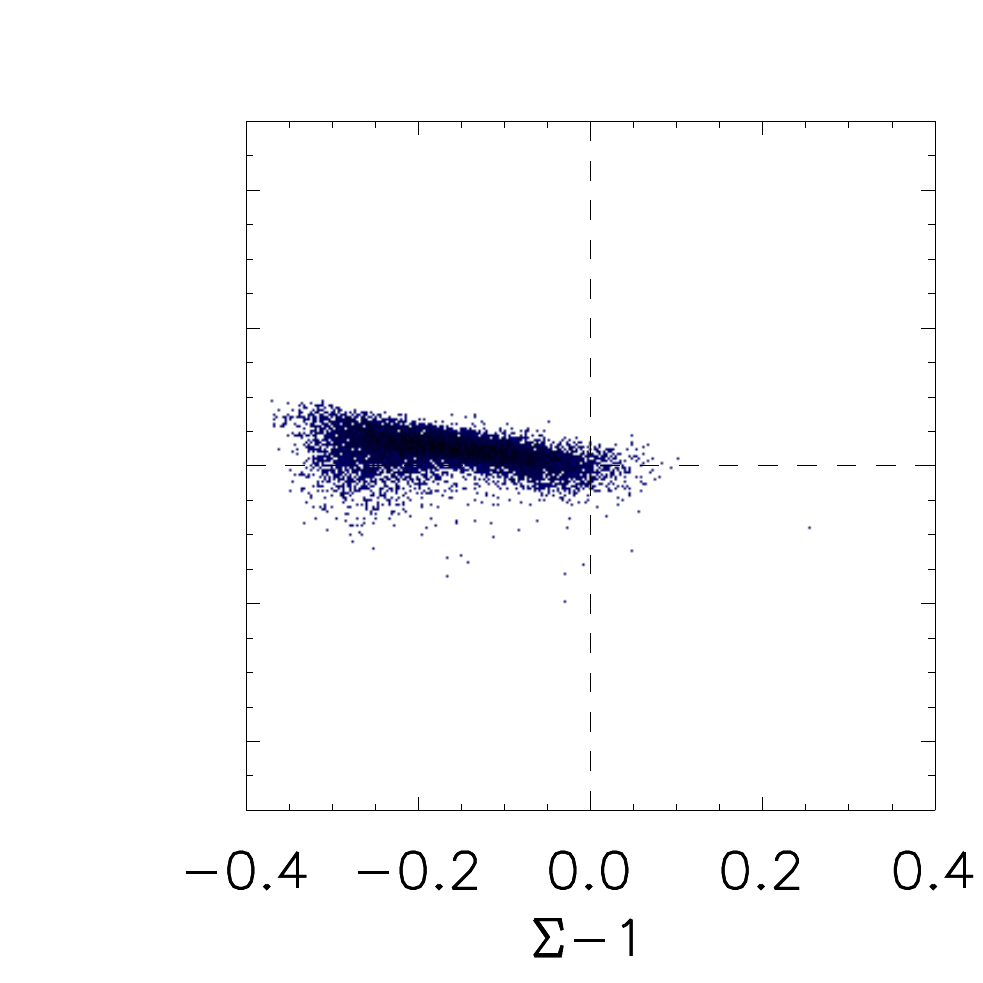}  \hskip-4mm
   \includegraphics[scale=0.27]{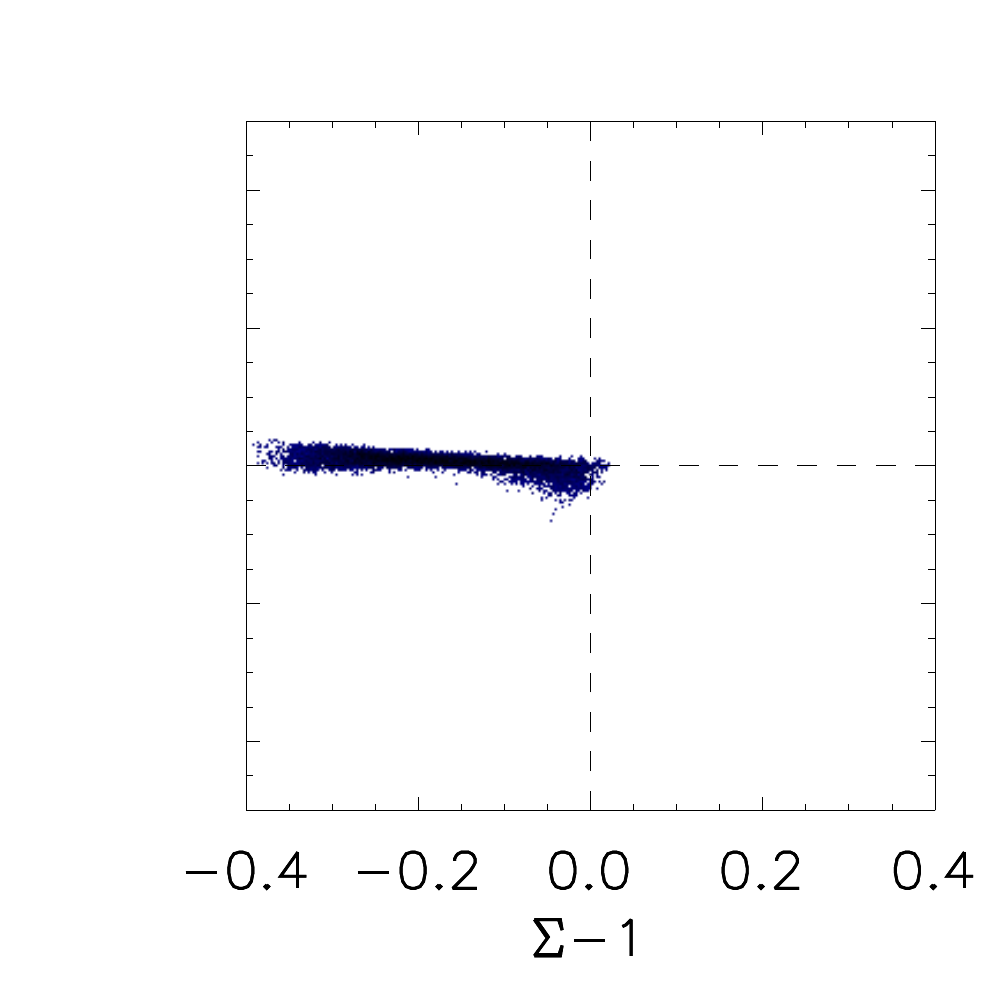}  \hskip-4mm
   \includegraphics[scale=0.27]{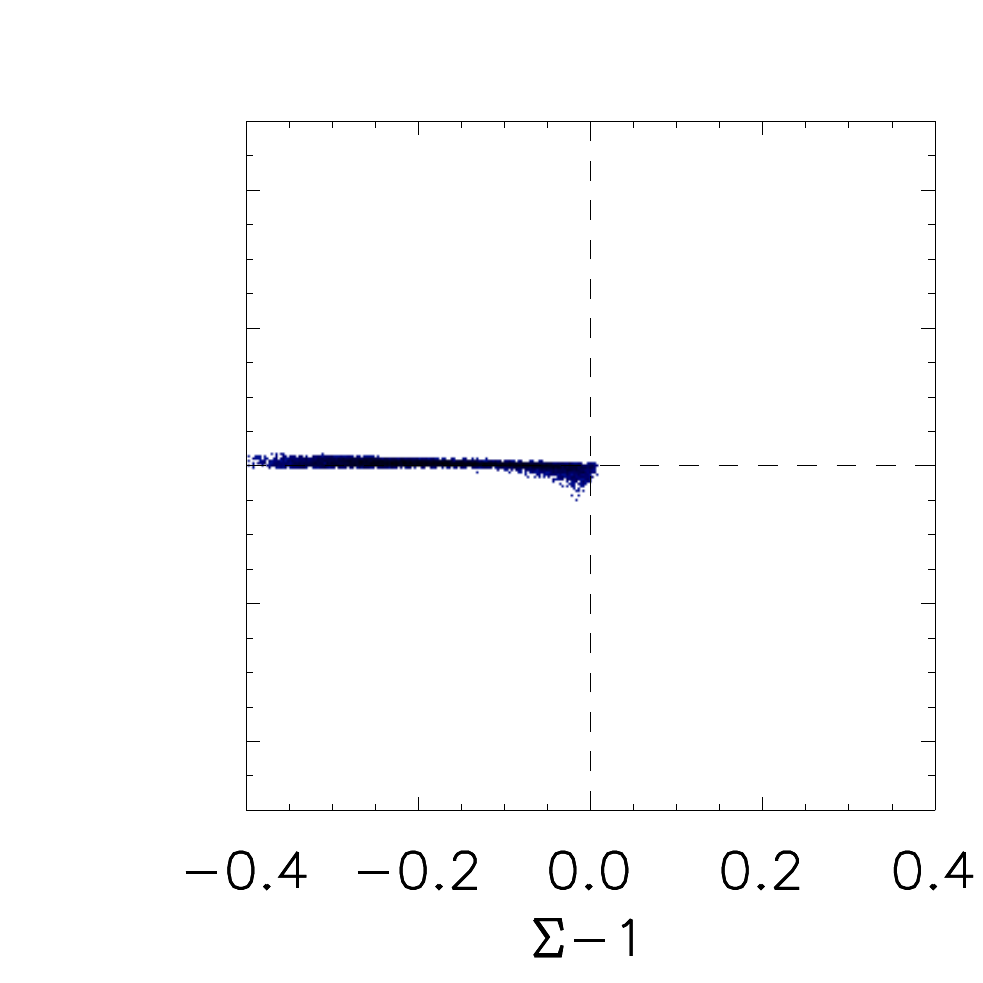}
   \vskip-2mm
   \includegraphics[scale=0.27]{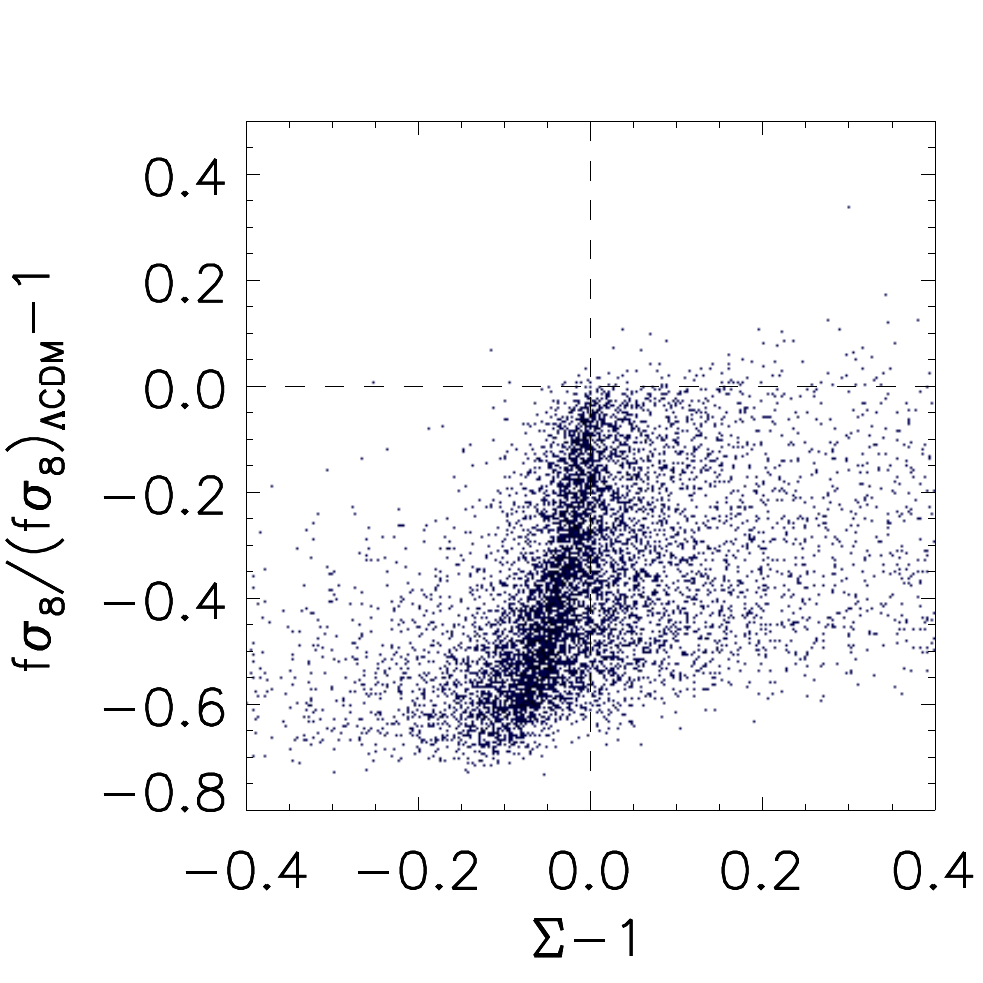}  \hskip-4mm
   \includegraphics[scale=0.27]{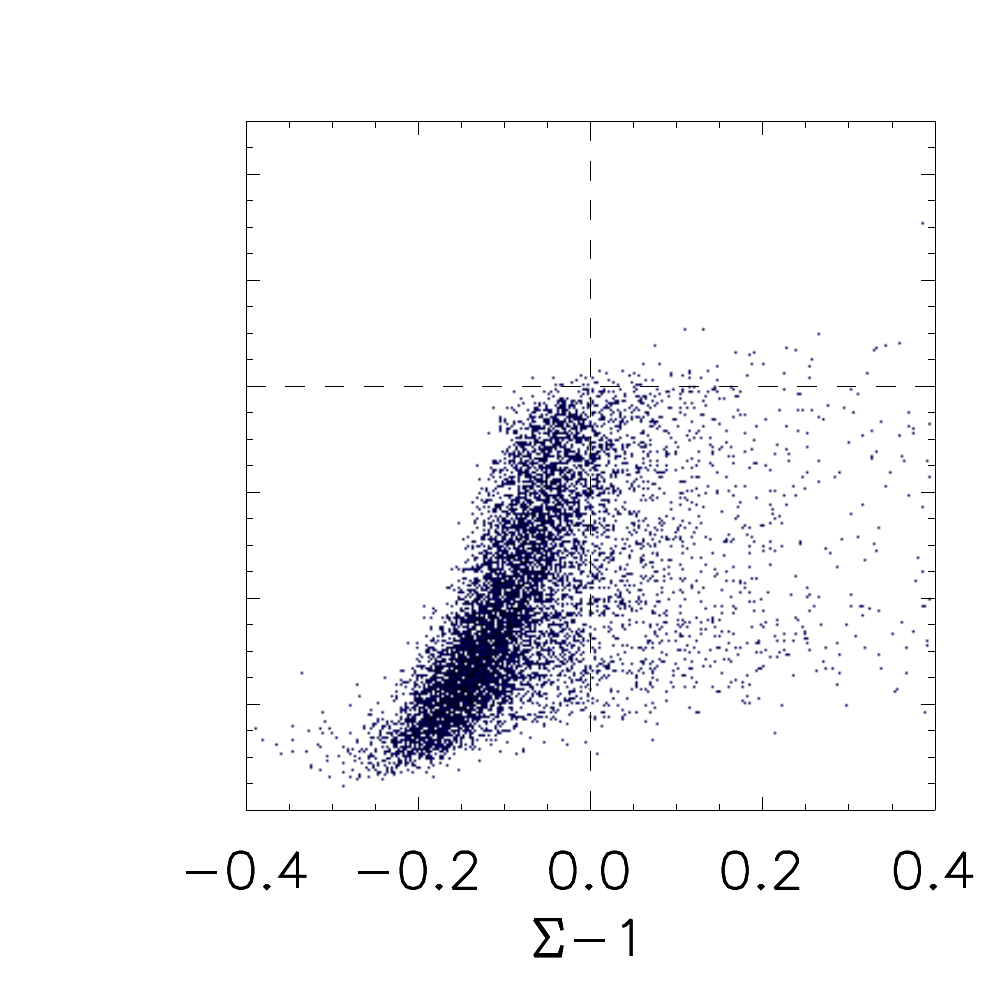}  \hskip-4mm
   \includegraphics[scale=0.27]{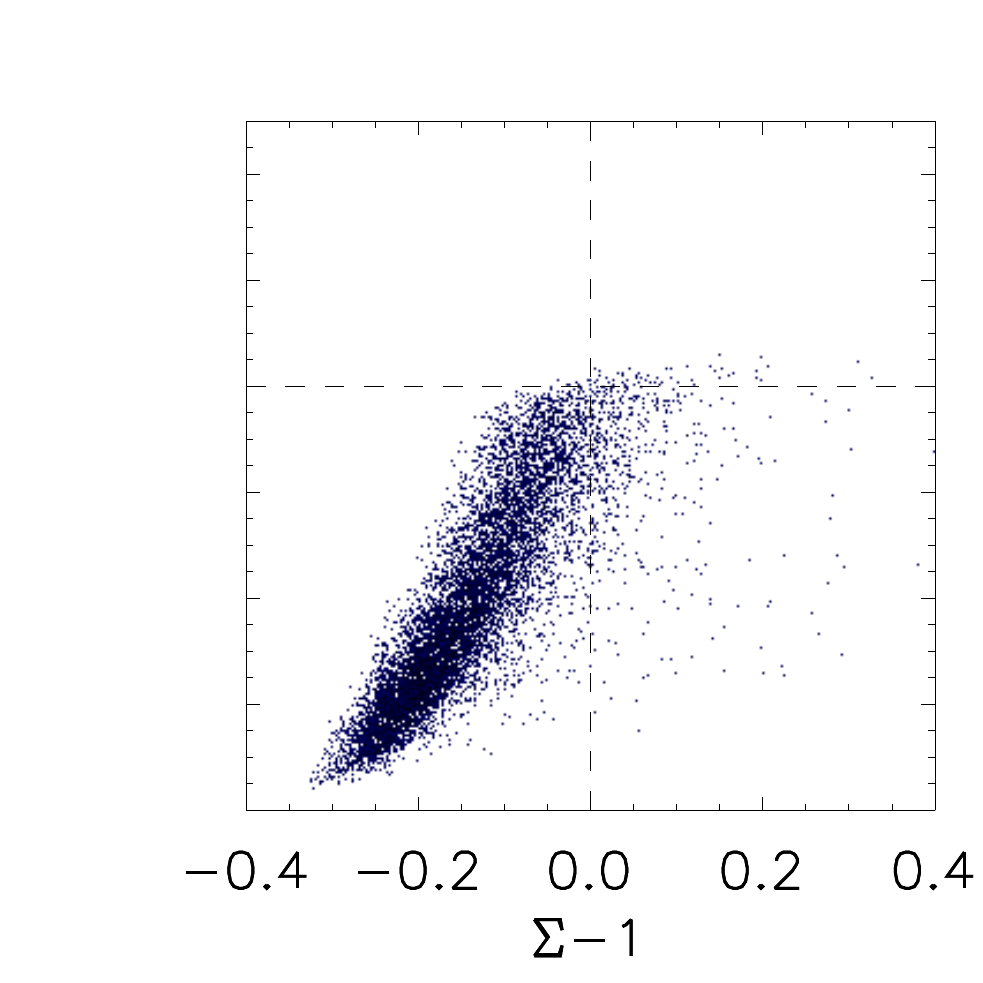}   \hskip-4mm
   \includegraphics[scale=0.27]{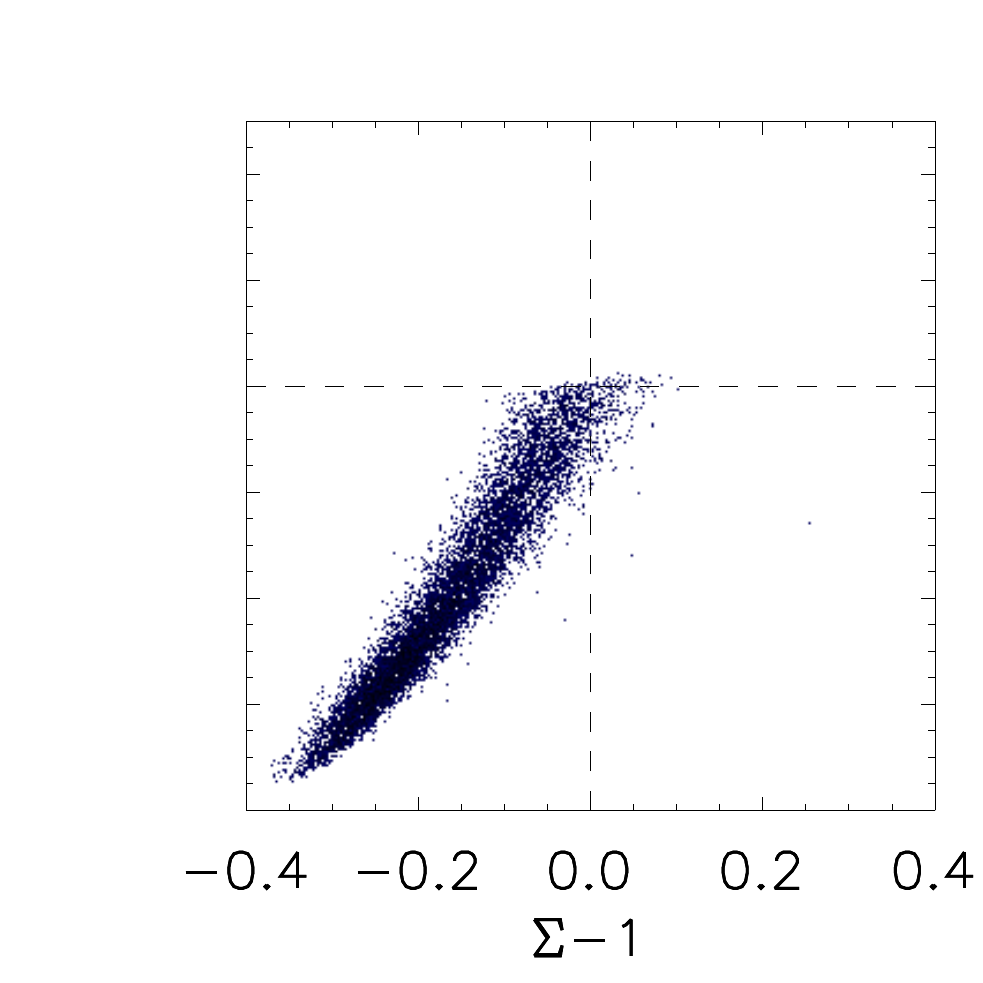}  \hskip-4mm
   \includegraphics[scale=0.27]{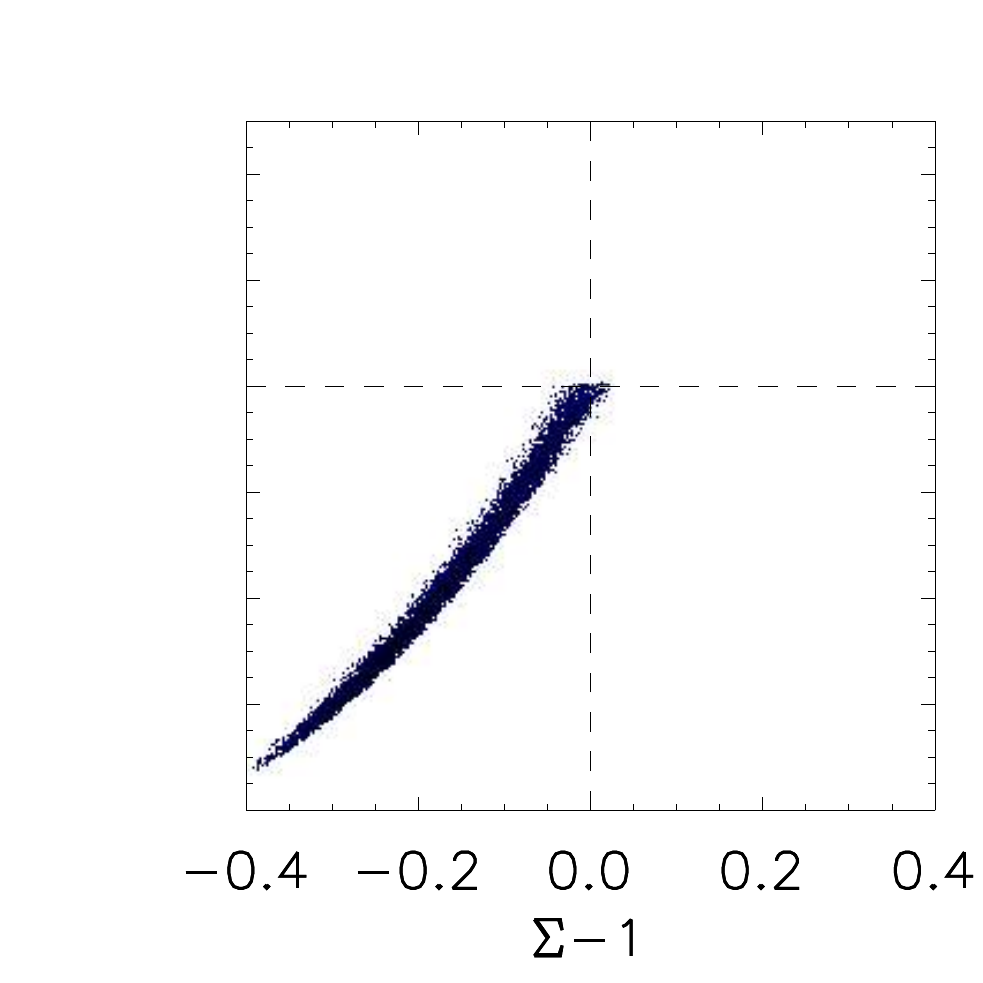}  \hskip-4mm
   \includegraphics[scale=0.27]{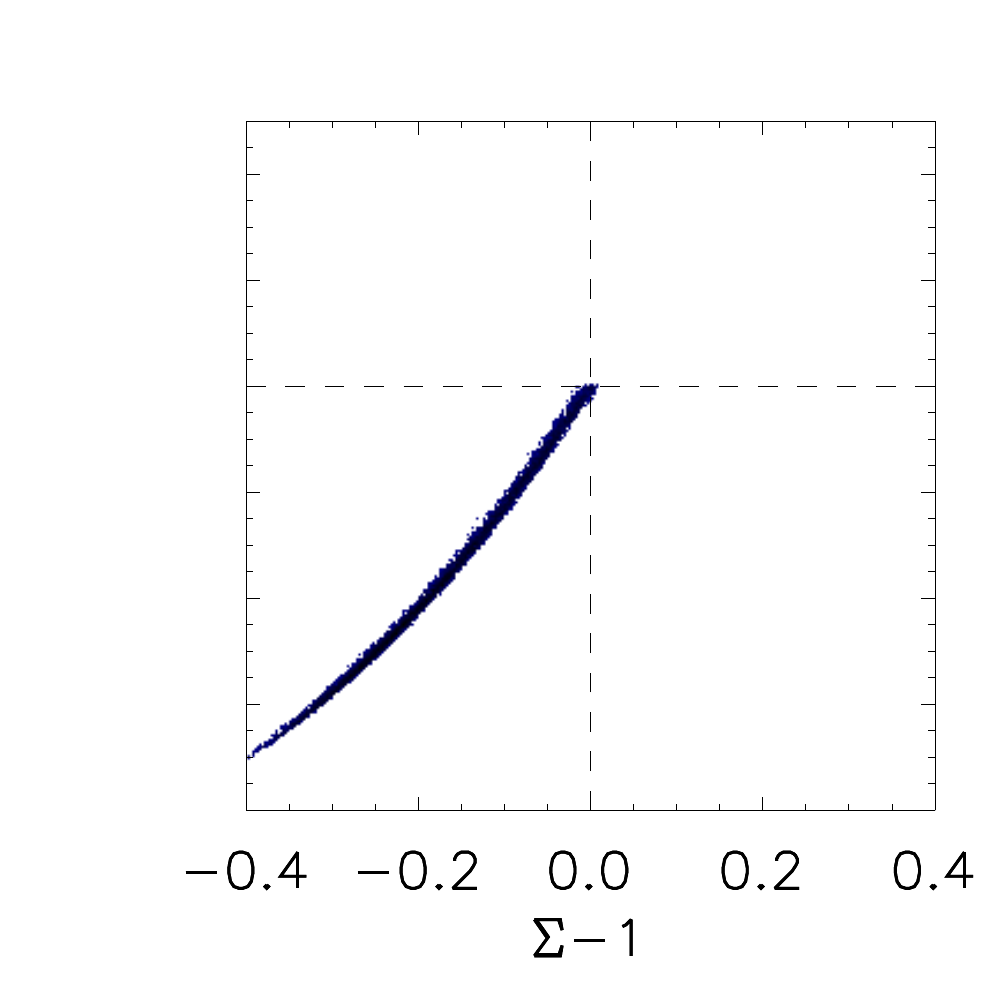}
\vskip1mm
- - - - - - - - - - - - - - - - - - - - - - - - - - - - - - - - - - - - - - - - - - - - - - - - - - - - - - - -	
\begin{flushleft} EMG : \end{flushleft}   \vskip-3mm
   \includegraphics[scale=0.27]{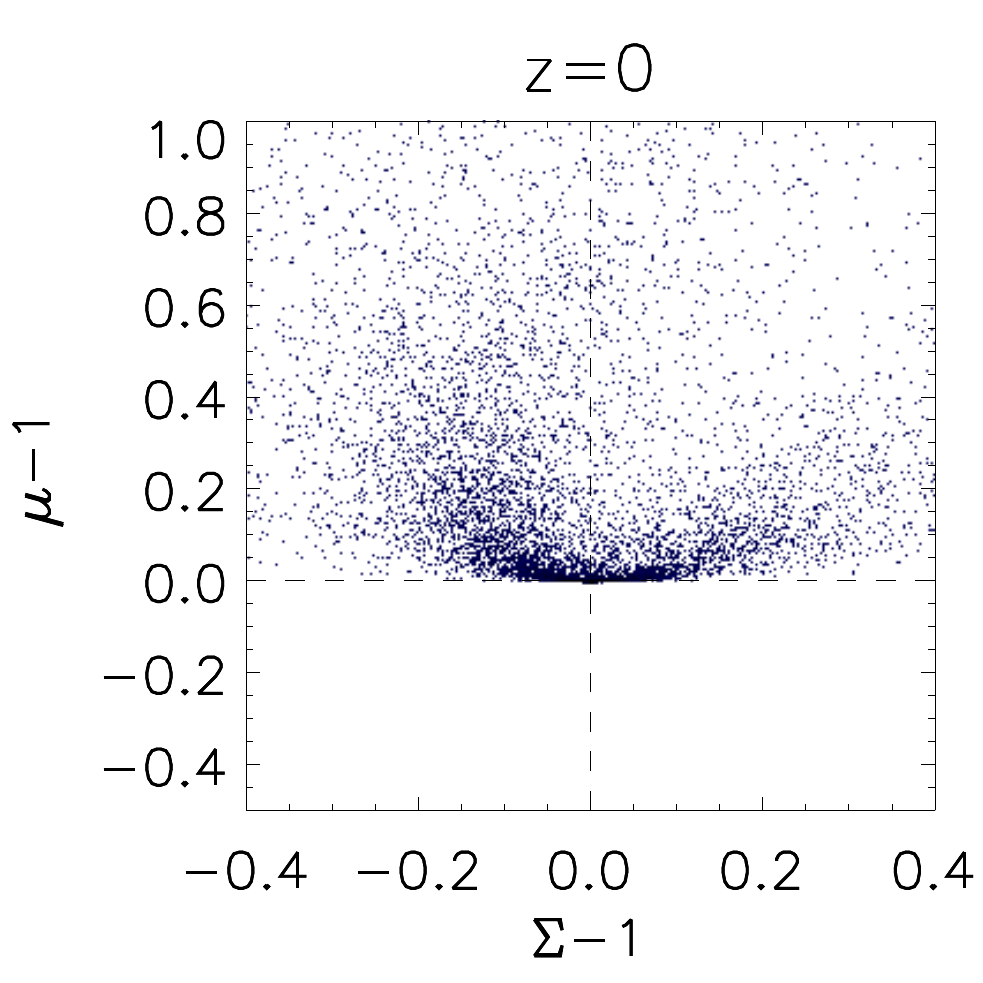}  \hskip-4mm
   \includegraphics[scale=0.27]{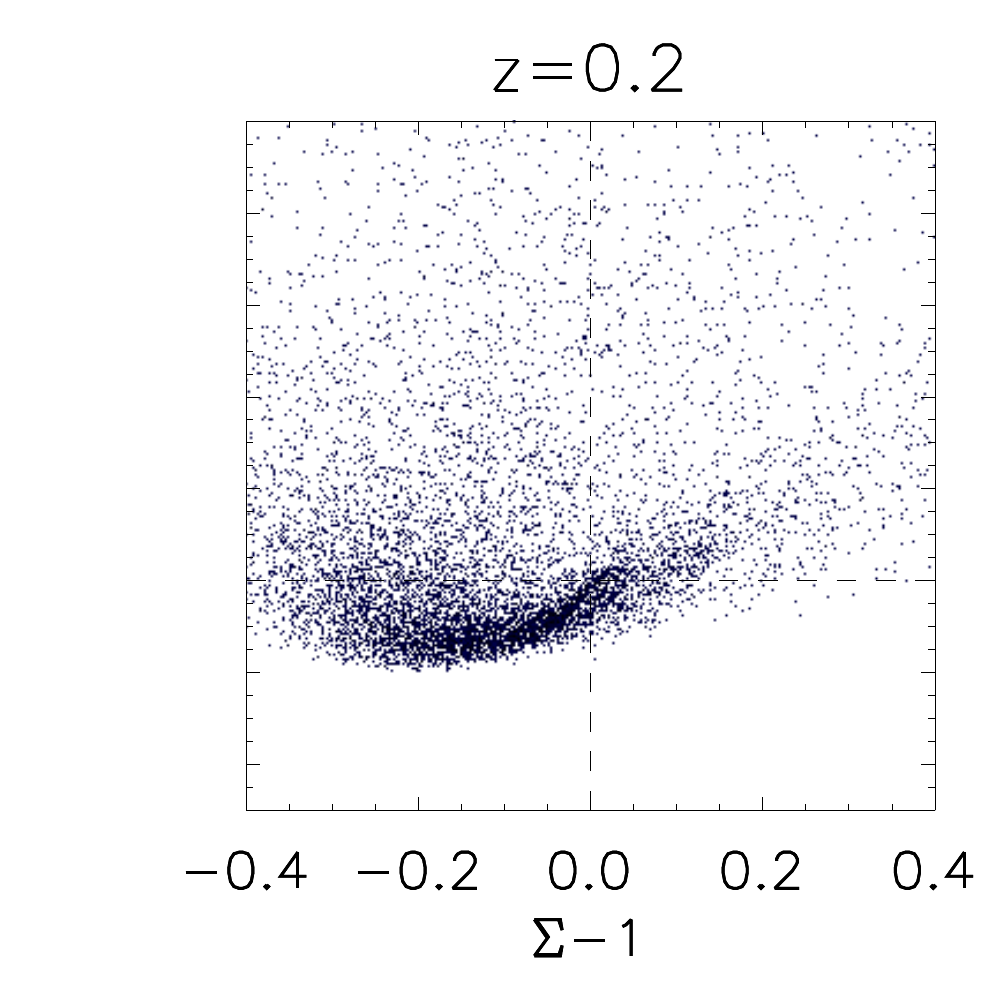}  \hskip-4mm
   \includegraphics[scale=0.27]{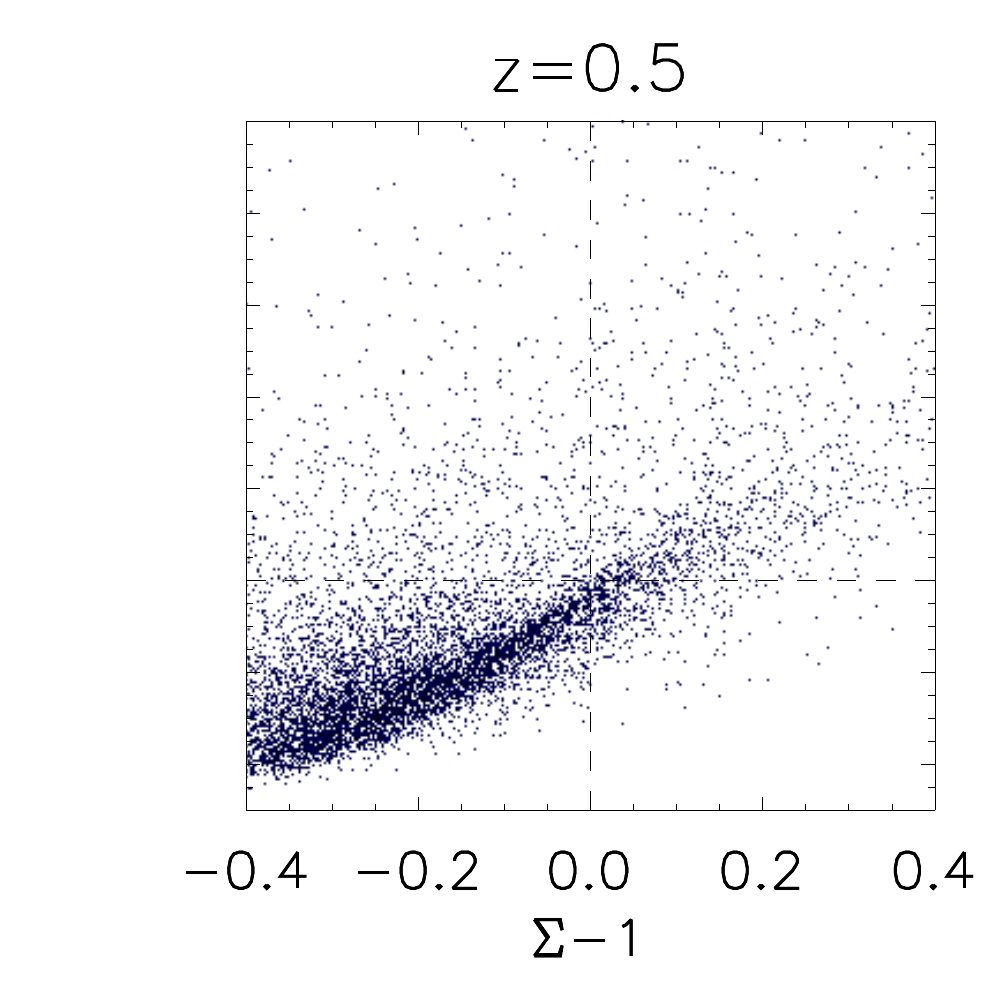}   \hskip-4mm
   \includegraphics[scale=0.27]{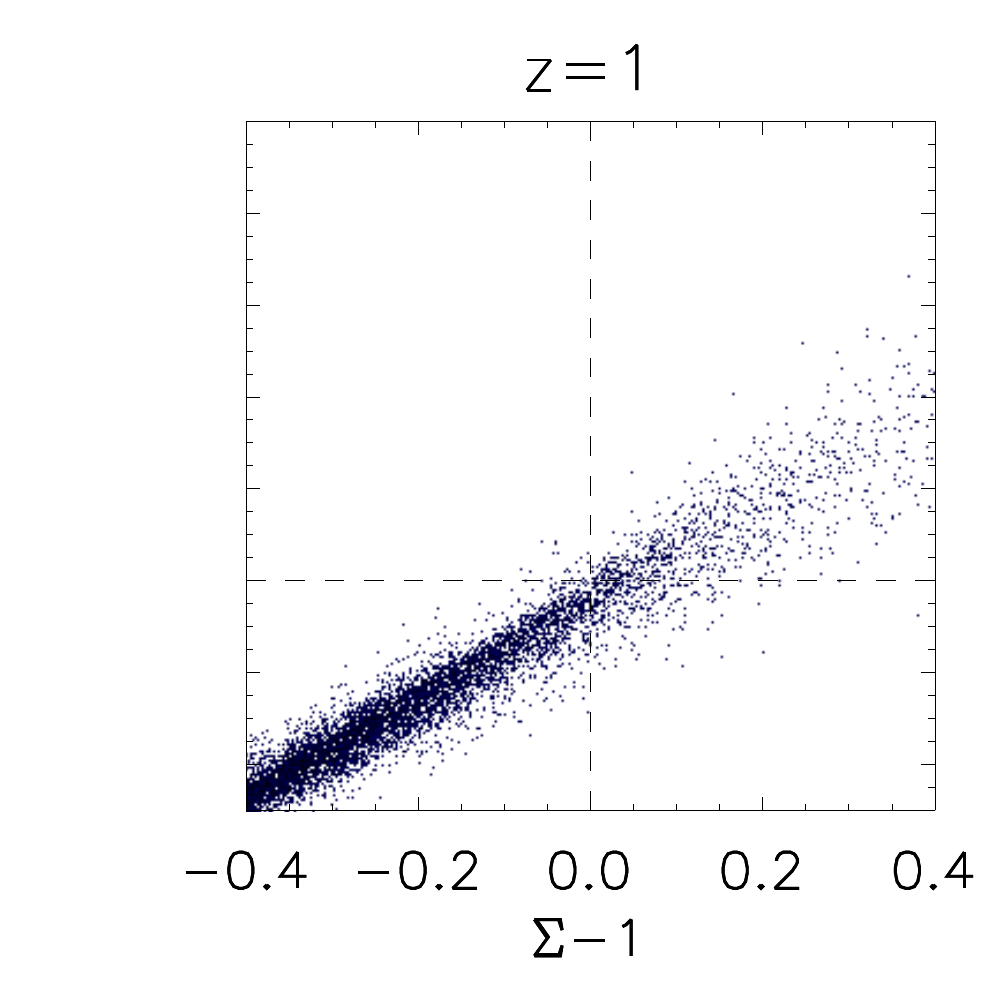}  \hskip-4mm
   \includegraphics[scale=0.27]{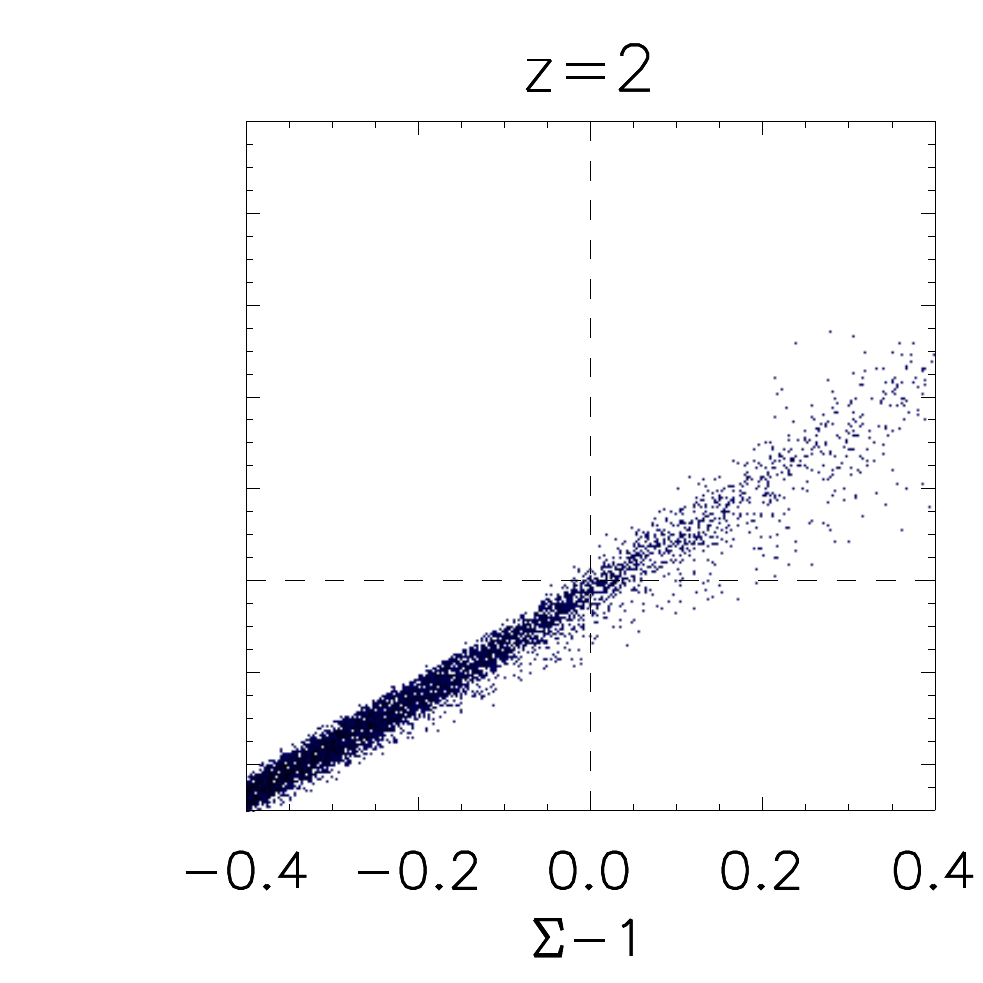}  \hskip-4mm
   \includegraphics[scale=0.27]{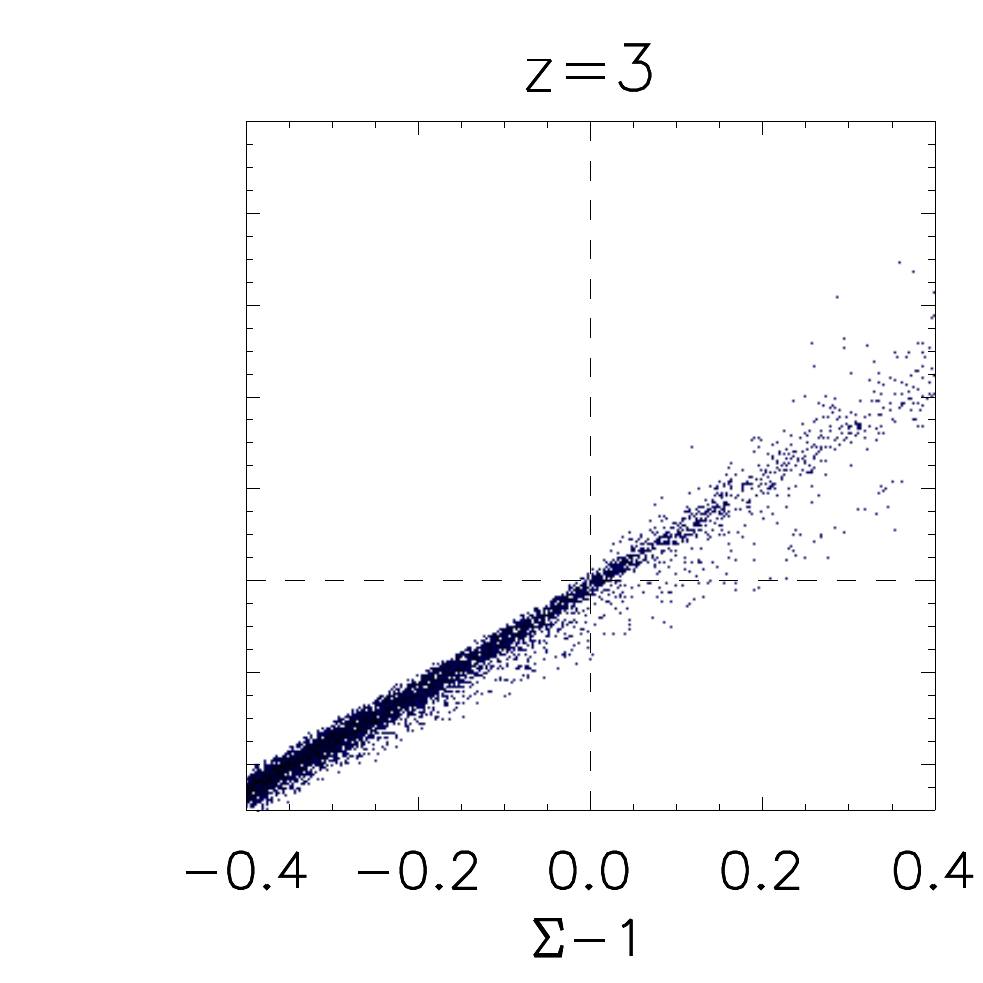}
   \vskip-2mm
   \includegraphics[scale=0.27]{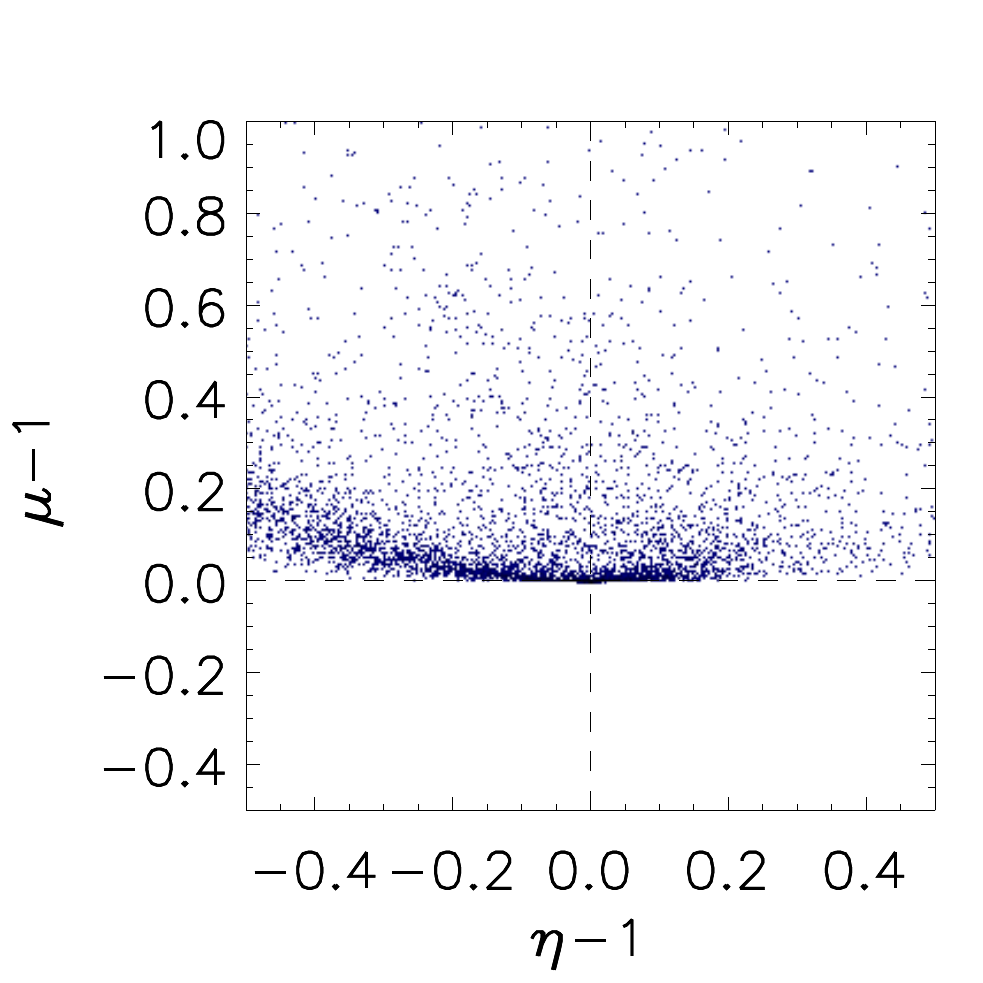}  \hskip-4mm
   \includegraphics[scale=0.27]{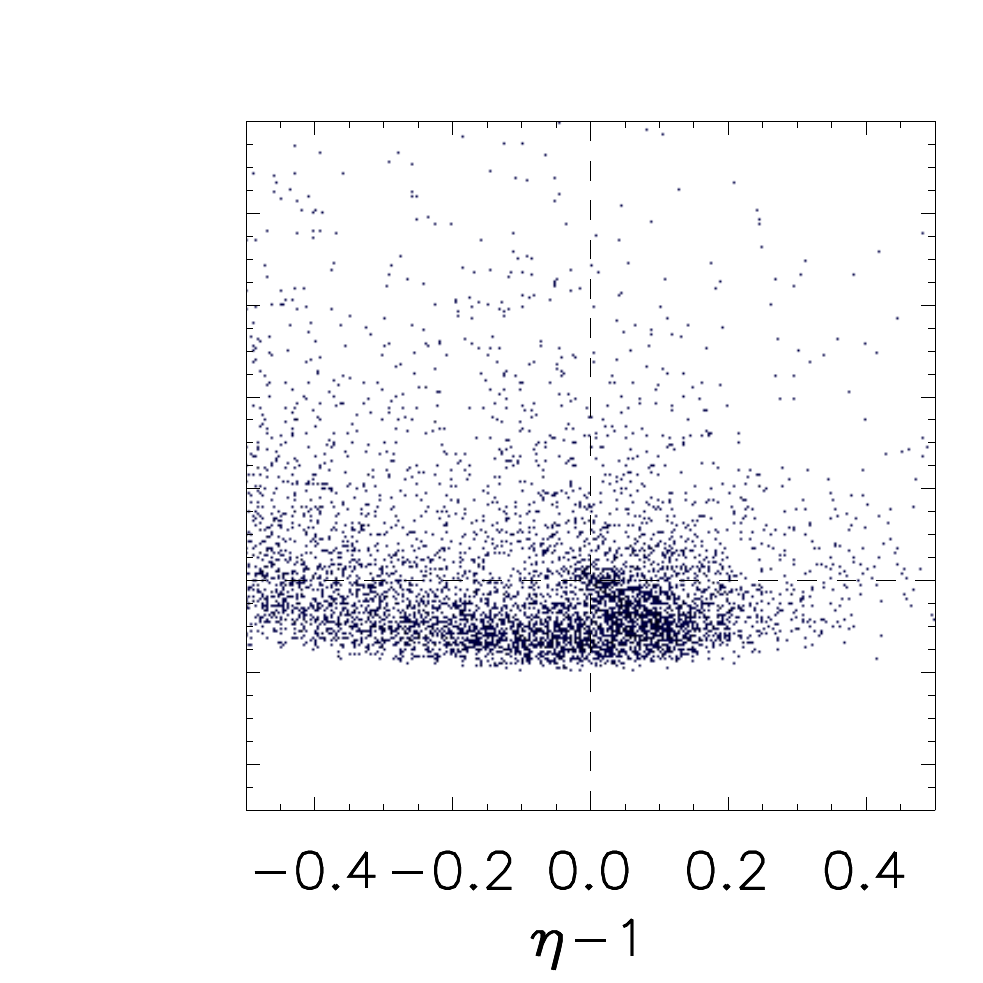}  \hskip-4mm
   \includegraphics[scale=0.27]{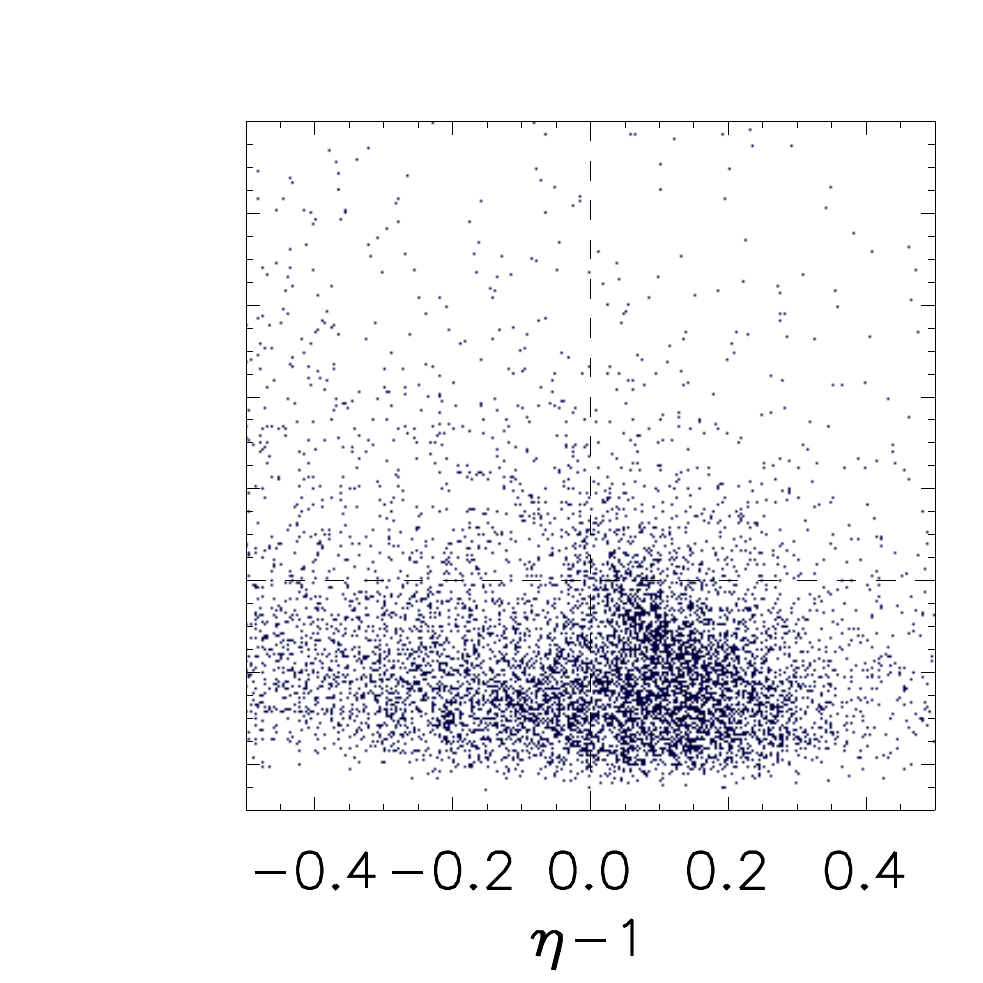}  \hskip-4mm 
   \includegraphics[scale=0.27]{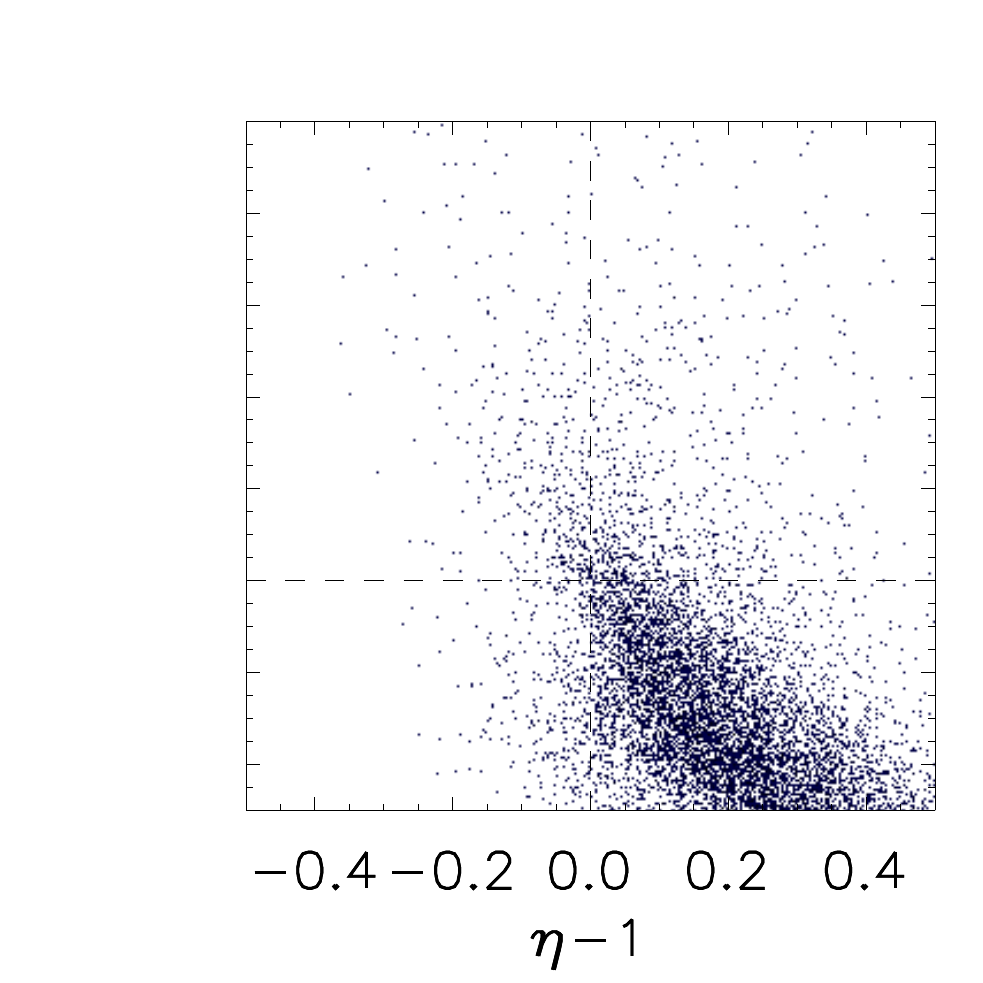}  \hskip-4mm
   \includegraphics[scale=0.27]{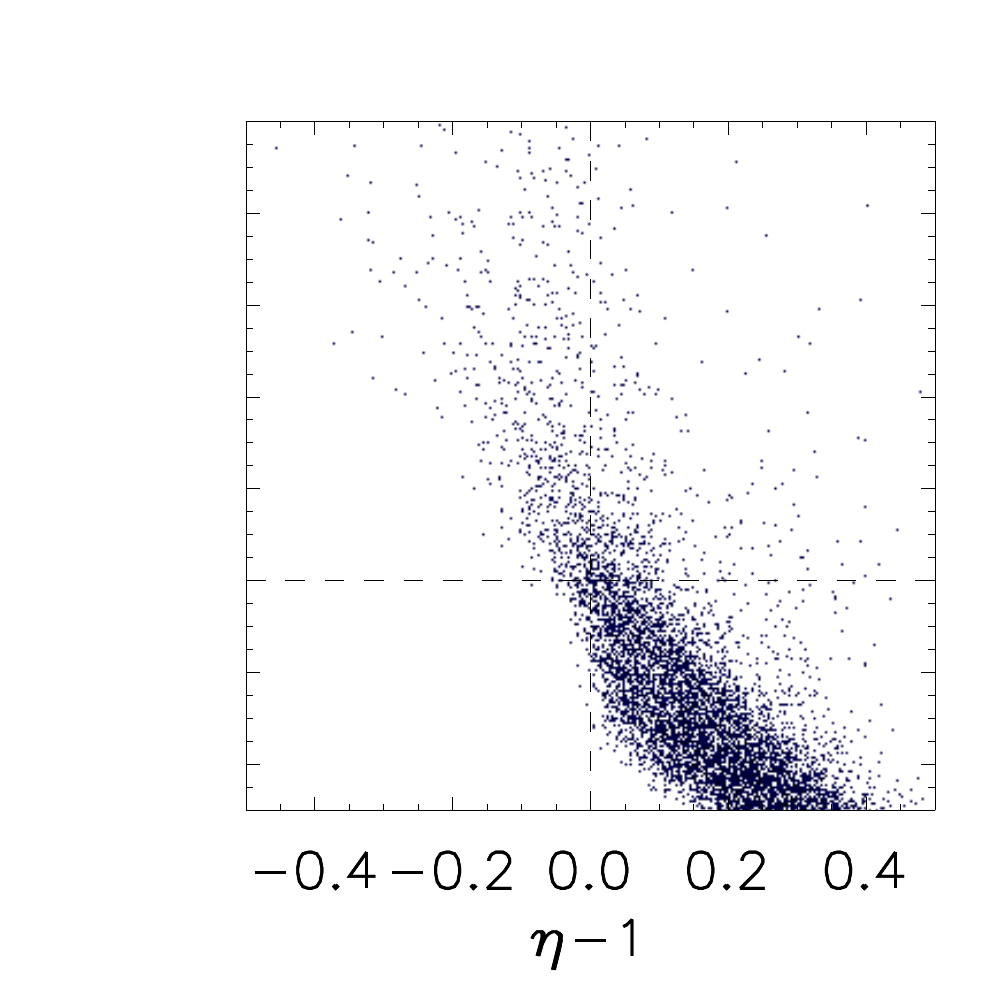}  \hskip-4mm
   \includegraphics[scale=0.27]{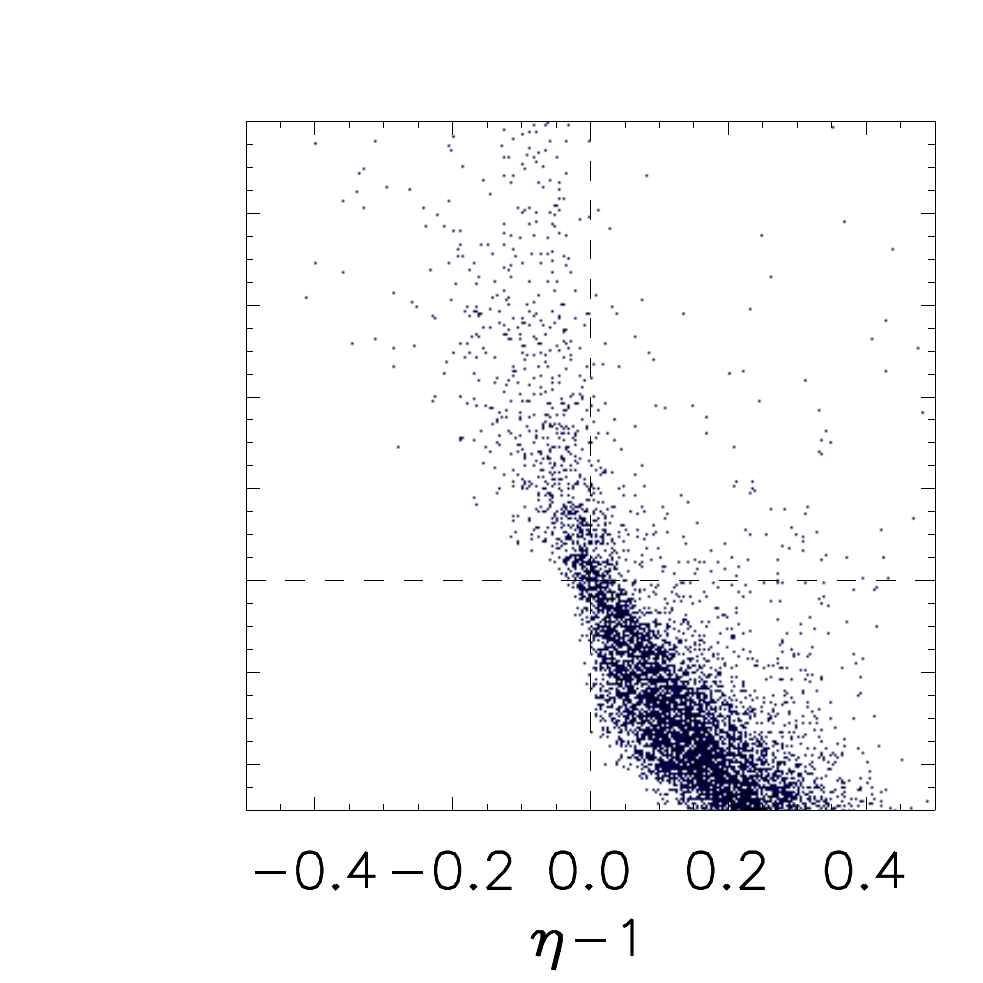}
   \vskip-2mm
   \includegraphics[scale=0.27]{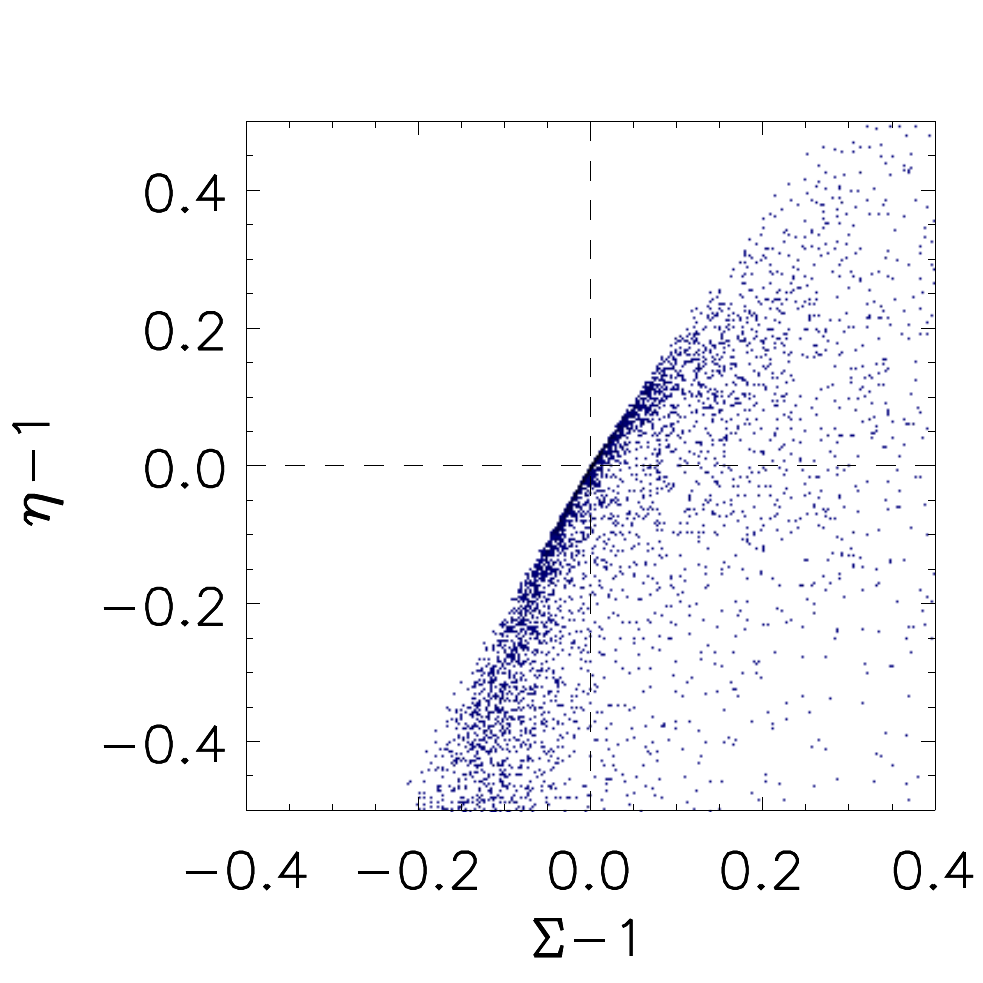}  \hskip-4mm
   \includegraphics[scale=0.27]{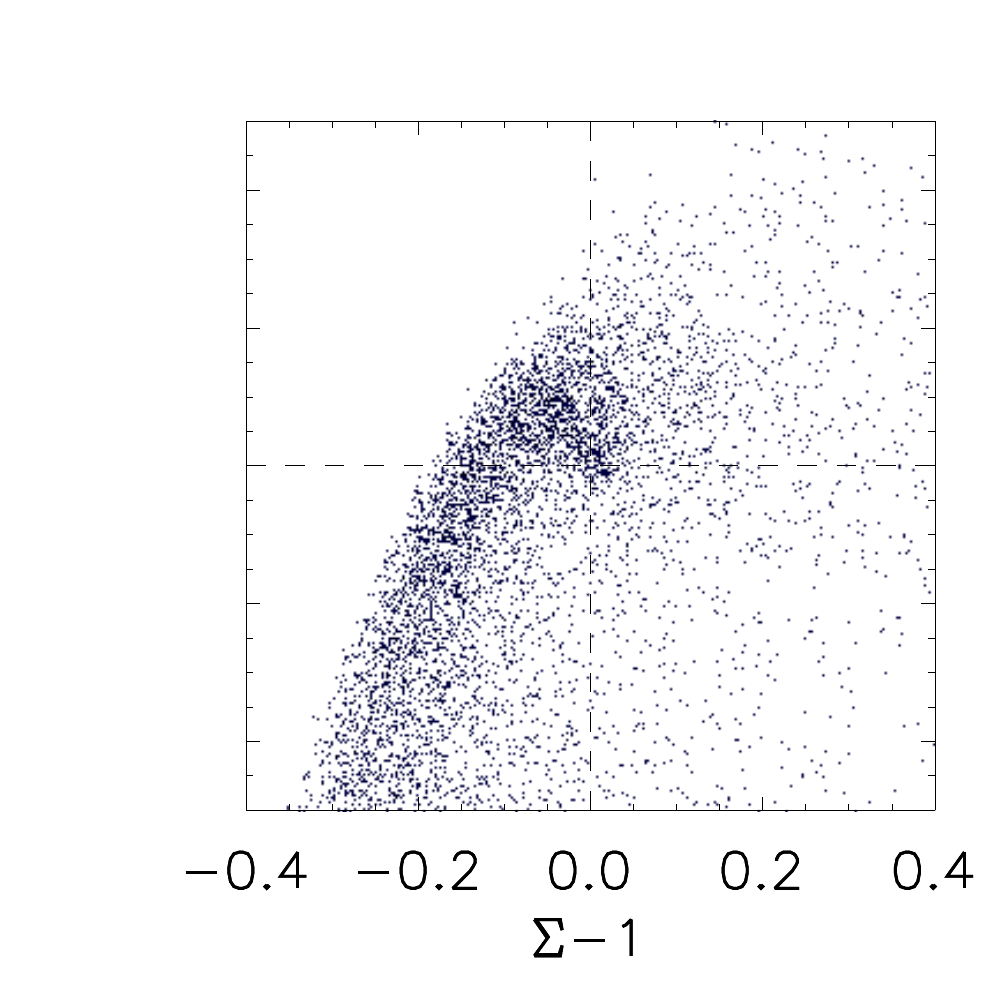}  \hskip-4mm
   \includegraphics[scale=0.27]{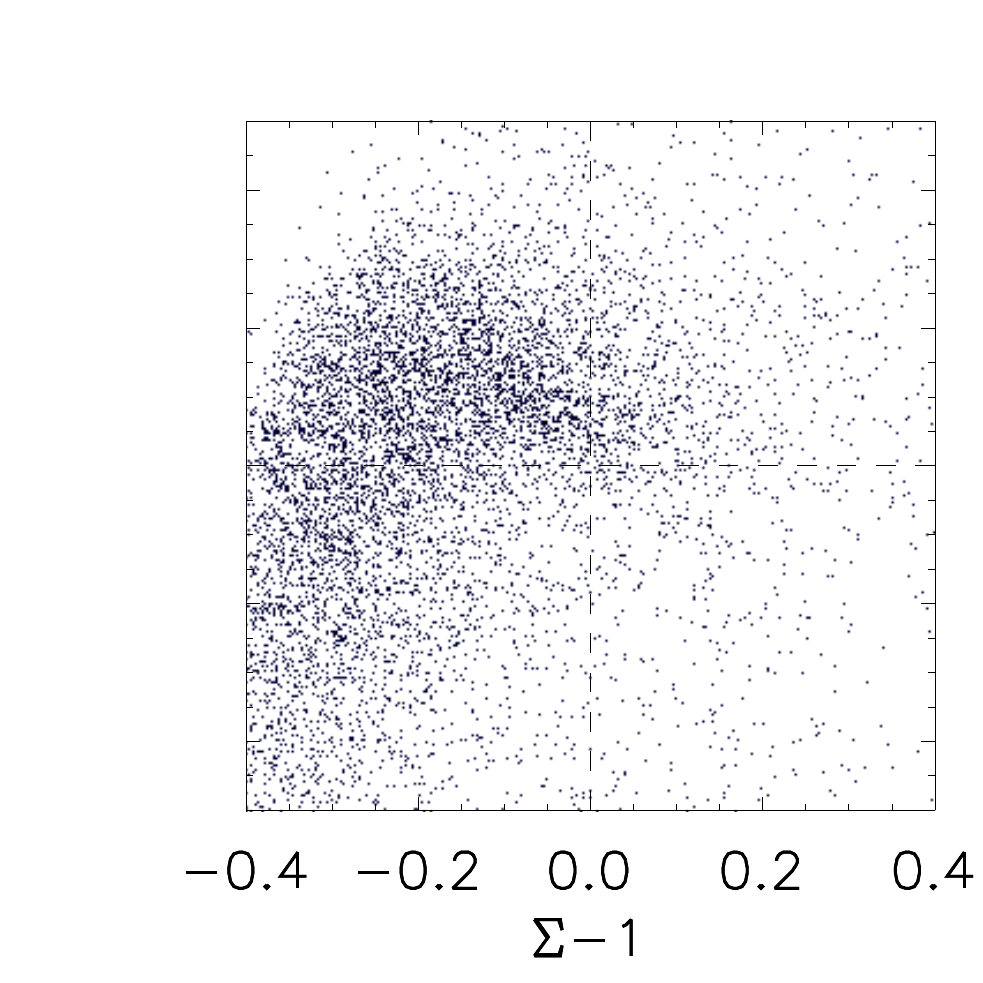}   \hskip-4mm
   \includegraphics[scale=0.27]{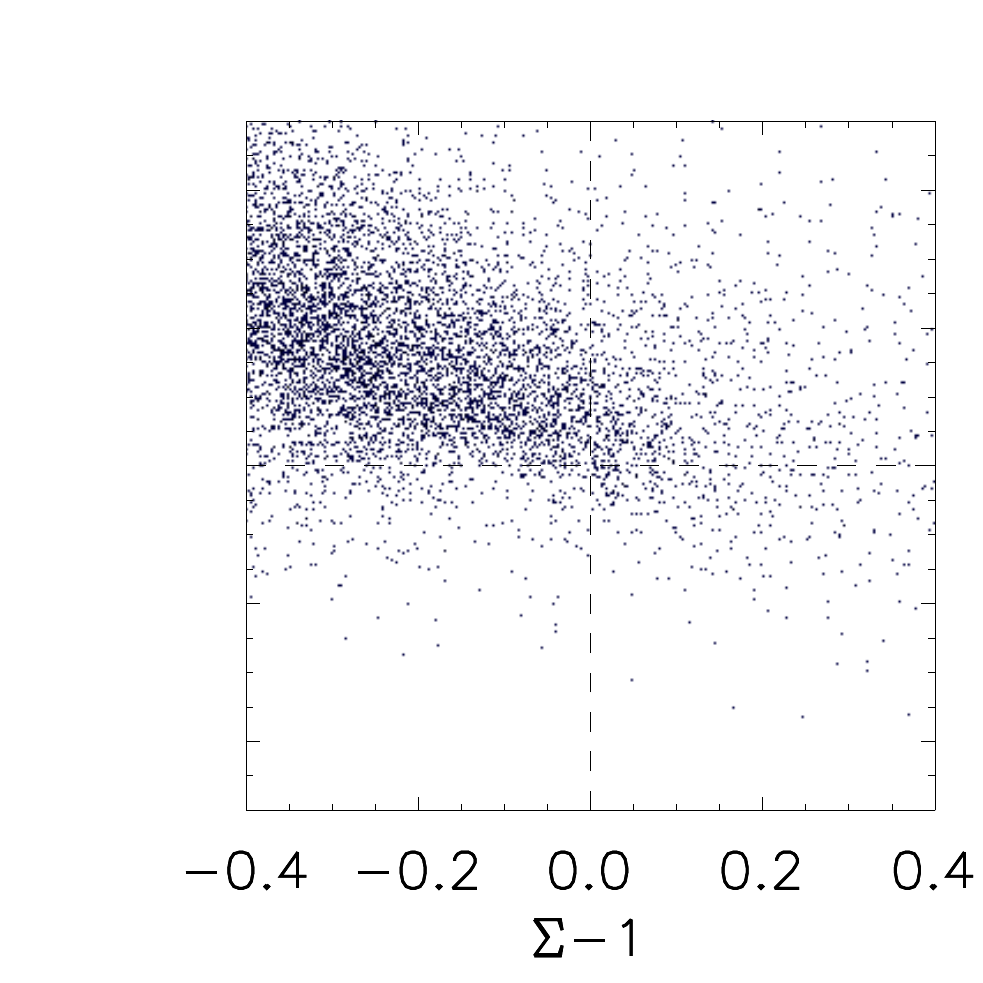}  \hskip-4mm
   \includegraphics[scale=0.27]{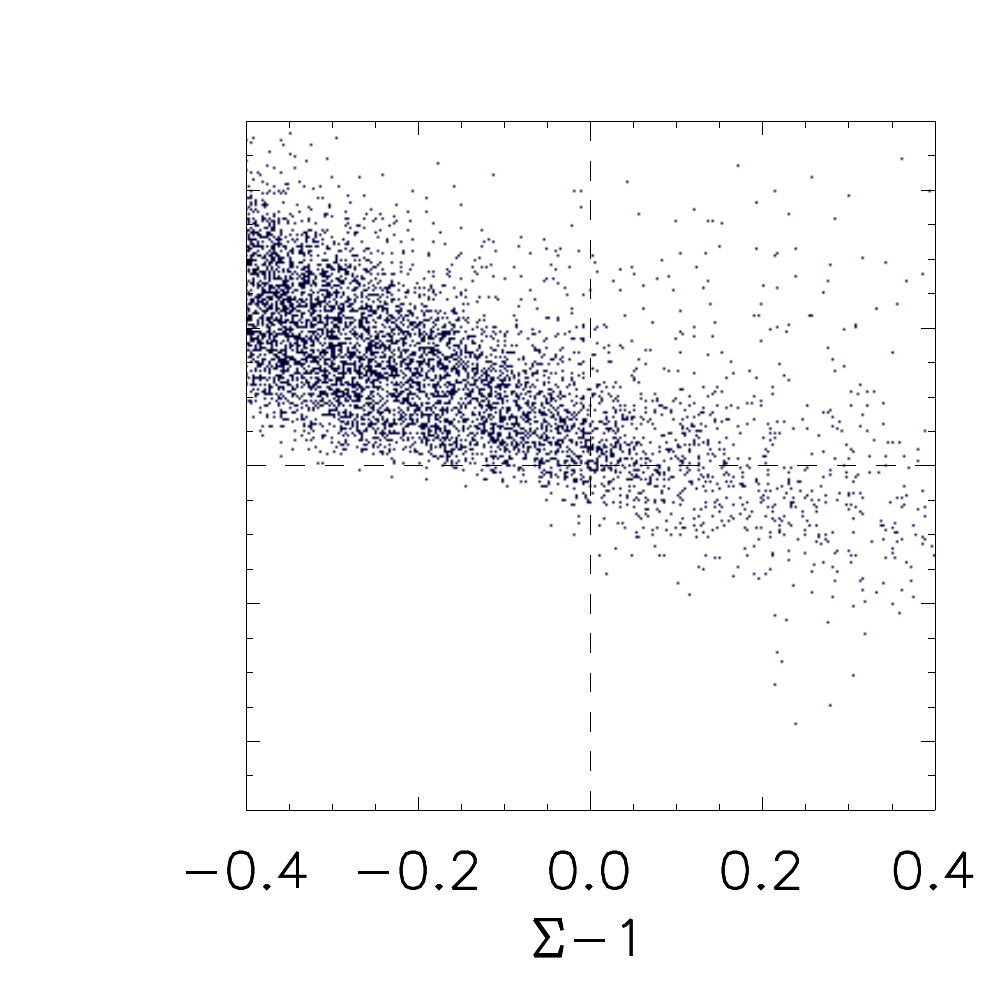}  \hskip-4mm
   \includegraphics[scale=0.27]{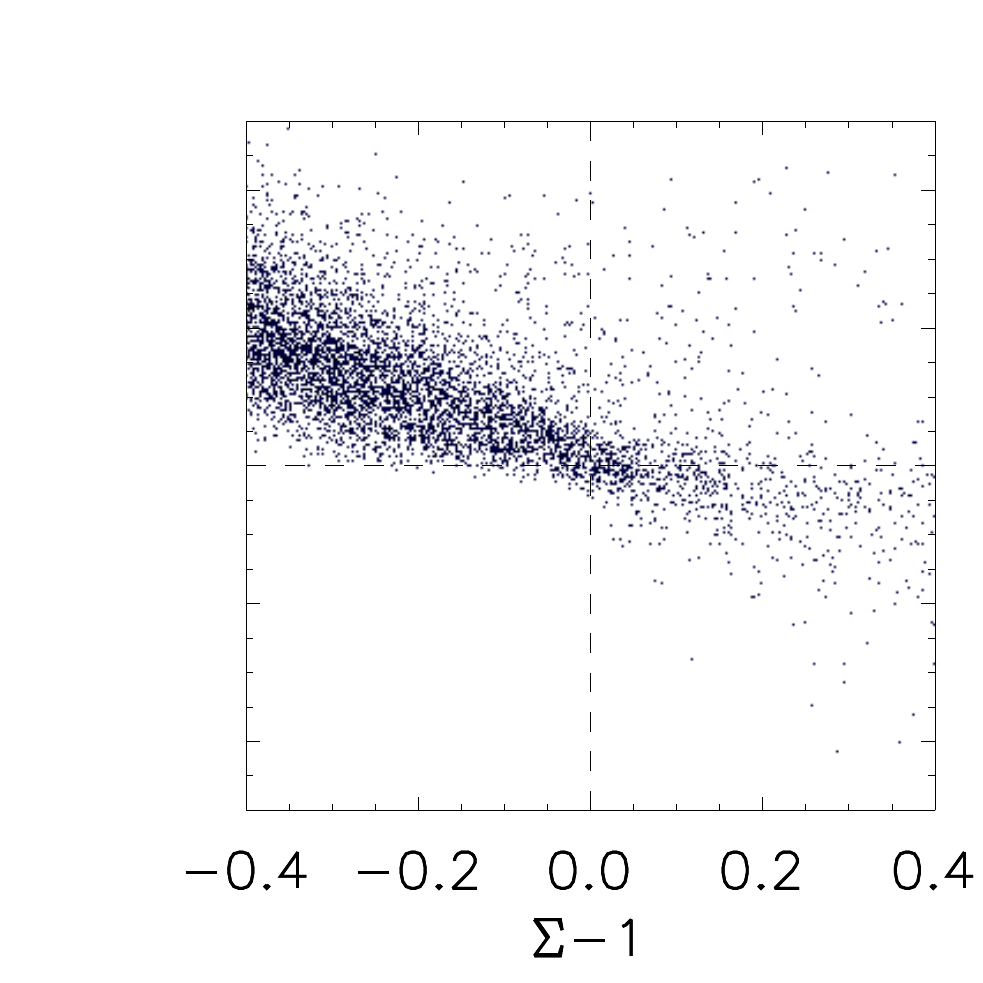}
   \vskip-2mm
   \includegraphics[scale=0.27]{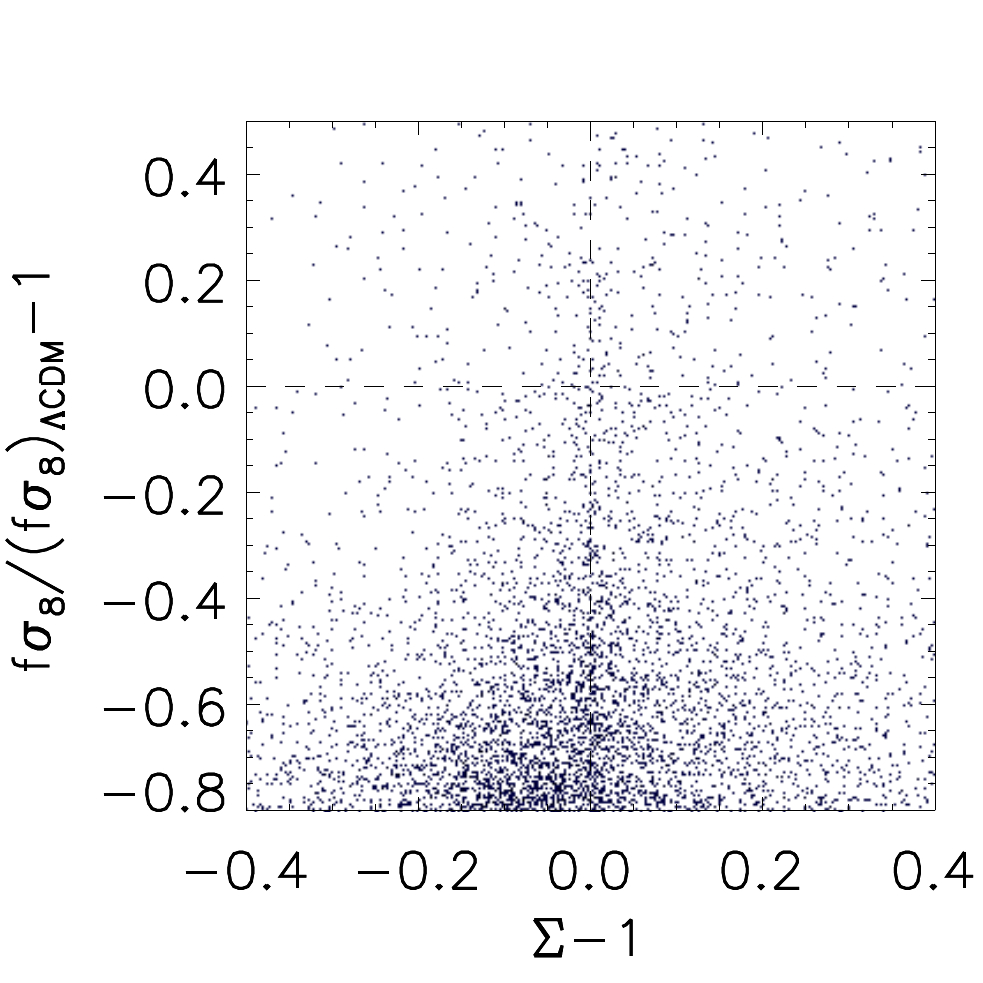}  \hskip-4mm
   \includegraphics[scale=0.27]{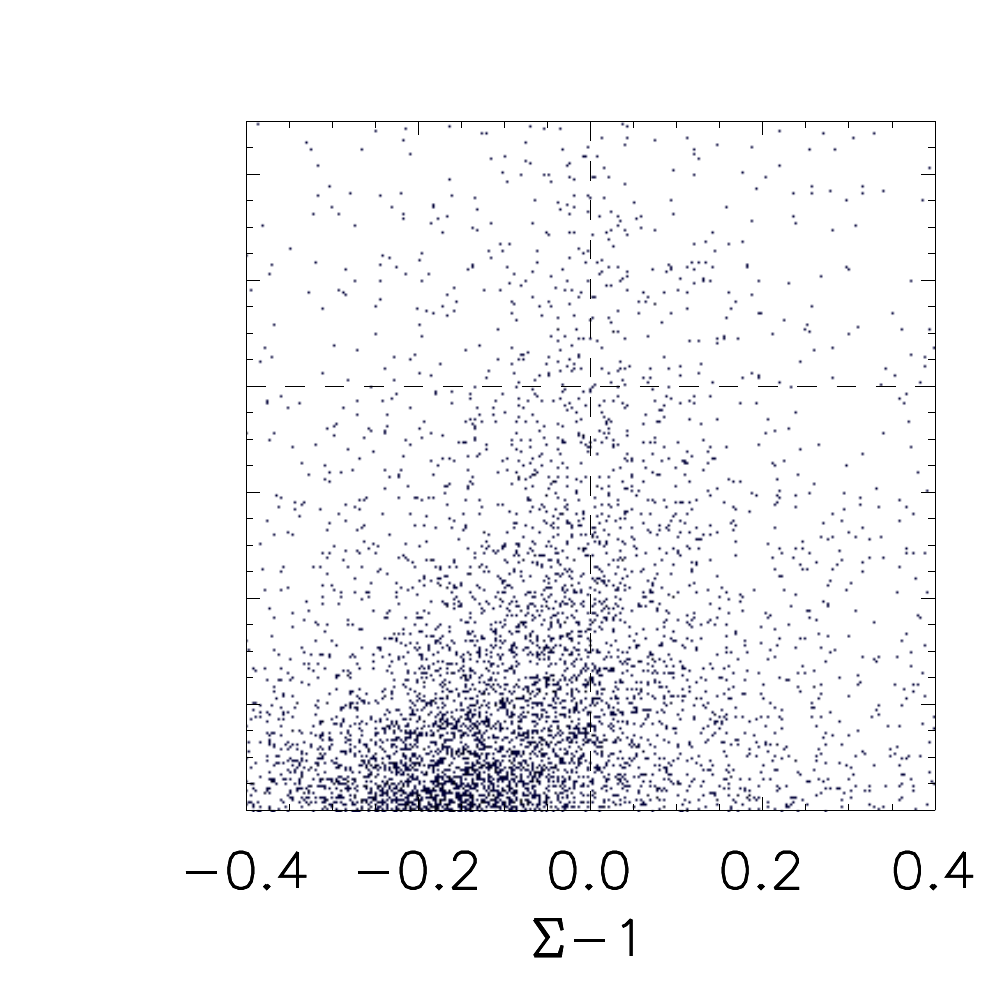}  \hskip-4mm
   \includegraphics[scale=0.27]{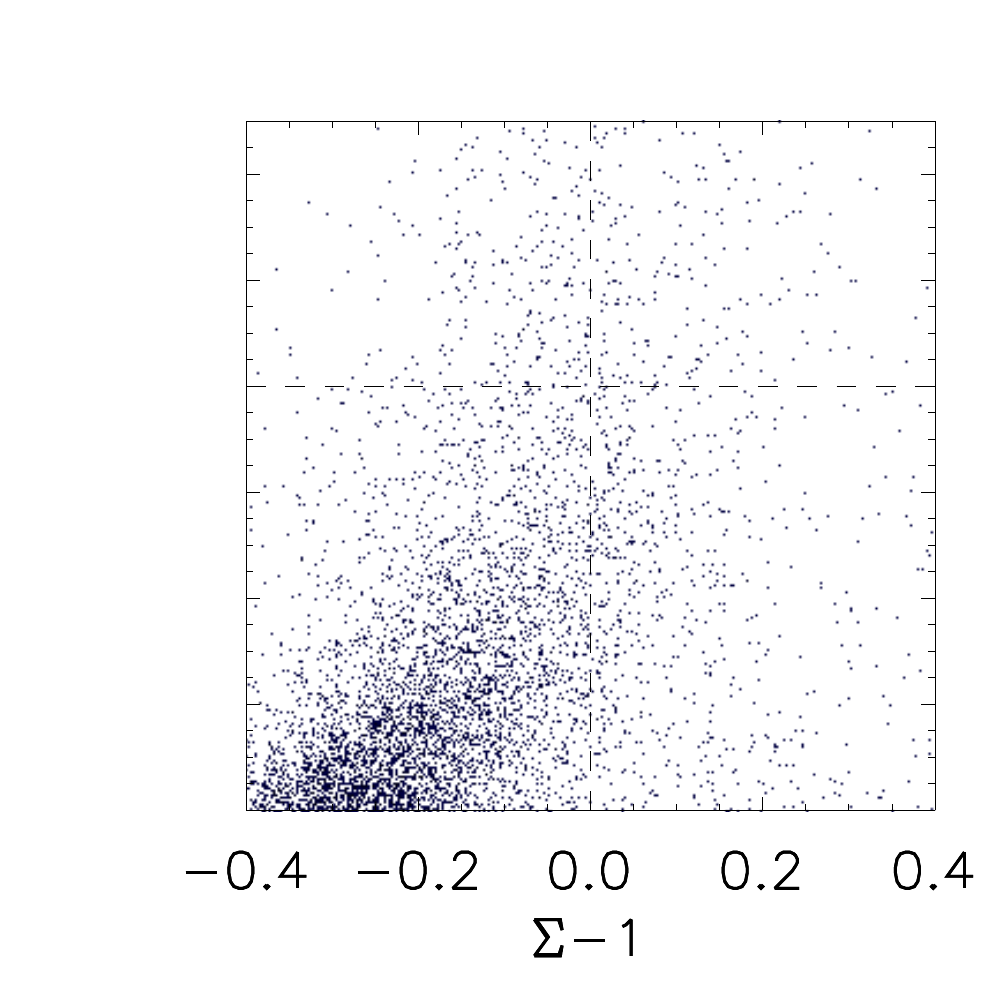}   \hskip-4mm
   \includegraphics[scale=0.27]{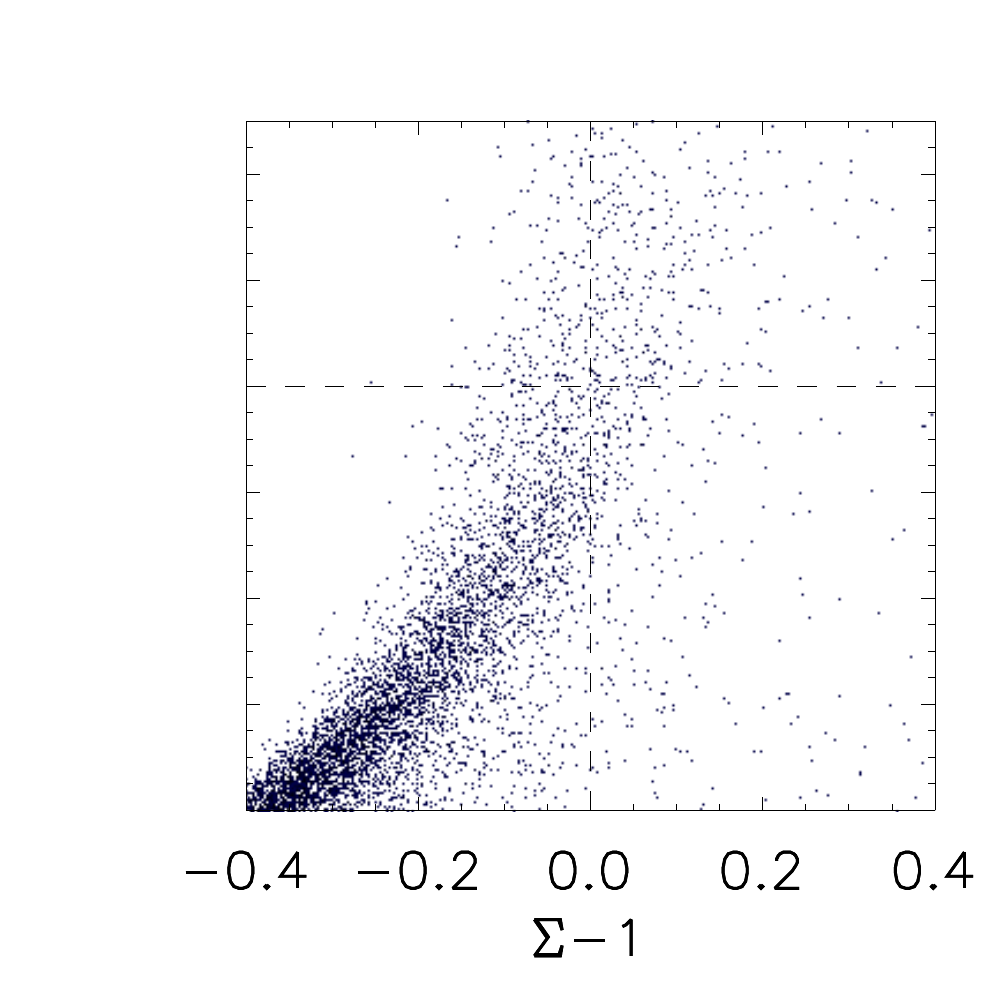}  \hskip-4mm
   \includegraphics[scale=0.27]{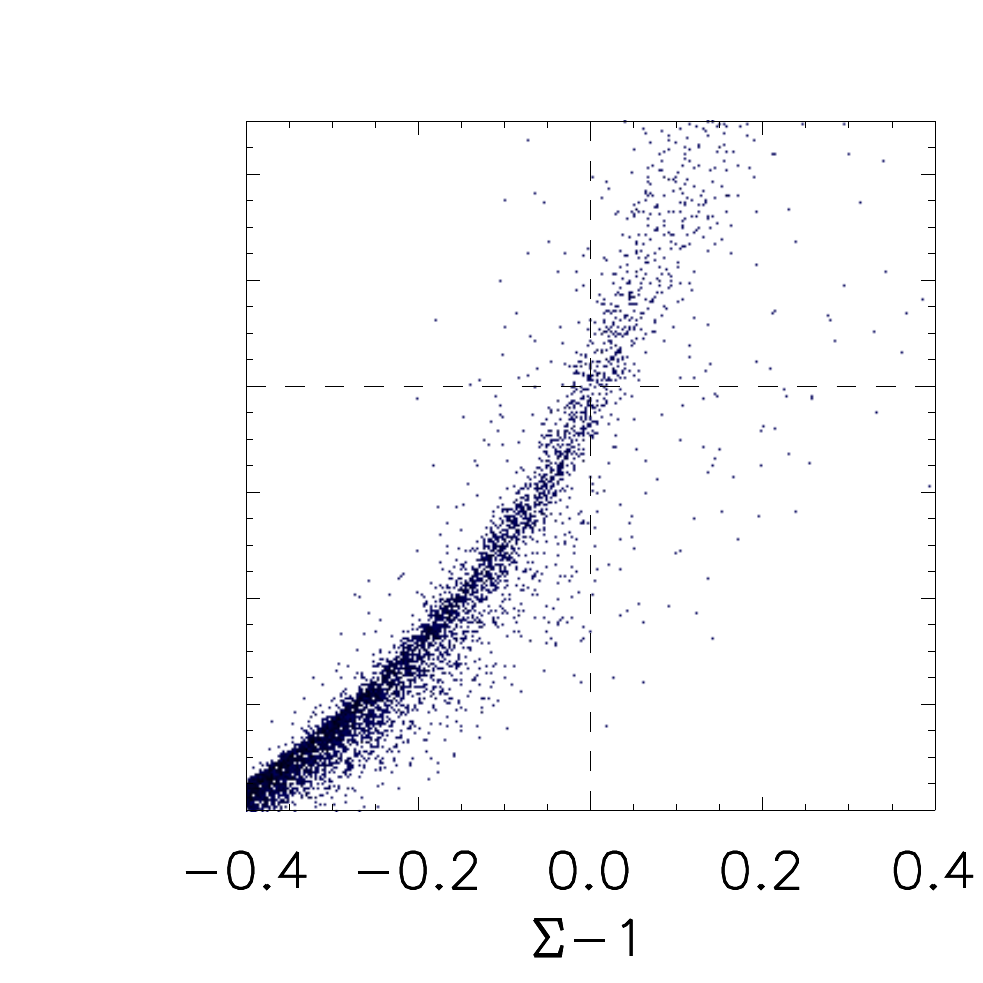}  \hskip-4mm
   \includegraphics[scale=0.27]{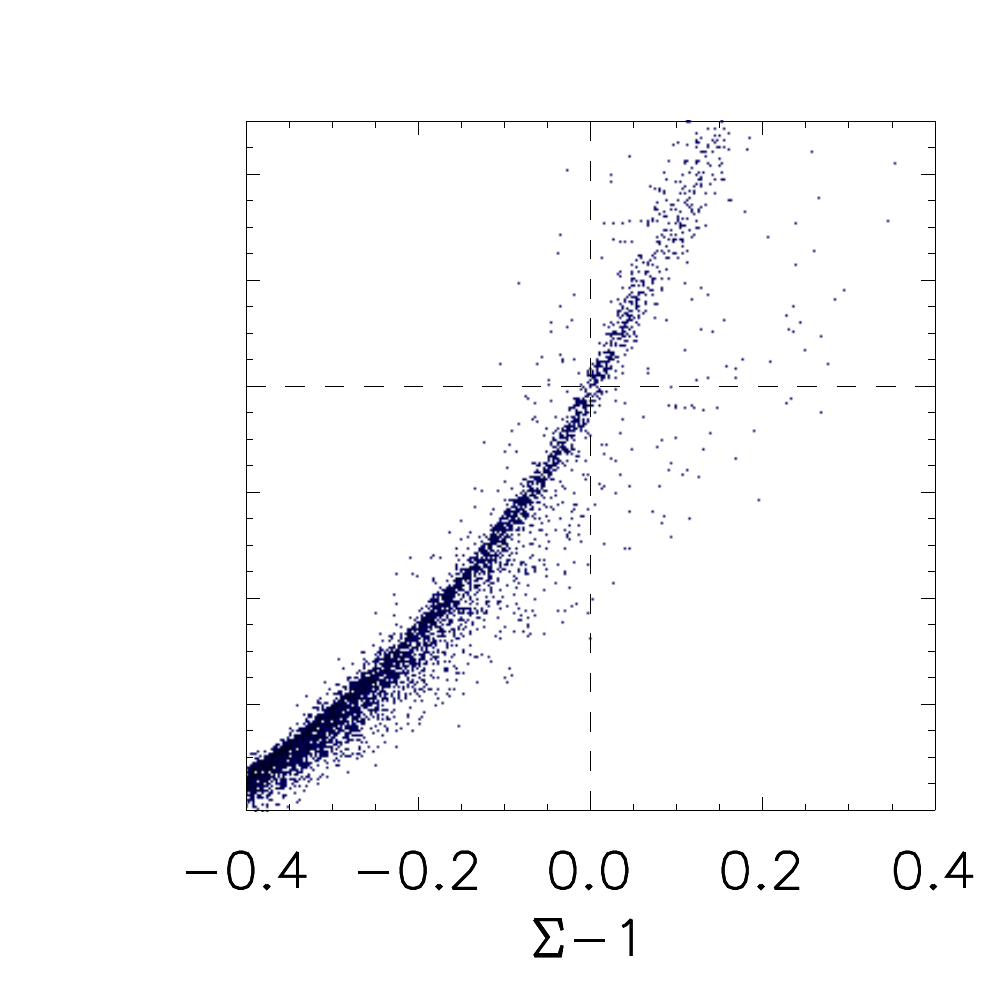}
   \caption{Same as in figure~\ref{fig_lde} but for $10^4$ models in the EDE scenario (top 4 rows) and for $10^4$ models in the EMG scenario (bottom 4 rows).}
      \label{fig_ede_emg}   
\end{center}
\end{figure}

Figure~\ref{fig_lde} shows that amplitude of $\fs$ expected in a $\Lambda$CDM  model is always minimal  if compared to Horndeski  expectations for  $z>1.5$. Interestingly, measurements of low $\fs$ amplitudes (with respect to the Planck-extrapolated value) provided by local redshift surveys seem to be quasi systematic, especially in analyses where the background is decoupled from the perturbation sector, see for instance 
\cite{Ruiz:2014hma,Kunz:2015oqa,Bernal:2015zom,Steigerwald:2016frc}. In parallel, recent observations at higher redshifts $z\sim 1.4$~\cite{Okumura:2015lvp}, seem to be suggestive of an early epoch with an excess of growth with respect to standard model predictions, although the error bar being too large does not yet allow to draw any meaningful interpretations.  
However, it serves as an illustration that, if such values were to hold up, they would effectively confirm a definitive prediction of Horndeski theories.

The remarkable tightness of the growth rate evolution of $\fs(z)$ also deserves a comment.  Despite $\mu$, $\gsp$ and $\Sigma$ spanning, especially at low redshift, a large range of values, absolute deviations of $\fs$ from the $\Lambda$CDM prediction are never larger than $0.2$ at all cosmic epochs investigated. The remarkably low theoretical dispersion, or equivalently, the poor sensitivity of $\fs$ to the variation of the Horndeski couplings, is not however prejudicial, from the observational side, for the purposes of  model identification. Indeed, it is as well remarkable that no single model displays both  $\fs<0$ and $\Sigma>0$ for any $z>1$. Measurements of $\fs$ from redshift surveys, when  combined with lensing estimations of $\Sigma$, provide thus an interesting diagnostic tool: evidences of even a single data point lying in the top right quadrant of the $\fs - \Sigma$ plane at redshift larger than 1 would definitely rule out LDE of the Horndeski type  as a viable candidate for theoretical interpretation.
\vv
Among the features emerging from Figure~\ref{fig_lde} is a strong positive correlation between $\mu$ and $\Sigma$ at high redshifts, or, even more telling, the lack of theories predicting  $\mu-1$ and $\Sigma-1$ of opposite sign as long as  $z>0.5$. When the behaviour of the gravitational slip parameter is closely scrutinized, the fact that $\gsp$ and $\mu$ cannot be both positive once $z>1$ also stands out.

The question is now whether any violation of these features is a smoking gun of the failure of only the LDE models, or,  more interestingly, if it can rule out even more general Horndeski scenarios. This issue is investigated in the next section.

\subsection{EDE and EMG scenarios}

The classification scheme proposed in Sec.~\ref{edeclass} contains the possibility that dark energy is present at early times, either in the energy momentum tensor (EDE) or also as early modified gravity (EMG). Figure~\ref{fig_ede_emg} shows that the presence of modifications of GR at early times alters the values of LSS observables even in the local universe. 

Irrespectively of the specific scenario, viability conditions favour theories with $\mu$ smaller than 1 for $z>0.5$. Despite we are now allowing initial values of $M^2$ different than $\mps$, the tendency of having $M^2>\mps$ survives. On the other hand, the EMG scenario is the only possibility to produce a small subset of models with $\mu > 1$ at early times. 

This can be understood by expressing the effective gravitational constant as
\begin{equation}\label{mudef2}
 \mu \ = \ \frac{M^2(x_0)[1+\epsilon_4(x_0)]^2}{ M^2(1+\epsilon_4)^2} \ \left[1+\frac{1+\epsilon_4}{B} \left(\dfrac{\mu_1 -\mu_3}{1+\epsilon_4}-(\mu_1 +\mathring{\epsilon}_ 4) \right)^2\right]. 
 \end{equation}
From stability requirements, $B\geqslant 0$ and $\epsilon_4\geqslant -1 $ ($c_T \geqslant 0$), $i.e$ respectively no gradient instabilities of scalar and tensor modes, the quantity contained in the squared brackets above is greater than or equal to 1. Therefore, allowing non vanishing $\epsilon_4$ and $\mu_3$ at $x=1$ pushes up the value of $\mu$ at early times.
The above expression shows that the value of $\mu$ at present time, $\mu(x_0)$, is always greater than or equal to unity whatever the DE scenario, a consequence of the definition eq.(14). Equation \eqref{mudef2} also illustrates  the competition between the two major physical mechanism that contribute to the amplitude of the  gravitational coupling : {\it (i)} the fifth force induced by the scalar field,  which must always be attractive, hence larger than unity, for a massless spin 0 field (embodied by the term in squared brackets) and  {\it (ii)}  the possibility of realising weaker gravity through the $1/(M^2(1+\epsilon_4)^2)$ component, which,  as pointed out in \cite{Perenon:2015sla}, is a term  related to the amount of self-acceleration a model produces. 
\vv
The behaviour of $\mu$ at early times affects the $\fs$ observable at late epochs. For instance, the amplitude of $\fs$ predicted in EDE scenarios is lower than the standard $\Lambda$CDM value for $z>0.5$, as opposed to LDE, for which models systematically flip over $\Lambda$CDM at $z\gtrsim 1.5$. Therefore, over almost all the most interesting epochs of the universe, the $\Lambda$CDM growth history appears as an extremum not only among the whole class of LDE models, but also when EDE scenarios are considered. Intriguingly, only EDE models manage to strongly suppress the amplitude of the  linear growth function at present time. On the opposite, the only models allowing for a faster growth than $\Lambda$CDM  (with $\fs$ more than $20\%$ higher), are the EMG scenarios. This is not surprising for, as we said, it is the only set-up for which $\mu>1$ at early times. 
\vv
The asymptotic value of the gravitational slip parameter $\gsp$ at $x=1$ is, by definition, 1 when the coupling functions $\mu_3$ and $\epsilon_4$ vanish. Therefore, only very mild differences arise between LDE and EDE at early times.  On the contrary, the redshift dependence of $\gsp$ is significantly affected in the EMG case, since $\mu_3$ and $\epsilon_4$ are  different from zero at all cosmic epochs. As far as the evolution of $\Sigma$ is concerned, the amplitude calculated in LDE and EDE scenarios is always lower  than $\Lambda$CDM for $z>1.5$. Once more, the standard model appears as an extremal model. 

EMG is the only mechanism  enabling $\Sigma$ to be grater than  unity also at high redshifts.

The marked positive correlation between $\mu$ and $\Sigma$ for $z>1.5$ persists in EDE models as it did in LDE models. In fact, since 
\begin{equation}\label{45cor}
\mu-1= \left(\frac{2}{1+\gsp}\right) \left(\Sigma-1 \right) \,-\,\frac{\gsp-1}{1+\gsp}\;,
\end{equation}
a clear $45^\circ$ correlation should be seen as long as $\gsp$ is close to unity. This stands out clearly, at high redshifts,  for LDE (see also figure 6 in \cite{Perenon:2015sla}) and for EDE scenarios, as opposed to the EMG case which displays slightly more dispersion, for it exhibits larger values of $\gsp$. Assuming a non-standard gravitational signal will be detected by future surveys, we can thus tentatively conclude from comparing Figs.~\ref{fig_lde} and~\ref{fig_ede_emg} that the Horndeski class of models would be ruled out by high redshit measurement if $\mu$ and $\Sigma$ have different sign for $z>1.5$. The same conclusion holds if future estimates should eventually converge on a local ($z=0$) value of the effective Newton constant lower than unity,  accordingly to the argument already given below eq.\eqref{mudef2}. Interestingly,  \cite{Pogosian:2016pwr} have argued that Horndeski theories are likely to display a sign agreement in the $\mu$--$\Sigma$ plane across all cosmic epochs. We suggest that these diverging conclusions arise because they seem to consider as representative only models displaying a gravitational slip parameter $\eta$ close to unity today. We observe that the range of possible $\eta$ values progressively broadens when a larger number of couplings is progressively switched on, and the generality of the scenarios is increased. We observe a systematic  increase in the scatter of the $\eta$ values, first when considering the $\m22$ parameter to be not zero at all times in the LDE case, and then when allowing more generic initial conditions such as in the EDE and EMG scenarios. On top of this,  Figure~\ref{fig_viab} also shows that the significance of the opposite sign statement is progressively lessened by relaxing some of our viability conditions. This was also highlighted in Figure 7 of \cite{Salvatelli:2016mgy}, where CMB data likelihood is plotted for the observables $\mu$ and $\Sigma$ calculated at $z=0$.

\begin{figure}[!]
\begin{center}
	\includegraphics[scale=0.49]{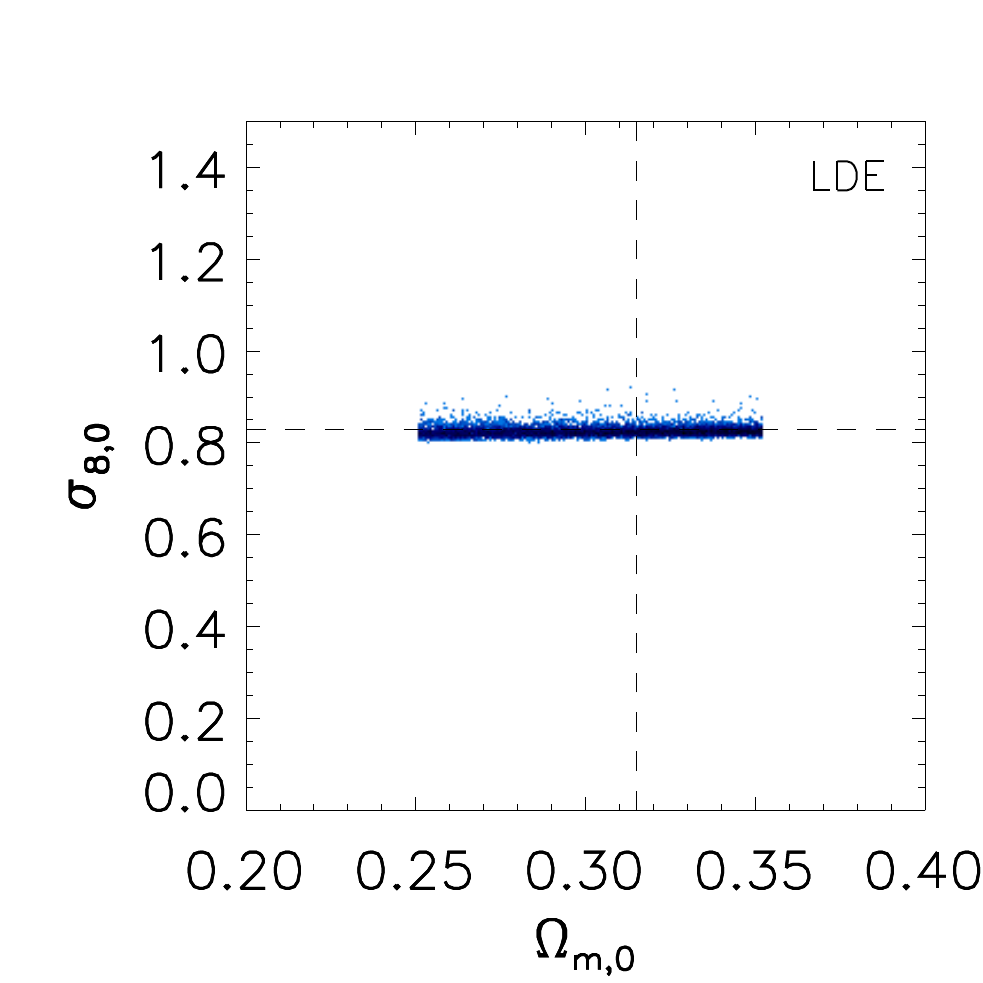}   \hskip4mm
	\includegraphics[scale=0.49]{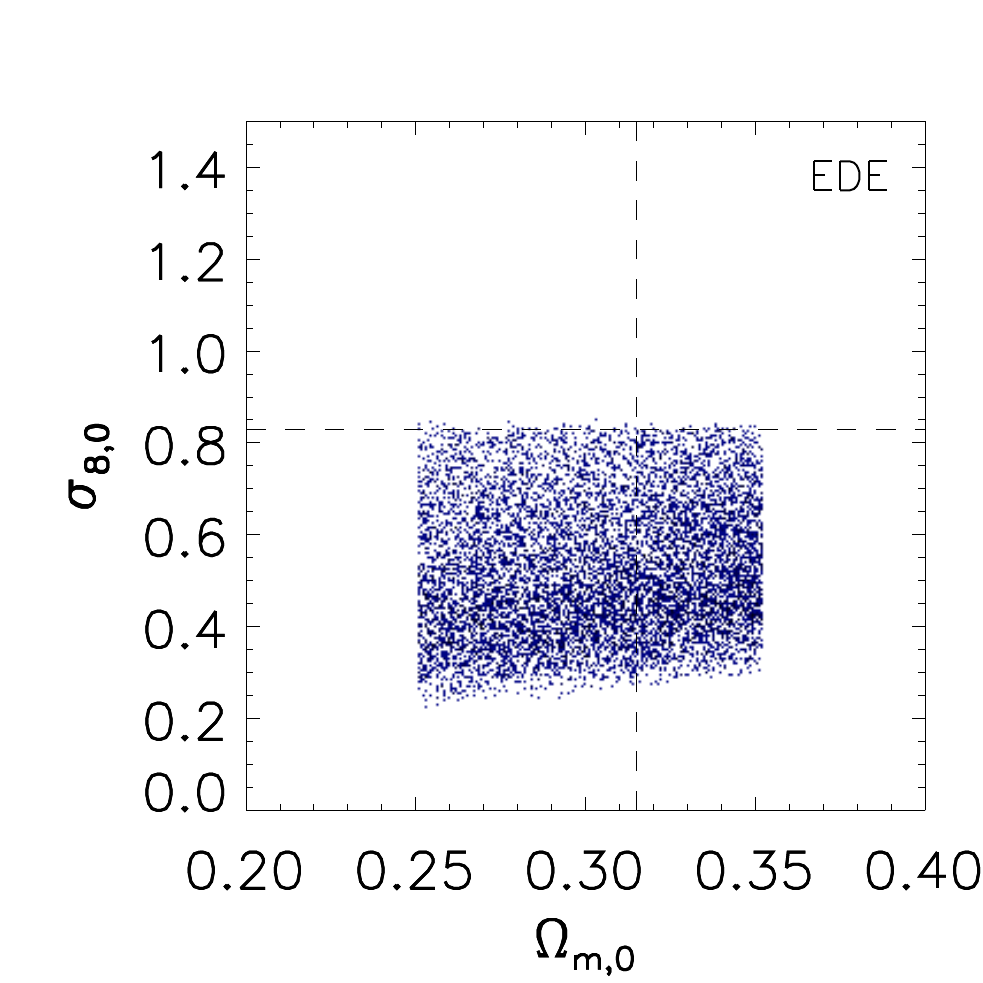} \hskip4mm
	\includegraphics[scale=0.49]{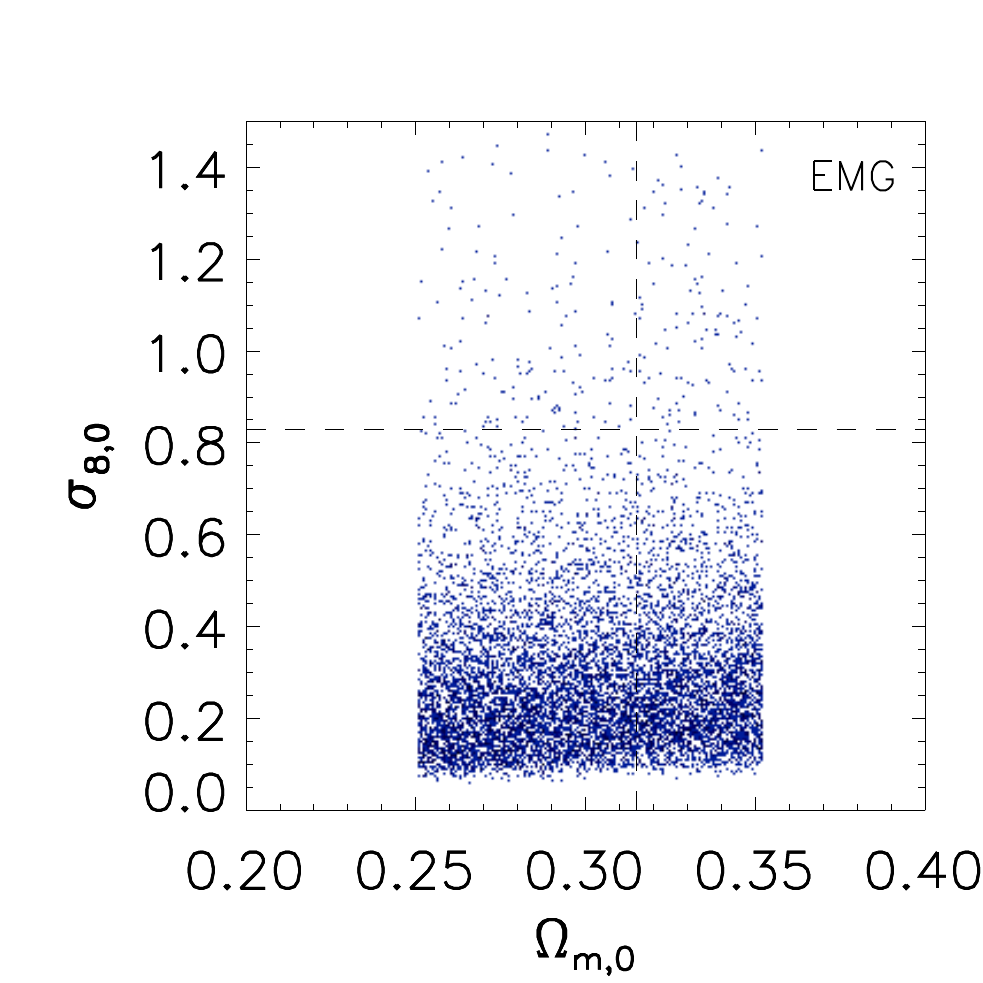}
   \caption{Present day value of $\sigma_8$ as a function of the fractional matter density today $\Omega_{m,0}$ of $10^4$ EFT models. Here, the value of $\Omega_{m,0}$ per model is left as free parameter and is randomly generated, as are the coefficients of the coupling functions. The background has been set to a flat $\Lambda$CDM background. The intersection of the dashed lines corresponds to the Planck measurements~\cite{Ade:2015xua}.}
	\label{fig_sigom} 
\end{center}
\end{figure}

\vv
In much the same way as the plane $\mu$--$\Sigma$ provides a diagnostic for the whole class of Horndeski models, the plane $\fs$--$\Sigma$ allows us to tell apart Horndeski dark energy sub-classes. For example, if $\fs > (f \sigma_{8})_{\Lambda CDM}$ at $z>1.5$, then EDE scenarios are ruled out. Similarly, LDE is not viable if both $\fs$ and $\Sigma$ have smaller amplitude than predicted by $\Lambda$CDM for $z>1.5$.
\vv
Identical conclusions follow from the analysis of the amplitude of the $\sigma_8$ observable alone. Figure~\ref{fig_sigom} shows the present-day linear extrapolation of the $rms$ fluctuations of the matter density field, which we assume to be normalized  by the Planck measurements at last scattering. Predictions closely reproduce the $\Lambda$CDM limit in all the LDE models. However, a local measurement of $\sigma_8$ showing large deviations from the $\Lambda$CDM extrapolation will be instrumental for disentangling EDE from EMG scenarios. The first would be definitely ruled out if observational evidences should indicate that $\sigma_{8,0} \gtrsim 0.9$.\footnote{We note that lower values of $\sigma_8$ are also found in the ``kinetic matter mixing" model considered in~\cite{D'Amico:2016ltd}.}
\vv
In summary, we find the effects of early modification of GR  to be conspicuous also at low redshift.  Joint measurements of the $\gsp$, $\Sigma$ and $\fs$ observables would give strong indications as to the type of DE required for a faithful description of cosmological perturbations. Indeed, complementing an analysis on the growth of structures with lensing observables increases substantially the discriminating power between models. 

\section{Discussion}\label{sec_4bis}

The next issue we tackle concerns the generality and robustness of our findings against changes in the specific settings and/or assumptions adopted in the analysis. We first explore whether our diagnostic predictions still stand out so clearly once some viability conditions about propagations speeds are relaxed. Then, we gauge the effects of considering non-standard evolution of the background expansion rate. Finally, we discuss additional checks on the generality of our results and the parametrisation of Horndeski theories.

\subsection{Constraining power of viability conditions}\label{sec_viab}

\begin{figure}[!]
\begin{center}
\begin{flushleft}
$S$ :
\end{flushleft}  \vskip-2mm
   \includegraphics[scale=0.27]{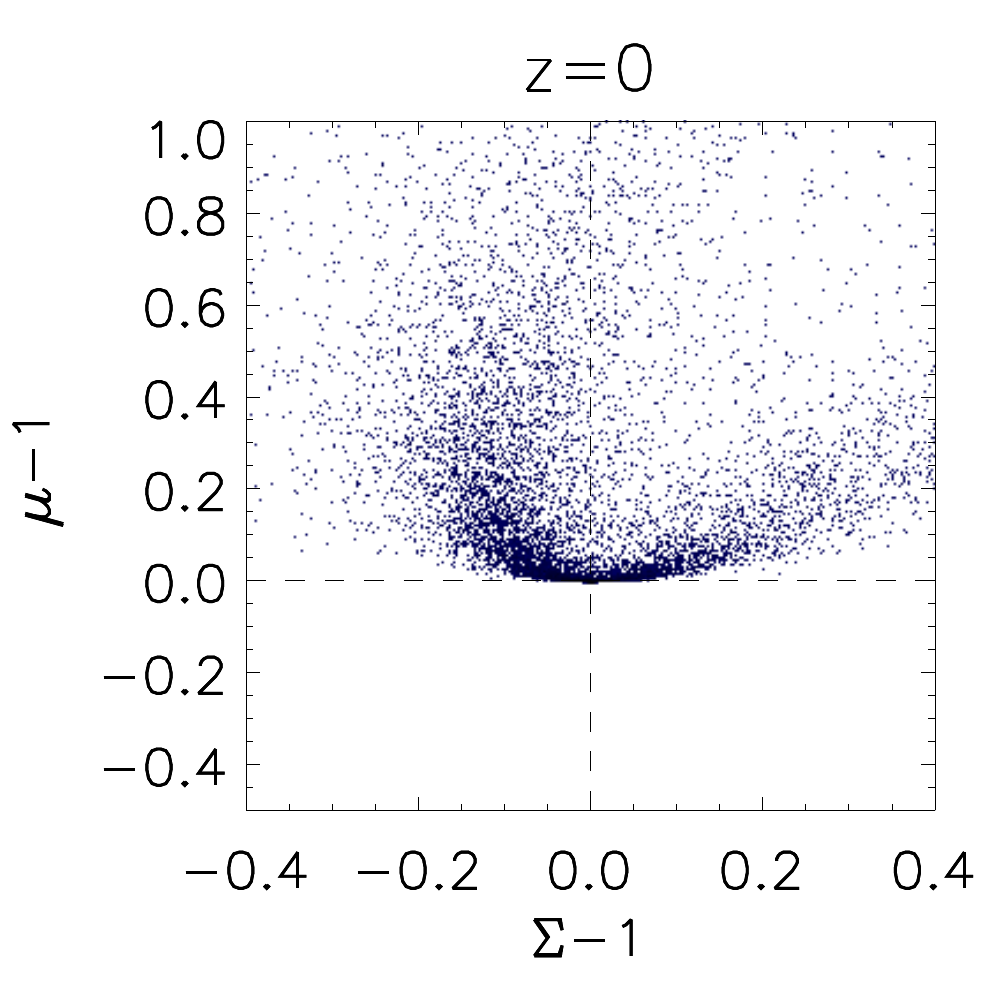}  \hskip-4mm
   \includegraphics[scale=0.27]{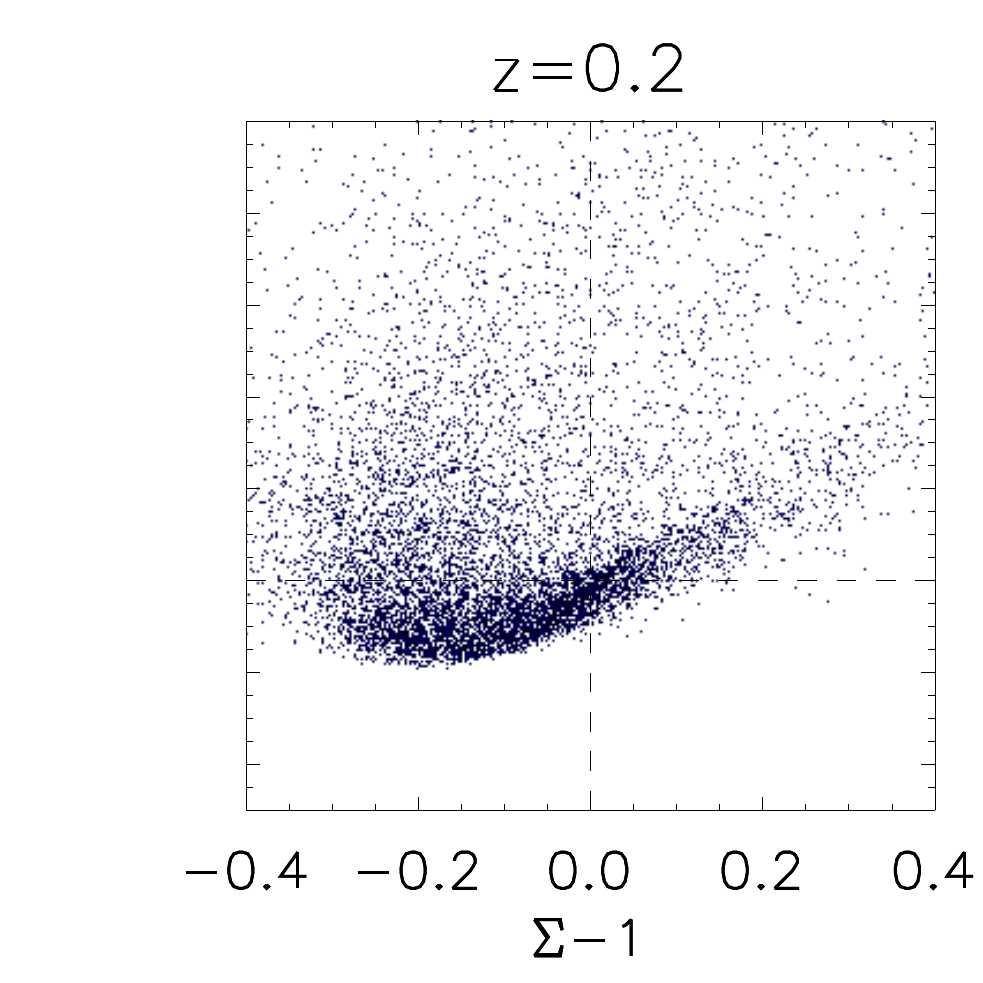}  \hskip-4mm
   \includegraphics[scale=0.27]{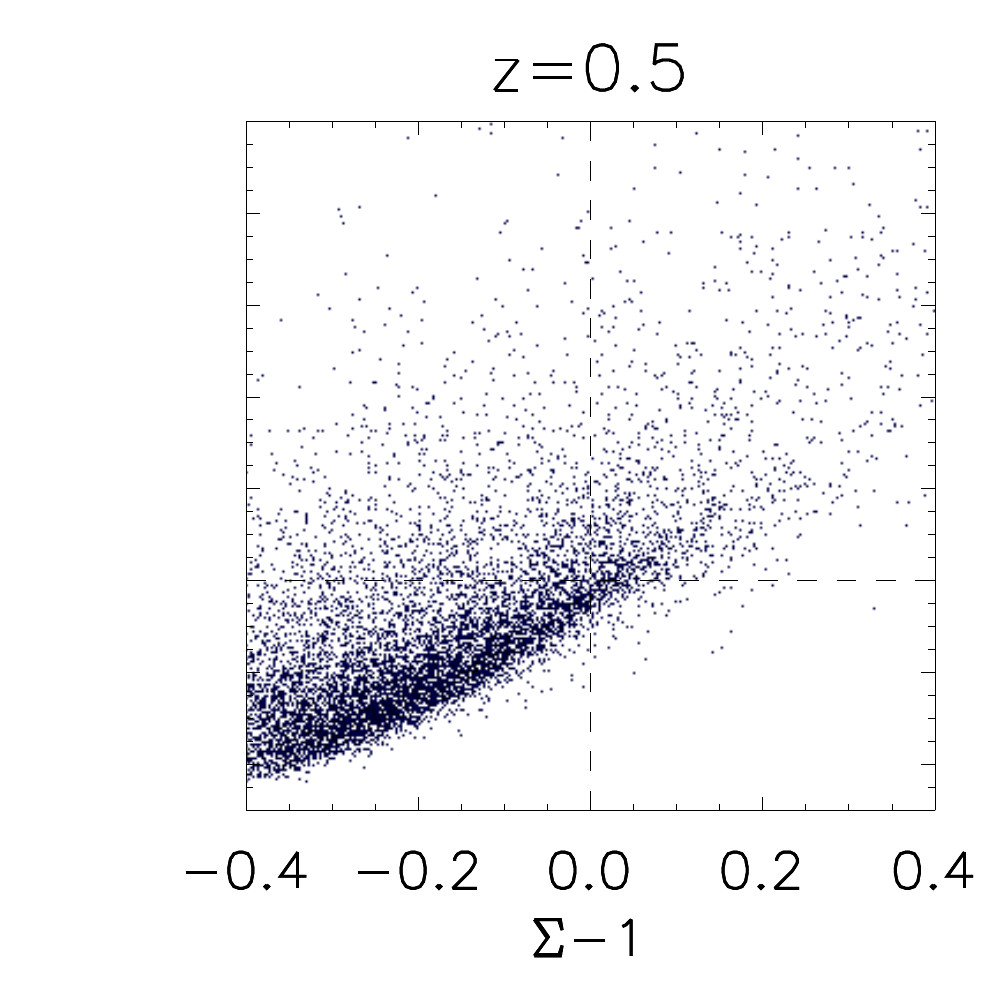}   \hskip-4mm
   \includegraphics[scale=0.27]{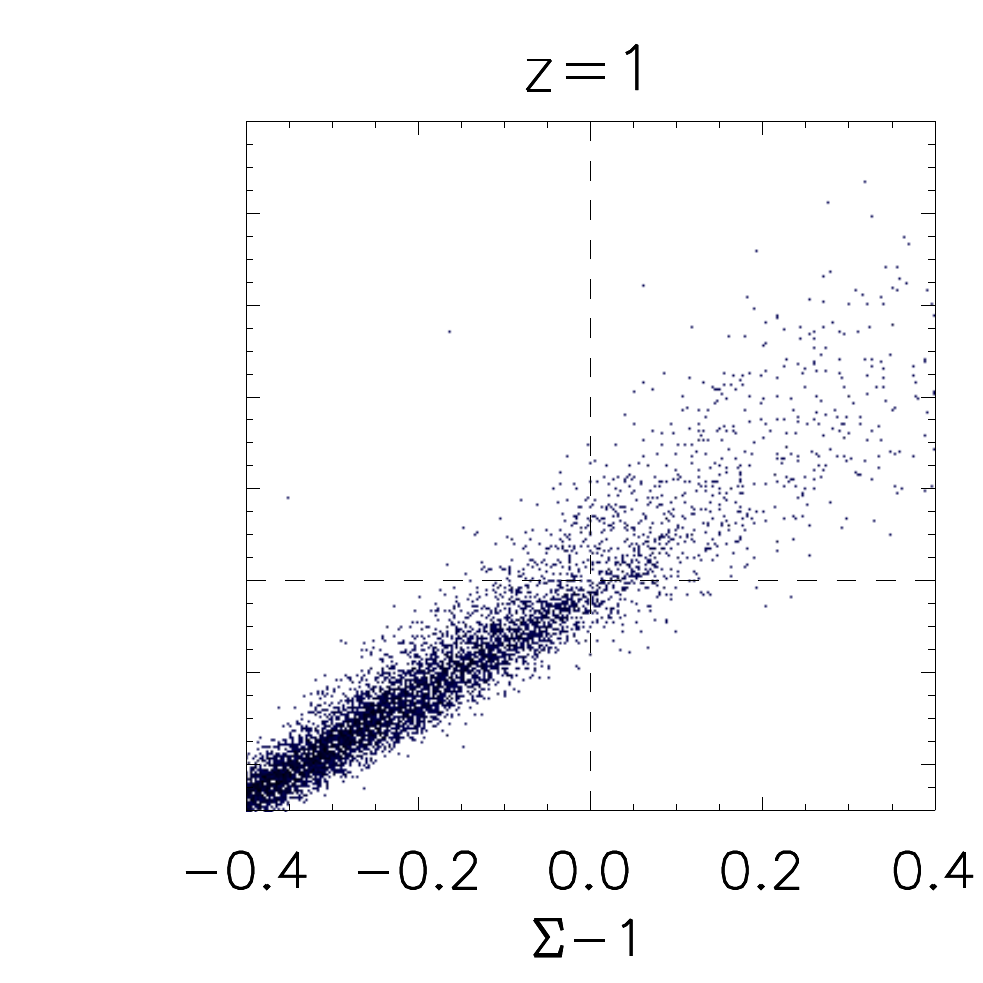}  \hskip-4mm
   \includegraphics[scale=0.27]{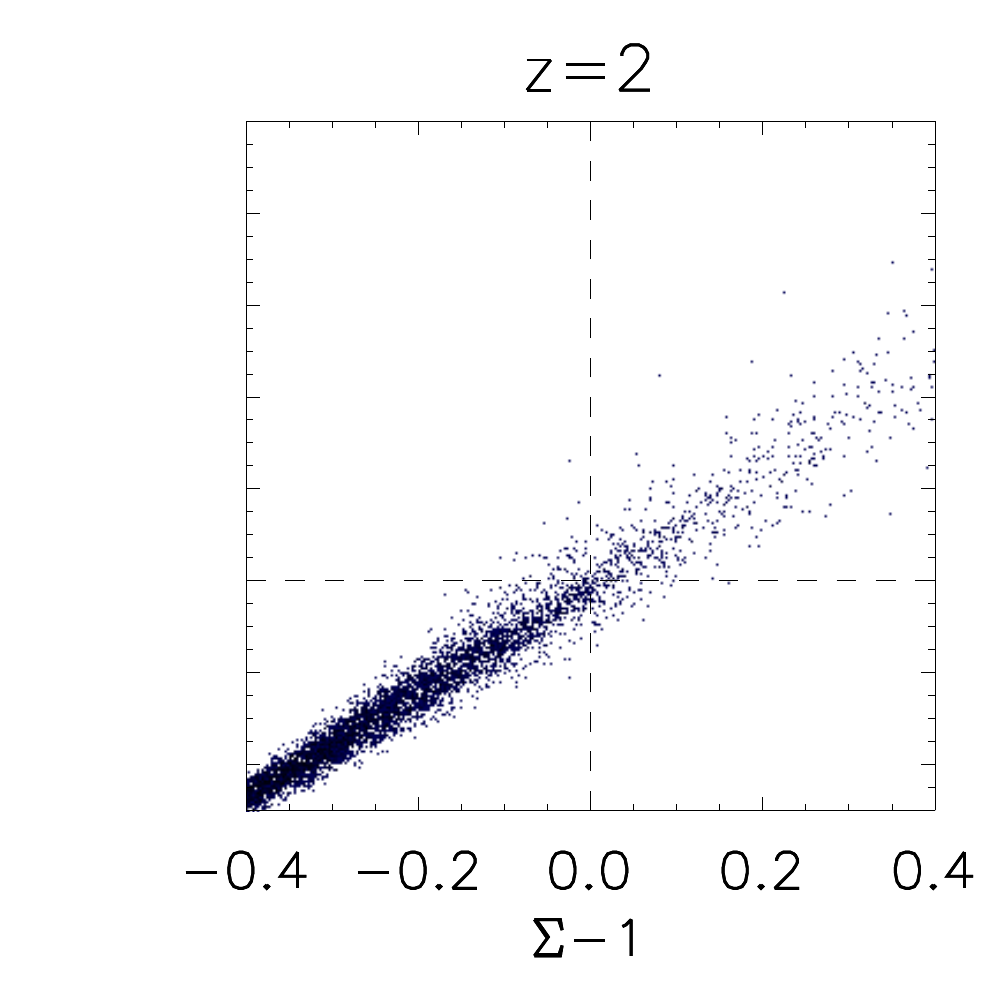}  \hskip-4mm
   \includegraphics[scale=0.27]{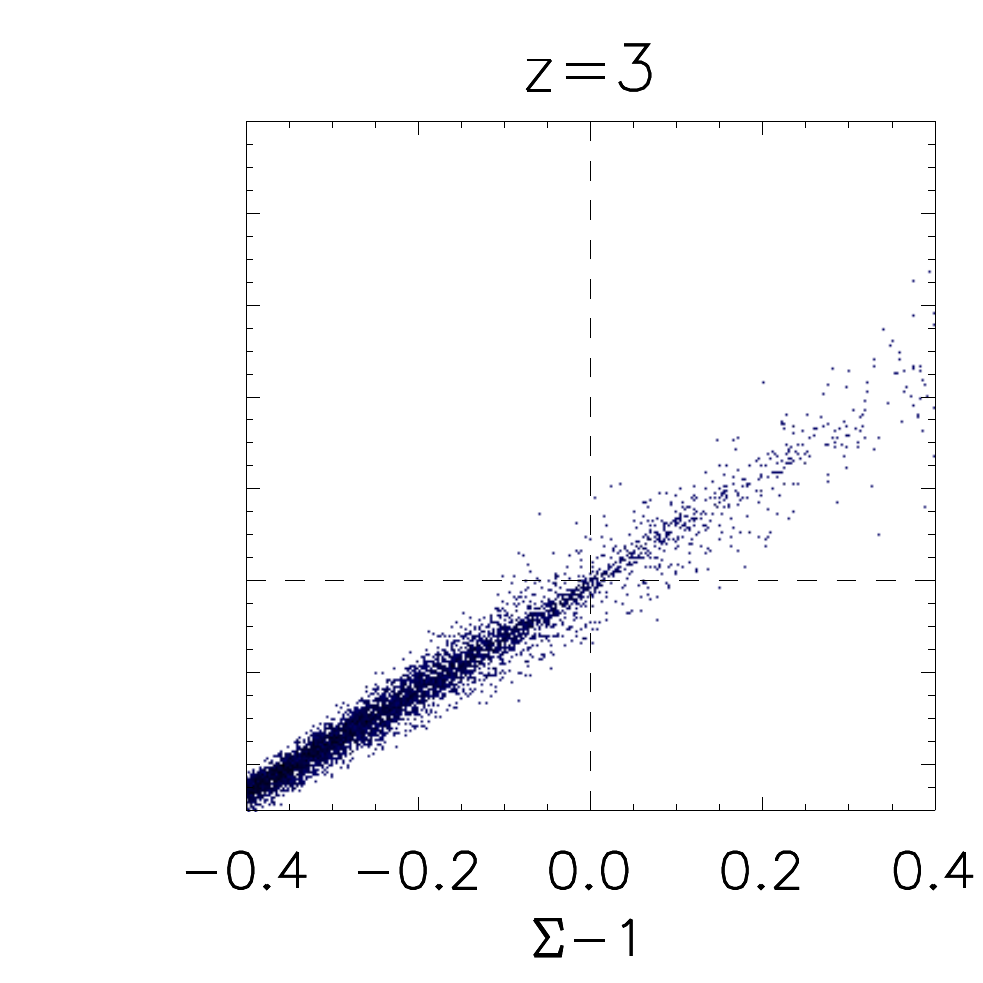}
   \vskip-2mm
   \includegraphics[scale=0.27]{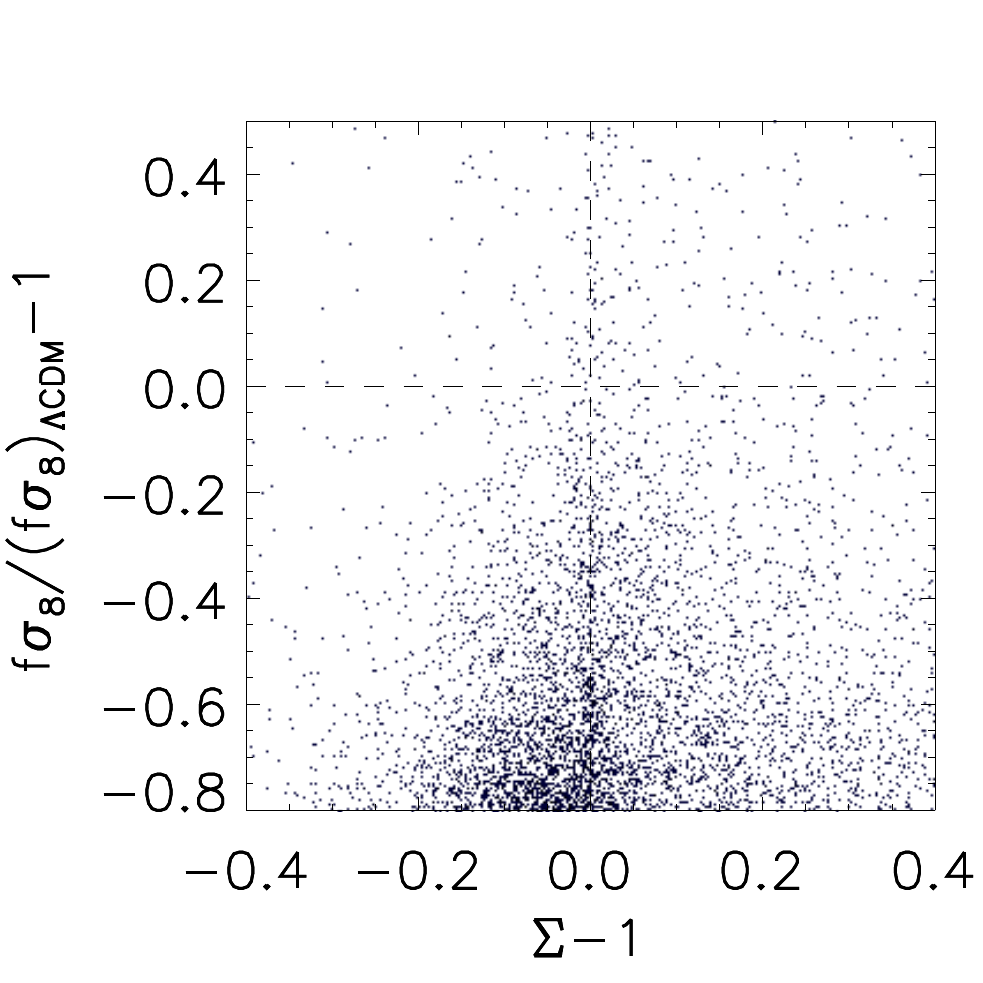}  \hskip-4mm
   \includegraphics[scale=0.27]{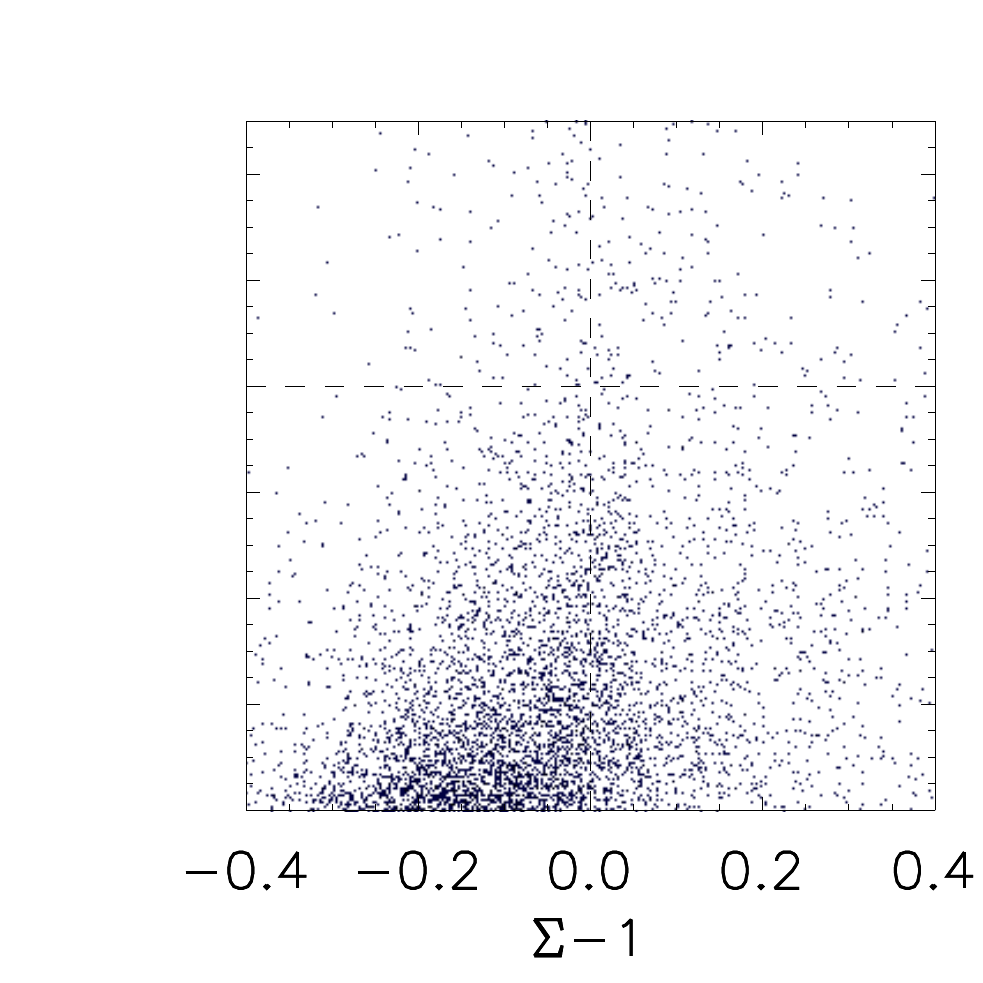}  \hskip-4mm
   \includegraphics[scale=0.27]{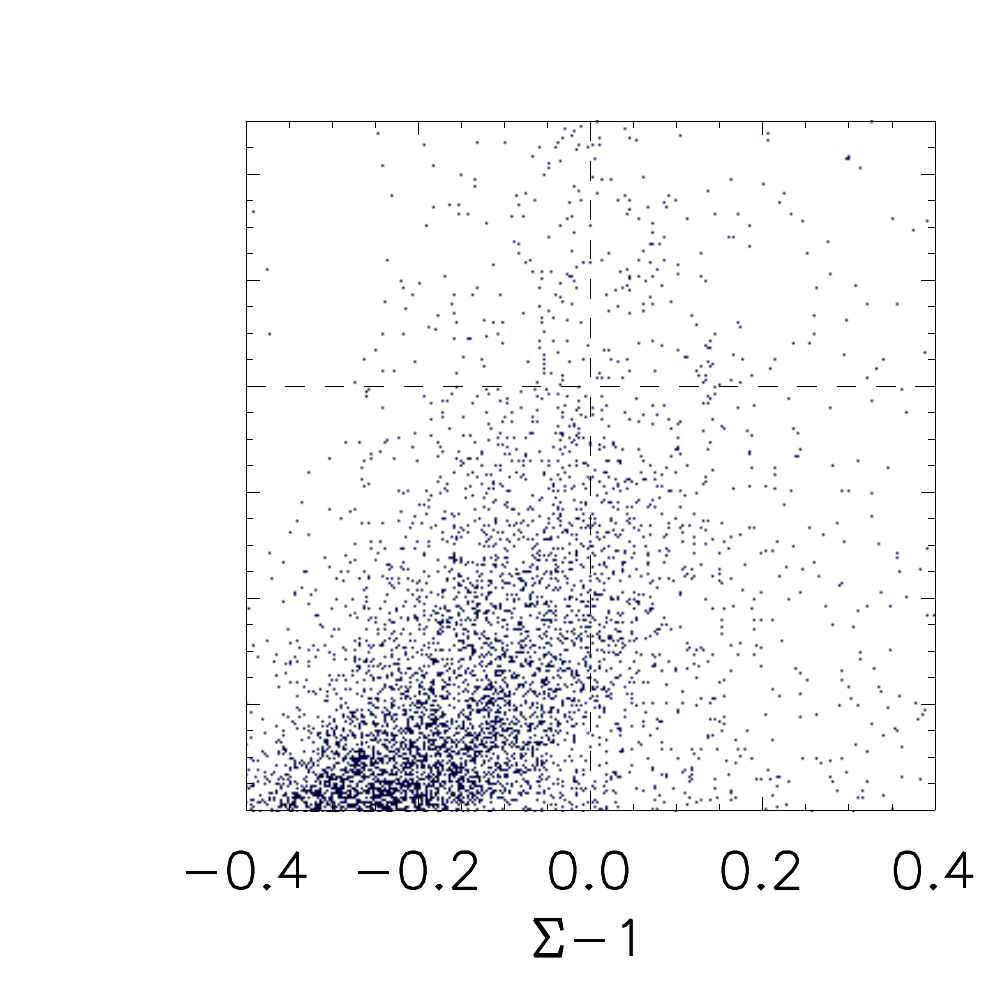}   \hskip-4mm
   \includegraphics[scale=0.27]{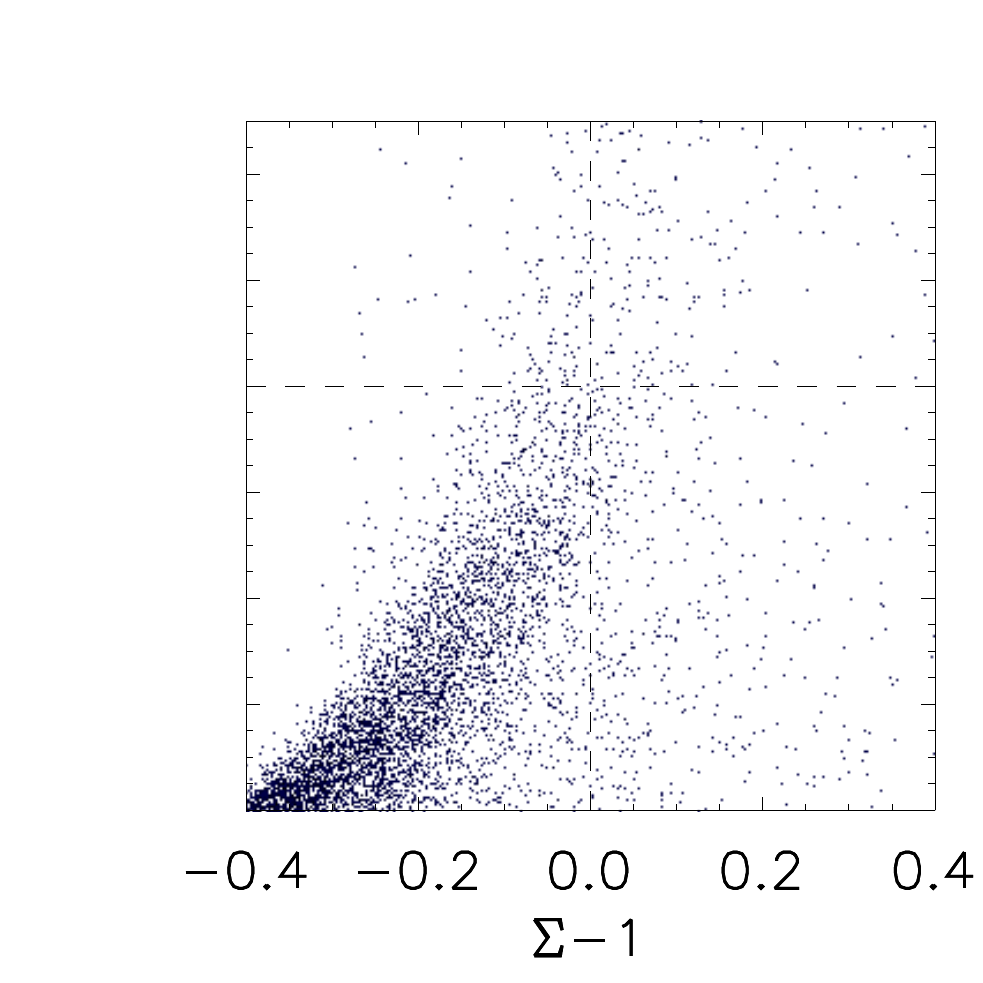}  \hskip-4mm
   \includegraphics[scale=0.27]{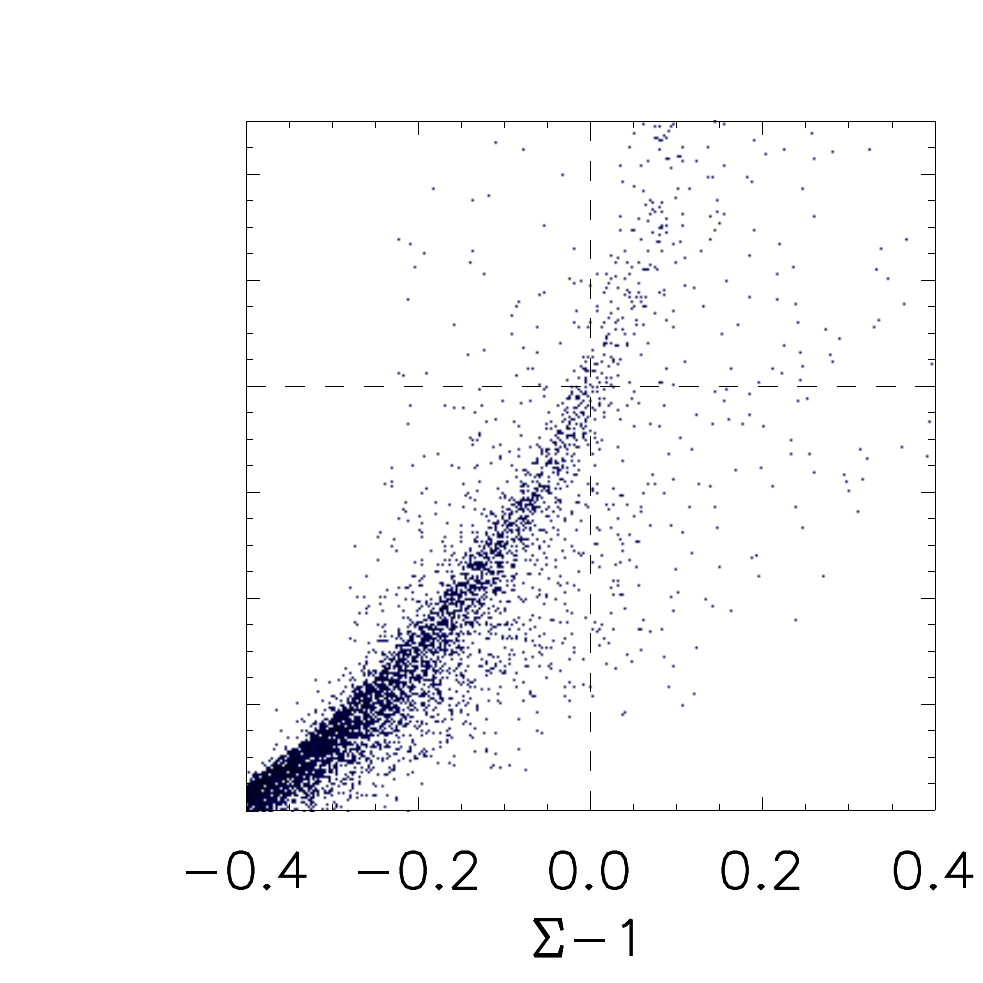}  \hskip-4mm
   \includegraphics[scale=0.27]{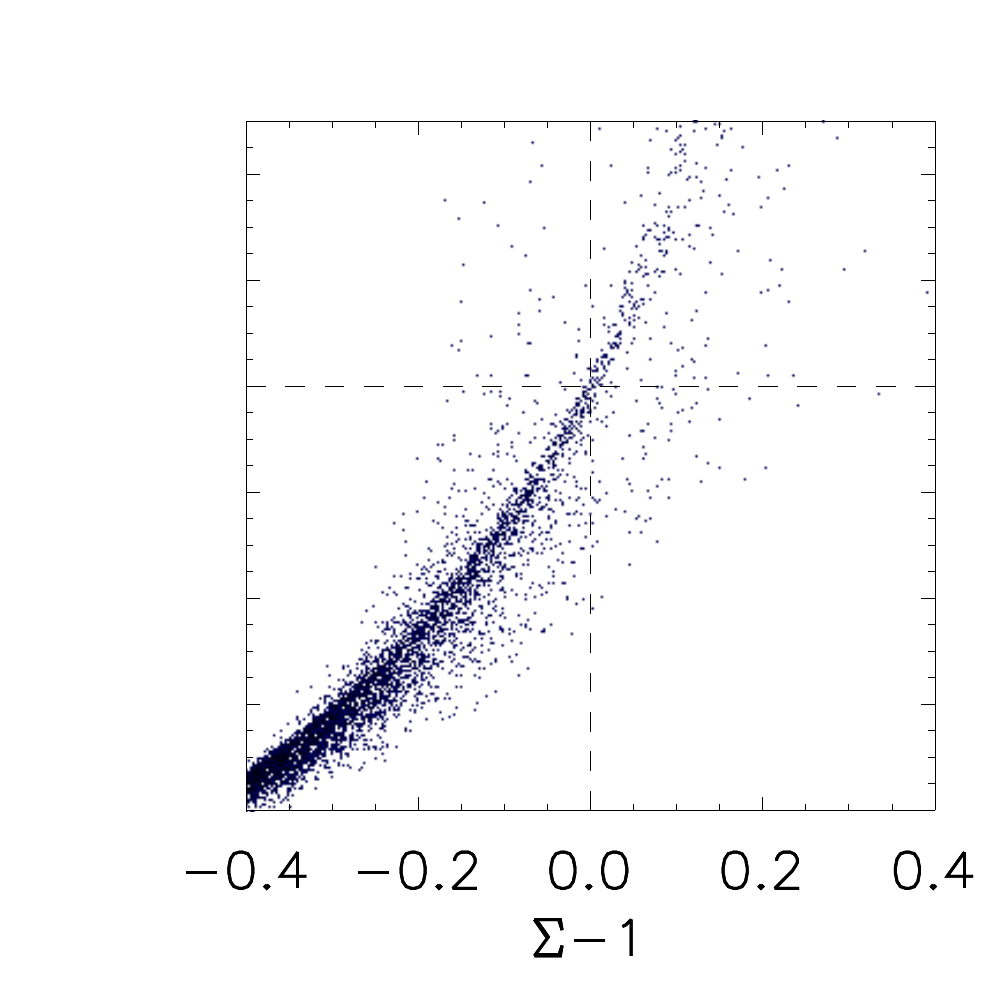}
%- - - - - - - - - - - - - - - - - - - - - - - - - - - - - - - - - - - - - - - - - - - - - - - - - - - - - - - -	
\begin{flushleft}
$S+c_s$ :
\end{flushleft}   \vskip-3mm
   \includegraphics[scale=0.27]{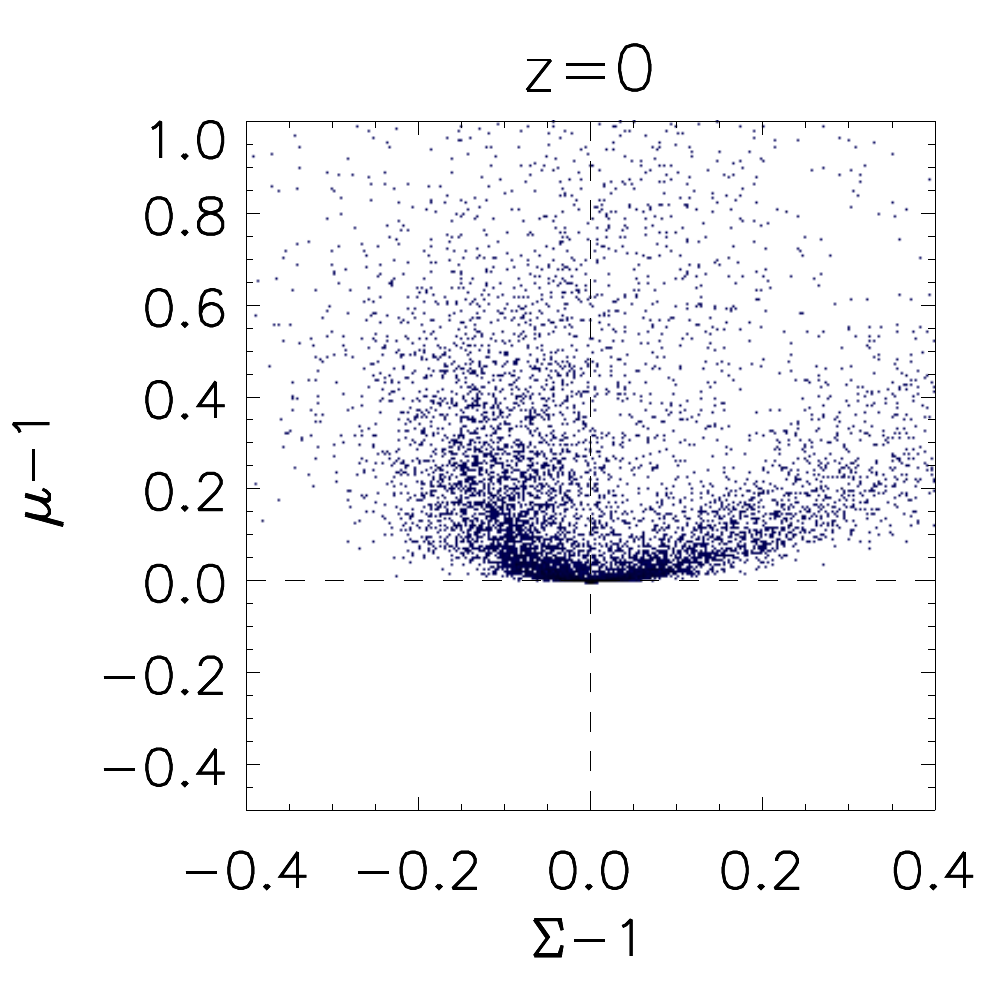}  \hskip-4mm
   \includegraphics[scale=0.27]{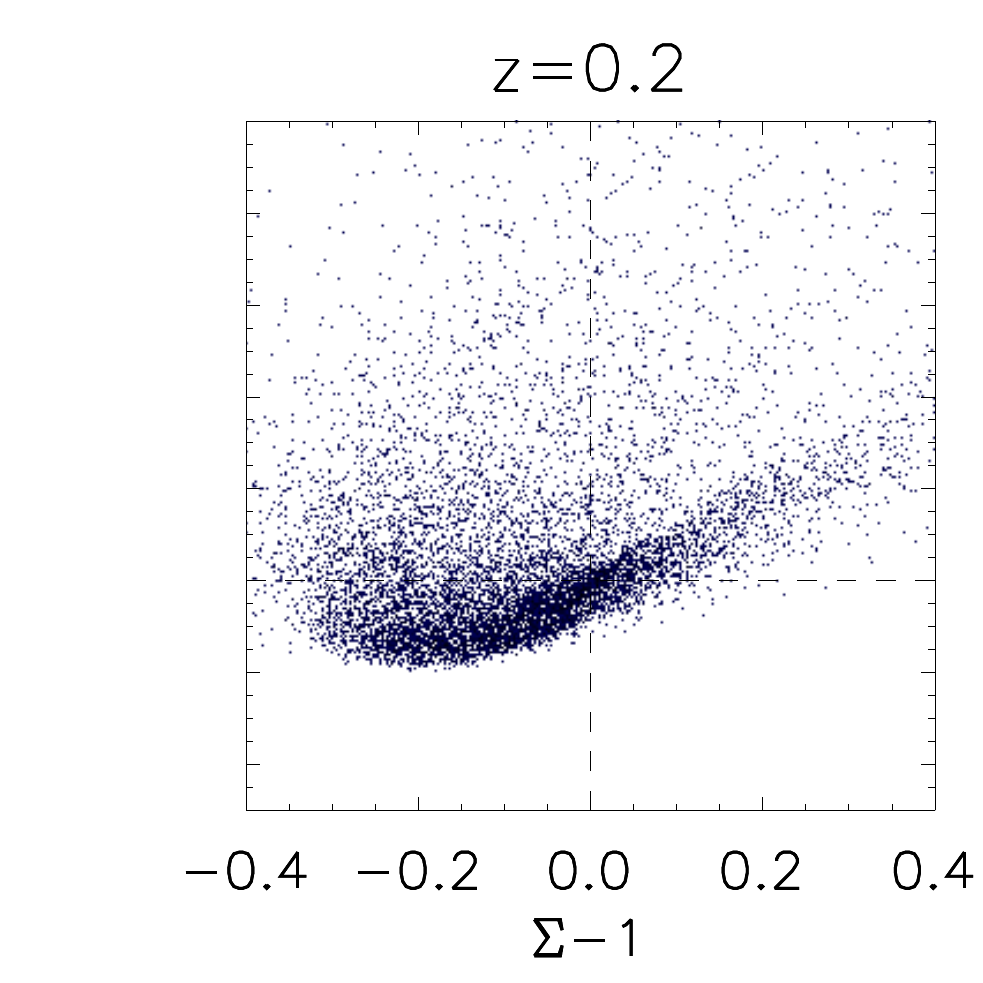}  \hskip-4mm
   \includegraphics[scale=0.27]{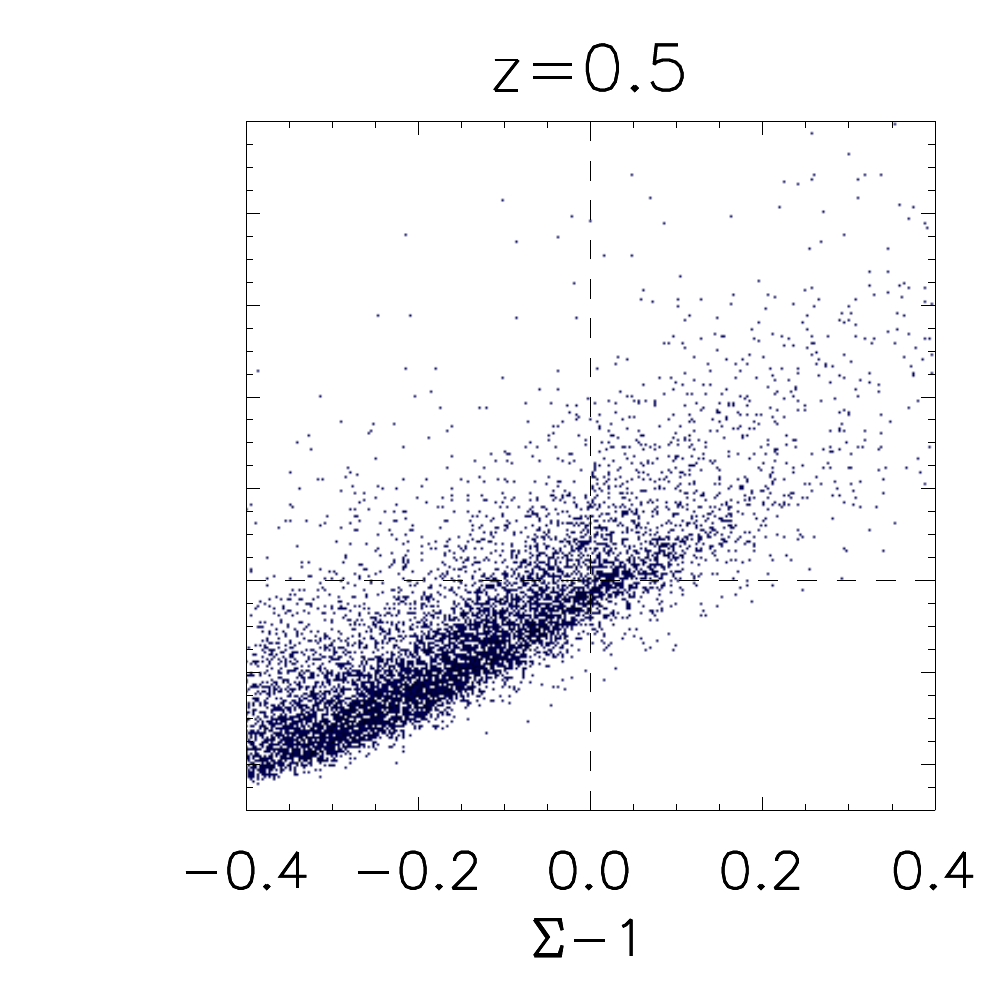}  \hskip-4mm 
   \includegraphics[scale=0.27]{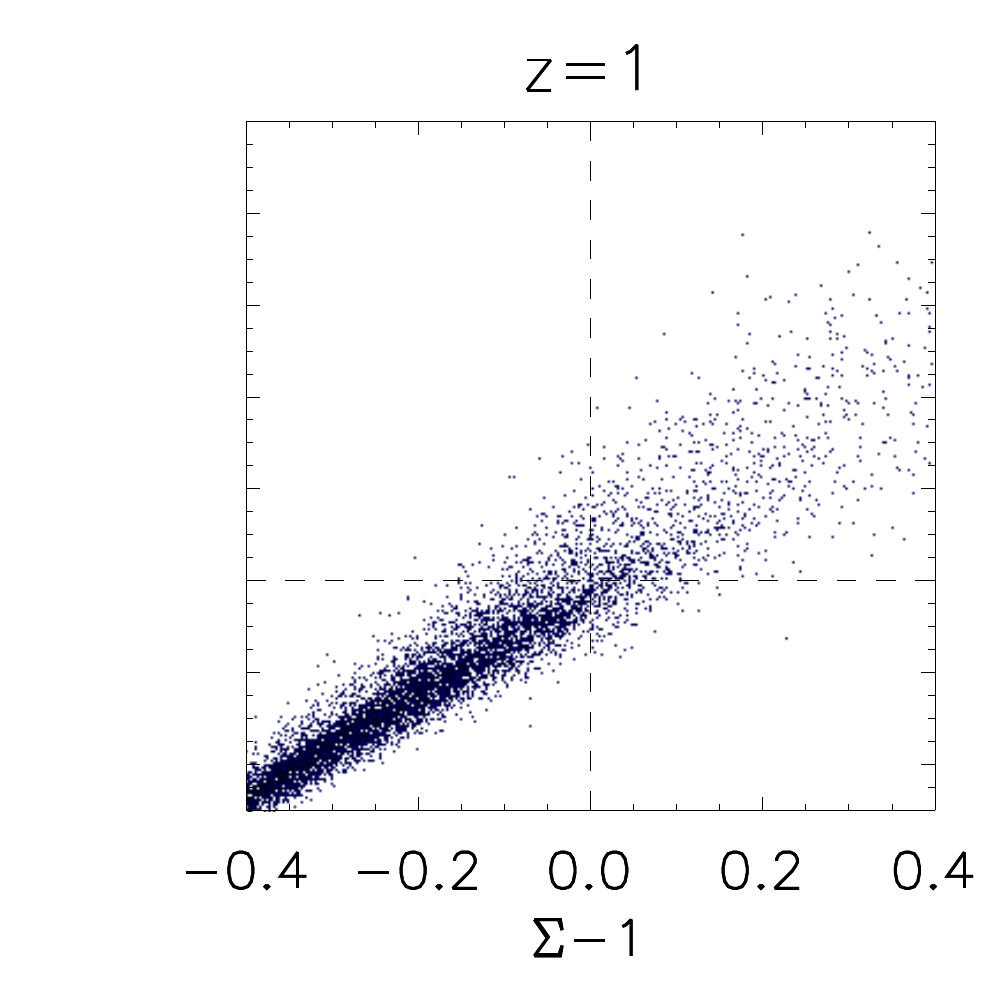}  \hskip-4mm
   \includegraphics[scale=0.27]{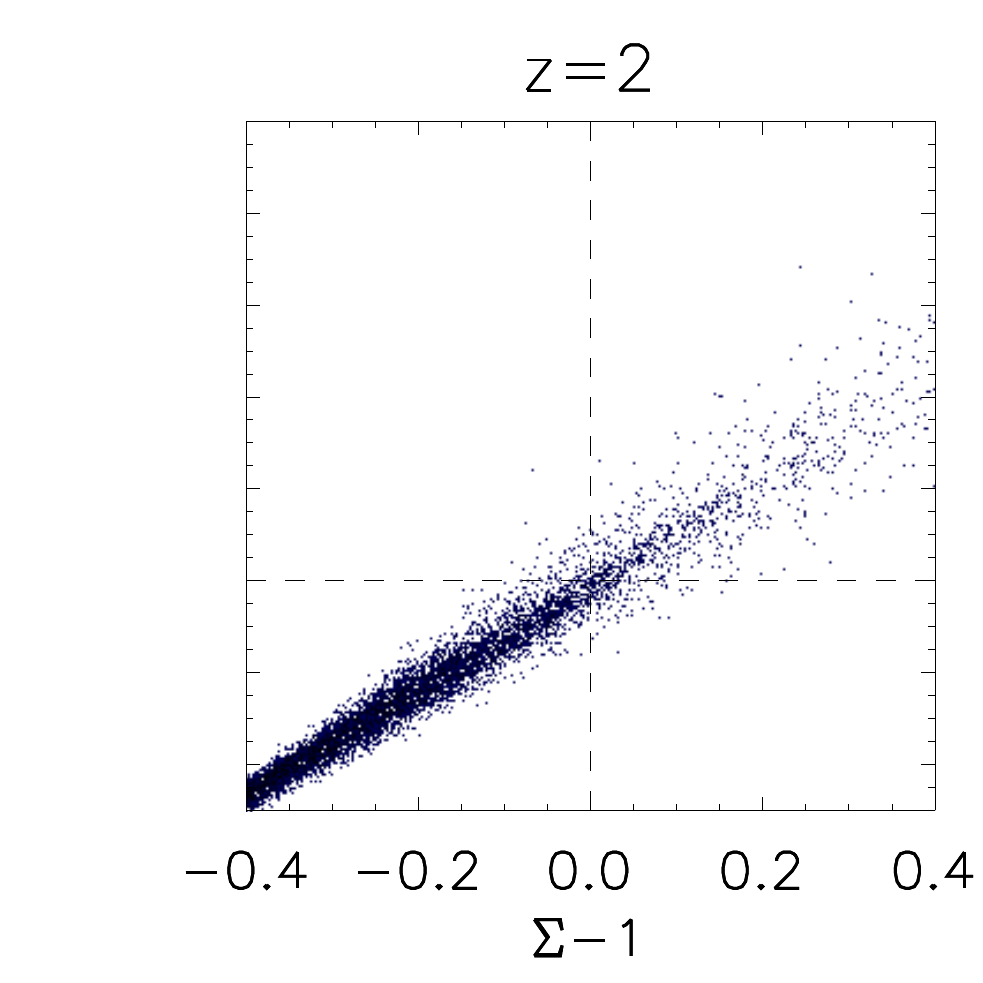}  \hskip-4mm
   \includegraphics[scale=0.27]{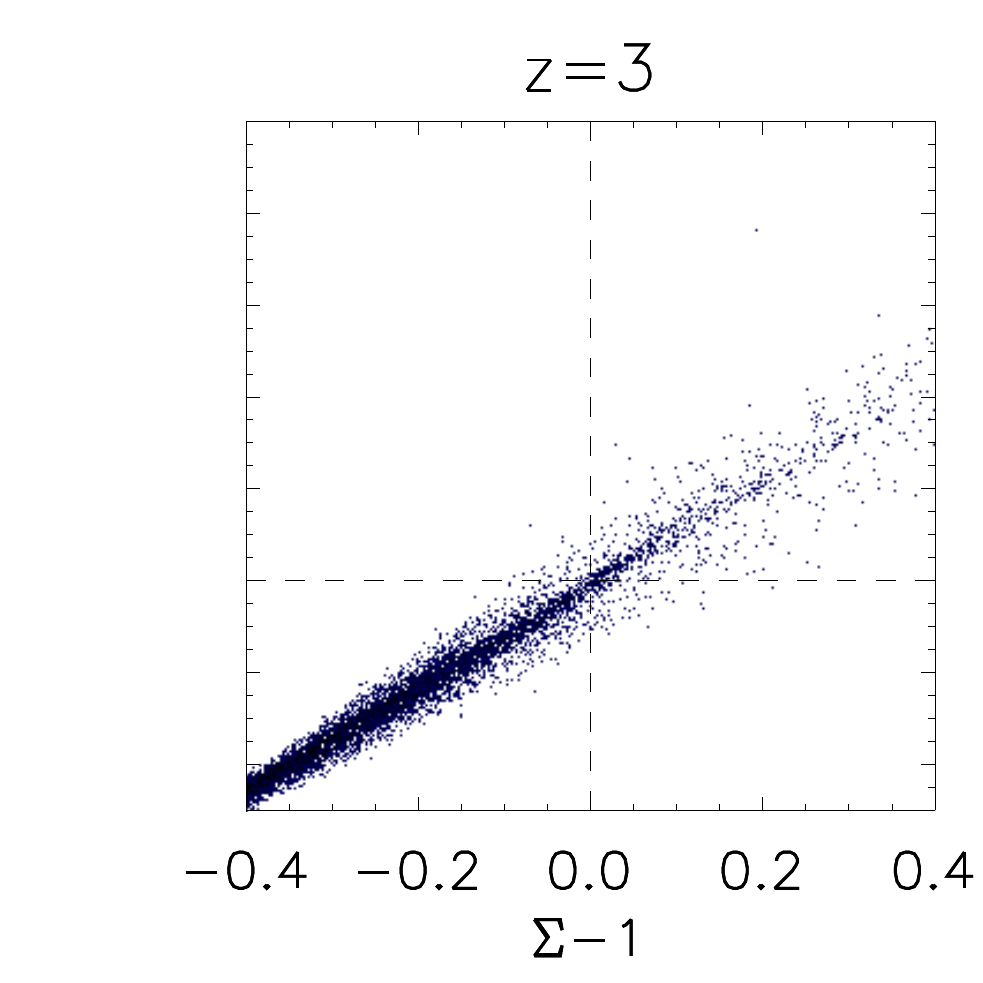}
   \vskip-2mm
   \includegraphics[scale=0.27]{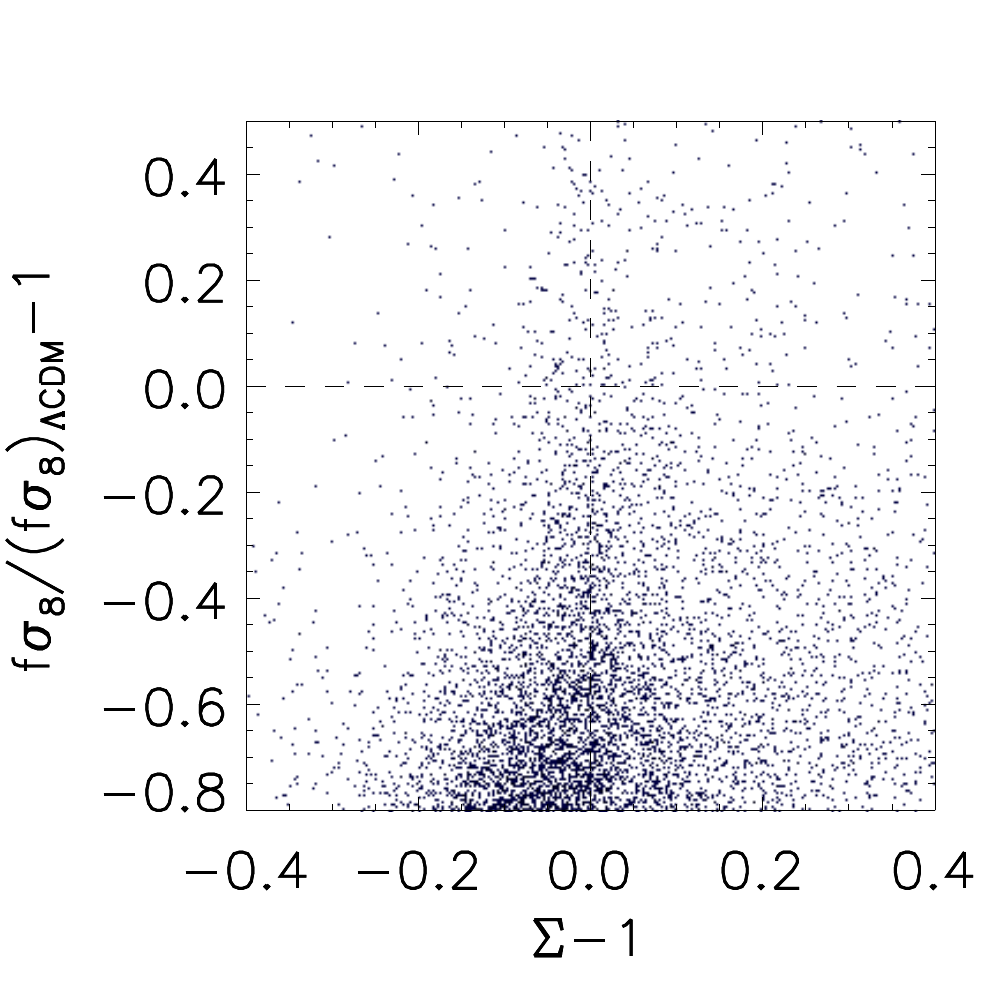}  \hskip-4mm
   \includegraphics[scale=0.27]{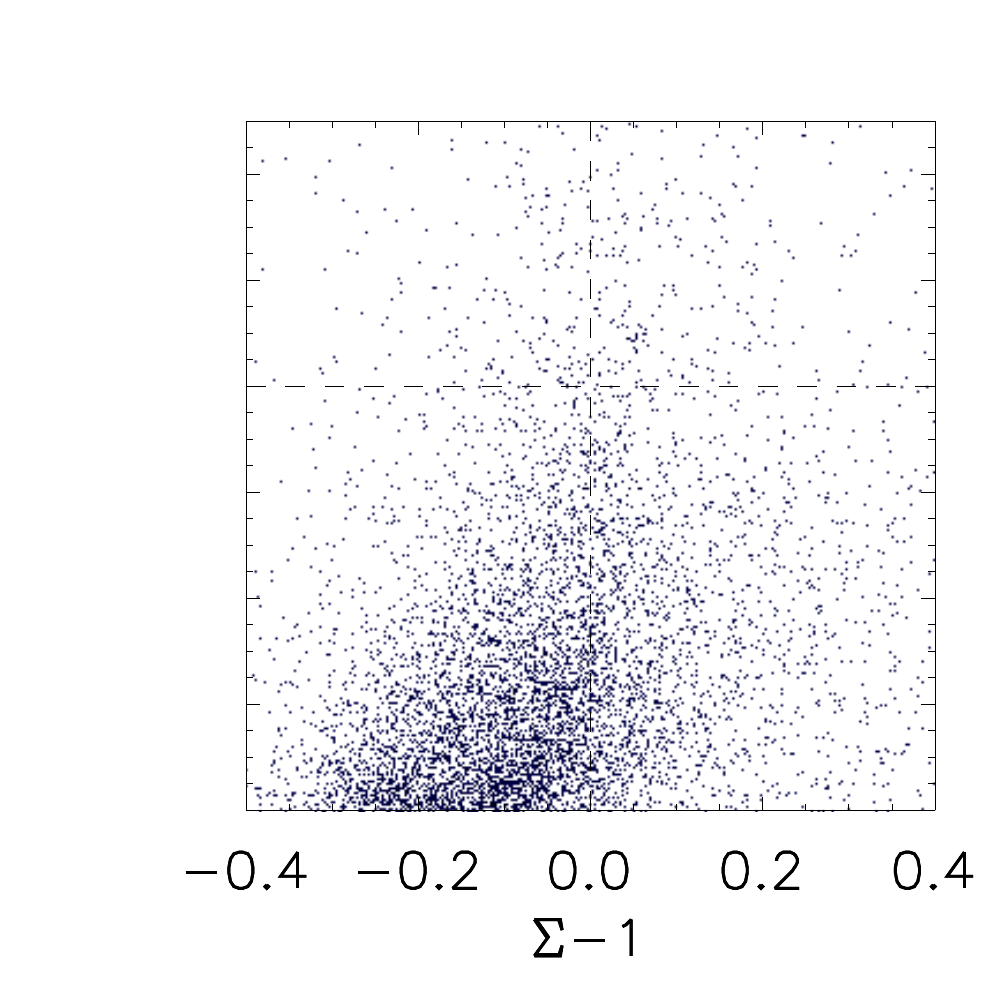}  \hskip-4mm
   \includegraphics[scale=0.27]{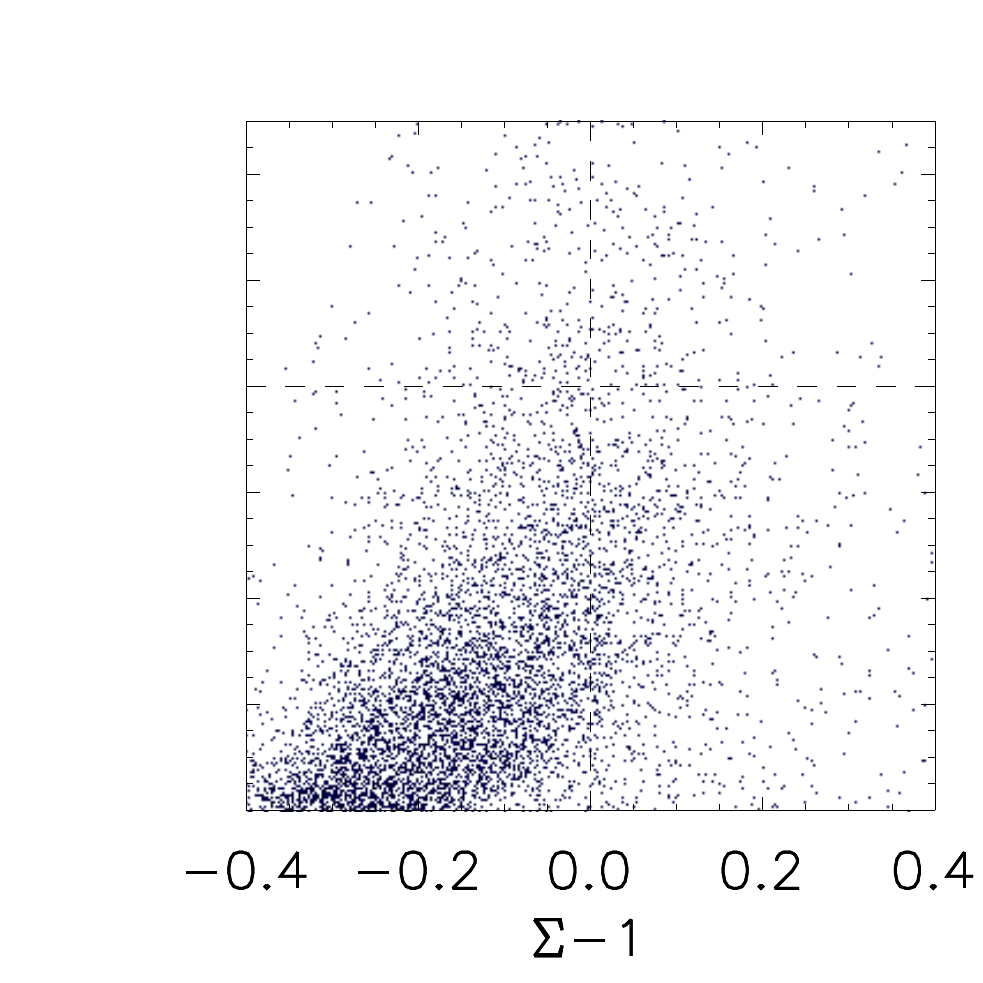}  \hskip-4mm 
   \includegraphics[scale=0.27]{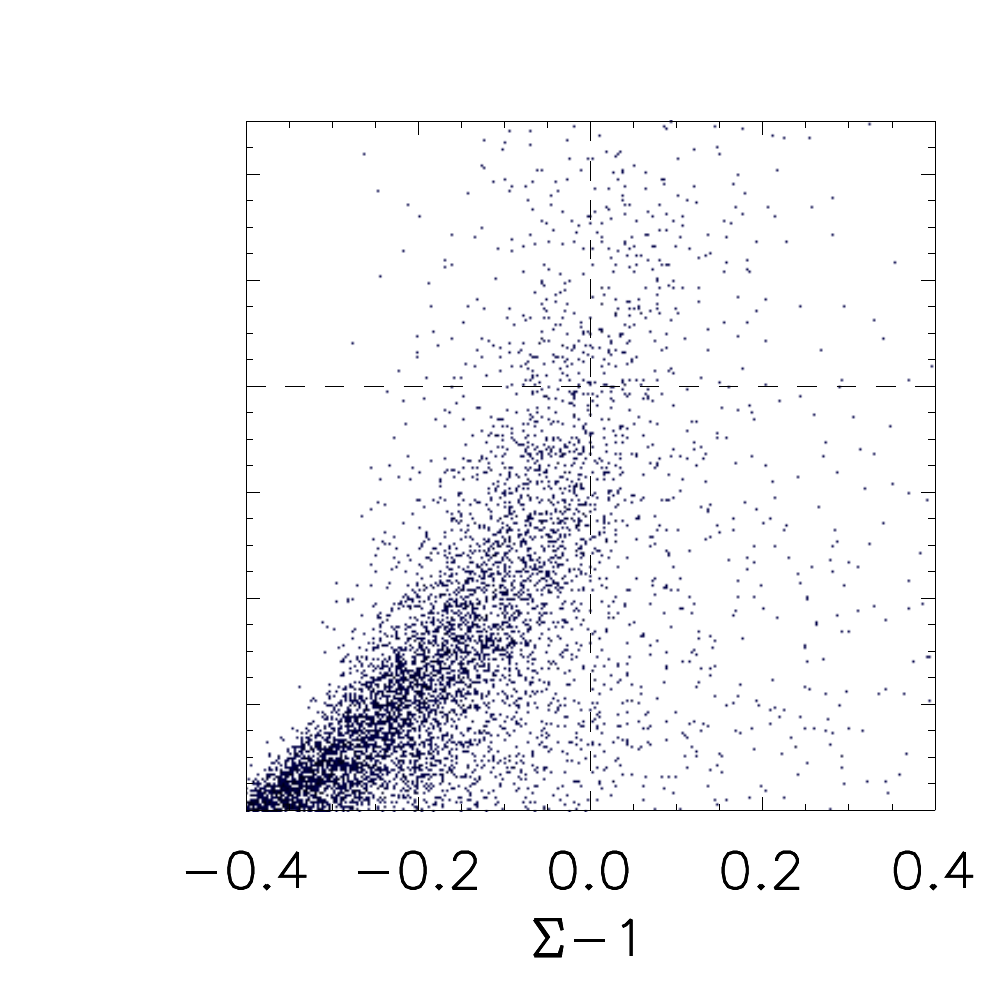}  \hskip-4mm
   \includegraphics[scale=0.27]{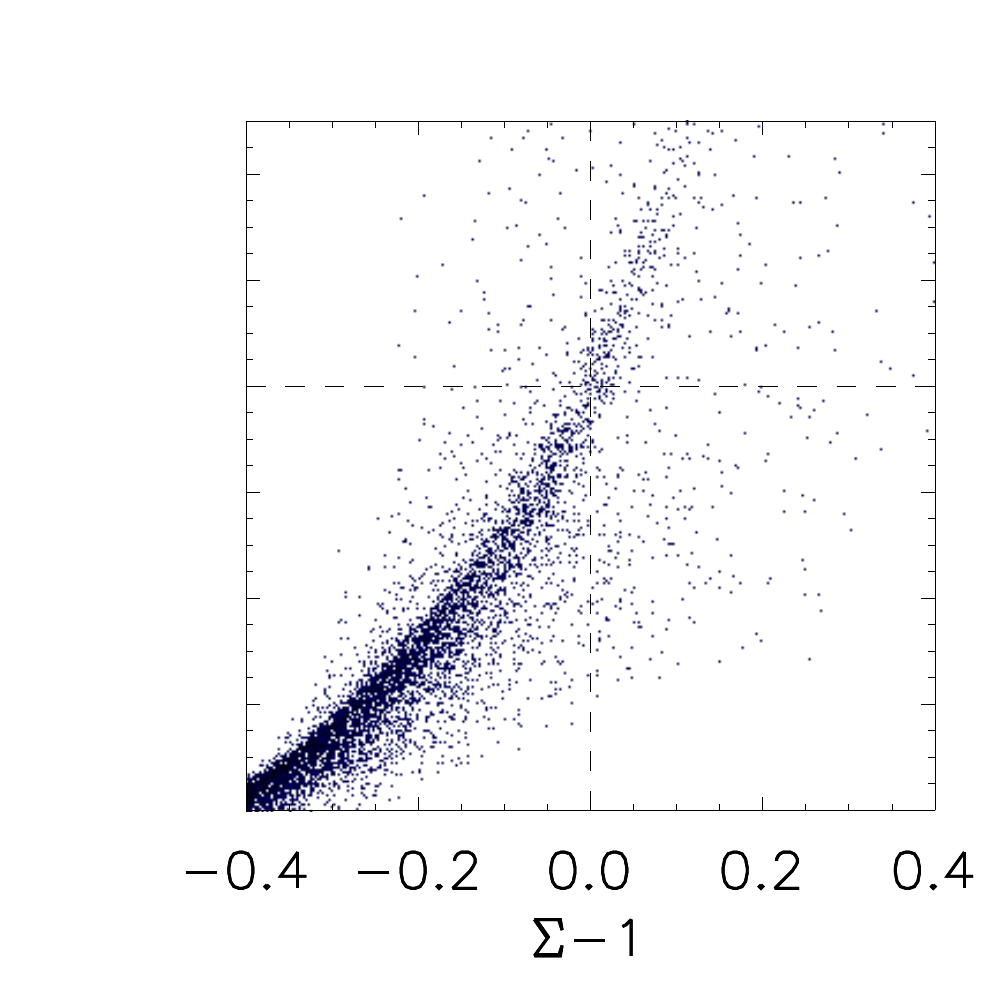}  \hskip-4mm
   \includegraphics[scale=0.27]{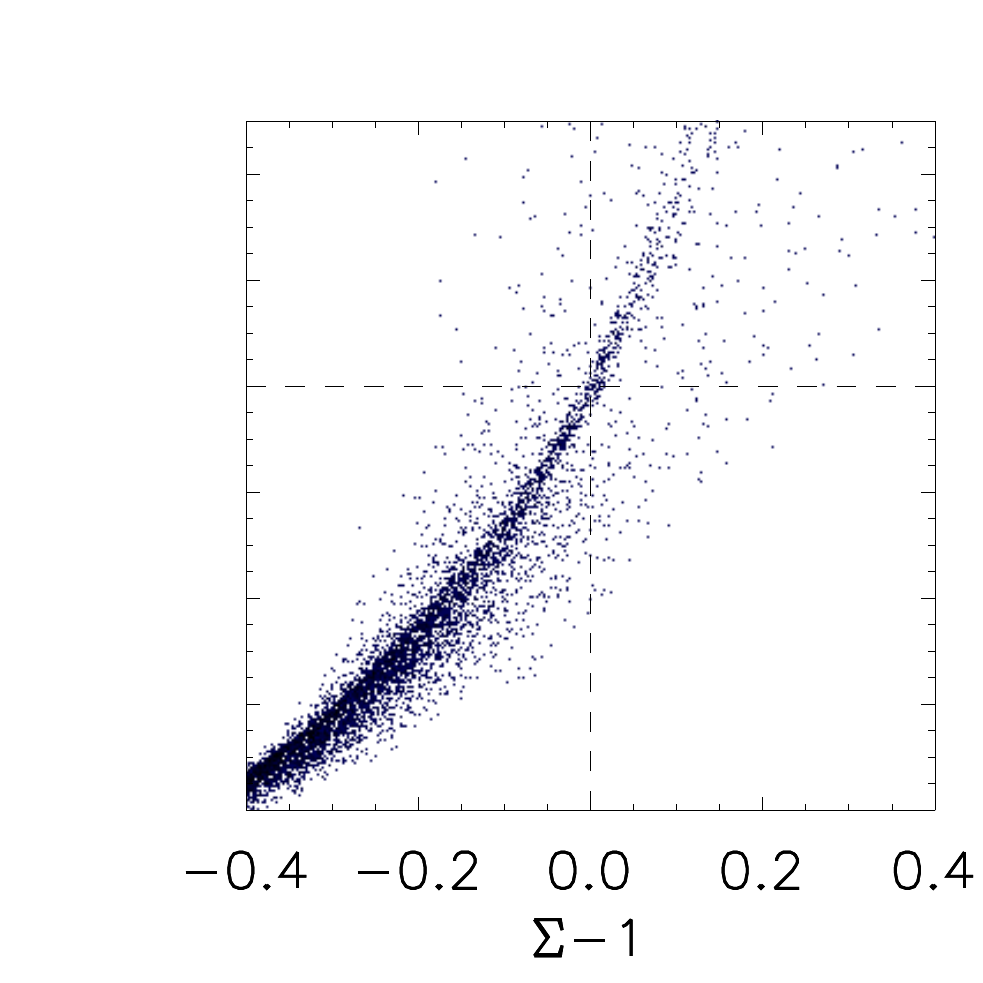}
%- - - - - - - - - - - - - - - - - - - - - - - - - - - - - - - - - - - - - - - - - - - - - - - - - - - - - - - -	
\begin{flushleft}
$S+c_T$ :
\end{flushleft}   \vskip-3mm
   \includegraphics[scale=0.27]{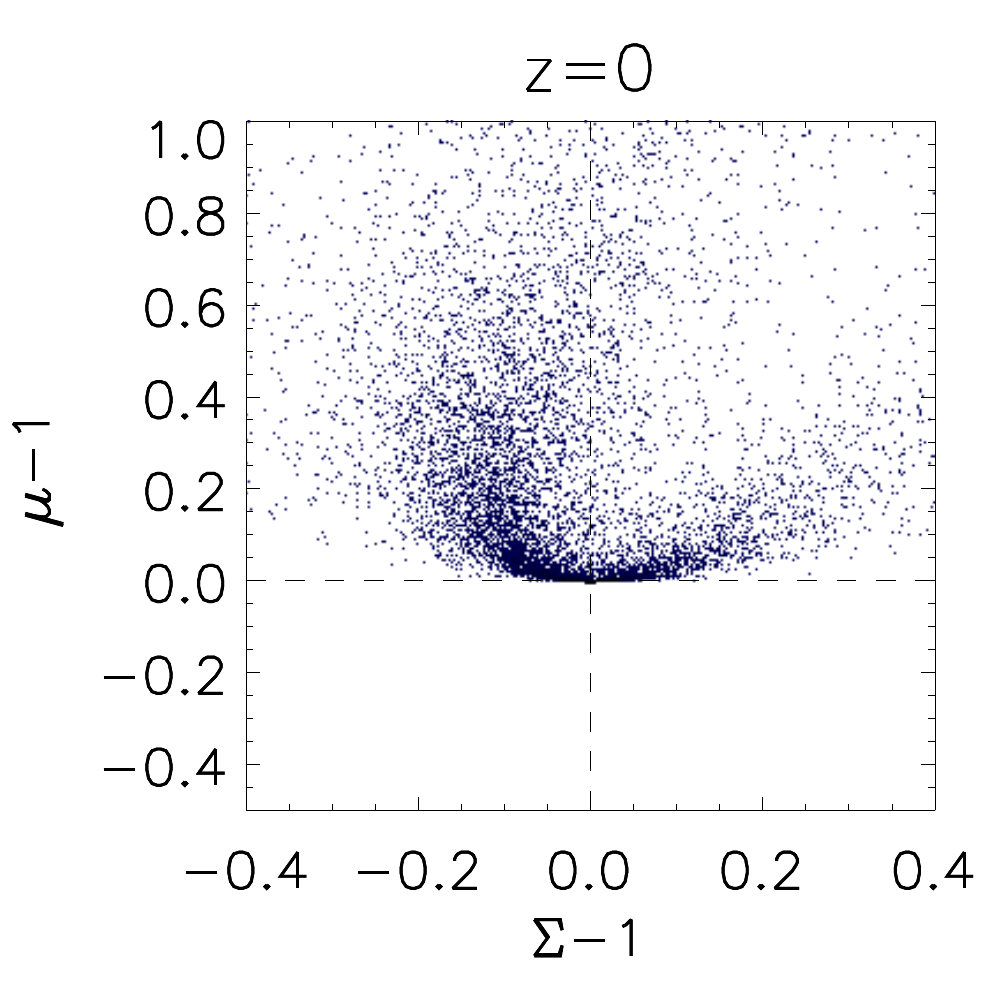}  \hskip-4mm
   \includegraphics[scale=0.27]{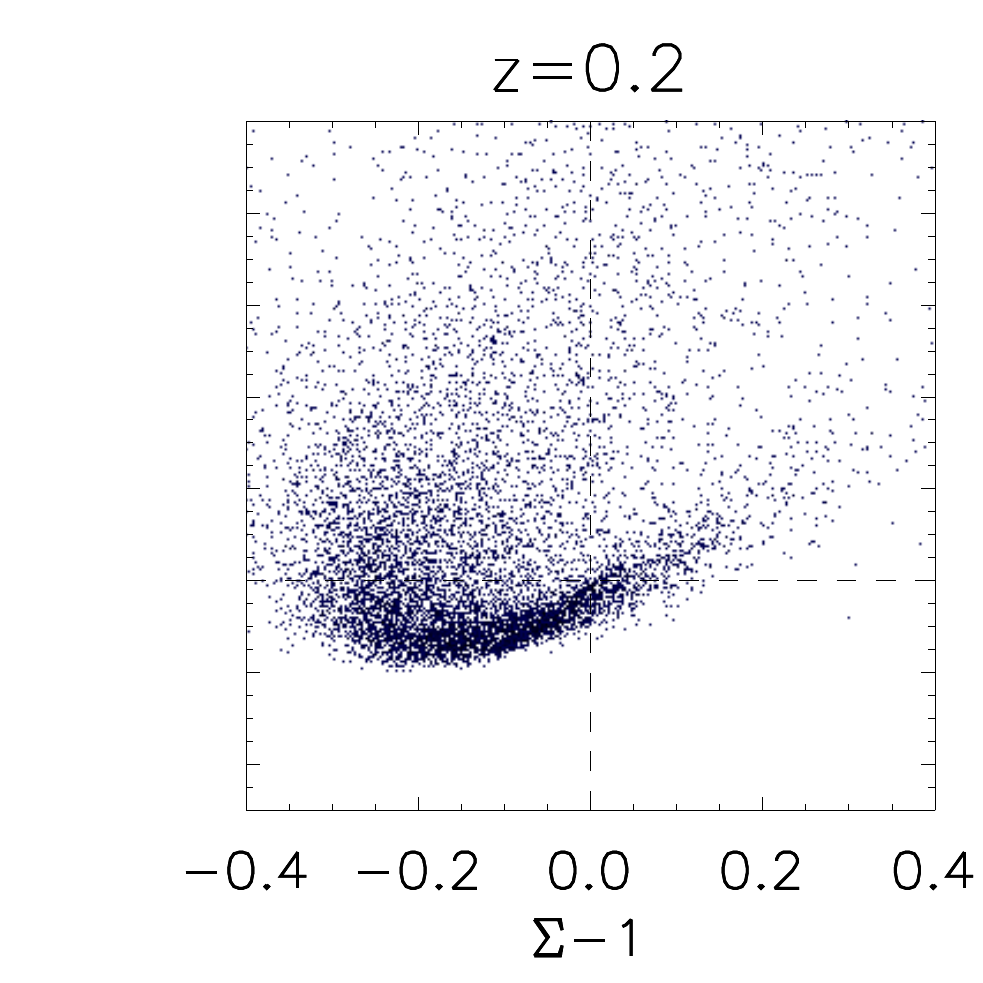}  \hskip-4mm
   \includegraphics[scale=0.27]{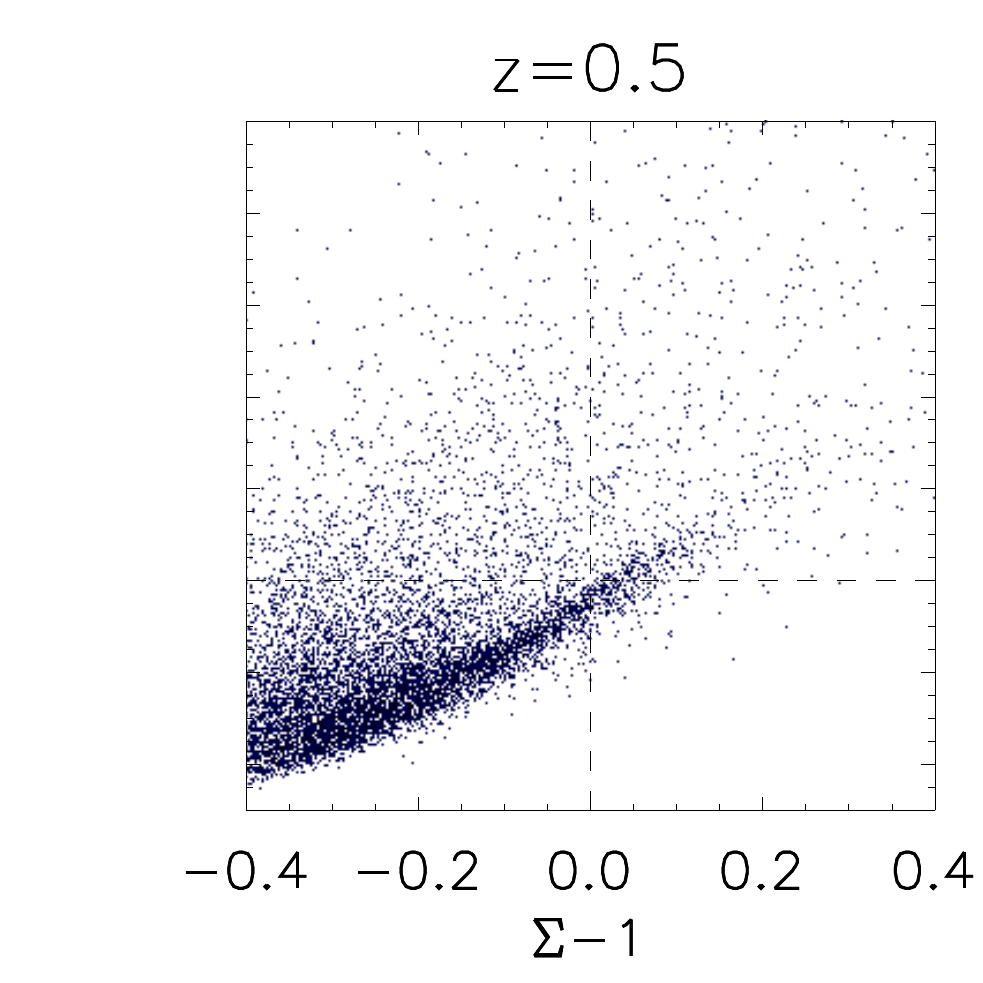}  \hskip-4mm 
   \includegraphics[scale=0.27]{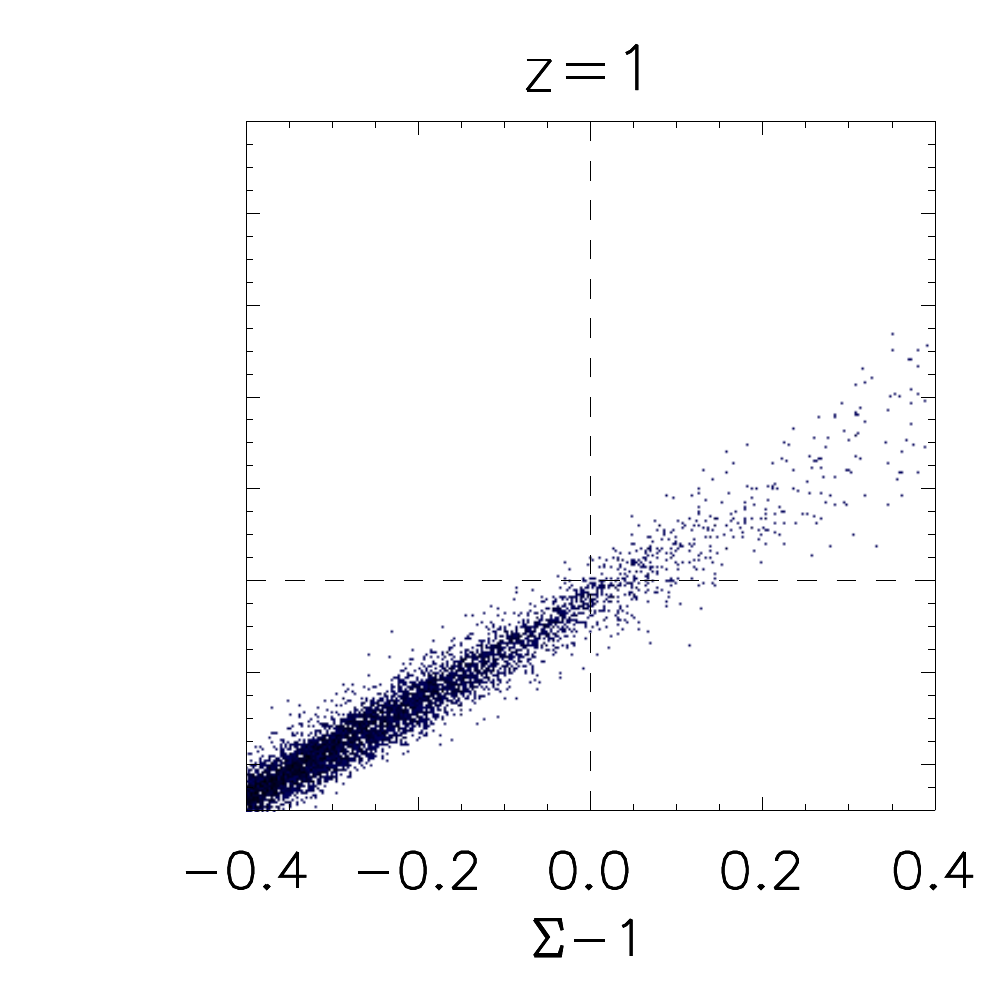}  \hskip-4mm
   \includegraphics[scale=0.27]{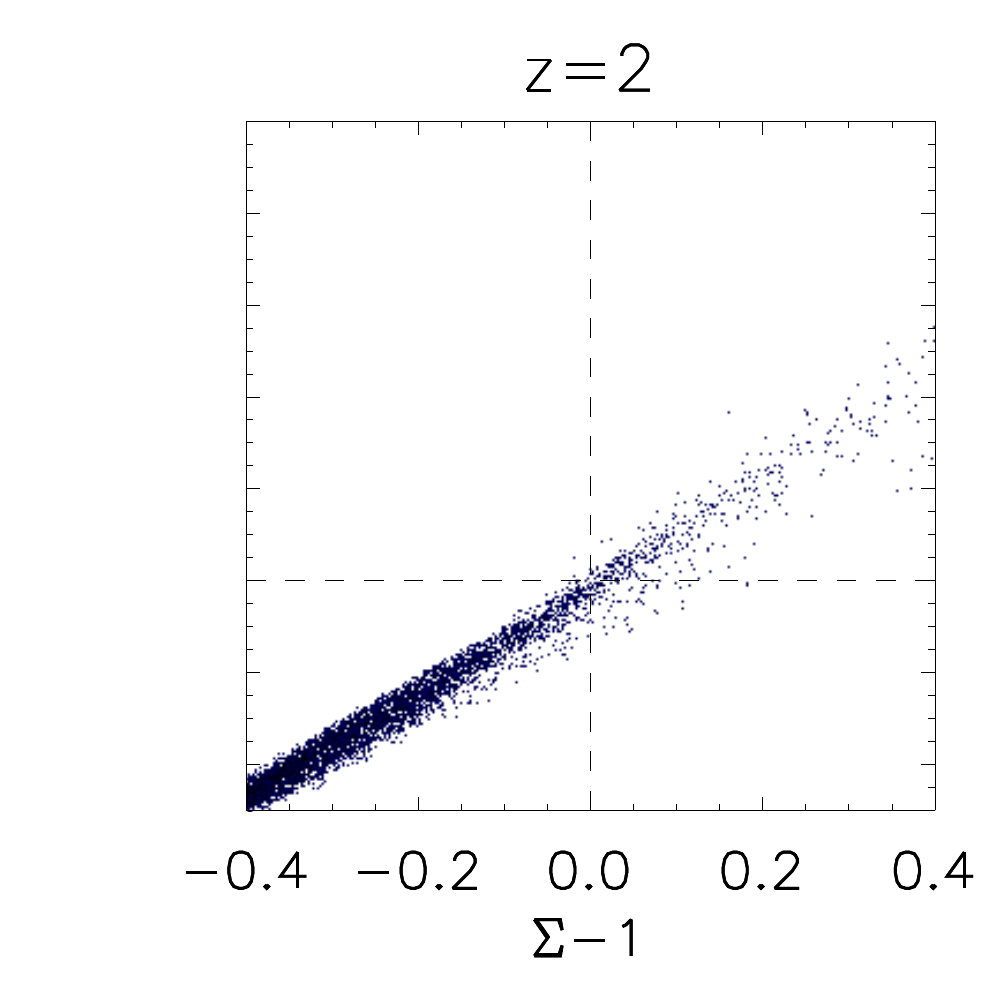}  \hskip-4mm
   \includegraphics[scale=0.27]{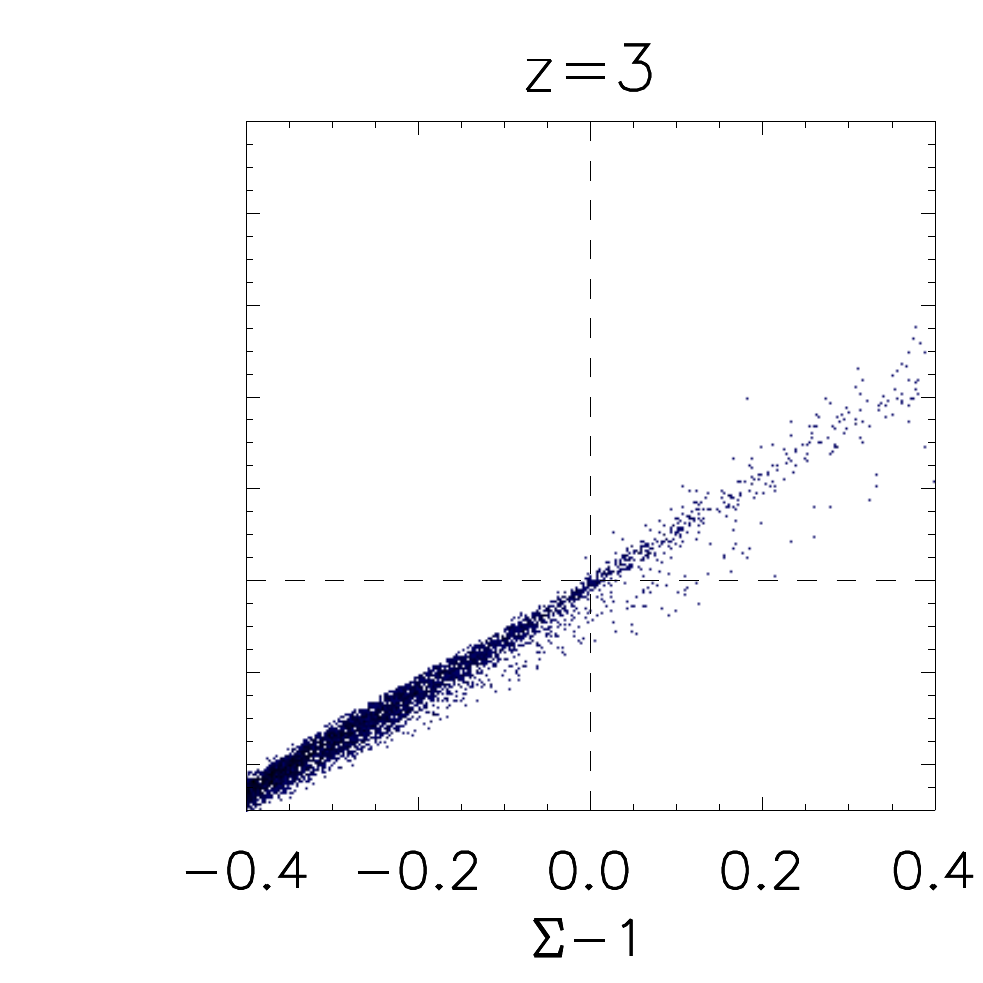}
   \vskip-2mm
   \includegraphics[scale=0.27]{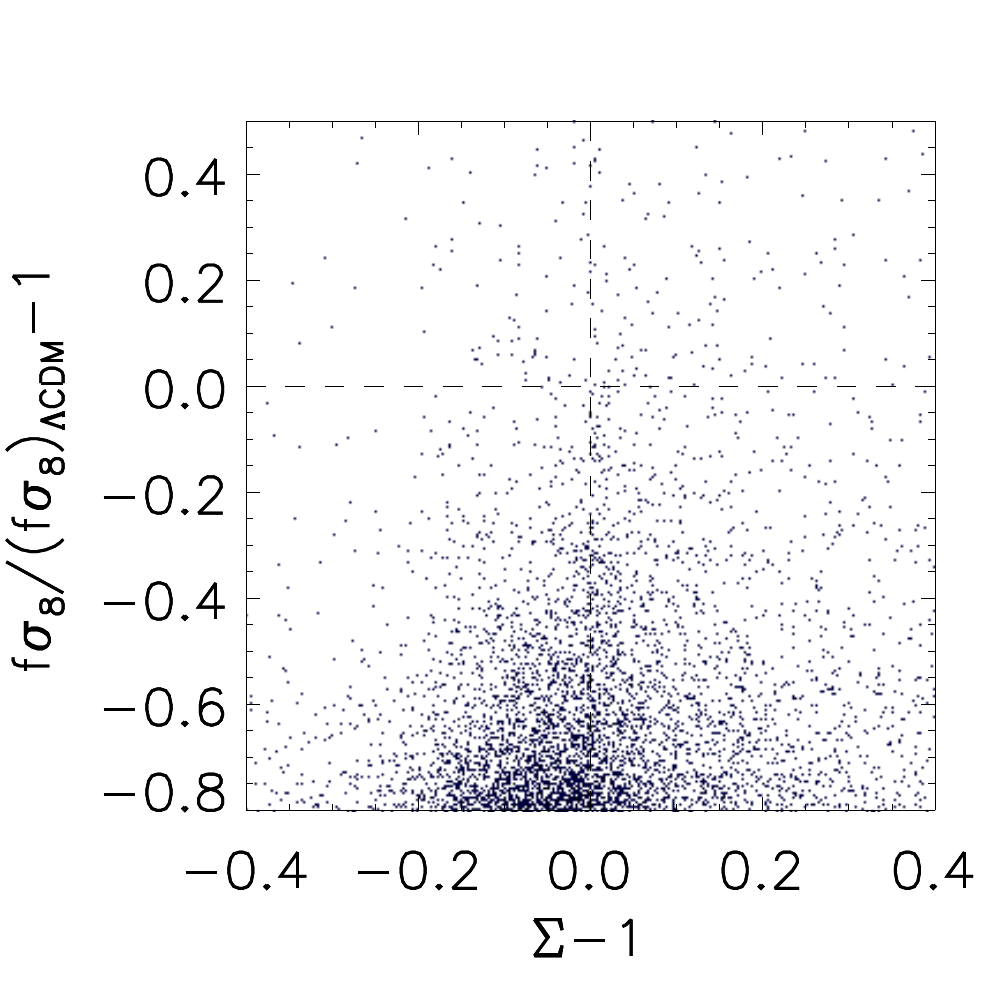}  \hskip-4mm
   \includegraphics[scale=0.27]{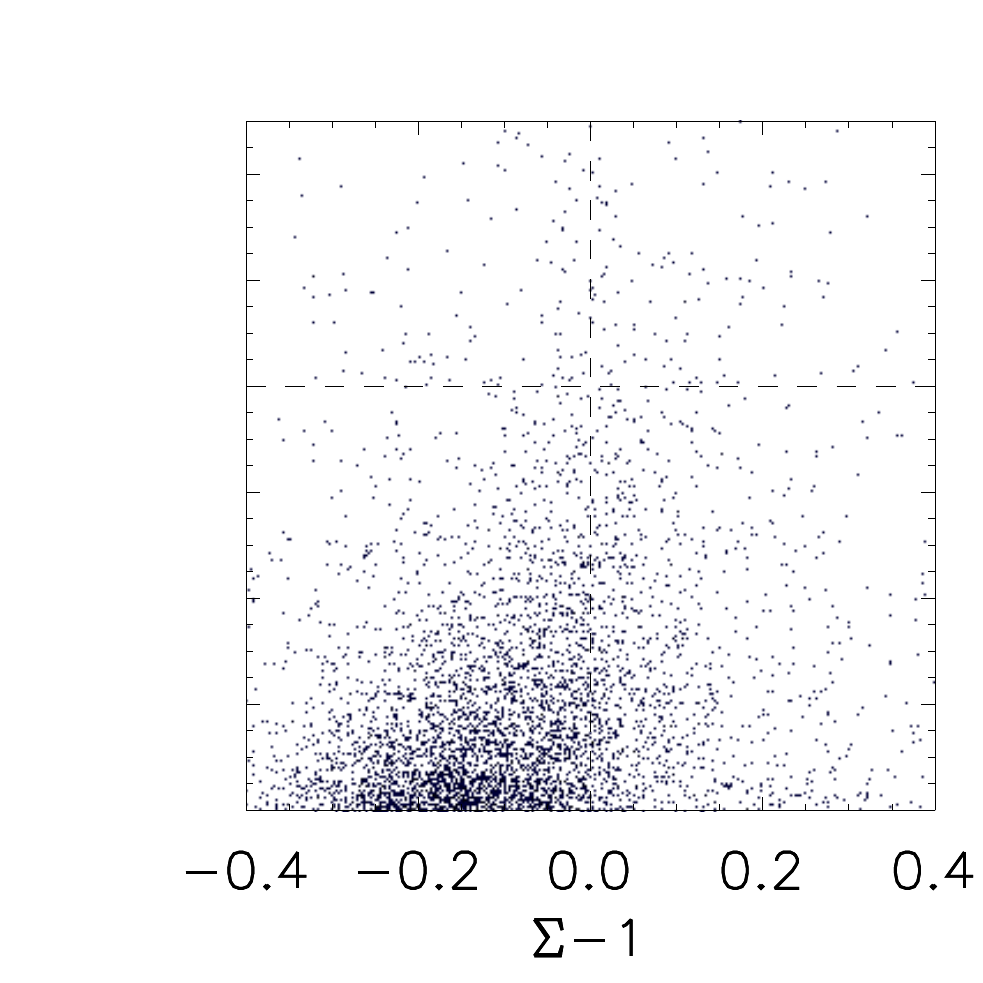}  \hskip-4mm
   \includegraphics[scale=0.27]{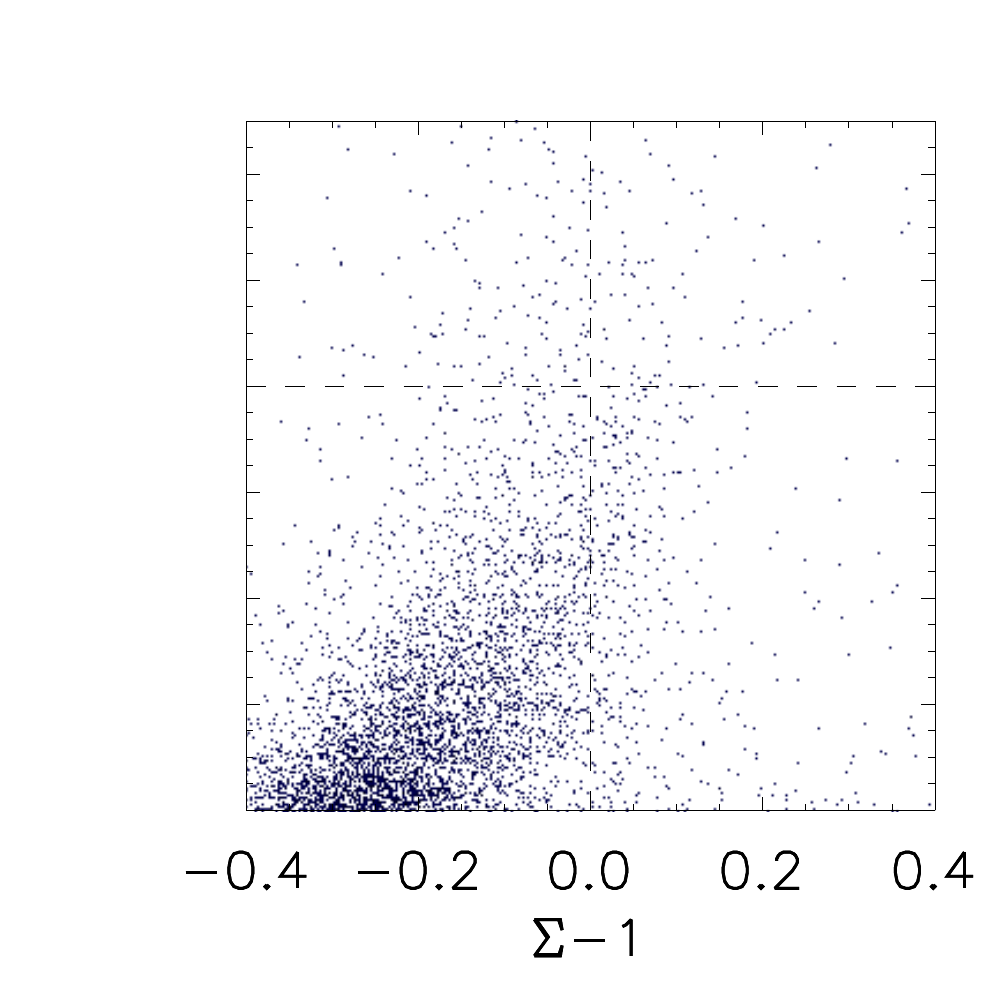}  \hskip-4mm 
   \includegraphics[scale=0.27]{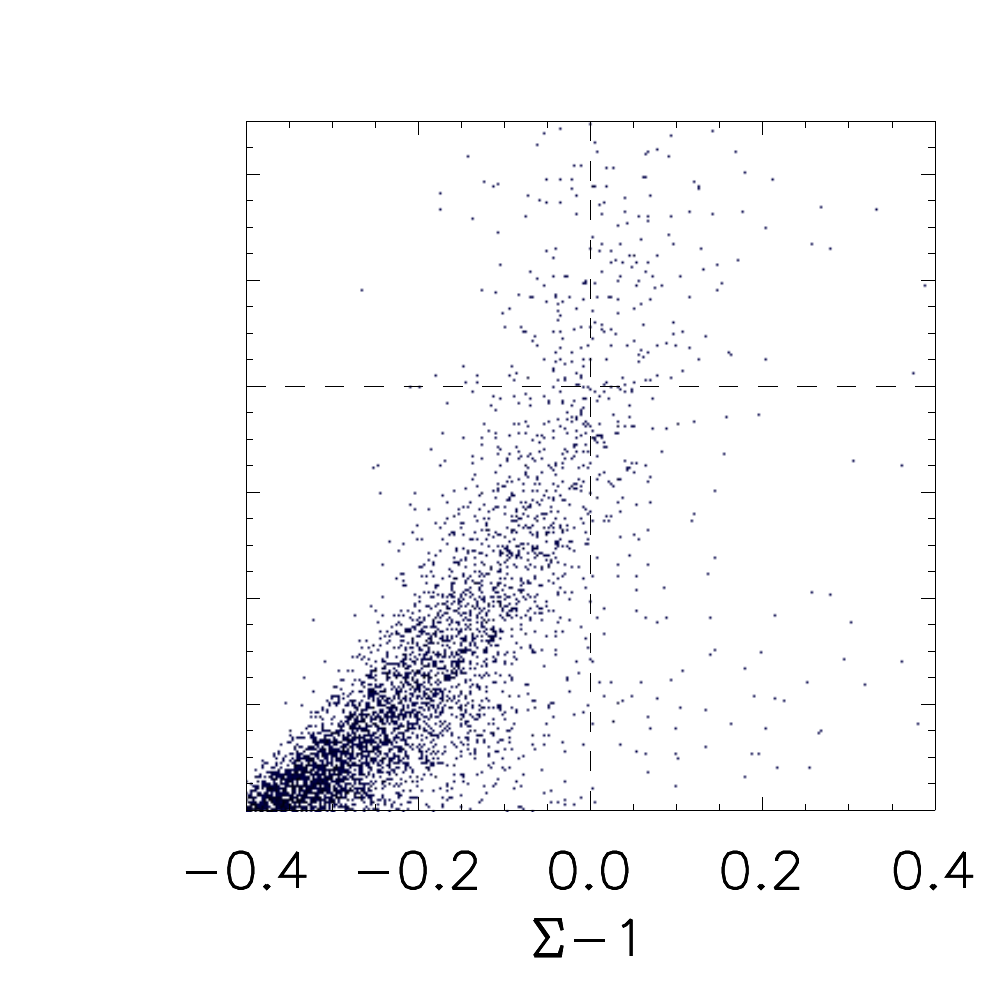}  \hskip-4mm
   \includegraphics[scale=0.27]{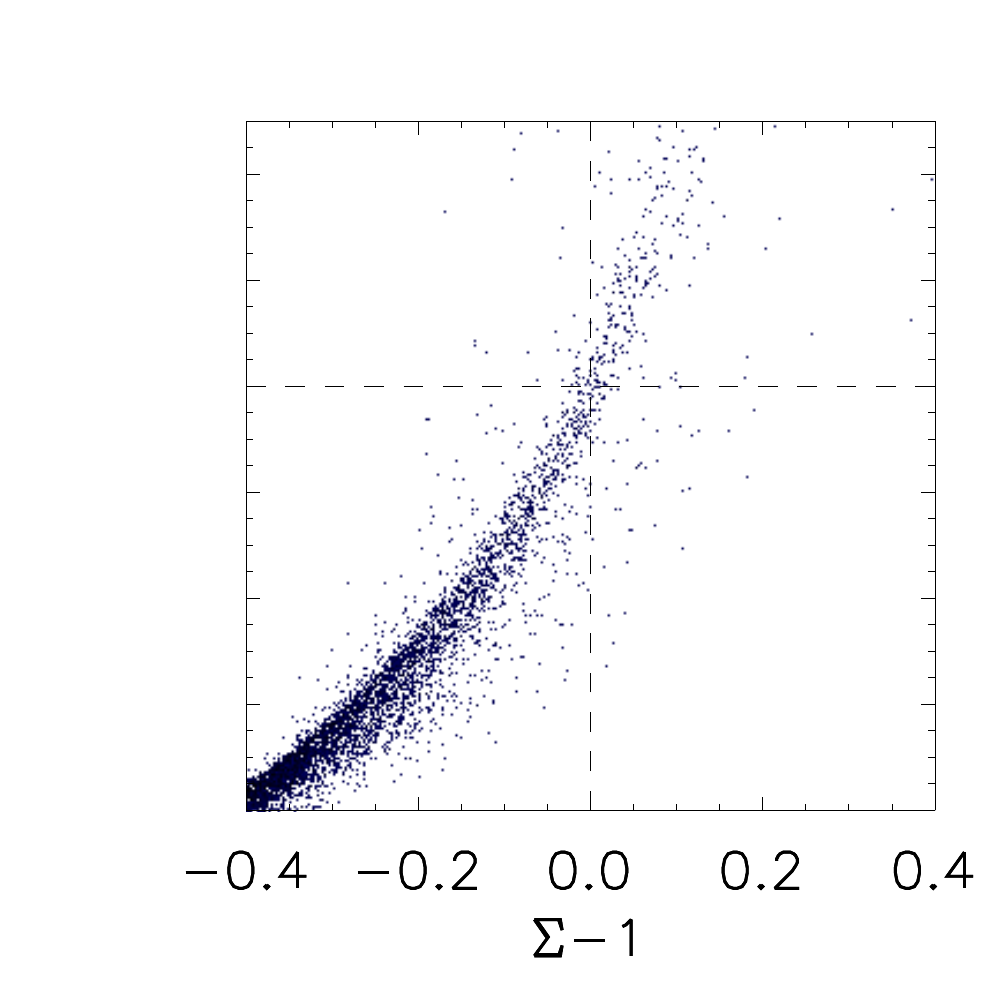}  \hskip-4mm
   \includegraphics[scale=0.27]{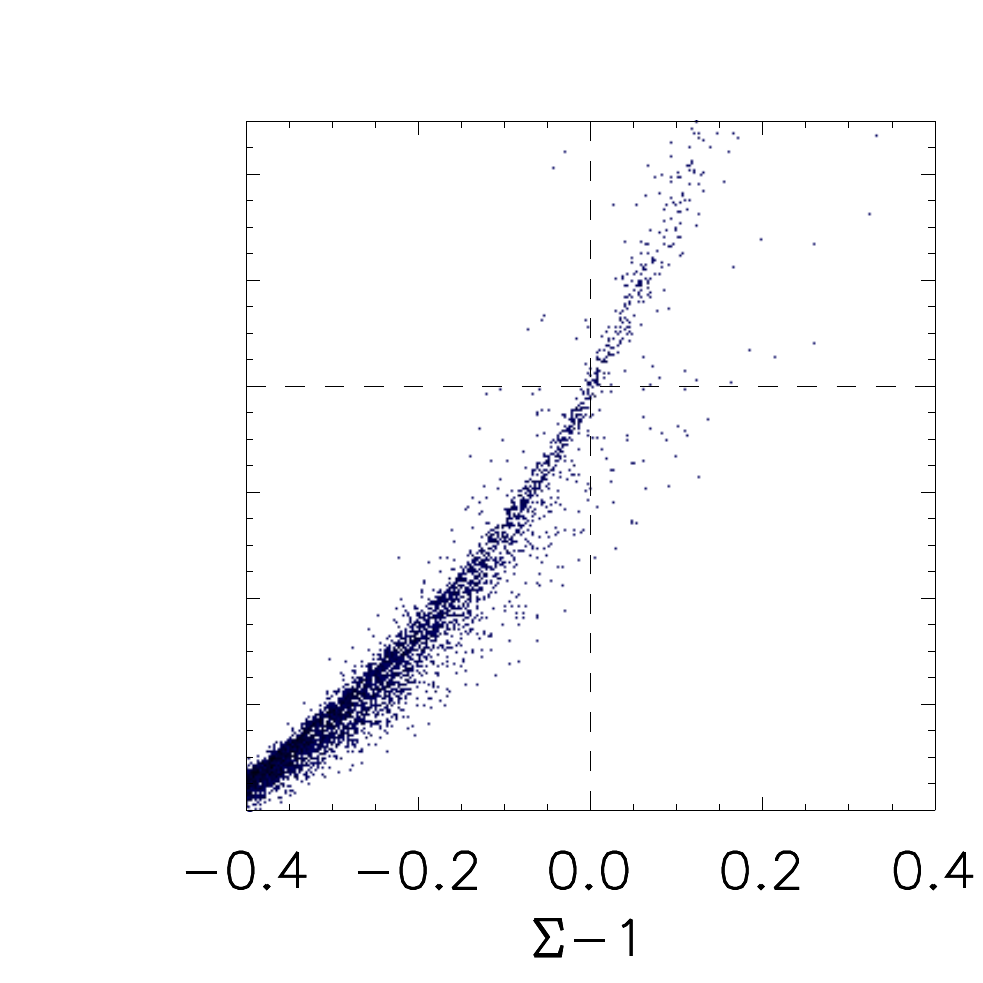}
   \caption{Correlations in the $\mu$--$\Sigma$ and $\fs$--$\Sigma$ planes for $10^4$ EMG models with the stability condition $S$ (first two rows), the stability condition $S$ and $c_s\leqslant 1$ (middle two rows) and the stability condition $S$ and $c_T\leqslant 1$ (bottom two rows). The background evolution has been set to a flat $\Lambda$CDM. The $\Lambda$CDM prediction stands at the intersection of the two dashed lines.} 
   \label{fig_viab}
\end{center}
\end{figure}

When dealing with the dark sector it is still debated if super-luminal propagation in a low-energy effective theory can be acceptable. We tend to see super-luminality as a serious pathology of a low-energy theory, following the reasoning in~\cite{Adams:2006sv}. However,  for the sake of generality, in Figure~\ref{fig_viab} we show the effects of relaxing the conditions on the propagation speeds of the scalar and the tensors, eqs.~\eqref{cs} and~\eqref{ct}.  We never intend to give up either ghost stability~\eqref{conditions} nor gradient stability~\eqref{conditions2}---gathered henceforth under the notation S.
It is worth noting that, in our formalism, a theory exhibiting $c_s>1$ can always be tuned back to $c_s \leqslant 1$ by using the parameter $\m22$. The latter, we recall, does not enter the expressions of the LSS observables. Therefore, switching on $\m22$ allows one to include Horndeski models with $c_s>1$ but $\m22=0$, in some sense. This is illustrated in Figure~\ref{fig_viab} where the predictions with the conditions $S$ and the condition $S+c_s$ are displayed. One can rightfully conclude that the selection criterion $c_s<1$ is useless once a non null $\m22$ is considered. 

On the other hand, by writing $\mu$ as $\mu= \left( \frac{c_T}{c_T(t_0)} \right)^4 \  \frac{M^2(t_0)}{M^2} \ \frac{a_0}{b_0}$ one already appreciates analytically how $c_T>1$ strengthens gravity at high redshifts. This is well highlighted by Figure~\ref{fig_viab}, the correlation lines of in the $\mu$--$\Sigma$ or $\fs$--$\Sigma$ planes are thinner once the $c_T\leqslant 1$ is implemented, the upper points being chopped out. The practical conclusion out of this analysis is if a data point was to be found at hight redshifts in the top left corner of either the $\mu$ --$\Sigma$ or $\fs$--$\Sigma$ plane, only a Horndeski model with a $c_T>1$ would be valid. More will be able to be said once $c_T$ is tightly constrained at large redshifts by future measurements of the electromagnetic counter part of gravitational wave emitting events~\cite{Nishizawa:2014zna,Bettoni:2016mij}.

\subsection{Effects of the background expansion history}

Does the evolution of perturbed sector observables depend on the acceleration of the background metric? 

\begin{figure}[!]
\begin{center}
\begin{flushleft} LDE : \end{flushleft} 
   \includegraphics[scale=0.27]{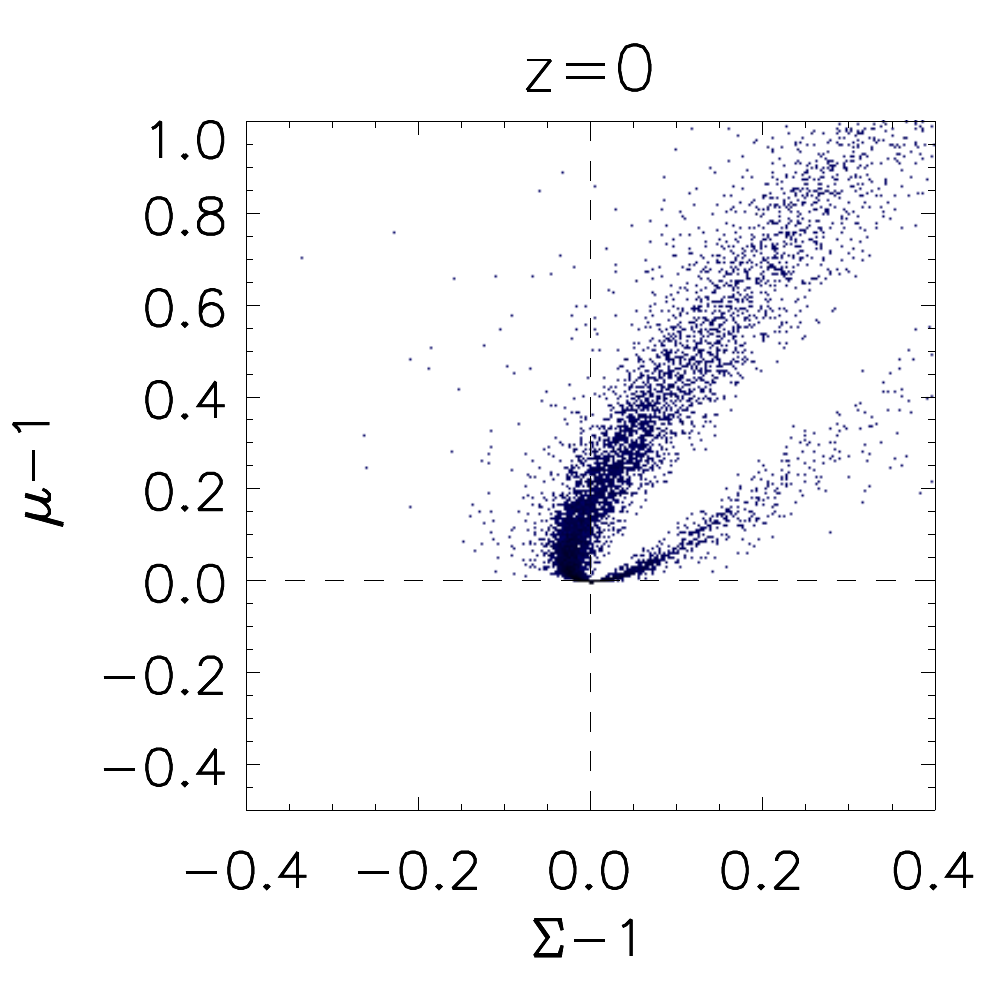}  \hskip-4mm
   \includegraphics[scale=0.27]{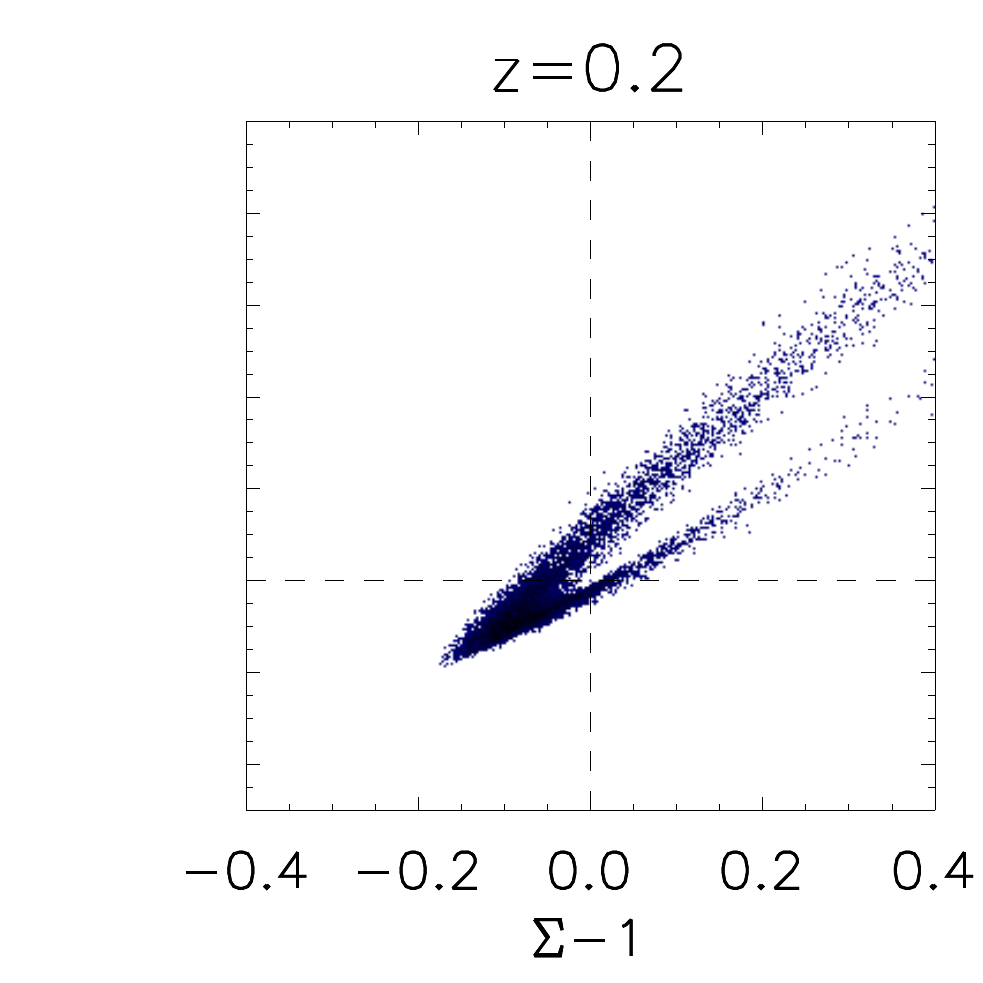}  \hskip-4mm
   \includegraphics[scale=0.27]{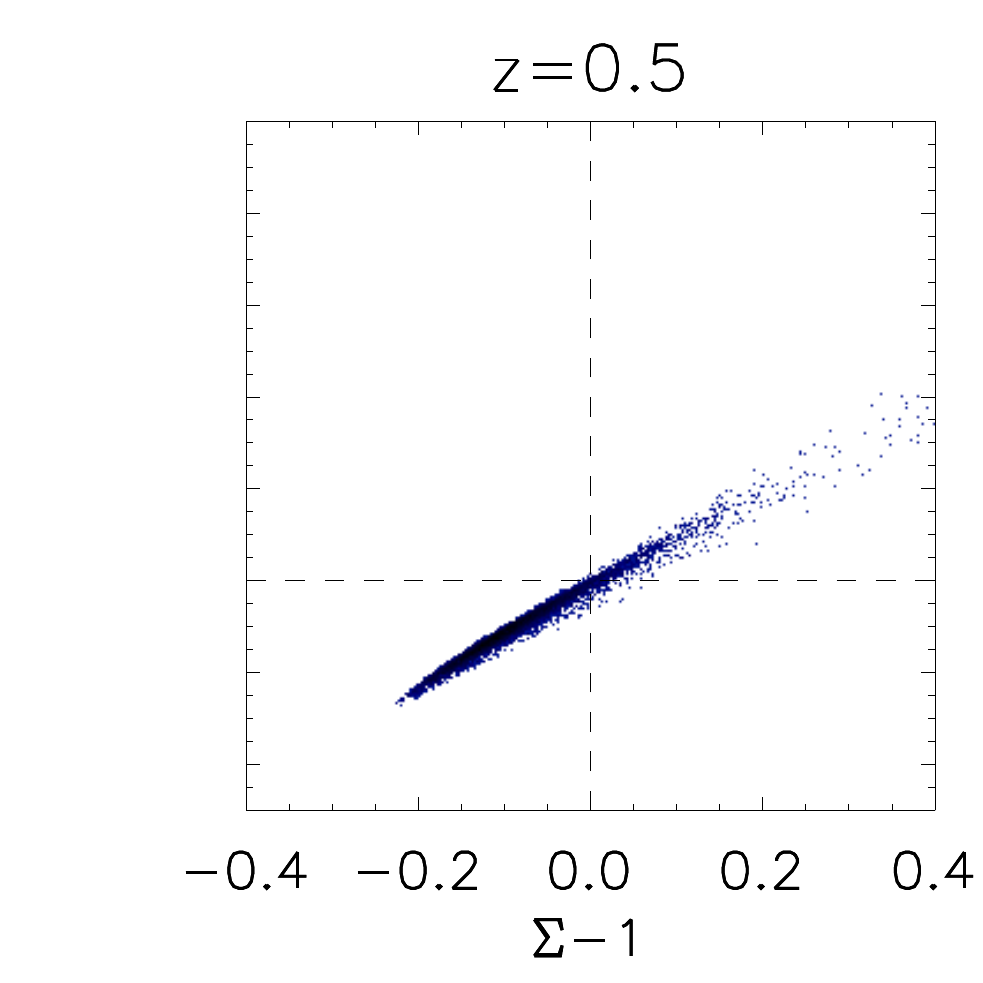}   \hskip-4mm
   \includegraphics[scale=0.27]{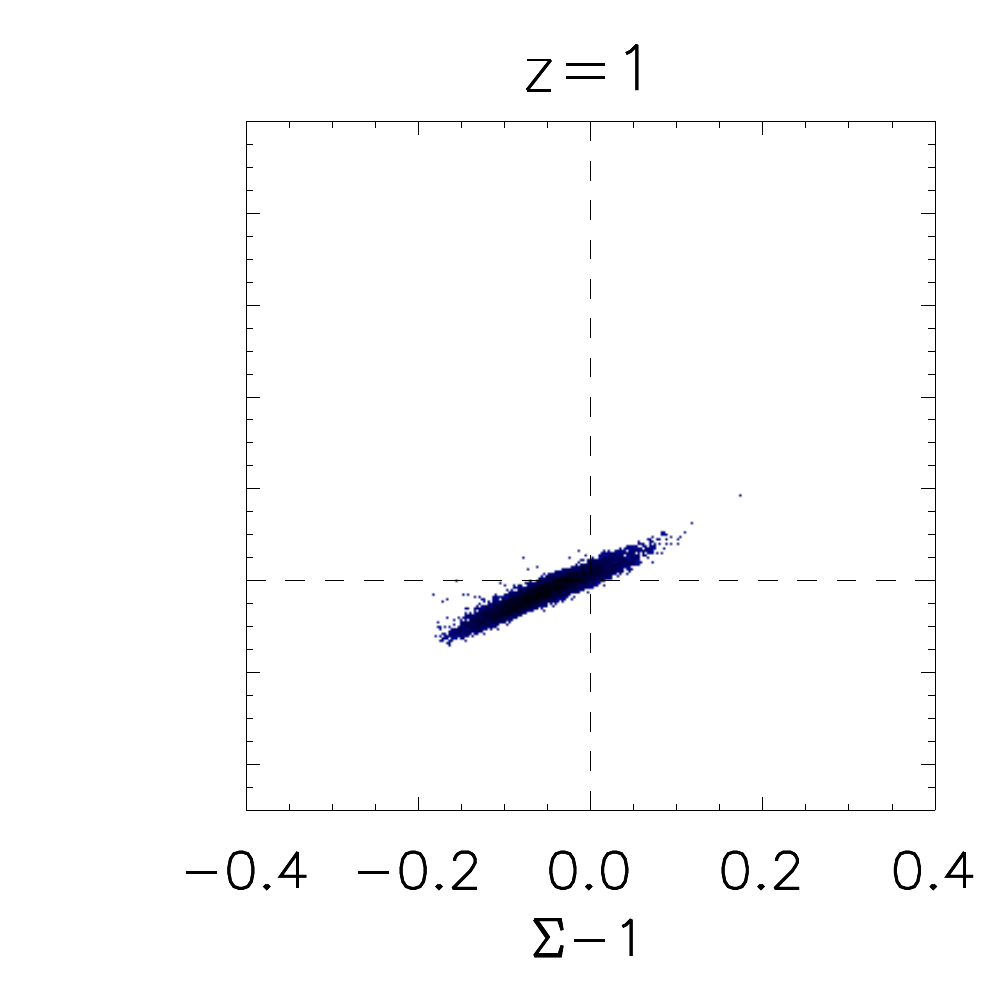}  \hskip-4mm
   \includegraphics[scale=0.27]{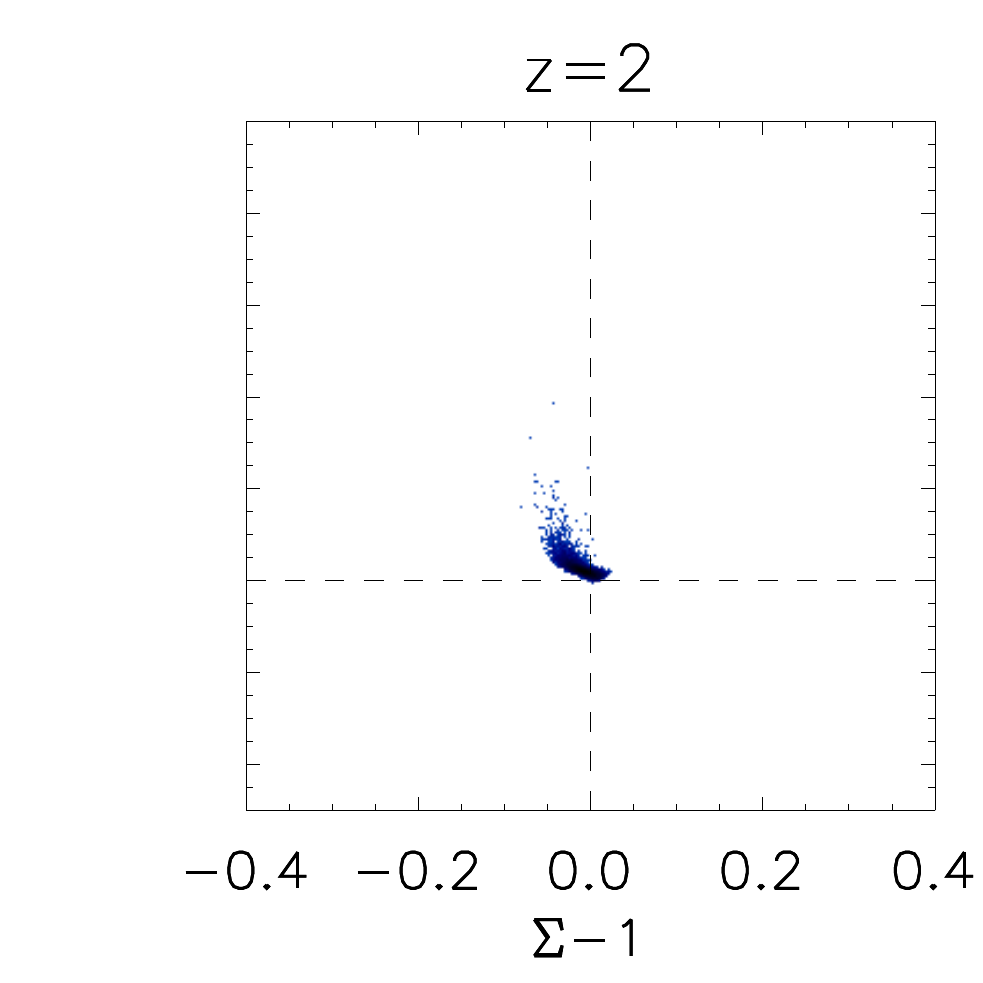}  \hskip-4mm
   \includegraphics[scale=0.27]{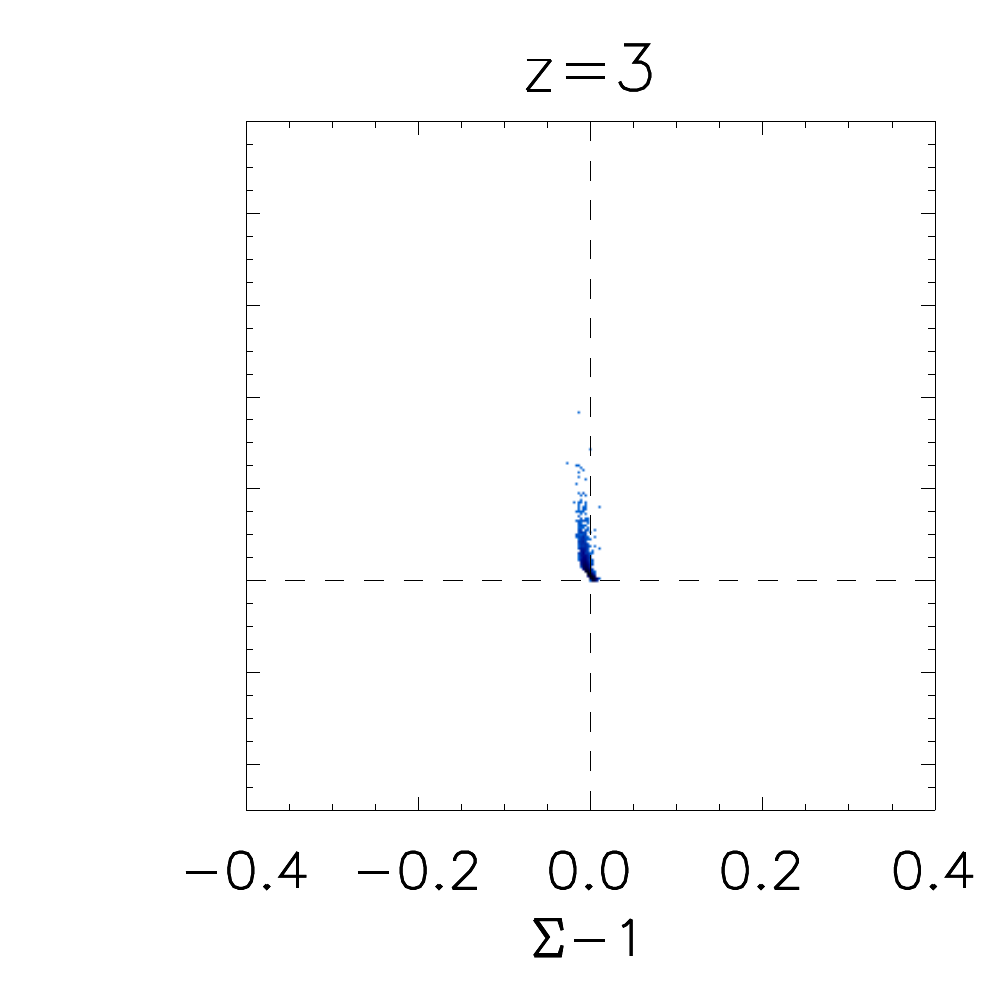}
   \vskip-2mm
   \includegraphics[scale=0.27]{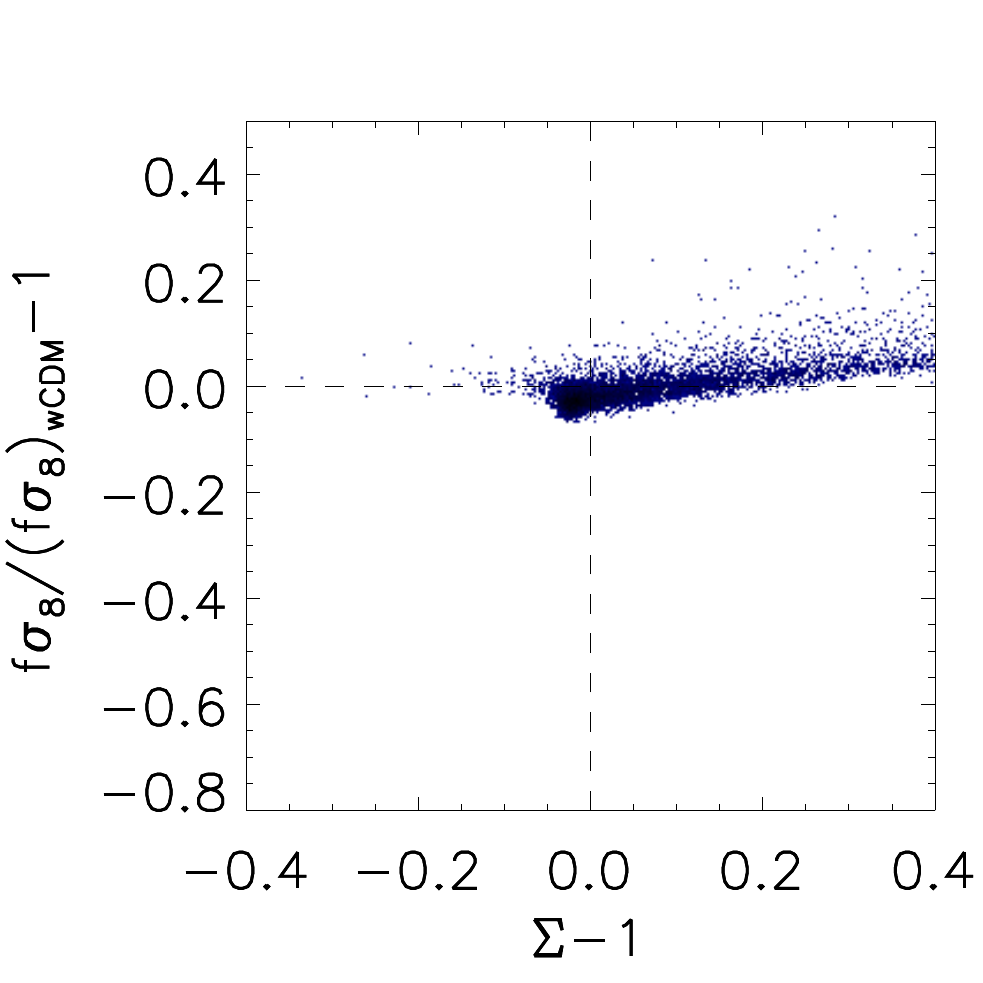}  \hskip-4mm
   \includegraphics[scale=0.27]{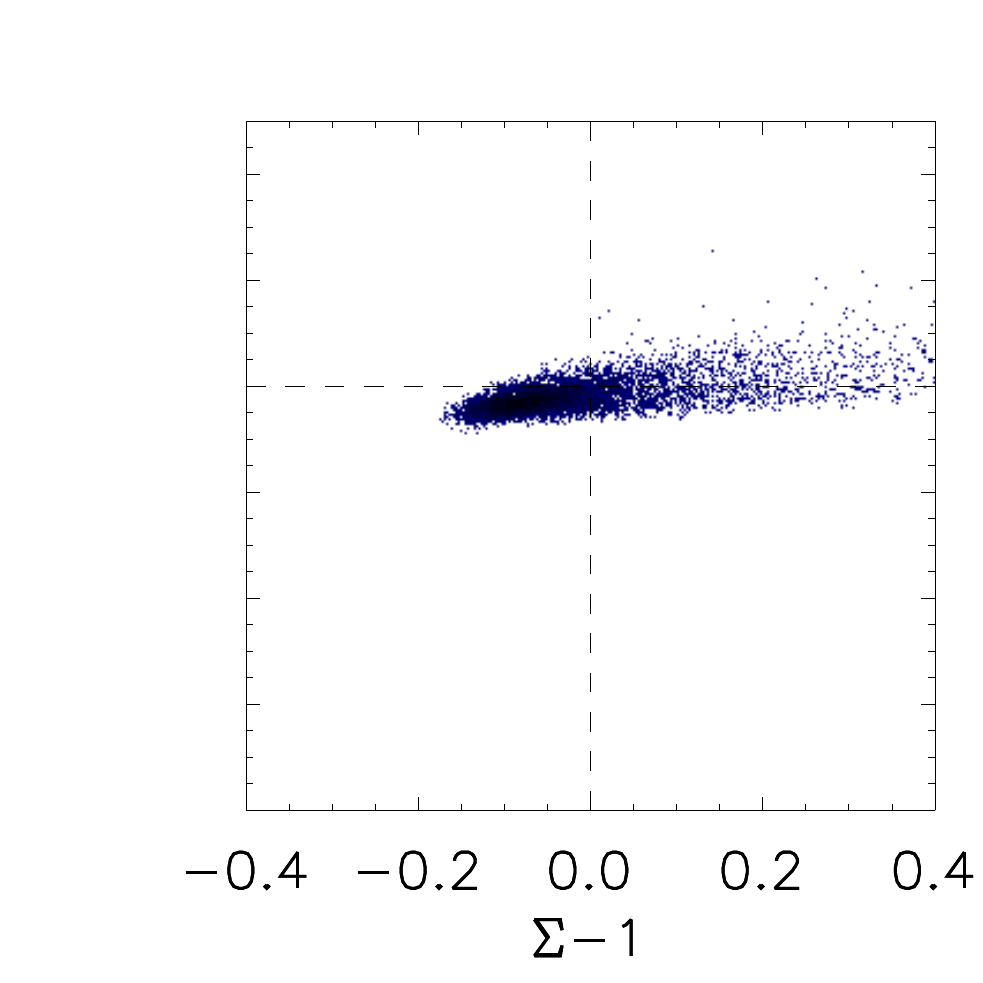}  \hskip-4mm
   \includegraphics[scale=0.27]{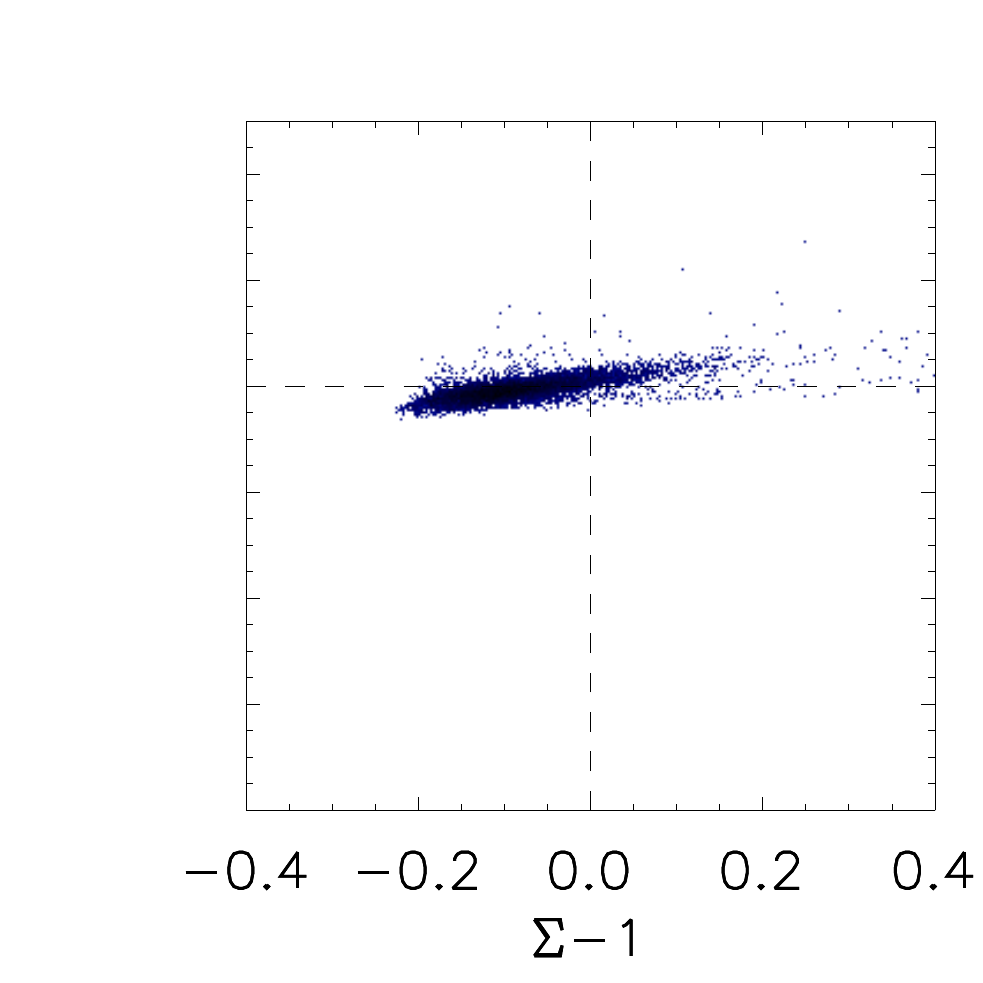}   \hskip-4mm
   \includegraphics[scale=0.27]{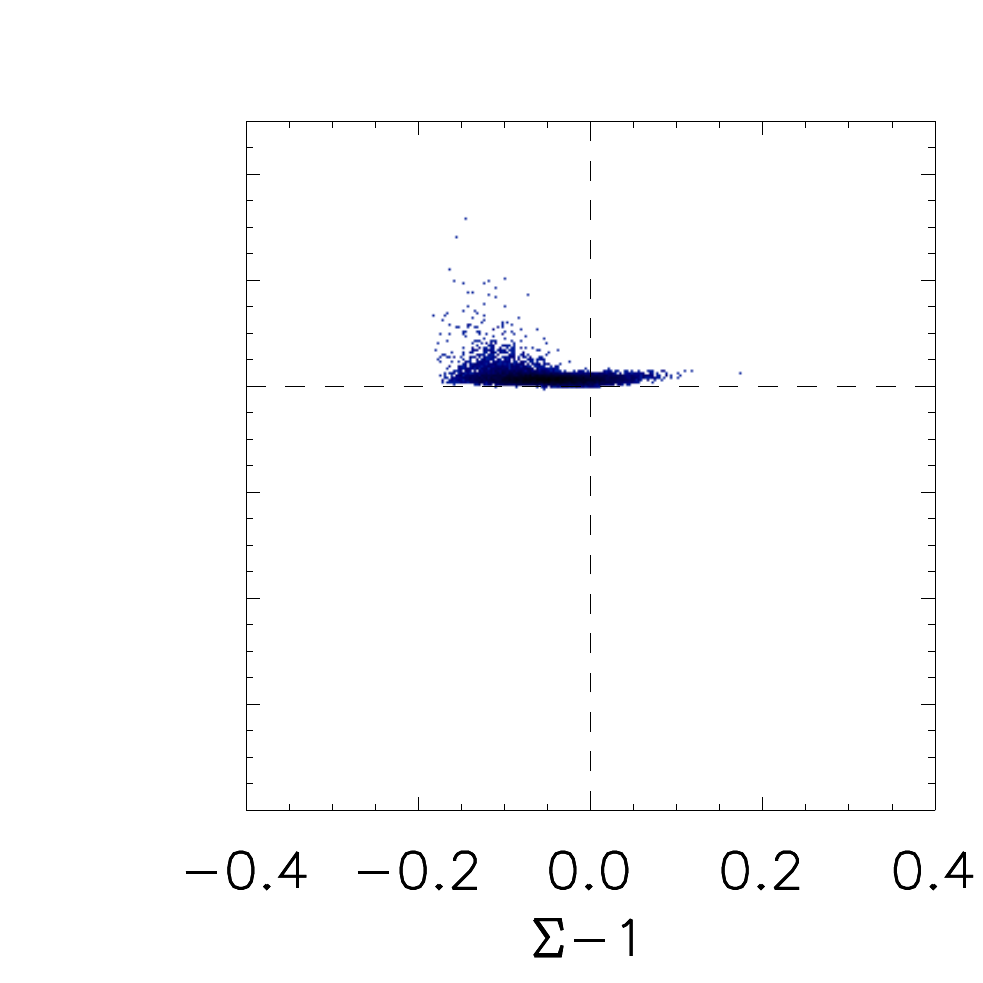}  \hskip-4mm
   \includegraphics[scale=0.27]{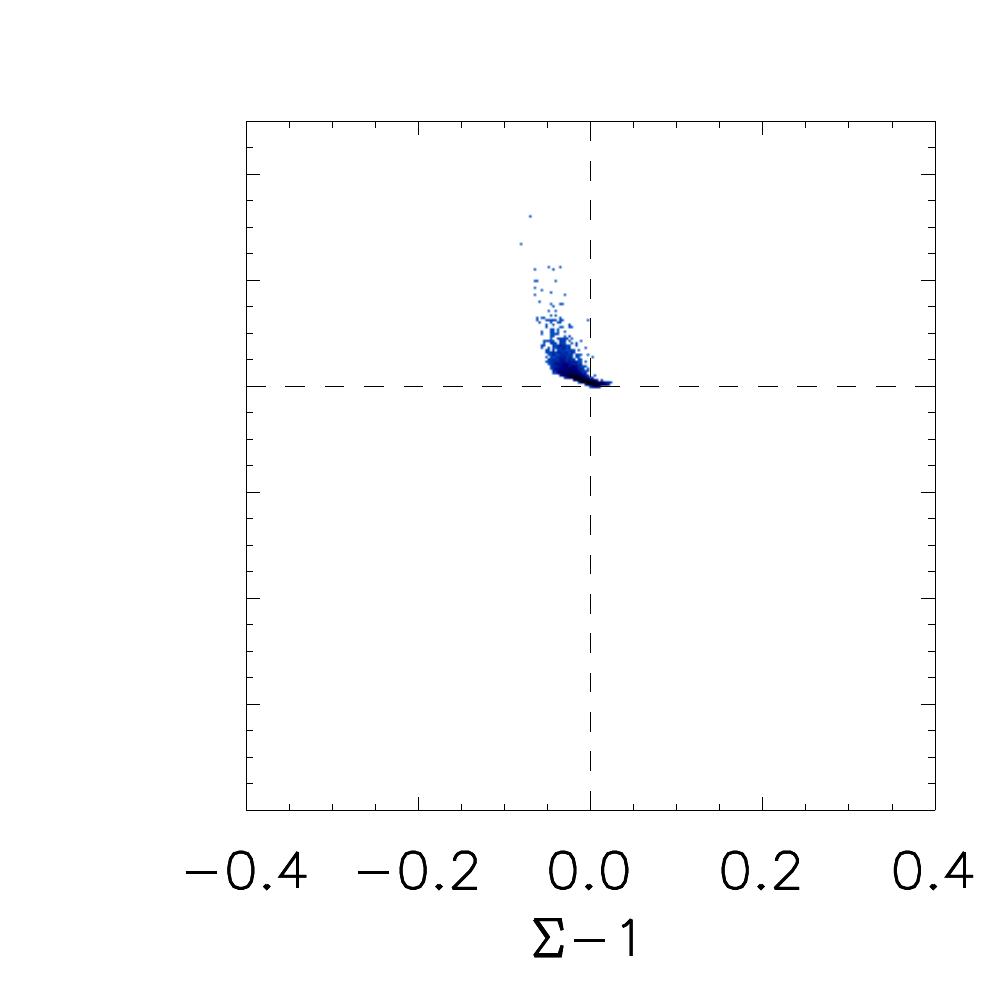}  \hskip-4mm
   \includegraphics[scale=0.27]{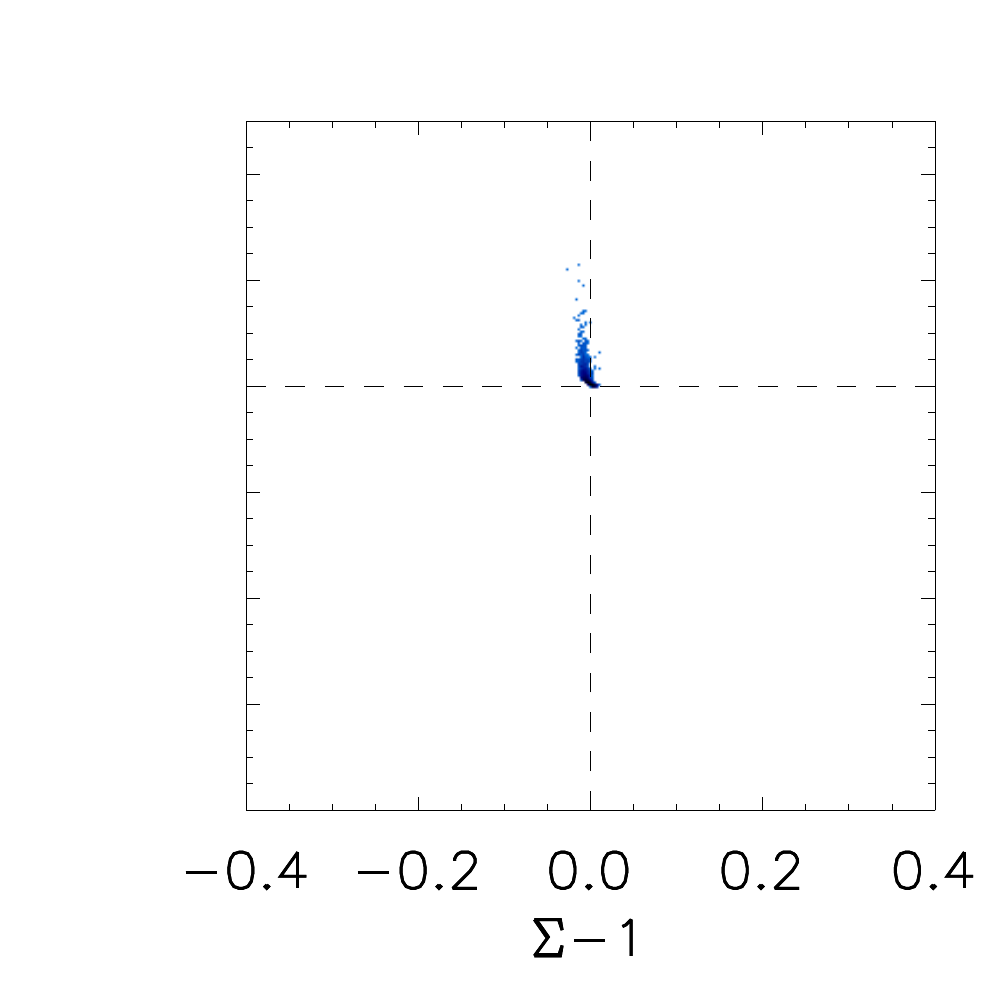}
%\vskip-2mm\vskip-2mm
%- - - - - - - - - - - - - - - - - - - - - - - - - - - - - - - - - - - - - - - - - - - - - - - - - - - - - - - -	
\begin{flushleft} EMG : \end{flushleft}   \vskip-3mm
   \includegraphics[scale=0.27]{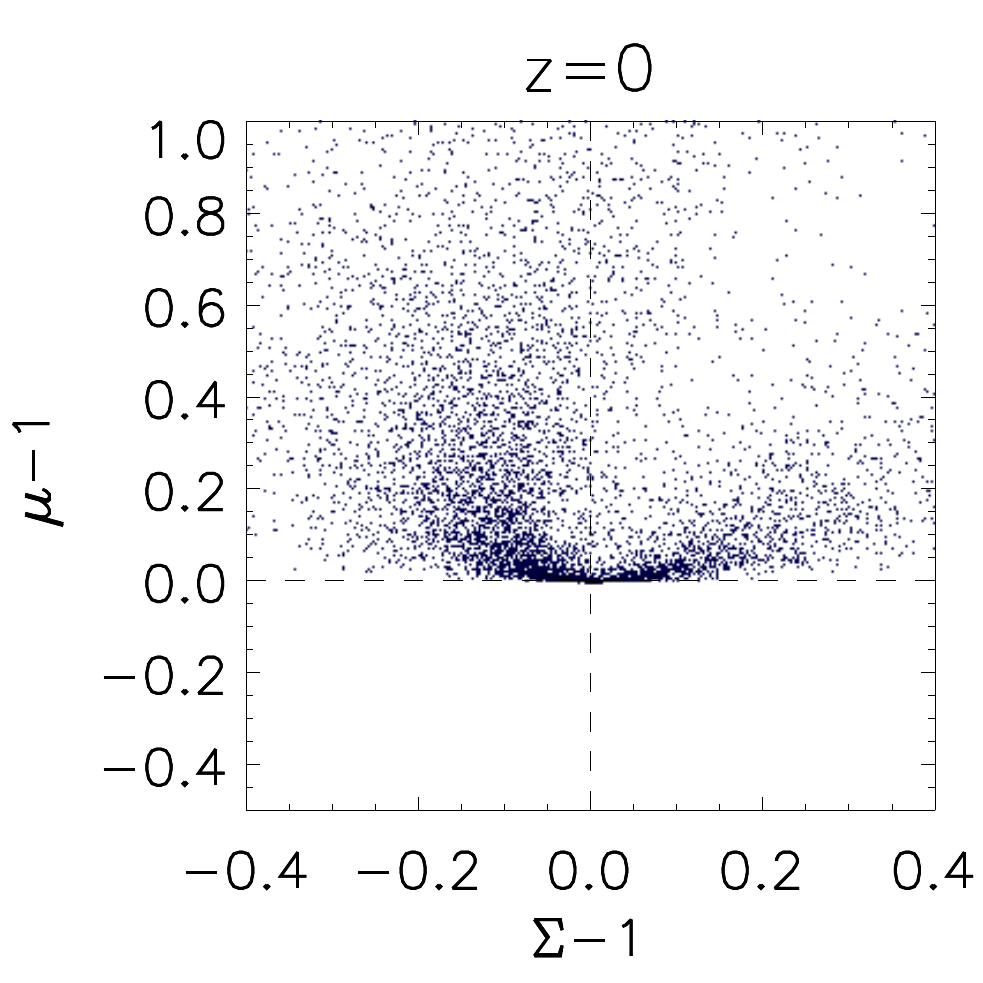}  \hskip-4mm
   \includegraphics[scale=0.27]{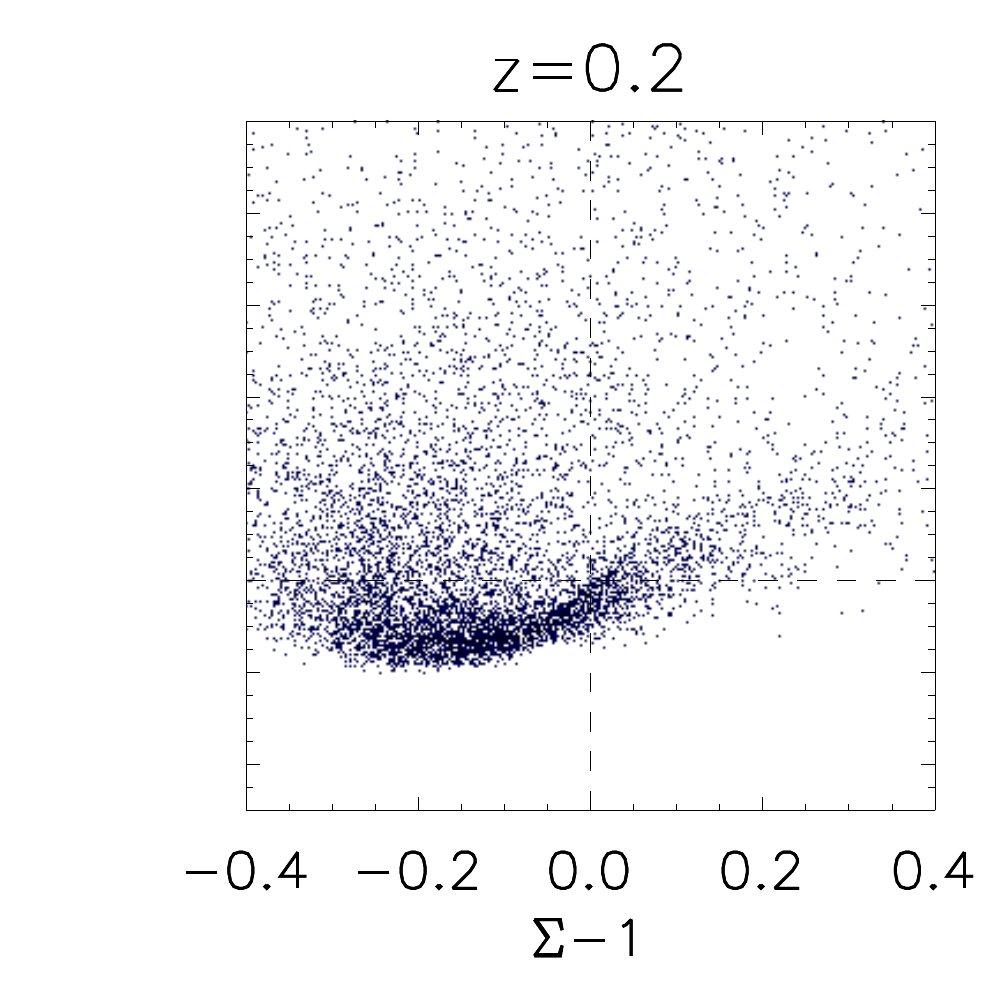}  \hskip-4mm
   \includegraphics[scale=0.27]{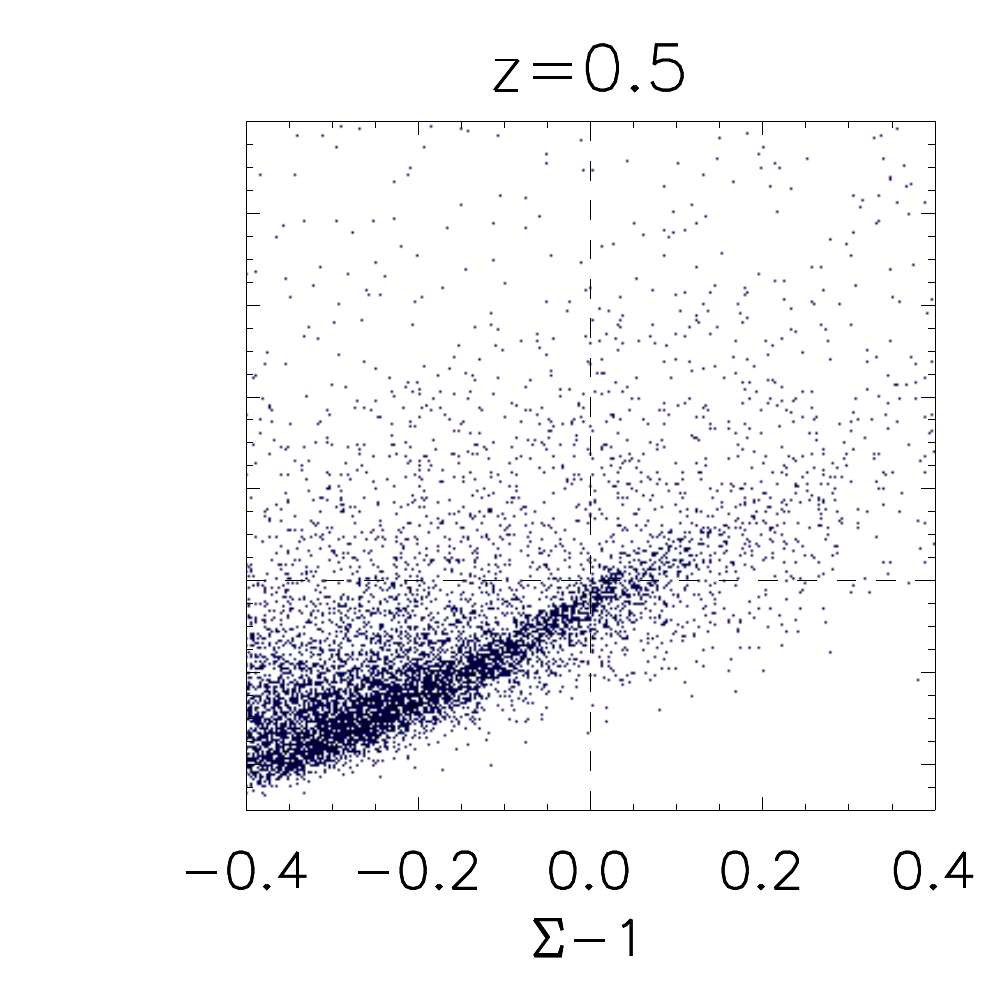}   \hskip-4mm
   \includegraphics[scale=0.27]{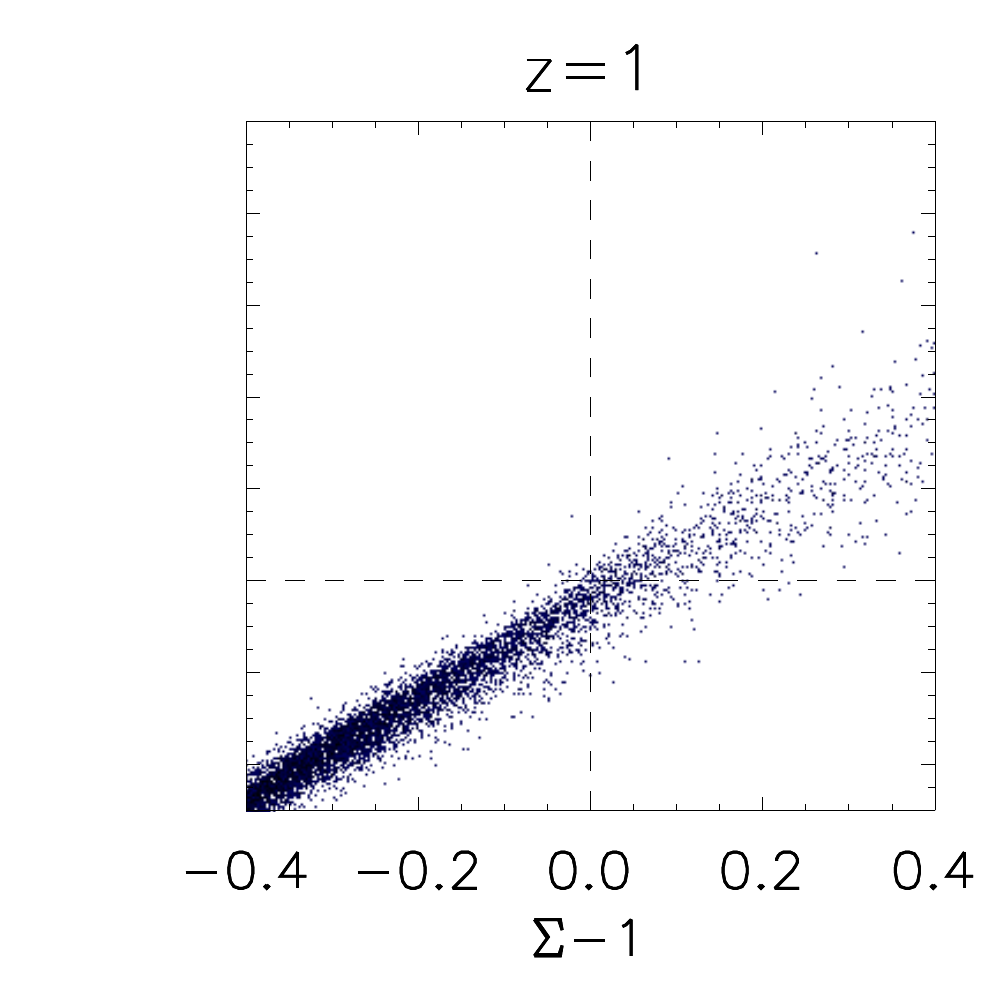}  \hskip-4mm
   \includegraphics[scale=0.27]{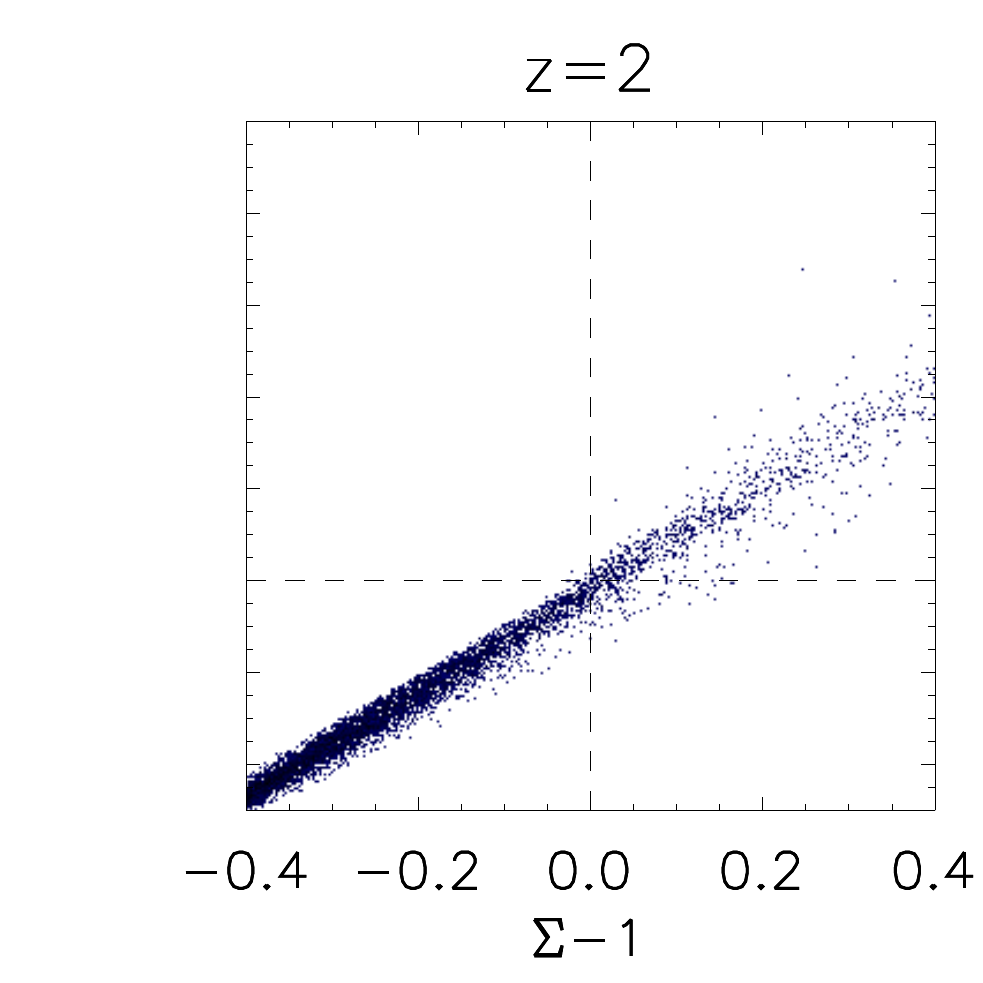}  \hskip-4mm
   \includegraphics[scale=0.27]{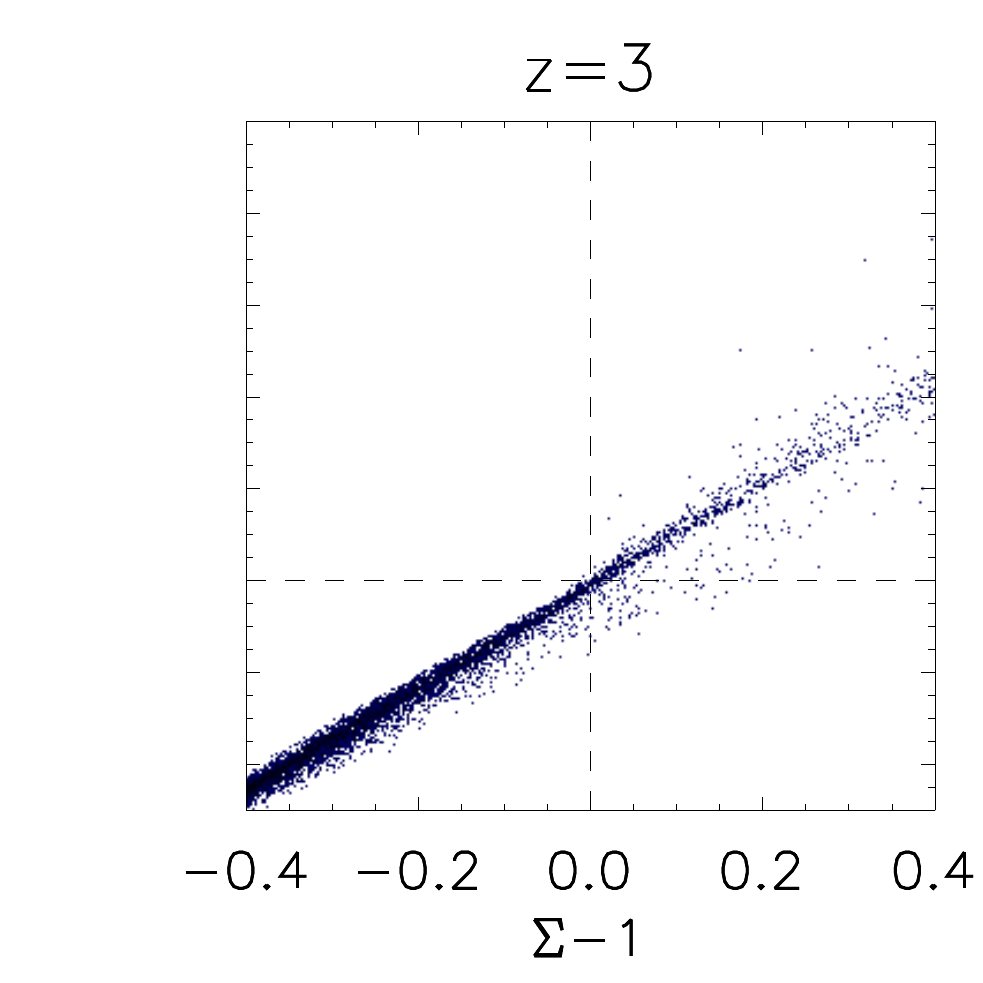}
   \vskip-2mm
   \includegraphics[scale=0.27]{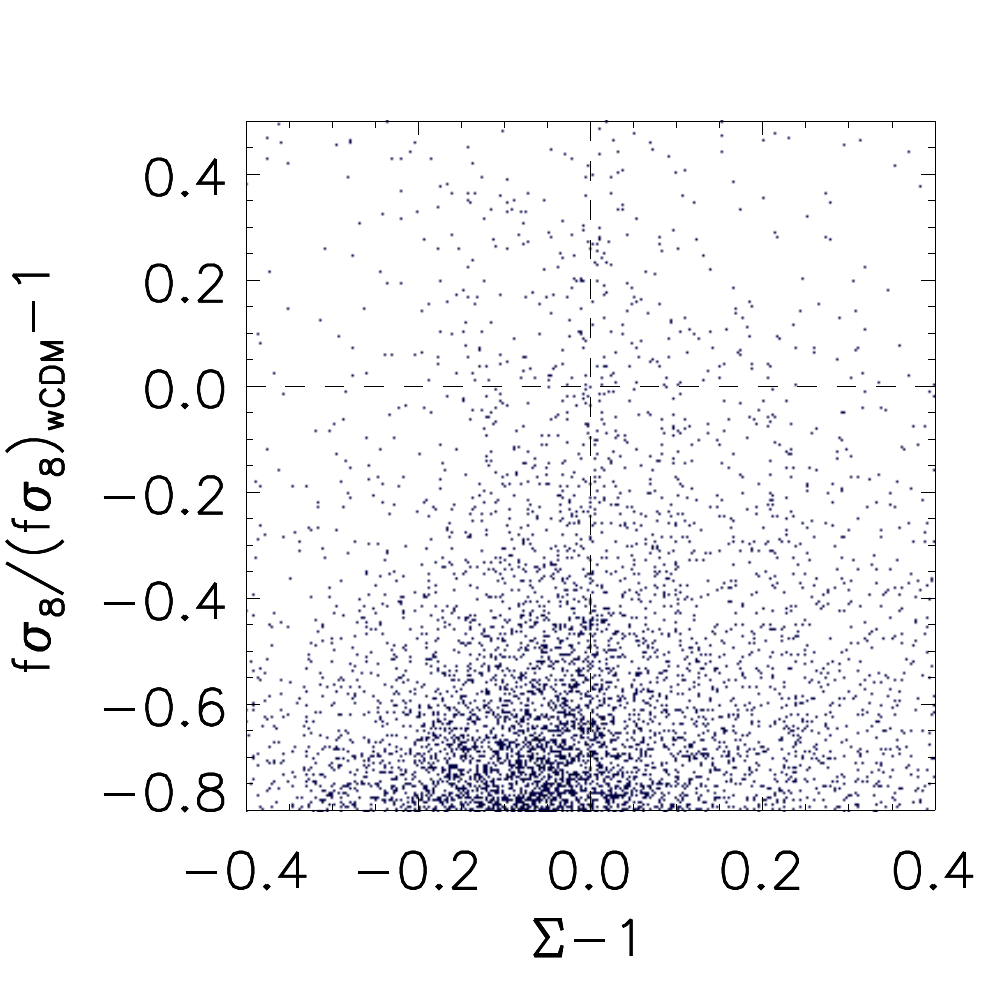}  \hskip-4mm
   \includegraphics[scale=0.27]{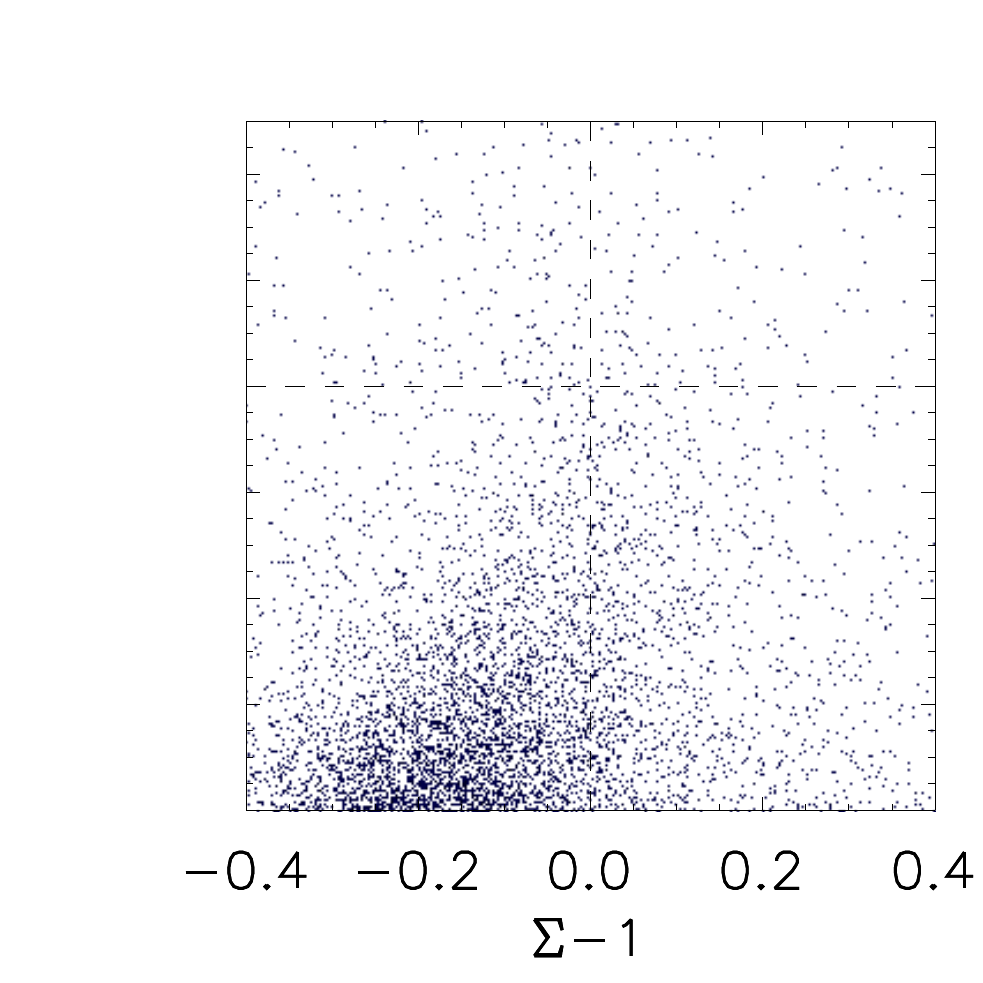}  \hskip-4mm
   \includegraphics[scale=0.27]{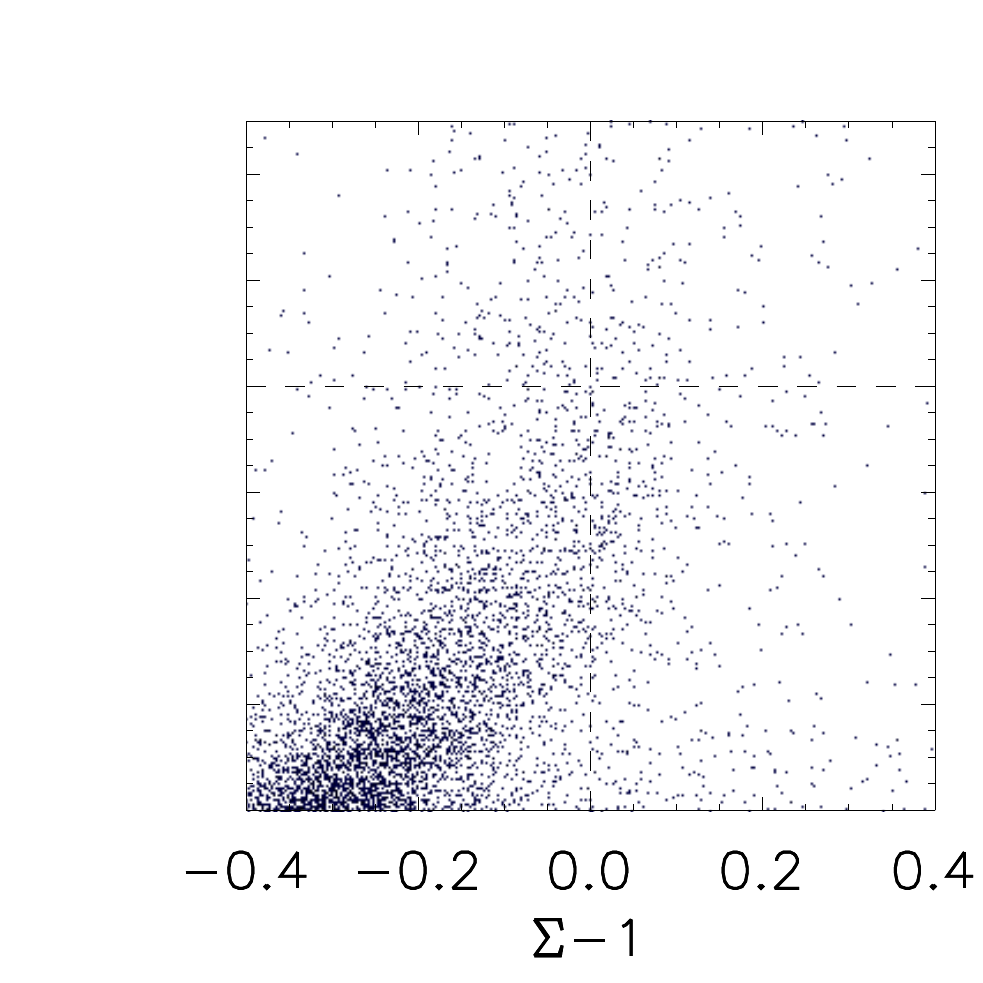}   \hskip-4mm
   \includegraphics[scale=0.27]{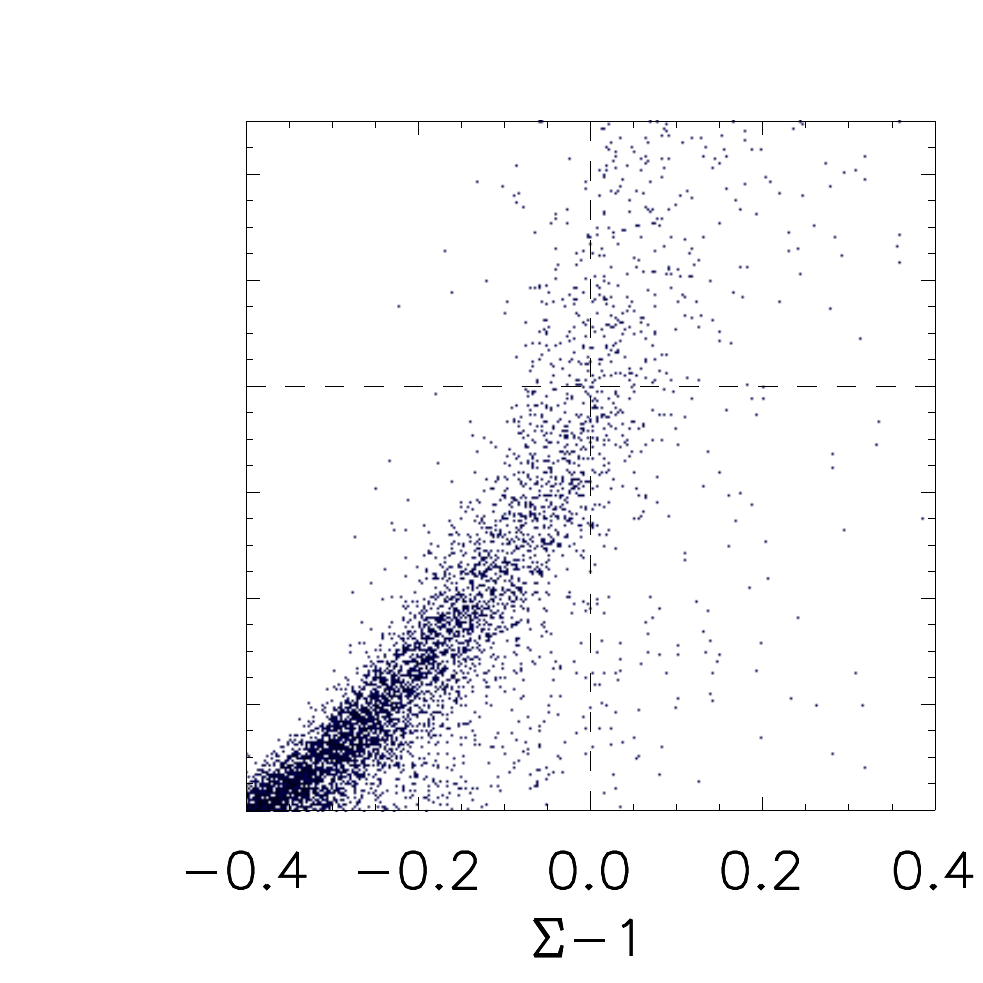}  \hskip-4mm
   \includegraphics[scale=0.27]{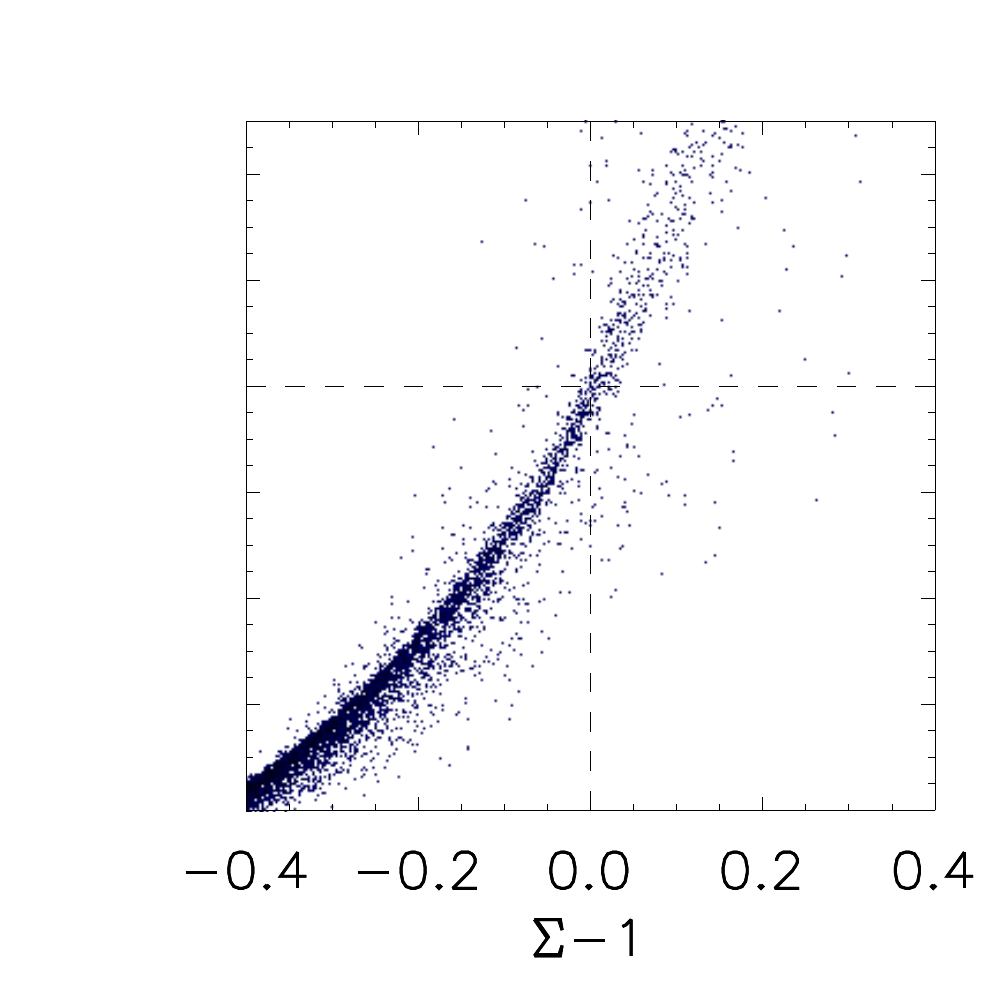}  \hskip-4mm
   \includegraphics[scale=0.27]{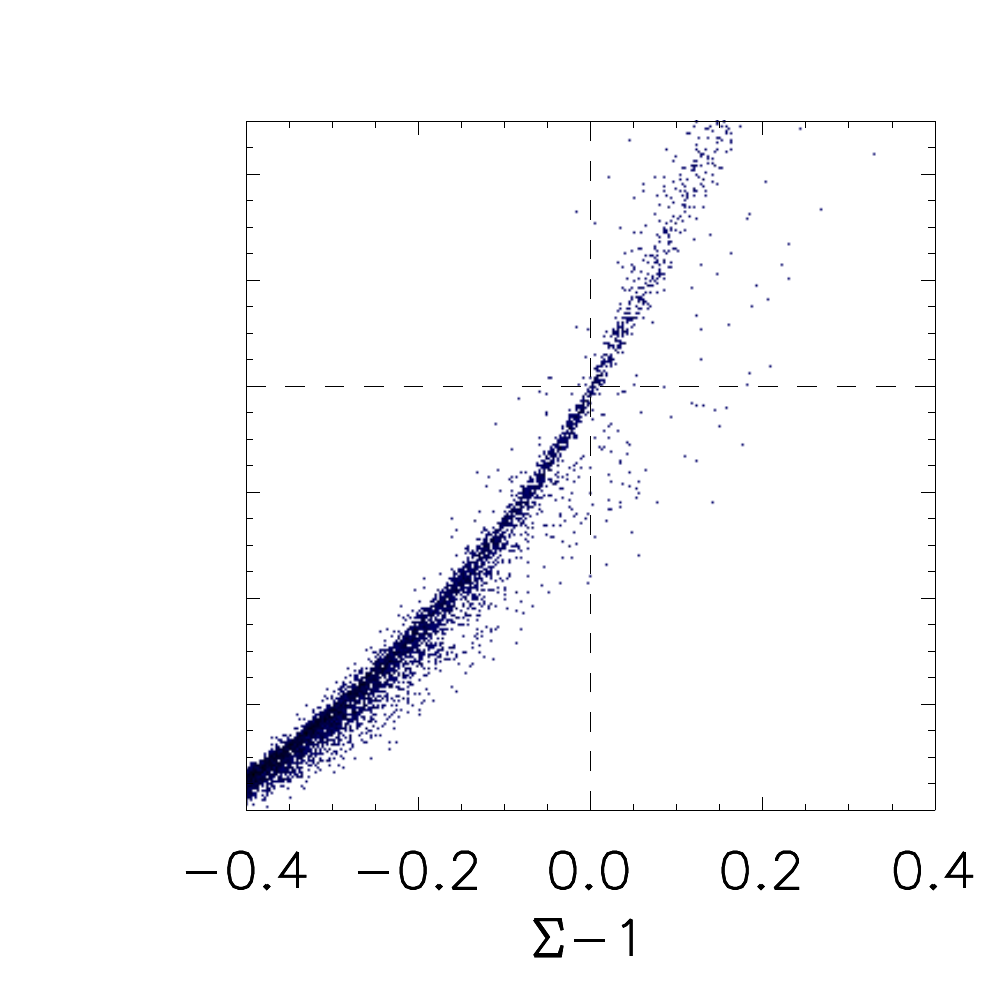}
   \caption{Correlations in the $\mu$--$\Sigma$ and $\fs$--$\Sigma$ for $10^4$ LDE models (two top rows) and $10^4$ EMG models (bottom two rows) with the full viability conditions requested but the background $e.o.s$ set to $\wb=-1.1$. The $(w=-1.1)$CDM prediction corresponds to the crossing of the two dashed lines.}  
   \label{fig_w}
\end{center}
\end{figure}
What we find is that setting $\wb=-0.9$ does not change the diagnostic described in the previous section---we thus do not display any plot for the sake of brevity. 

The effect of lowering the dark energy equation of state below $\wb= -1$ is also very mild, but worth commenting. Crossing the $\wb = -1$ means considering violations of the null energy condition. As known, such violation can be produced in a stable theory only by switching on the non-minimal gravitational couplings (see Figs.~1 and 2 of Ref.~\cite{Piazza:2013pua}), and therefore in a region of the space of theories that is ``far" from $\Lambda$CDM where all couplings vanish. Nevertheless, the effects of modified gravity are not significantly amplified as can be seen by the comparison between Figure~\ref{fig_w} and Figs.~\ref{fig_lde} and Figure~\ref{fig_ede_emg} (lower panel). The amplification effect is effectively compensated by the wide variations of the couplings and by the large volume in the theory space that we are considering, even in the $\wb = -1$ case. In summary, we do not find distinctive definite features of LSS phenomenology related to stable violations of the null energy condition.

\subsection{Consistency and robustness checks}

To assess  the generality of our results we have conducted two consistency checks. From the analysis undertaken in this paper and in~\cite{Perenon:2015sla,Piazza:2013pua,Salvatelli:2016mgy}, the $\Lambda$CDM paradigm stands out as an extremal model among EFT models. However, one could wonder whether our randomly generated coefficients within a uniform distribution could end up favouring models that are very far from the standard paradigm. However, we have checked that the correlations we find are unchanged even if the  coefficients are picked from Gaussian distributions centred around 0 (the $\Lambda$CDM value) and with a standard deviation of 1. 

As a second check, we have noted that our diagnostic is also unchanged under using a different parametrisations of the couplings. In particular we have considered an alternative choice of the coupling functions, the so called "$\alpha$-parametrisation" first proposed by~\cite{Bellini:2014fua} and extended by~\cite{Bellini:2015wfa,Bellini:2015xja,Gleyzes:2014rba,Gleyzes:2015pma,Gleyzes:2015rua},   as opposed to the "$\mu$-parametrisation" presented in this paper.
Appendix~\ref{alpha} contains a dictionary to switch between the two parametrisations.

\section{Conclusions} \label{sec_5}

What Horndeski theories have to say about early dark energy? This is the original question motivating the analysis presented in this paper. 
Early modifications of GR  are found to have non-negligible, lasting and potentially detectable effects in the LSS observables of the local and recent universe. 

\begin{figure}[H]
\begin{center}
	\hskip4mm
	\includegraphics[scale=0.55]{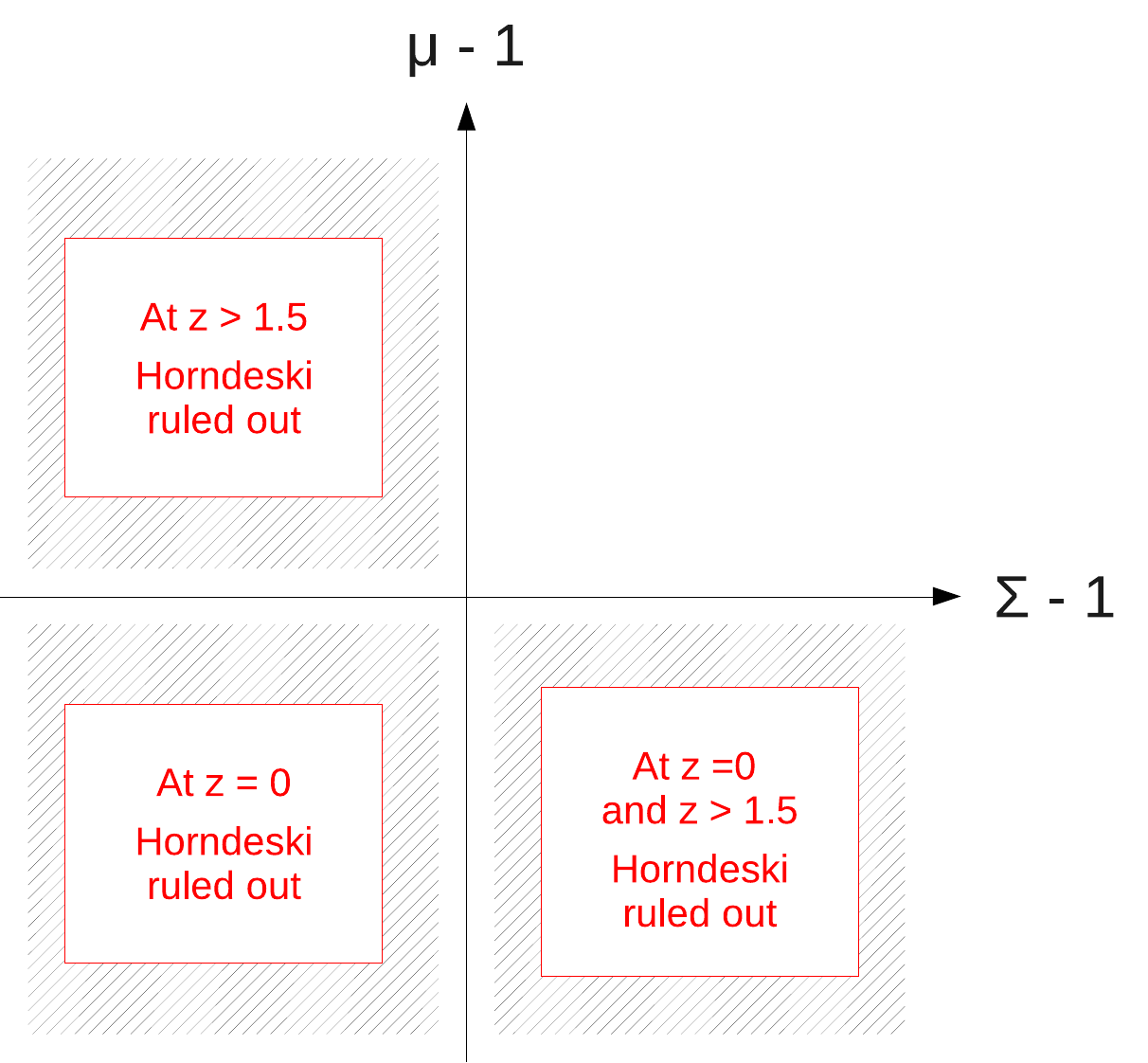}
	\hskip8mm
	\includegraphics[scale=0.55]{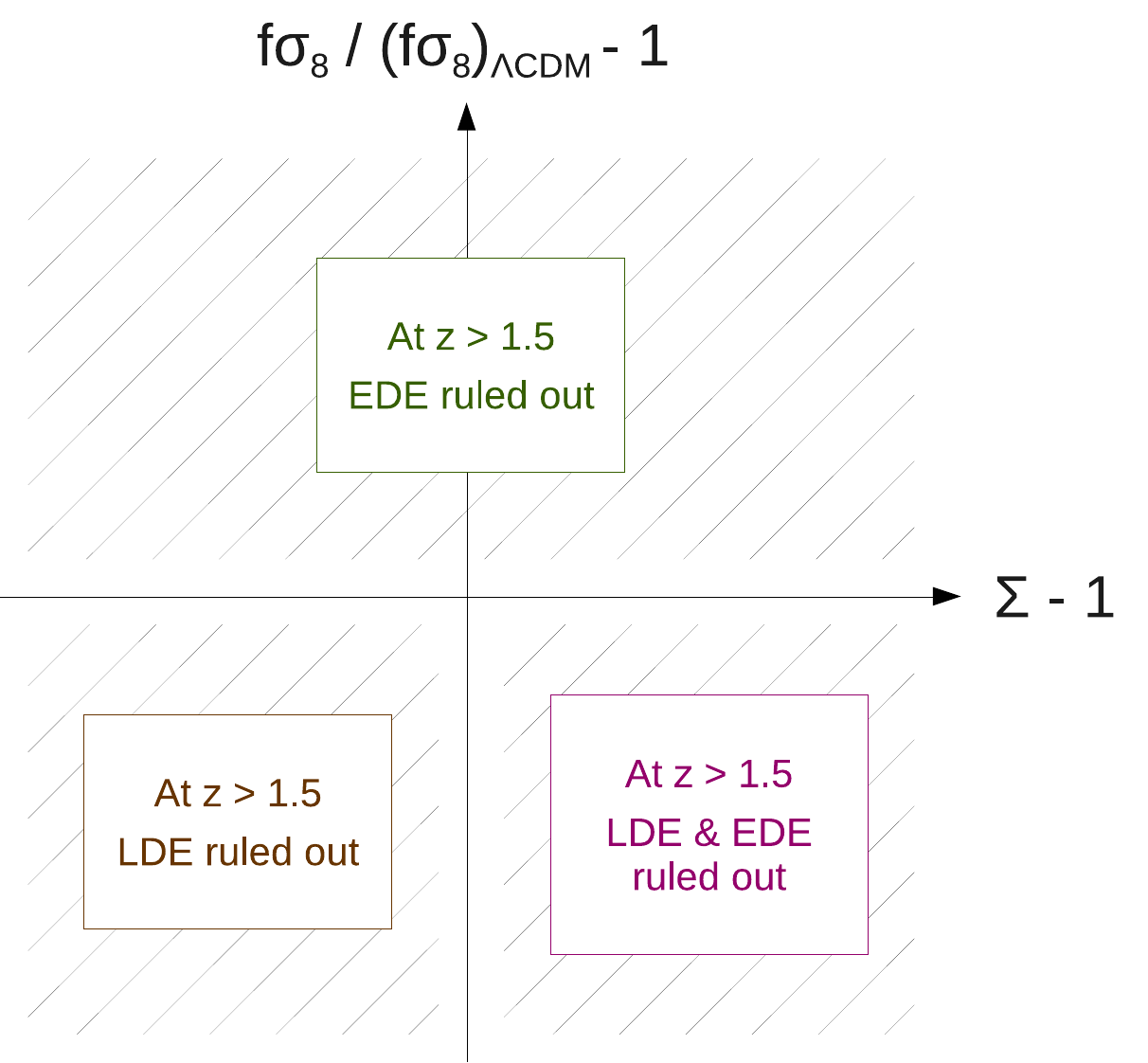}
\end{center}
\caption{Schematics of the fundamental observable planes allowing to discard Horndeski theories (left diagram) and the type of dark energy embedded (right diagram).}
\label{fig_diag}
\end{figure}

In Figure~\ref{fig_diag} we summarize our main findings. By tracing the time evolution, from early epochs ($z=100$) down to present day, of fundamental LSS observables such as the reduced effective Newton constant $\mu$, the gravitational slip parameter $\eta$, the lensing parameter $\Sigma$ and the linear growth function of LSS $\fs$ we have found that GR extensions contemplating an additional scalar degree of freedom with second order equations of motion can be definitely ruled out
if one of the following conditions apply (Figure~\ref{fig_diag}, left panel): 
\begin{itemize}
\item The observables  $\mu$ and $\Sigma$ have opposite sign for $z>1.5$\, 
\item $\mu<1$  at $z=0$\, 
\end{itemize}
Specific sub-classes of such theories in which the modified gravity effects are limited to late times could be discriminated if data at redshift $z>1.5$ eventually become available for both redshift and lensing surveys. 
Indeed, we  find that above that critical redshift (Figure~\ref{fig_diag}, right panel): 
\begin{itemize}
\item  LDE  will be ruled out if  $\fs < (f \sigma_{8})_{\Lambda CDM}$ at $z>1.5$
\item  EDE  will be ruled out if  $\fs > (f \sigma_{8})_{\Lambda CDM}$ at $z>1.5$ or $\fs > (f \sigma_{8})_{\Lambda CDM}$ and $\Sigma>1$ at $z>1.5$ 
\end{itemize}
These results  are  insensitive to the background dark energy $e.o.s$ parameter within the  reasonable range $\wb \in [-1.1,-0.9]$. Indeed, we have found the diagnostic tool does not lose predictability when  progressively less constraining requirements are imposed. 
Two complementary strategies  allow to estimate the likelihood of data given the Horndeski class of theories. 
A model-dependent analysis is optimal if one is  to  exploit  theoretical priors about the physical viability of specific
Horndeski models to complement the discriminatory power of data. Indeed,  \cite{Salvatelli:2016mgy} have shown  that  by this approach the region
of the parameter space that is not rejected by observations is  significantly reduced. 
An orthogonal approach consists in implementing model-independent likelihood analysis, parametrising LSS observables 
in a  purely phenomenological way, blindly of any gravitational theory.  The  diagnostics developed in this paper are meant to  
facilitate theoretical interpretation in these cases.  Interestingly,  model-independent analyses have  been pursued 
for example in~\cite{Simpson:2012ra} and more recently in~\cite{Ade:2015rim} and preliminary  results are  suggestive of a negative value of  $\mu-1$  at redshift z = 0.  Should  future, higher precision data  strengthen the statistical significance of  these  findings,  the Horndeski landscape would face hard times. 
\vv
Exploring beyond standard GR, and notably the functional space of scalar-field extension of GR will eventually become  less disorienting than previously suspected. However, much must still be accomplished and a number of improvements would be desirable. We have focused on scales much smaller than the Hubble radius in this paper. As data improve on ever larger scales, our analysis should be extended to include possible scale dependent effects coming from mass terms of the scalar field  that are of the order of Hubble. On the contrary, it would be interesting to evaluate,  on small scales,  how many models survive once solar system tests are applied. Lastly, it would be useful to study to which level our diagnostic plots are stable to the inclusions of more general scenarios in which, for instance, the scalar field is allowed to satisfy larger than second order equation of motions (the so called beyond Horndeski theories \cite{Gleyzes:2014qga,Gleyzes:2014dya}, see also~\cite{Zumalacarregui:2013pma} for early studies in this direction), or when conformal-disformal couplings of matter to gravity are considered (the so called effective field theory of interacting dark energy~\cite{Gleyzes:2015pma}).

\section*{Acknowledgments} 

We acknowledge useful discussions with Jose Beltr\`an Jiménez, Julien Bel, Lam Hui, Stephane Ili\'c, Levon Pogosian, Valentina Salvatelli, Alessandra Silvestri, Filippo Vernizzi and Miguel Zumalacarregui. F.P. warmly acknowledges the financial support of A*MIDEX project (n$^{o}$ ANR-11-IDEX-0001-02) funded by the ``Investissements d'Avenir" French Government program, managed by the French National Research Agency (ANR). We are grateful for support from project funding of the Labex OCEVU. This article is based upon work from COST Action CANTATA CA15117, supported by COST (European Cooperation in Science and Technology). We acknowledge financial support from ``Programme National de Cosmologie and Galaxies" (PNCG) of CNRS/INSU, France.

%**************************************************************************************************
%                                            APPENDIX
%**************************************************************************************************
\appendix
\section{Parametrisation of the couplings}\label{param}

In this appendix we present how we parametrise the EFT coupling functions in the three different DE scenarios.
\vskip4mm
\textbf{\ LDE and EDE :}
\begin{align}
\mu\left(x\right)        \ &=\ H   \ \, \, (1-x)  \left(p_{11}+p_{12} \left(x -x_0\right)+p_{13} \left(x -x_0\right)^2 \right) \;,\\[1mm]
\mu^2_2\left(x\right)    \ &=\ H^2 \,      (1-x)  \left(p_{21}+p_{22} \left(x -x_0\right)+p_{23} \left(x -x_0\right)^2 \right) \;,\\[1mm]
\mu_3\left(x\right)      \ &=\ H   \ \, \, (1-x)  \left(p_{31}+p_{32} \left(x -x_0\right)+p_{33} \left(x -x_0\right)^2 \right) \;,\\[1mm]
\epsilon_4\left(x\right) \ &=\ \ \ \, \, \, \, \, (1-x) \left( \qquad \ \ p_{42} \left(x -x_0\right)+p_{43} \left(x -x_0\right)^2 \right) \;.
\end{align}

For the LDE case the constrain $p_{12} = \frac{p_{11} \log (\Omega_m^0)-6\log\left(1 + (1- \Omega_m^0) p_{41}\right)}{1 -\Omega_m^0+ \Omega_m^0 \log (\Omega_m^0)}$ must be imposed to enforce $M^2(x\rightarrow 1)\rightarrow\mps$, see~\cite{Perenon:2015sla} for details.
\vskip4mm

\textbf{\ EMG :}
\begin{align}
\mu\left(x\right)        \ &=\ H   \ \, \, (1-x)  \left(p_{11}+p_{12} \left(x -x_0\right)+p_{13} \left(x -x_0\right)^2 \right) \;,\\[1mm]
\mu^2_2\left(x\right)    \ &=\ H^2 \,        \left(p_{21}+p_{22} \left(x -x_0\right)+p_{23} \left(x -x_0\right)^2 \right) \;,\\[1mm]
\mu_3\left(x\right)      \ &=\ H   \ \, \,   \left(p_{31}+p_{32} \left(x -x_0\right)+p_{33} \left(x -x_0\right)^2 \right) \;,\\[1mm]
\epsilon_4\left(x\right) \ &=\ \ \; \, \, \, \, \,  \left( \qquad \ \ p_{42} \left(x -x_0\right)+p_{43} \left(x -x_0\right)^2 \right) \;.
\end{align}

\section{Links with the $\alpha$-parametrisation}\label{alpha}

The EFT action of Horndeski theories for linear perturbations about an FLRW background can be also parametrised by
\begin{equation}
\begin{split}
S = \ \int d^3x dt a^3\, \frac{M_*^2}2\bigg[ &\delta K_{ij }\delta K^{ij}-\delta K^2  +\R\, \delta N \\
&  +(1+\alpha_T) \bigg( \R \frac{\delta \sqrt{h}}{a^3} \bigg) + \alpha_K H^2 \delta N^2+4\alpha_B H \, \delta K\, \delta N \bigg]\, ,
\end{split}
\end{equation}
with $N$ being the lapse function and $S_m$ the action of matter perturbations in the Jordan frame. 

In this action the EFT couplings stand as a running planck mass $M_*$, the excess speed of gravitational waves, the kineticity $\alpha_K$ and the braiding $\alpha_M$. It customary to redefine the running of the planck mass trough the non minimal coupling $\alpha_M=\frac{1}{H}\frac{{\rm d\,ln}M^2_*}{{\rm d\,ln}t}$. 
The alpha parametrisation has the benefit of attaching the evolution coupling functions to clear physical effects. However, the theory-friendly view point is lost, subsets of Horndeski theories, such as Brans-Dicke for instance, are described by more involved combinations of the $\alpha$ functions as compared to the $\mu$'s (see table 1 in~\cite{Piazza:2013pua}). In parallel, the $\mu$-parametrisation has the benefit of displaying a Lagrangian expanded in small perturbations where the couplings are expected to be of order 1.
\vv
The mapping with the $\mu$-characterisation is the following. 
\begin{align}
M_* & =  M \sqrt{1+\epsilon_4}      \\
\alpha_M & =  \frac{\dot{\epsilon_4}}{H(1+\epsilon_4)}+ \frac{\mu_1}{H}  \\
\alpha_K & =  \frac{2\mathcal{C}+4\m22}{H^2(1+\epsilon_4)}  \\
\alpha_B & =  \frac{\mu_1-\mu_3}{2H(1+\epsilon_4)} \\
\alpha_T & =  -\frac{\epsilon_4}{1+\epsilon_4}  \\
\end{align}
From this, the DE scenarios are characterised by:

\begin{align}
{\rm \bf LDE} &: \left\{\frac{M_*^2}{\mps} \rightarrow  1 \;,\; \alpha_M  \rightarrow 0 \;,\; \alpha_K \rightarrow 0 \;,\; \alpha_B\rightarrow 0 \;,\; \alpha_T \rightarrow 0 \right\}\underset{x \rightarrow 1} \;\;\;,\nonumber\\[2mm]
{\rm \bf EDE} &: \left\{\frac{M_*^2}{\mps} \rightarrow  const. \;,\; \alpha_M  \rightarrow 0 \;,\; \alpha_K \rightarrow const. \;,\; \alpha_B\rightarrow 0 \;,\; \alpha_T \rightarrow 0 \right\}\underset{x \rightarrow 1} \;\;\;,\nonumber\\[2mm]
{\rm \bf EMG} &: \left\{\frac{M_*^2}{\mps} \rightarrow  const. \;,\; \alpha_M  \rightarrow 0 \;,\; \alpha_K \rightarrow const. \;,\; \alpha_B\rightarrow const. \;,\; \alpha_T \rightarrow const.\right\}\underset{x \rightarrow 1} \;\;\;.\nonumber\
\end{align}

%\newpage
\section{A covariant description of the DE scenarios}\label{cov}

A simplified Lagrangian in the Horndeski class which encompasses our DE scenarios is, for example,
\begin{equation}\label{cov-lag}
\mathcal{L}= \frac{M_*^2(\phi)}{2}R - \frac{1}{2} Z(\phi)\partial_\mu \phi \partial^\mu \phi -V(\phi) + \frac{A(\phi)}{\l33} \partial_\mu \phi \partial^\mu \phi \square \phi \;,
\end{equation}
where $M_*^2$, $Z$, $V$, and $A$ are functions of the scalar field $\phi$. The energy scale for modelling cosmic acceleration is typically $\Lambda^3_3 \sim H_0^2 M_{\rm Pl}$.

Now, say that we are given in the EFT formalism some coupling functions of the time $\mu_1(t)$ and $\mu_3(t)$. In this appendix, it is easier to work with the proper time variable $t$. Switching to the variable $x$ defined in \eqref{xdef} is straightforward. The example that we consider here will allow us to implement models with $\epsilon_4=0$.  

A given background expansion history will be specified through the scale factor $a(t)$ as a function of the time. From this, one can define a Hubble parameter $H$ and a non-relativistic matter density for any given background evolution with some equation of state parameter $w_{\rm eff}$. In summary, we consider the following input functions: 
\begin{equation}
H^{w_{\rm eff}}(t), \qquad \rho_m^{w_{\rm eff}}(t),\qquad \mu_1(t), \qquad \mu_3(t).
\end{equation}
The first three functions above allow us to define the following two quantities, by using eq.~\eqref{candlambda},

\begin{align}\label{cccc}
\cb(t) & =  \dfrac12 \left( H^{w_{\rm eff}} \mu_1-\dot\mu_1-\mu_1^2 \right) - {\dot H}^{w_{\rm eff}} -\dfrac{\rho_m^{w_{\rm eff}}}{2M^2} \ , \\
\lambda(t) & =  \dfrac12 \left( 5 H^{w_{\rm eff}}\mu_1+\dot\mu_1+\mu_1^2 \right) + {\dot H}^{w_{\rm eff}} +3(H^{w_{\rm eff}})^2-\dfrac{\rho_m^{w_{\rm eff}}}{2 M^2 } \ ,\, 
\end{align}
where $M^2(t)$ will be calculated through $d \ln M^2(t)/dt = \mu_1(t)$.
\vv
The contributions of the terms in the Lagrangian~\eqref{cov-lag} to the EFT operators can be read off  the dictionary provided in App. C of Ref.~\cite{Gleyzes:2013ooa},

\begin{align}
M^2(t) & = M_*^2(\phi) \\
\lambda(t) & = \frac{1}{M_*^2(\phi)}\left[\dot{\phi}^2 \left(\ddot{\phi}+3H\dot{\phi} \right)\frac{A(\phi)}{\l33} +V(\phi) \right] \\
\mathcal{C}(t) & = \frac{1}{M_*^2(\phi)}\left[-\dot{\phi}^2 \left(\ddot{\phi}+3H\dot{\phi} \right)\frac{A(\phi)}{\l33} +\frac{\dot \phi^4}{\l33}\frac{dA(\phi)}{d\phi} +\frac{\dot\phi^2}{2}Z(\phi)\right] \\
\mu_3(t) & = \frac{2}{M_*^2(\phi)}\,\dot{\phi}^3 \,\frac{A(\phi)}{\l33} 
\end{align}
Under a field redefinition $\phi \rightarrow \tilde \phi (\phi)$, the Lagrangian~\eqref{cov-lag} does not change its structure but just the defining functions $M_*^2, Z, V$ and $A$. Therefore we assume to work directly with the field $\phi$ that is proportional to the time coordinate: $\phi(t)= c t$, $\dot \phi = c$. Then, one can express $A(\phi)$ as 
\begin{equation}
A(\phi) = \frac{\l33}{2 c^3} \;\mu_3(\phi/c)\;M^2(\phi/c) \;.
\end{equation}
The potential  follows,
\begin{equation}
V(\phi) = M^2(\phi/c)\left(\lambda(\phi/c) -\frac{3}{2}H^{w_{\rm eff}}(\phi/c)\mu_3(\phi/c)\right)                   \;,
\end{equation}
and, finally,
\begin{equation}
Z(\phi) = \frac{M^2(\phi/c)}{c^2}\left(2\mathcal{C}(\phi/c) -3H(\phi/c) \mu_3(\phi/c)-\mu_1(\phi/c)\mu_3(\phi/c)-\dot{\mu}_3(\phi/c)\right)                   \;,
\end{equation}
where $\mathcal{C}$ is defined in~\eqref{cccc}.

%**************************************************************************************************
%                                            BIBLIOGRAPHY
%**************************************************************************************************

\small
\label{Bibliography}
\bibliography{References}
\bibliographystyle{JHEP}
\end{document}